\documentclass[a4paper]{article}
\pdfoutput=1
\usepackage {amssymb}
\usepackage {amsmath}
\usepackage{amsfonts}
\usepackage{amsthm}
\usepackage{arydshln}
\usepackage{pdflscape}

\usepackage{graphicx}
\usepackage{hyperref}

\makeatother%
\newdimen\theight

\newcommand{\be}{\begin{equation}}
\newcommand{\ee}{\end{equation}}
\newcommand{\bi}{\begin{itemize}}
\newcommand{\ei}{\end{itemize}}

\newcommand{\sn}{\mbox{sn}}
\newcommand{\cn}{\mbox{cn}}
\newcommand{\dn}{\mbox{dn}}
\newcommand{\ba}{\begin{array}}
\newcommand{\ea}{\end{array}}
\newcommand{\bea}{\begin{eqnarray}}
\newcommand{\eea}{\end{eqnarray}}

\newcommand{\arcsinh}{\mbox{arcsinh}}
\newcommand{\arccosh}{\mbox{arccosh}}

\newcommand{\diag}{\mathop{\mathrm{diag}}}

\begin{document}

\begin{center}
{\Large{\bf Lie Algebra Contractions and Separation

\vspace{0.3cm}
of Variables on Two-Dimensional Hyperboloids.

\vspace{0.3cm}
Coordinate Systems}}

\vspace{0.9cm}

{\bf G.S.~Pogosyan$^{(1)}$
and A.~Yakhno$^{(2)}$
}

\vspace{0.3cm}
{\it Departamento de Matematicas, CUCEI, Universidad de Guadalajara, \\
Guadalajara, Jalisco, Mexico}

\vspace{0.6cm}
$^{(1)}${\it pogosyan@theor.jinr.ru}
\vspace{0.3cm}
$^{(2)}${\it alexander.yakhno@cucei.udg.mx}
\end{center}

\vspace{0.9cm}

\begin{abstract}
In this work the detailed geometrical description of all possible orthogonal and nonorthogonal systems of coordinates, which allow separation of variables of two-dimensional Helmholtz equation is given as for two-sheeted (upper sheet) $H_2$, either for one-sheeted ${\tilde H}_2$ hyperboloids. It was proven that only five types of orthogonal systems of coordinates, namely: pseudo-spherical, equidistant, horiciclic,  elliptic-parabolic and elliptic system cover one-sheeted ${\tilde H}_2$ hyperboloid completely. For other systems on ${\tilde H}_2$ hyperboloid, well defined In\"on\"u--Wigner contraction into pseudo-euclidean plane $E_{1,1}$ does not exist. 
Nevertheless, we have found the
relation between all nine orthogonal and three nonorthogonal separable systems of coordinates on the one-sheeted hyperboloid 
and eight orthogonal plus three nonorthogonal ones on pseudo-euclidean plane $E_{1,1}$. We could not identify the counterpart
of parabolic coordinate of type II on $E_{1,1}$ among the nine separable coordinates on hyperboloid ${\tilde H}_2$, but we have 
defined one possible candidate having such a property in the contraction limit. In the light of contraction limit we have understood 
the origin of the existence of an additional invariant operator which does not correspond to any separation system of coordinates 
for the Helmholtz equation on pseudo-euclidean plane $E_{1,1}$. 

Finally we have reexamine all contraction limits from the nine separable 
systems on two-sheeted $H_2$ hyperboloid to Euclidean plane $E_2$ and found out some previously unreported transitions.

\end{abstract}

\today

\newpage
\tableofcontents

\newpage

\section{Introduction}

The are many limiting processes which connect the global physical theories as relativistic and non relativistic
physics including the quantum and classical mechanics. In the base of existence of such limit procedures between
physical theories there is an idea that every new physical theory must have appropriate limit when the old one can be
recovered. Segal \cite{SEGAL} was the first mathematician who noted the correlation between limit processes for physical
theories with corresponding Lie algebras. Two years later, In\"on\"u and Wigner \cite{INONU1} using the
idea of Segal, introduced the concept of limit process in physics, now called In\"on\"u--Wigner contraction
(IW contraction), which induces a transition from Poincare algebra to the Galilean one where the velocity of light
$c \to \infty$. Similarly, de Sitter and anti de Sitter spaces with their $SO(3,2)$ and $SO(4,1)$ isometry groups were contracted to flat Minkowski space 
with its Poincare isometry group $P(3,1)$ with the limit $R\to \infty$.

The standard In\"on\"u and Wigner contractions \cite{GILMOR, INONU1, SALETAN} can be viewed  as singular changes of basis
in a given Lie algebra. Indeed, let us consider basis $\{X_1, X_2, ... X_n\}$ of Lie algebra $L$. Introducing a
new basis $\tilde{X}_i = u_{ik} (\varepsilon) X_k$ (hereafter the summation on repeating indices is assumed), where matrix $U(\varepsilon) = (u_{ik})$ is responsible for this transformation between basis, depends on some parameter $\varepsilon$
and it is nonsingular for $\varepsilon \not= 0$, $|\varepsilon|< \infty$ and is such that at $\varepsilon = 1$ we have
$u_{ik}(1) = \delta_{ik}$, and for $\varepsilon \to 0$ matrix $U(\varepsilon)$ become singular i.e. $|U(0)|=0$. In this limit the commutation 
relations of algebra $L$ are continuously changed into new commutation relations which determine non-isomorphic algebra $\tilde{L}$ to algebra $L$. 

Later, the class of IW contractions was extended and
the {\it generalized or singular IW contractions} were introduced in literature. This contraction is generated by the diagonal
matrix in the form of the integer power of contraction parameter $\varepsilon$  \cite{DOEBNER, INONU2, LOHMUS1, TALMAN}, see also articles
\cite{WW1,WW2,WW3}. As an example we can mention the contraction
from group $SU(2)$ into Heisenberg--Weyl group \cite{APW2, APW1, LOHMUS1, TALMAN}.

The future extension of the IW contractions is connected with so-called {\it graded contractions}, introduced firstly in articles \cite{PATERA3,PATERA2,PATERA1} and developed in many works \cite{MPT, HMOS,PPW, TOLART}.
These contractions belong to the type of discrete contractions and are quite natural from mathematical point of view. The idea of graded
contractions can be formulated as follows. Multiplying the structure constants of graded Lie algebra $L$ by some parameters $p_i$
in such a way that the structure constants keep the same grading and then taking the limit when these parameters go to zero.
Let us also mention that the reader can find different kinds of the properties of contractions in the following works
\cite{BGHOS, BHOS1, CASTANOS, CGT, GROMOV1,  HB, KMPOST, LOHMUS2, POPOWIZ} and in references therein.

Some aspects of the theory of Lie group and Lie algebra contractions in the context of the separation of variables of Laplace-Beltrami
equation in the homogeneous spaces, namely:
{\it the relation between separable coordinates systems in curved and flat spaces, related by the contraction of their isometry groups}
have been presented in a series of articles \cite{ IZPOG1, IZM1, IZM3,IZM5, IZM2,   KMP, KALYAKHPOG1, PSW1, PSWW,YAKHPOG1,YAKHPOG2, YAKHPOG3}.
The approach makes use of specific realizations of IW contractions, called the {\it analytic contractions}. The contractions are analytic
because contraction parameter $\varepsilon = 1/R$, where $R$ is the radius of sphere or pseudo-sphere, appears in the separable systems of coordinates, in the operators of Lie algebra, in the eigenvalues and eigenfunctions of Laplace-Beltrami operator and not only
in the structure constants.  Using this method, for instance, it is possible to observe the contraction limit $\varepsilon \to 0$
($R\rightarrow\infty$) at all the levels: the level of the Lie algebra as realized by vector fields, the Laplace-Beltrami operators in the
two-dimensional four homogeneous spaces (sphere or hyperboloid on one hand and Euclidean or pseudo-Euclidean spaces on the other), the
second order operators in the enveloping algebras, characterizing separable systems, the separable coordinate systems themselves, the
separated (ordinary) differential equations, the separated eigenfunctions of the invariant operators and finally the interbasis expansions.
In particular, in articles \cite{IZM1,IZM3} the method of analytic contractions demonstrated on the two simple homogeneous spaces: the
two-dimensional
sphere $S_2$$\sim O(3)/O(2)$ and the two-sheeted hyperboloid $H_2$$\sim O(2,1)/O(2)$, where all the types of separable coordinates where
considered.  For example, contractions of $O(3)$ to $E(2)$ relate elliptic coordinates on $S_2$ to elliptic and parabolic coordinates on the Euclidean plane $E_2$.
They also relate spherical coordinates on $S_2$ to polar and Cartesian coordinates on $E_2$. Similarly, all 9 coordinate systems on the $H_2$
hyperboloid can be contracted to at least one of the four systems on $E_2$  \cite{ IZM3, KMP}.  The tree-dimensional sphere was considered in paper \cite{YAKHPOG1}. 
In paper \cite{IZM5} the dimension of the space was arbitrary, but only the simplest types of coordinates were considered, namely subgroup ones.
The analytic contractions from the rotation group $O(n+1)$ to the Euclidean group $E(n)$ are used to obtain asymptotic relations for matrix
elements between the eigenfunctions of the Laplace-Beltrami operator corresponding to separation of variables in the subgroup--type coordinates
on $S_n$ \cite{IZM2}. The contraction for non subgroup coordinates have been described in \cite{KMP}. The case of the contractions for
non-subgroup basis on $H_2$ have been recently considered in the article \cite{KALYAKHPOG1}.

Up today one can distinguish two principal approaches of separation of variable for Laplace-Beltrami and Helmholtz equations on homogeneous spaces with isometry groups $SO(p,q)$ and $E(p,q)$. Historically, the first approach uses the methods of differential geometry. In such a way the problem of separation of variables for Helmholtz, Hamilton-Jacobi and Schrodinger equations on two- and three-dimensional spaces of constant curvature (including the sphere and two-sheeted hyperboloid) was firstly solved by Olevskii \cite{OLEV}. A similar problem for two- and three-dimensional Minkowski spaces was discussed in detail in the work by Kalnins \cite{Kalnins1} and for hyperboloid with isometry group $SO(2,2)$ and complex sphere $SO(4,C)$ in works by  Kalnins and Miller \cite{KALMIL77,KALMIL78} (see also the article \cite{BOYER-Kalnins}). 
Later the graphic procedure of the construction of an orthogonal coordinate systems on $n$-dimensional Euclidean space and on $n$-dimensional sphere was presented in \cite{Kalnins-SE1} and in book \cite{KAL-BOOK}. Each of the separating coordinate systems is characterized by a complete set of mutually commuting symmetry operators belonging to the enveloping algebra of the isometry group of the space for Helmholtz equation. For example, the problem of classification of complete sets of symmetry operator corresponding to the separation of variables of Helmholtz equation on two-sheeted hyperboloid with Lorenz isometry group $SO(3,1)$ was solved in work \cite{SMORTUGOV}. A similar problem for the Helmholtz equation in the case of other spaces of constant curvature (two- and three-dimensional)
was solved in \cite{KALMIL77, KMW, KALTHOM, LUKA1, LUKA2, MILPATWIN,  PW1, WINLUKSMOR}. 

The other, a purely algebraic approach to the problem of separation of variables is based on an isometry group of the space. The classification of second-order operators on nonequivalence classes permits the construction of all possible orthogonal coordinate systems admitting the separation of variables for Helmholtz equation on the spaces of constant curvatures. However, in contrast to the direct approach when the separation of variables in a given coordinate system is uniquely identified by the full set of symmetry operators, the inverse problem, namely the construction of coordinate systems where the given set of commuting operators can be diagonalized does not look so trivial and consistently has done only for some two-dimensional spaces of constant curvature in work \cite{PSW1}. Moreover, there is an example (for instance on plane $E_{1,1}$) when the set of symmetry operators does not correspond to any coordinate system at all. A detailed discussion of this problem for three-dimensional spaces can be found in \cite{MILPATWIN}. Here we show that this phenomena can be understand in frames of contraction procedure. 

In spite of the wide bibliography devoted to the investigation of various aspects (separation of variable, integrals of motion, wave functions, 
spectrum, etc.) of  Helmholtz equation in homogeneous spaces  the analogous investigation (including the contractions) in the case of one-sheeted hyperboloids or more general transitive surfaces of $SO(p,q)$ $(p,q \geq 1)$ group has not been done with the same amount of detail as the other spaces of constant curvature.  Actually, there are many peculiarities which differ two transitive surfaces of Lorentz group $SO(2,1)$ in Minkowski space. 
Whereas the distance $r$ between two fixed points $x=(x_0,x_1,x_2)$ and $y=(y_0,y_1,y_2)$ of two-dimensional two-sheeted hyperboloid $[x, x] = x_0^2 - x_1^2 - x_2^2 = R^2$ given by the formula $\cosh kr = [x, y]/R^2 \geqslant 1$ ($k=1/R$) is nonnegative real, for the one-sheeted hyperboloid we have $\cosh kr = [x, y]/R^2 \geqslant 0$, where now $[x, x] = x_0^2 - x_1^2 - x_2^2 = - R^2$ and therefore the distance $r$ is nonnegative real for $\cosh kr \geqslant 1$ and is imaginary when $0 \leqslant \cosh kr < 1$ (see \cite{GG}). For example $[x,y] = 0$ for two points $x=(0,0,R)$ and $y=(0,R,0)$ of one-sheeted hyperboloid. Let us note that in these cases all diametrically opposed points of hyperboloids are considered to be equivalent.

This is the reason why on the two-sheeted hyperboloid only continuous representation of $SO(2,1)$ group is realized, while on the one-sheeted hyperboloid both continuous and discrete representations are realized. Another consequence of this fact is that many separable systems of coordinate for Helmholtz equation are piecewise defined and not even all of them completely cover the surface of one-sheeted hyperboloid and thus only some of coordinates on two-sheeted hyperboloid correspond in a natural way to the one-sheeted hyperboloid.  

Apparently for the first time pseudo-spherical functions as a solutions of Helmholtz equation on the one-sheeted hyperboloid  were introduced in the 
article by Zmuidzinas \cite{ZMUIDZINAS} for Lorenz group $SO(3,1)$ in the frame of expansion theorem of Titchmarsh \cite{TITCHMARSH}, and independently using the reduction to the canonical bi pseudo-spherical coordinates for the more general group $SO(p,q)$ in series of articles by Lumic, Niederle and Raczka \cite{LUMNOCI, LUMNOC}. Following the
approach of integral geometry developed by Gelfand-Graev \cite{GG}, Kuznetsov and Smorodinski in \cite{SMORODIN}, Verdiev in \cite{VERDIEV2} 
and Kalnins and Miller \cite{KALMIL-1} various separable bases for the Laplace-Beltrami equation on the one-sheeted hyperboloid embedded in Minkowski space have been studied. We also should mention works \cite{KALMIL77,KALTHOM} where group $SO(2,2)$  was considered. More recently the wave functions on the two-dimensional one-sheeted hyperboloid in subgroup parametrization were presented in \cite{VERDIEV,REM,YAKHPOG2}.

Therefore, the precise form of separable coordinate systems, square integrable solutions for Helmholtz
equation and the overlap functions (at least for subgroup bases) for two- and three-dimensional spheres and two-sheeted hyperboloid (see for
example articles \cite{KAMIb, KALMIL-1, KMW, PAWI} and books \cite{MILLER, VILENKIN}) is more or less well known, but our knowledge about the one-sheeted 
hyperboloid is restricted mainly to the cases of subgroup separable coordinates and bases and obviously is not complete \cite{VERDIEV,YAKHPOG2,YAKHPOG3}.

The present work pretends to fill this gap and in the ideological  sense is a continuation of a series of papers 
\cite{IZPOG1, IZM1, IZM3, IZM5, IZM2, KMP, KALYAKHPOG1, PSW1, YAKHPOG1, YAKHPOG2, YAKHPOG3}.   
Here we restricted ourselves to consideration of contractions of the separable coordinate systems for the Helmholtz equation on the two-dimensional one-sheeted 
hyperboloid 
${\tilde H_2} \sim O(2,1)/O(1,1)$ into pseudo-Euclidean space (two-dimensional Minkovski space) $E_{1,1}$. This type of contractions has never been discussed in the literature.
We also reexamined the procedure of contraction for separable systems on two-dimensional two-sheeted hyperboloid and found out some new types of connection between systems 
on $H_2\sim O(2,1)/O(2)$ and $E_2$. Let us note that all obtained results can be generalized for higher dimensions.

The importance of this study is dictated by many factors. From the physical point of view one-sheeted hyperboloid is a model of imaginary Lobachevsky space 
(if we identify the diametrically opposed points \cite{GG}), de Sitter $\text{dS}^{1+1}$ or anti de Sitter $\text{AdS}^{1+1}$ spaces, which is of great use not only in modern field theory \cite{WARD}, cosmology \cite{PEEBLES}
 or elementary particle physics \cite{AGAMAL}, but also in quantum physics (see for instance \cite{GAZEAU}). 
From the mathematical point of view the analytic contractions are motivated by numerous applications to the theory of 
superintegrable systems \cite{CALZADA,VINITSKI,HEREDERO} and special functions \cite{APW2, APW1, MILLER-POST, PSW1}.

The paper is organized in the following way. In section \ref{sec:2} we shortly present some known aspects of separation of variables for the Helmholtz equation on the spaces with constant curvature (more detailed exposition is done in \cite{PSW1}) and we give a description of a group theoretical procedure for classification of all separable systems of coordinates for two-dimensional Helmholtz equation into the nonequivalent classes (or orbits) of symmetry algebra $so(2,1)$, which only were designated but not resolved in article \cite{PSW1}. We also mention the subgroup systems 
of coordinates which come from the first order symmetry operators  (see for general case \cite{PW1}). 
Then, using the connection between classical and quantum mechanics we have constructed the non-subgroup systems of coordinates coming from the second order symmetry operator. We have determined various equivalent forms of the separable systems of coordinates, which are most convenient in the sense of contractions.
 Section \ref{sec:3} is devoted to the contractions of  the separable systems of coordinates for Laplace-Beltrami equation and  the  symmetry  operators on  the  two-sheeted hyperboloid. Some of these  contractions are known from papers \cite{IZM3,PSW1},  
but we present them here for completeness. In section \ref{sec:4}  the contractions  on  one-sheeted hyperboloid have been presented.  
As far as we know all these results are new ones. Many of  these results  are listed in Table \ref{tab:3}.  Finally in section \ref{sec:5} 
we have discussed some aspects of contractions. Let us note, that some part of the material of the present paper was published in proceedings \cite{YAKHPOG2}.  

The classification of the separable systems of coordinates for two-dimensional Helmholtz equation on pseudo Euclidean space $E_{1,1}$ and Euclidean space $E_{2}$ is well known (see  \cite{Kalnins1,PSW1}). We just summarize these results in Tables \ref{tab:6} and \ref{tab:7}.

\section{Separation of Variables on Two Dimensional Hyperboloids}
\label{sec:2}

The manifold we are interested in is the two-dimensional hyperboloid
\begin{eqnarray}
\label{HYPER-01}
u \cdot u = G_{\mu \nu} u_\mu u_\nu = - u_0^2 + u_1^2 + u_2^2  = - \epsilon R^2,
\qquad \epsilon = \pm 1, \qquad \mu,\nu =0,1,2,
\end{eqnarray}
where $R$  is a pseudo-radius, the case when $\epsilon = 1$ corresponds to the two-sheeted hyperboloid $H_2$ and $\epsilon = - 1$ to 
one-sheeted hyperboloid ${\tilde H_2}$. Cartesian coordinates $u_\mu$ and metric $G_{\mu\nu} = \diag (-1, 1, 1)$ belong to the three-dimensional ambient Minkowski space $E_{2,1}$.  Lie group $SO(2,1)$ is the isometry group of hyperboloid (\ref{HYPER-01}). The standard basis for
the Lie algebra $so(2,1)$ we take in the form 
\begin{eqnarray}
K_1 = - (u_0 \partial_{u_2} + u_2 \partial_{u_0}), \qquad K_2 =
-(u_0 \partial_{u_1} + u_1 \partial_{u_0}), \qquad L =  u_1
\partial_{u_2} - u_2 \partial_{u_1},
\label{algebra_basis}
\end{eqnarray}
that satisfies to commutation relations:
\begin{eqnarray}
\label{NONCOMPACT1} 
\left[K_1,K_2\right] = - L,\,\,\,\,\,
\left[L,K_1\right] =   K_2,\,\,\,\,\, \left[K_2,L\right] =
K_1.
\end{eqnarray}
The Casimir operator of $so(2,1)$ algebra is
\begin{equation}
\label{ALGEBRA3} {\cal C} =   K_1^2 + K_2^2 - L^2
\end{equation}
and it is proportional to Laplace-Beltrami operator $\Delta_{LB} = {\cal C}/R^2$, which in the local curvilinear coordinates
$\xi=(\xi^1, \xi^2)$ on the hyperboloid (\ref{HYPER-01}) has the form
\begin{equation}
\label{LAPLACE1}
 \Delta_{LB} =
\frac{1}{\sqrt{g}}
\, \frac{\partial }{\partial
\xi^i} \sqrt{g} \, g^{ik} \, \frac{\partial }{\partial \xi^k},
\qquad\qquad\qquad i,k =1,2,
\end{equation}
where the metric tensor $g_{ik}$ is given by
\begin{equation}
\label{METRIC-01} 
ds^2 = g_{ik} d\xi^i d\xi^k, \qquad g^{ij} g_{jk} = \delta^{i}_{k}, \qquad g=|\det (g_{ik})|, \qquad i,k
=1,2.
\end{equation}
The relation between metric tensor $G_{\mu\nu}$ in the ambient space and  $g_{ik}(\xi)$ of Eqs. (\ref{LAPLACE1}) and 
(\ref{METRIC-01}) looks as follows
\begin{equation}
\label{METRIC-001} 
g_{ik}(\xi) = \epsilon G_{\mu\nu} \frac{\partial u_\mu}{\partial\xi^i} \frac{\partial u_\nu}{\partial
\xi^k}.
\end{equation}
The Helmholtz equation on manifold (\ref{HYPER-01}) has the form
\begin{equation}
\label{02-HELM} 
\Delta_{LB} \Psi = \mathcal{E} \Psi,
\end{equation}
where $\mathcal{E}$ is nonzero complex or real constant. The orthogonal separable systems of coordinates on hyperboloids (\ref{HYPER-01}) are associated  with the second order 
operator $S^{(2)}_{\alpha}$  in the enveloping  algebra of $so(2,1)$ (where index $\alpha$ marks the separated system of coordinates) commuting  with operator 
$\Delta_{LB}$. Let us recall that for an orthogonal system its metric tensor has diagonal form 
$g_{ik} = \delta_{ik} H_k^2$, where 
$H_k^2$ are the Lame coefficients \cite{LAME}.

The classification of all  nonequivalent  operators $S^{(2)}_\alpha$ from  the set  of  second order polynomials  in terms of the elements of the algebra $so(2,1)$:
\be 
\label{FIRST-OPER-01} 
S^{(2)}_\alpha = 
m_{ik}^{(\alpha)} X_i X_k,
\qquad  m_{ik}^{(\alpha)} = m_{ki}^{(\alpha)},  
\qquad  m_{ik}^{(\alpha)} = \text{const.}, 
\qquad i,k = 1,2,3
\ee
($X_1 = K_1, X_2 = K_2, X_3 = L$) into orbits under the action of group  $SO(2,1)$ lies  in  the  basis of  algebraic approach  of  the classification of  separable systems of 
coordinates for Helmholtz equation (\ref{02-HELM}).  
 The task  is  reduced  to  the classification  of  symmetric matrices $M =  \left(m_{ki}^{(\alpha)}\right)$ into nonequivalent classes under the transformation  $M^\prime = \alpha_1 A^{T} M A + \alpha_2 A_{\mathcal{C}}$,  
where  $A^{T}$  is transposed matrix and  $A$ is an element of the group of inner automorphisms of  $so(2,1)$; $\alpha_1 \ne 0$, $\alpha_2$ are the real constants; $A_{\mathcal{C}}$ is a matrix corresponding to Casimir operator (\ref{ALGEBRA3}).

The next procedure is the construction of the local orthogonal curvilinear coordinates $\xi=(\xi^1, \xi^2)$ which simultaneously diagonalize two second order operators  
$\Delta_{LB}$ and $S^{(2)}_{\alpha}$. The corresponding separable 
solutions $\Psi_{\mathcal{E} \mu} (\xi^1, \xi^2) = Y_{\mathcal{E} \mu} (\xi^1) \, \Phi_{\mathcal{E} \mu} (\xi^2)$  of the Helmholtz equation (\ref{02-HELM}) are the common functions for 
two differential equations:  (\ref{02-HELM})  and   $S^{(2)}_\alpha \Psi = \mu \Psi$,   where $\mu$  is a separation constant.  

If operator $S^{(2)}_{\alpha}$ is the square of the first order operator $X_i$ (or the square of the sum of some two operators $X_i$ as in the case of 
horicyclic coordinates), then it has to be of the subgroup type \cite{PW1}. 
In fact, the subgroup coordinates can be classified through the first order operators
 of $so(2,1)$ algebra
\be 
S^{(1)} =  
m_i X_i,
\qquad\qquad m_{i} = \text{const.}, \qquad i,k = 1,2,3.
\ee
Note that the classification based on first order operators is not unique because in general not only orthogonal coordinate systems can be obtained. Nevertheless, as it is shown in \ref{sec:system_first_order}, using some arbitrariness in the definition of coordinate systems one can reduce them to orthogonal form.

In work \cite{PSW1} in the frame of algebraic approach, the problem of the classification of orthogonal coordinate systems for two-dimensional 
Helmoltz equation on the spaces of a constant curvature was discussed. The case of the separation of variables for Minkovski space $E_{1,1}$ was 
investigated in detail whereas the solving of similar problems on the two-dimensional hyperboloids ($H_2$ and $\tilde{H_2}$) was only designated. Below in this section we fill this gap and solve  
the same problem of separable systems of Helmoltz equation for the case of one- and two-sheeted hyperboloids.

\subsection{First-order symmetries and separation of variables}
\label{sec:2.1}

\subsubsection{Classification of first-order symmetries}

Let us consider the vector of general position $\textbf{v} = (a,b,c)$ of an
arbitrary element
\be
\label{FIRSTSYM}
{S}^{(1)} (a,b,c) = a K_1 + b K_2 + c L
\ee
of Lie algebra $so(2,1)$. Under the action of group $A_3$ of
inner automorphisms (\cite{OVSYANNIKOV}, Chapter 4) of this algebra the
coordinates of vector $\textbf{v}$ are transformed in such a
way:
\begin{eqnarray}
K_1:\ {\overline{\textbf{v}}} &=& (a,b,c)\cdot
\left( {{\begin{array}{*{20}c}
 1 & 0 & 0 \\
 0 & \cosh a_1 & \sinh a_1 \\
 0 & \sinh a_1 & \cosh a_1
\end{array} }} \right)
=\textbf{v} \cdot A_{K_1},
\label{A_{K_1}}
\\[3mm]
K_2:\ \overline{\textbf{v}} &=& (a,b,c)\cdot
\left( {{\begin{array}{*{20}c}
 \cosh a_2 & 0 & \sinh a_2 \\
 0 & 1 & 0 \\
 \sinh a_2 & 0 & \cosh a_2
\end{array} }} \right)
=\textbf{v} \cdot A_{K_2},
\label{A_{K_2}} \\[3mm]
L:\ \overline{\textbf{v}} &=& (a,b,c)\cdot
\left( {{\begin{array}{*{20}c}
 \cos a_3 & - \sin a_3 & 0 \\
 \sin a_3 & \cos a_3 & 0 \\
  0 & 0 & 1
\end{array} }} \right)
=\textbf{v} \cdot A_{L},    \label{A_{L}}
\end{eqnarray}
where $a_1$, $a_2$, $a_3$ are the group parameters, $\overline{\textbf{v}}
= (\bar{a},\bar{b},\bar{c})$  is a transformed  vector.  We will also include three reflections into classification scheme: 
\be
\label{reflections}
R_0:   (u_0, u_1, u_2)  \to  (-u_0, u_1, u_2), 
\
R_1:  (u_0, u_1, u_2) \to (u_0, - u_1, u_2), 
\
R_2:  (u_0, u_1, u_2)  \to  (u_0, u_1, - u_2),
\ee
which leave the Laplace-Beltrami operator $\Delta_{LB}$ to be invariant. 

Let us note, that there is the following correspondence between transformations of operators and coordinates $(u_0,u_1,u_2)$:
\begin{eqnarray}
\label{relation_coords_operators}
\left( K_1^\prime, K_2^\prime, L^\prime\right)^T &=& A_{K_1}^{-1}\left(K_1, K_2, L\right)^T  \sim \left(u_0^\prime, u_1^\prime, u_2^\prime \right)^T = \left( {{\begin{array}{*{20}c}
  \cosh a_1 & 0 & -\sinh a_1 \\
 0 &  1 & 0 \\
 -\sinh a_1 & 0 & \cosh a_1
\end{array} }} \right) \left(u_0, u_1, u_2 \right)^T,  \nonumber \\
[3mm]
\left( K_1^\prime, K_2^\prime, L^\prime\right)^T &=&  A^{-1}_{K_2} \left(K_1, K_2, L\right)^T \sim \left(u_0^\prime, u_1^\prime, u_2^\prime \right)^T = \left( {{\begin{array}{*{20}c}
 \cosh a_2 & \sinh a_2 & 0 \\
 \sinh a_2 &  \cosh a_2 & 0 \\
 0 & 0 & 1
\end{array} }} \right) \left(u_0, u_1, u_2 \right)^T, \\
[3mm]
\left( K_1^\prime, K_2^\prime, L^\prime\right)^T &=& A_{L}^{-1} \left(K_1, K_2, L\right)^T  \sim \left(u_0^\prime, u_1^\prime, u_2^\prime \right)^T = \left( {{\begin{array}{*{20}c}
1 & 0& 0\\
0& \cos a_3 & -\sin a_3 \\
0& \sin a_3 &  \cos a_3 
\end{array} }} \right) \left(u_0, u_1, u_2 \right)^T. \nonumber
\end{eqnarray}
The superposition of trigonometric rotation counter-clockwise through angle $a_3 = \pi/2$ about the $u_0$-axis with reflection $R_1$ we well call the permutation of $u_1$ and $u_2$: $u_1 \leftrightarrow u_2$. Let us note that such permutation transforms operators: $K_1^\prime = K_2$, $K_2^\prime = K_1$, $L^\prime = - L$.

The general invariant of transformations (\ref{A_{K_1}}) -- (\ref{A_{L}}) has the form
\be
I = a^2 + b^2 - c^2,
\ee
and in order to classify all subalgebras to non-conjugate classes we need to consider the following three cases.

\vspace{0.4cm}
{\bf A.} \,\, {$I=0$, or $c^2 = a^2 + b^2$.}
If $a \neq 0$, then $c^2 > b^2$ and acting by $A_{K_1}$ to $\textbf{v}$ with $a_1 = - {\rm arctanh\, } ({b}/{c})$, and taking into account that $I=0$, we can reduce coordinate
$b$ to zero and to get vector $(a,0,|a|)$. Dividing $(a, 0, |a|)$ by $a$, finally we have vector $(1,0, \pm 1)$ that corresponds to the symmetry
\be
\label{HOR-01}
{S}^{(1)} (1,0, \pm 1) = K_1 \pm L.
\ee
If $a = 0$, then $|c|=|b| \ne 0$ and dividing $(0, b, c)$ by $b$ we obtain $(0, 1, \pm 1)$ or
\be
\label{HOR-02}
{S}^{(1)} (0, 1, \pm 1) = K_2 \pm L.
\ee
Note that these two cases are connected by rotation $A_L$ (we make $\bar{a} \neq 0$), and the sign in (\ref{HOR-01}) can be taken positive due to discrete symmetry $R_2$  (\ref{reflections}). 

\vspace{0.4cm}
{\bf B.} \,\, {$I > 0$ or $c^2 < a^2 + b^2$.}
In this case $a\ne 0$ or $b\ne 0$. If $a\ne 0$, then using $A_L$ with $a_3 = -{\rm arccot\, } (b/a)$ we get $\bar{a} = 0$, so $\bar{c}^2 < \bar{b}^2$ and there is $a_1$ such that by $A_{K_1}$ we can vanish $c$ and we finally have the vector $(0,1,0)$, corresponding to the symmetry
\be
\label{HOR-03}
{S}^{(1)} (0,1,0) = K_2.
\ee
Following the same procedure we can construct  vector $(1,0,0)$ that gives
\be
\label{HOR-04}
{S}^{(1)} (1,0,0) = K_1.
\ee

\vspace{0.4cm}
{\bf C.}\,\, {$I < 0$.}
In this case $c^2 > a^2 + b^2$. If $b\ne 0$ and $a\ne 0$, then using of $A_L$ we can make  $\bar{a} = 0$, so $\bar{c}^2 > \bar{b}^2$. Then by $A_{K_1}$ we can vanish $\bar{b}$ and obtain vector $(0,0,1)$. If $b = 0$ and $a \ne 0$, then by $A_{K_2}$ one can vanish $\bar{a}$ and we get the same vector $(0,0,1)$ that corresponds to the symmetry
\be
\label{HOR-05}
{S}^{(1)} (0,0,1) = L.
\ee

Thus, operator ${S}^{(1)}(a,b,c)$ (\ref{FIRSTSYM}) has been transformed into the three nonequivalent forms, each of them 
corresponds to the subgroup type coordinates, which is the consequence of the existence of three one-parametric subalgebras  $o(2)$, 
$o(1,1)$ and $e(1)$ of $so(2,1)$ algebra \cite{PW1}.  Let us note that operators (\ref{HOR-01}) and (\ref{HOR-02}) as well as 
(\ref{HOR-03}) and (\ref{HOR-04})  describe the equivalent systems of coordinates.

\subsubsection{Subgroup coordinate systems}
\label{sec:system_first_order}

Now we will construct the subgroup separable systems of coordinates connected to each of the first order symmetry 
operators $S^{(1)}$ obtained in the previous subsection. We require that the operator $S^{(1)}$ presented in one of the 
forms (\ref{HOR-01}) -- (\ref{HOR-05}) or 
\be
\label{INT-MOTION-01}
S^{(1)} = \xi_0 \partial_{u_0} + \xi_1  \partial_{u_1}
+ \xi_2 \partial_{u_2},
\qquad\qquad 
\xi_i \equiv \xi_i (u_0, u_1,u_2), 
\ee
in some of the local curvilinear coordinates $\xi(u_0, u_1,u_2)$, $\eta(u_0, u_1,u_2)$ has the canonical or 
diagonalized form
\be
\label{0-INT-MOTION-01}
S^{(1)} = \frac{\partial}{\partial \eta}.
\qquad
\ee
The solution of this problem is equivalent to the common solution of the two first order partial differential 
equations   
\be
\label{01-INT-MOTION-01}
\xi_0 \frac{\partial \xi}{\partial u_0} + \xi_1 \frac{\partial \xi}{\partial u_1} +  \xi_2 \frac{\partial \xi}{\partial u_2} = 0,
\qquad
\xi_0 \frac{\partial \eta}{\partial u_0} + \xi_1 \frac{\partial \eta}{\partial u_1} +  \xi_2 \frac{\partial \eta}{\partial u_2} = 1
\ee
and equation (\ref{HYPER-01}).  Let us now run for each case separately.


\paragraph{1. ${S}^{(1)} = L.$} To reduce symmetry operator $L = u_1 \partial_{u_2} - u_2 \partial_{u_1}$
to canonical form ${L} = \partial_{\eta}$ we must solve the following system of equations:
\begin{eqnarray}
\label{SEC-ORD-01}
u_1 \frac{\partial \xi}{\partial u_2} -
u_2 \frac{\partial \xi}{\partial u_1} = 0,
\qquad
u_1 \frac{\partial \eta}{\partial u_2} -
u_2 \frac{\partial \eta}{\partial u_1} = 1.
\end{eqnarray}
To solve these partial differential equations we have to construct the equation for
characteristics in form:
\begin{eqnarray}
\label{SEC-ORD-02}
\frac{du_1}{u_2}= - \frac{du_2}{u_1} = \frac{d \xi}{0},
\qquad
\frac{du_1}{u_2}= - \frac{du_2}{u_1} = \frac{d \eta}{1}.
\end{eqnarray}
From the above equations we have
\begin{equation}
\label{ff}
f^2(\xi,R) = u^2_1 + u^2_2, \qquad
\eta = \arcsin \frac{u_2}{f(\xi)} + g(\xi,R),
\end{equation}
so $u_2 = f(\xi,R) \sin (\eta - g(\xi,R))$.  Relation  (\ref{ff}) gives
$u_1 = \sqrt{f^2(\xi,R) - u^2_2} = f(\xi,R)\cos (\eta - g(\xi,R))$.
The third coordinate $u_0$ is determined from equation (\ref{HYPER-01}). 

\begin{figure}
\begin{center}
\begin{minipage}[t]{0.45\linewidth}
\includegraphics[scale=0.3]{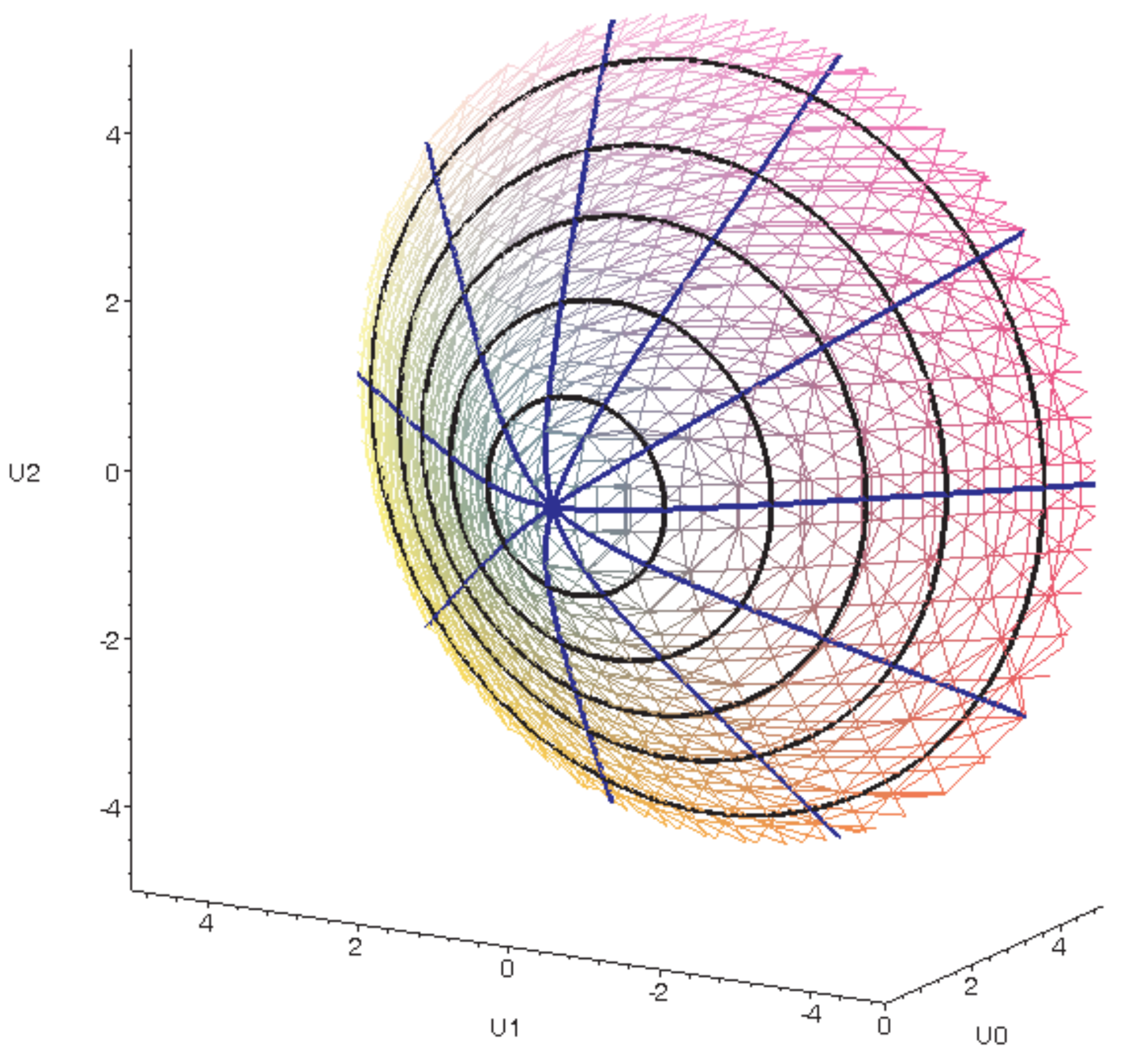}
\begin{center}
\caption{Spherical system}
\label{fig:2}
\end{center}
    \end{minipage}
    \hfill
    \begin{minipage}[t]{0.45\linewidth}
 \includegraphics[scale=0.3]{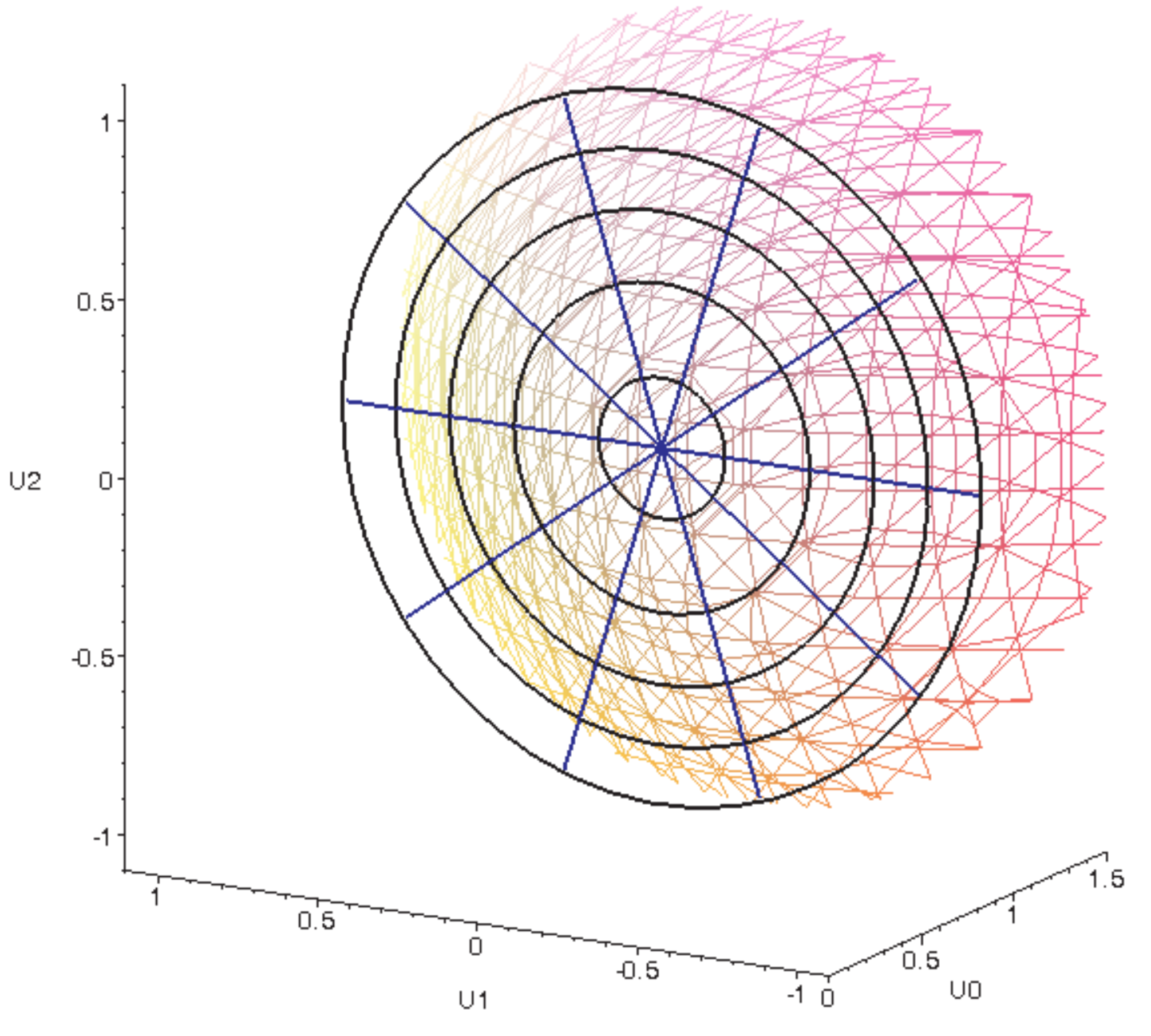}
\caption{Projective plane for spherical system is formed by circles and straight lines
passing through the center}
\label{fig:3}
\end{minipage}
\end{center}
\end{figure}

{\bf 1a}. For two-sheeted hyperboloid $\epsilon = 1$ and we consider $u_0 = \sqrt{f^2(\xi,R) + R^2} > 0$.
Choosing now for $f(\xi,R) = R \sinh \xi$ and $g(\xi,R) = 0$, denoting
instead of $(\xi, \eta)$ new coordinates $(\tau, \varphi)$ we obtain the orthogonal 
(pseudo-) spherical system of coordinates, corresponding to the upper sheet of 
two-sheeted hyperboloid $H_2$:
\begin{eqnarray}
\label{PSEUDO-SPHER-01}
u_0 = R \cosh \tau,
\qquad
u_1 = R \sinh \tau \cos \varphi,
\qquad
u_2 = R \sinh \tau \sin \varphi
\end{eqnarray}
with $\tau > 0$ and $\varphi \in [0,2\pi)$ (see Fig. \ref{fig:2}).

Let us note, if one takes $g(\xi,R)$ such that $g^\prime_\xi \ne 0$, than one obtains nonorthogonal spherical system, which admits the separation on coordinates. In particular, we will consider nonorthogonal spherical coordinates of the form ($g = - R\xi/\alpha$, $\alpha$ is nonzero constant):
\begin{eqnarray}
\label{PSEUDO-SPHER-01_NO}
u_0 = R \cosh \tau,
\qquad
u_1 = R \sinh \tau \cos \left(\varphi + R\tau/\alpha\right),
\qquad
u_2 = R \sinh \tau \sin \left(\varphi + R\tau/\alpha\right).
\end{eqnarray}
Let us note, that above system tends to orthogonal one as $\alpha \to \infty$. 

{\bf 1b}. In case of one-sheeted hyperboloid $\epsilon = -1$ we obtain from (\ref{ff}) that 
$u_0 = \pm \sqrt{f^2(\xi,R)-R^2}$. Thus  putting $f(\xi,R) = R\cosh\xi$, $g(\xi,R) = 0$ and introducing $\xi = \tau$, $\eta =\varphi$ we obtain the orthogonal pseudo-spherical system of coordinates on ${\tilde H_2}$ (see Fig. \ref{fig:19}):
\begin{eqnarray}
\label{pspherical}
u_0 = R \sinh \tau ,
\qquad
u_1 = R \cosh \tau \cos \varphi,
\qquad
u_2 = R \cosh \tau \sin \varphi
\end{eqnarray}
with $\varphi \in [0,2\pi)$, $\tau \in \mathbb{R}$. 

In the same manner, taking $g(\xi, R) = -R \xi/\alpha-\pi/2$ we consider nonorthogonal system:
\begin{eqnarray}
\label{pspherical_NO}
u_0 = R \sinh \tau ,
\qquad
u_1 = - R \cosh \tau \sin \left(\varphi + R \tau/\alpha\right),
\qquad
u_2 = R \cosh \tau \cos \left(\varphi + R \tau/\alpha\right).
\end{eqnarray}

Note that pseudo spherical systems of coordinates completely cover one-sheeted hyperboloid
and the upper sheet of two-sheeted hyperboloid.

\paragraph{2. ${S}^{(1)} = K_2.$}
\noindent
Let us take operator $K_2$ in terms of separable coordinates ($\xi, \eta$):
\be
- K_2 = u_0  \frac{\partial}{\partial u_1}  +
u_1  \frac{\partial}{\partial u_0} =
\left( u_0 \frac{\partial \xi}{\partial u_1} +
u_1 \frac{\partial \xi}{\partial u_0} \right) \frac{\partial}{\partial \xi}
+ \left( u_0 \frac{\partial \eta}{\partial u_1}
+ u_1 \frac{\partial \eta}{\partial u_0} \right)
\frac{\partial}{\partial \eta}
\ee
and let us require that it transforms into canonical form (\ref{0-INT-MOTION-01}).  
Hence, we have
\be
\label{K2_1}
u_0 \frac{\partial \xi}{\partial u_1}
+ u_1 \frac{\partial \xi}{\partial u_0} = 0,
\qquad
u_0 \frac{\partial \eta}{\partial u_1} +
u_1 \frac{\partial \eta}{\partial u_0} = 1.
\ee
The characteristic equations of the above ones are:
\begin{eqnarray}
\label{CARAC}
\frac{d u_1}{u_0}=  \frac{d u_0}{u_1} = \frac{d \xi}{0},
\qquad
\frac{du_1}{u_0}= \frac{du_0}{u_1} = \frac{d \eta}{1}.
\end{eqnarray}
Relations (\ref{K2_1}) imply
\be
\label{25}
f(\xi,R) = u^2_0 - u^2_1,
\qquad
\eta = \sinh^{-1}\frac{u_1}{\sqrt{f(\xi,R)}} + g(\xi,R).
\ee

\begin{figure}[htbp]
\begin{center}
\begin{minipage}[t]{0.45\linewidth}
\includegraphics[scale=0.3]{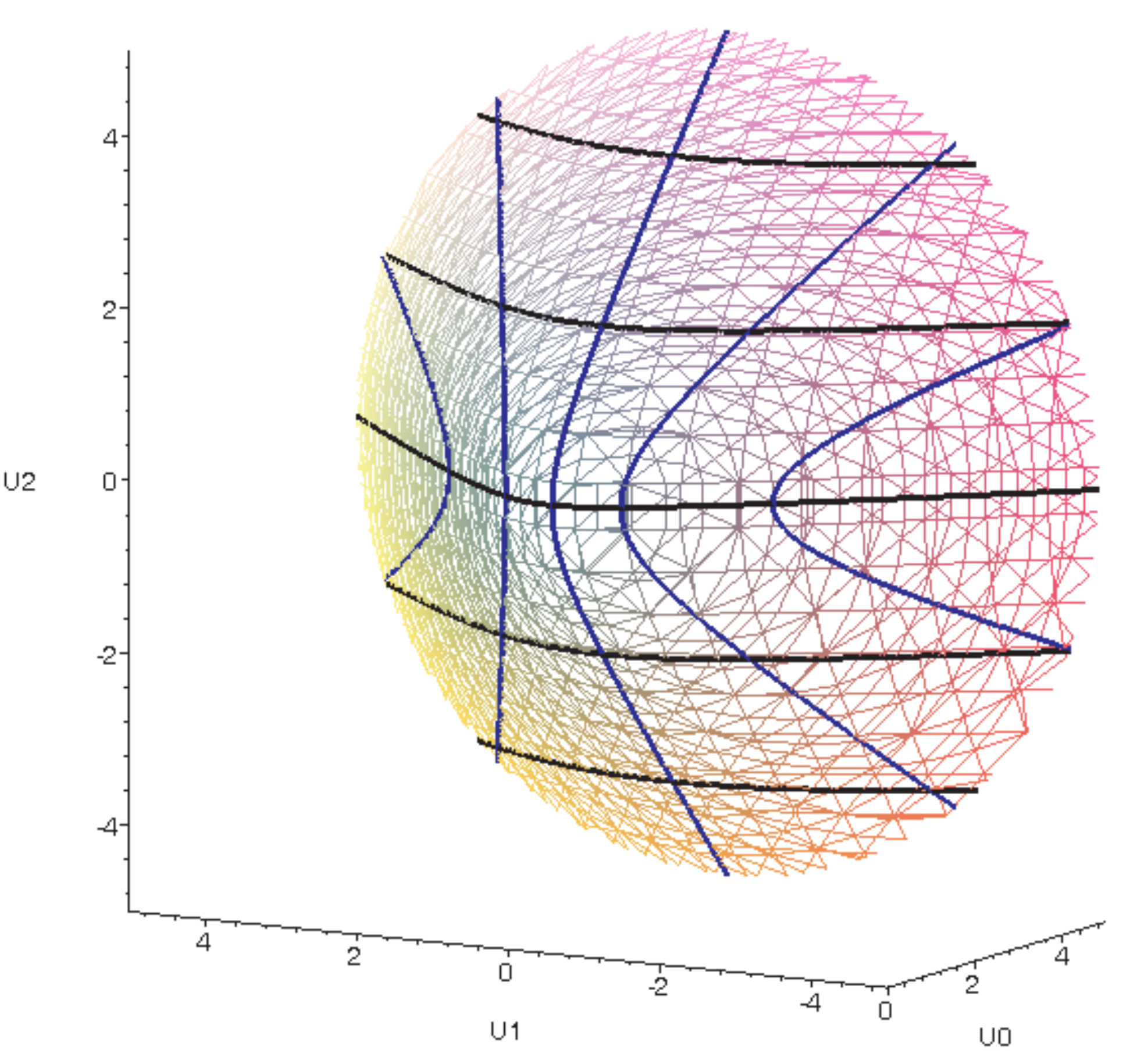}
\begin{center}
\caption{Equidistant system}
\label{fig:4}
\end{center}
    \end{minipage}
    \hfill
    \begin{minipage}[t]{0.45\linewidth}
 \includegraphics[scale=0.3]{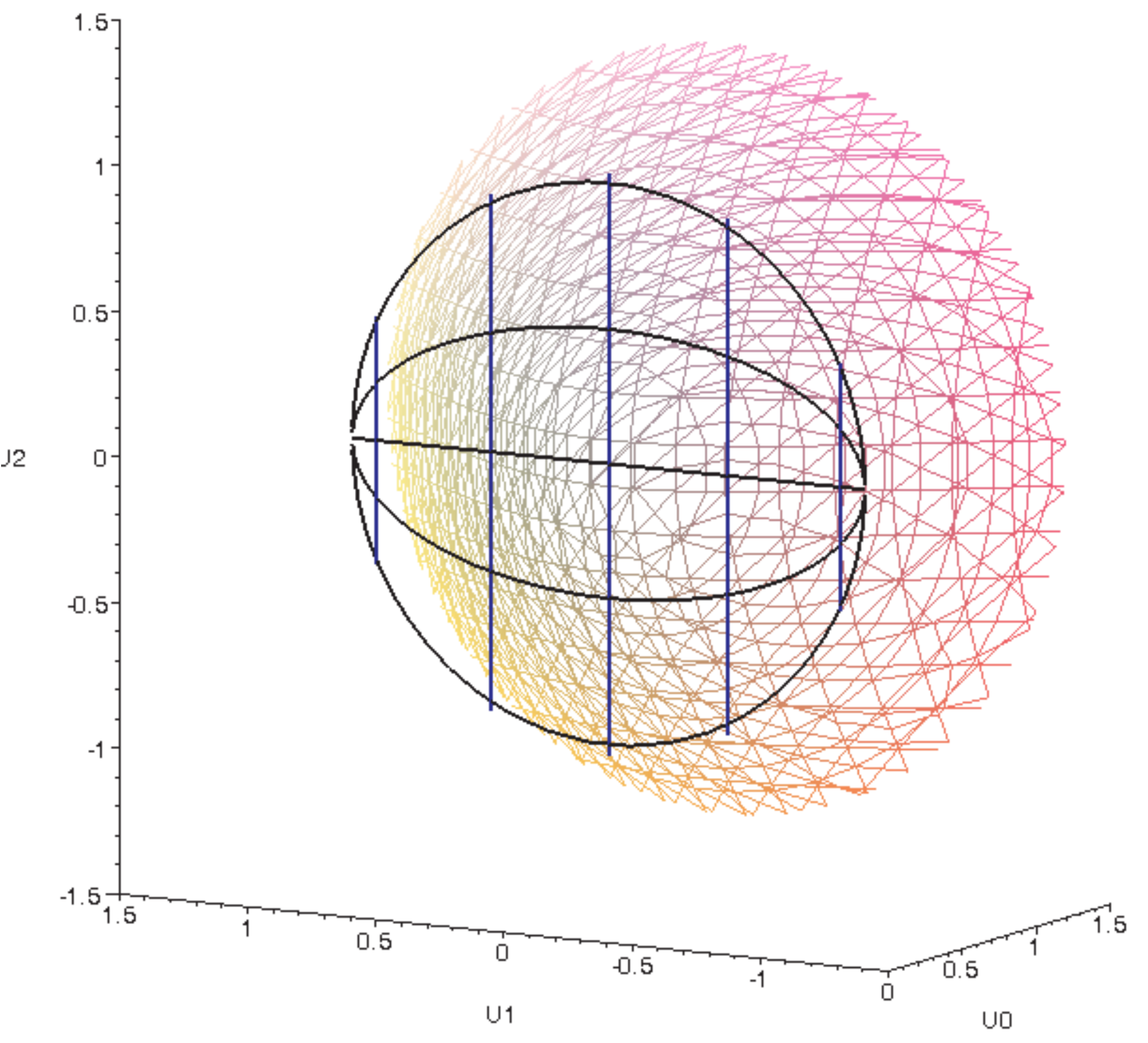}
\caption{Projective plane for equidistant system is formed by equidistant and parallel straight lines}
\label{fig:5}
    \end{minipage}
  \end{center}
\end{figure}

{\bf 2a.} 
In the case of two-sheeted hyperboloid $f(\xi,R) > 0$ and we get that 
\[u_0 = \sqrt{f(\xi,R) + u^2_1} = \sqrt{f(\xi,R}) \cosh(\eta - g(\xi,R)).\] 
The third coordinate is determined from equation $u_2 = \pm \sqrt{f^2(\xi,R) - R^2}$.
Taking now function $f(\xi,R) = R^2 \cosh^2 \xi$, putting $g(\xi,R)=0$ and choosing new notations $\tau_1$ and $\tau_2$ instead of $\xi$ and$\eta$, we obtain the orthogonal equidistant 
system of coordinates:
\be
\label{sys_equi}
u_0 = R \cosh \tau_1 \cosh \tau_2,
\quad
u_1 = R \cosh \tau_1 \sinh \tau_2,
\quad
u_2 = R \sinh \tau_1,
\ee
where $\tau_1, \tau_2 \in \mathbb{R}$ (see Fig. \ref{fig:4}).

If we take $g(\xi,R) = -\xi/\alpha$, than nonorthogonal equidistant system has the form:
\be
\label{sys_equi_NO}
u_0 = R \cosh \tau_1 \cosh (\tau_2 + R \tau_1/\alpha),
\quad
u_1 = R \cosh \tau_1 \sinh (\tau_2 + R \tau_1/\alpha),
\quad
u_2 = R \sinh \tau_1.
\ee

{\bf 2b.}  
On the one-sheeted hyperboloid ${\tilde H_2}$ function $f(\xi,R)$ can be positive or negative. When $f(\xi,R) > 0$ ($|u_2|\geq R$) we have 
again $u_0 = \sqrt{f(\xi,R}) \cosh(\eta - g(\xi,R))$, but now $u_2 = \pm \sqrt{f(\xi,R) + R^2}$. 

Taking ${f(\xi,R}) = R^2 \sinh^2 \xi$ 
and $g(\xi,R) = 0$, renaming coordinates $(\xi, \eta)$ as $(\tau_1, \tau_2)$, we obtain the orthogonal equidistant coordinate system of type Ia that only covers part $|u_2|\geq R$ on ${\tilde H_2}$ (see Fig. \ref{fig:21}):
\be
\label{sys_equi_1}
u_0 = R \sinh \tau_1 \cosh \tau_2,
\quad
u_1 = R \sinh \tau_1 \sinh \tau_2,
\quad
u_2 = \pm R \cosh \tau_1,
\ee
where $\tau_1, \tau_2 \in \mathbb{R}$.

In the same manner, taking
$g(\xi,R) = \ln[R \xi/\alpha]$
one can consider nonorthogonal  $EQ$ system of Type Ia ($\tau_1 \ne 0$):
\be
\label{sys_equi_1_NO}
u_0 = R \sinh \tau_1 \cosh \left( \tau_2 - \ln[R \tau_1 /\alpha]\right),
\quad
u_1 = R \sinh \tau_1 \sinh \left( \tau_2 - \ln[R \tau_1 /\alpha] \right),
\quad
u_2 = \pm R \cosh \tau_1.
\ee

In case when $u_0^2 - u_1^2 \leq 0$ or $|u_2|\leq R$, the relations in 
formula (\ref{25}) take the form
\be
\label{100}
f(\xi,R) = u_1^2 - u_0^2 > 0,
\qquad
\eta = \cosh^{-1}\frac{u_1}{\sqrt{f(\xi,R)}} + g(\xi,R).
\ee
By analogy with the previous case, but taking $f(\xi,R)=R^2 \sin^2\xi$, we come to the equidistant coordinate system which covers part $|u_2|\leq R$ of one-sheeted hyperboloid ${\tilde H_2}$ (see Fig. \ref{fig:21}):
\be
\label{sys_equi_11}
u_0 = R \sin \varphi \sinh \tau,
\quad
u_1 = R \sin \varphi \cosh \tau,
\quad
u_2 =  R \cos \varphi,
\ee
where  $\tau \in \mathbb{R}$, $\varphi \in [0, 2\pi)$. We will call this system equidistant  coordinates of type Ib.   
As for nonorthogonal $EQ$ system of type Ib, we take the same form 
$g(\xi,R) = \ln  [R \xi/\alpha]$
and obtain ($\varphi \ne 0$):
\be
\label{sys_equi_11_NO}
u_0 = R \sin \varphi \sinh \left( \tau - \ln[R\varphi/\alpha] \right),
\quad
u_1 = R \sin \varphi \cosh \left( \tau - \ln[R\varphi/\alpha] \right),
\quad
u_2 =  R \cos \varphi.
\ee

Let us  also note,  that system (\ref{sys_equi_11}) can be obtained 
from (\ref{sys_equi_1}) through change:  $\tau_1 = - i\varphi$,  $\tau_2 = \tau + i\pi/2$. 
Moreover, the difference between systems of type Ia and Ib  is  manifested by  the different contraction limits showed later.

Together with two equidistant coordinate systems of type Ia and Ib one can introduce the coordinate systems of  IIa ($|u_1| \geqslant R$) and IIb ($|u_1| \leqslant R$) types, which diagonalize operator $S^{(1)} = K_1$. As a result we come to the same systems (\ref{sys_equi_1}), (\ref{sys_equi_11}) up to the permutation $u_1 \leftrightarrow u_2$. 

As for nonorthogonal equidistant coordinates IIb, we can consider, for example, the following ones:
\be
\label{sys_equi_11b_NO}
u_0 = - R \cos \varphi \sinh \left( \tau + R \varphi /\alpha \right),
\quad
u_1 = R \sin \varphi,
\quad
u_2 =  - R \cos \varphi \cosh \left( \tau + R \varphi /\alpha \right).
\ee


\paragraph{3. ${S}^{(1)} = K_1 + L$.}
\noindent
Let us consider operator $K_1 + L  = (u_1 - u_0) \partial_{u_2} - u_2
\partial_{u_0} - u_2 \partial_{u_1}$. To reduce it to canonical form $\partial_{\eta}$
we need to solve the following system
\begin{eqnarray}
(u_1 - u_0)\frac{\partial \xi}{\partial u_2} -
u_2 \frac{\partial \xi}{\partial u_0} - u_2 \frac{\partial \xi}{\partial u_1} =0,
\qquad
(u_1 - u_0)\frac{\partial \eta}{\partial u_2} - u_2 \frac{\partial \eta}{\partial u_0}
- u_2 \frac{\partial \eta}{\partial u_1} = 1.
\end{eqnarray}
Then,
\be
f(\xi,R) = u_0- u_1,\qquad u_2 = g(\xi,R) - \eta f(\xi,R).
\ee

\begin{figure}[htbp]
 \begin{center}
    \begin{minipage}[t]{0.45\linewidth}
      \includegraphics[scale=0.3]{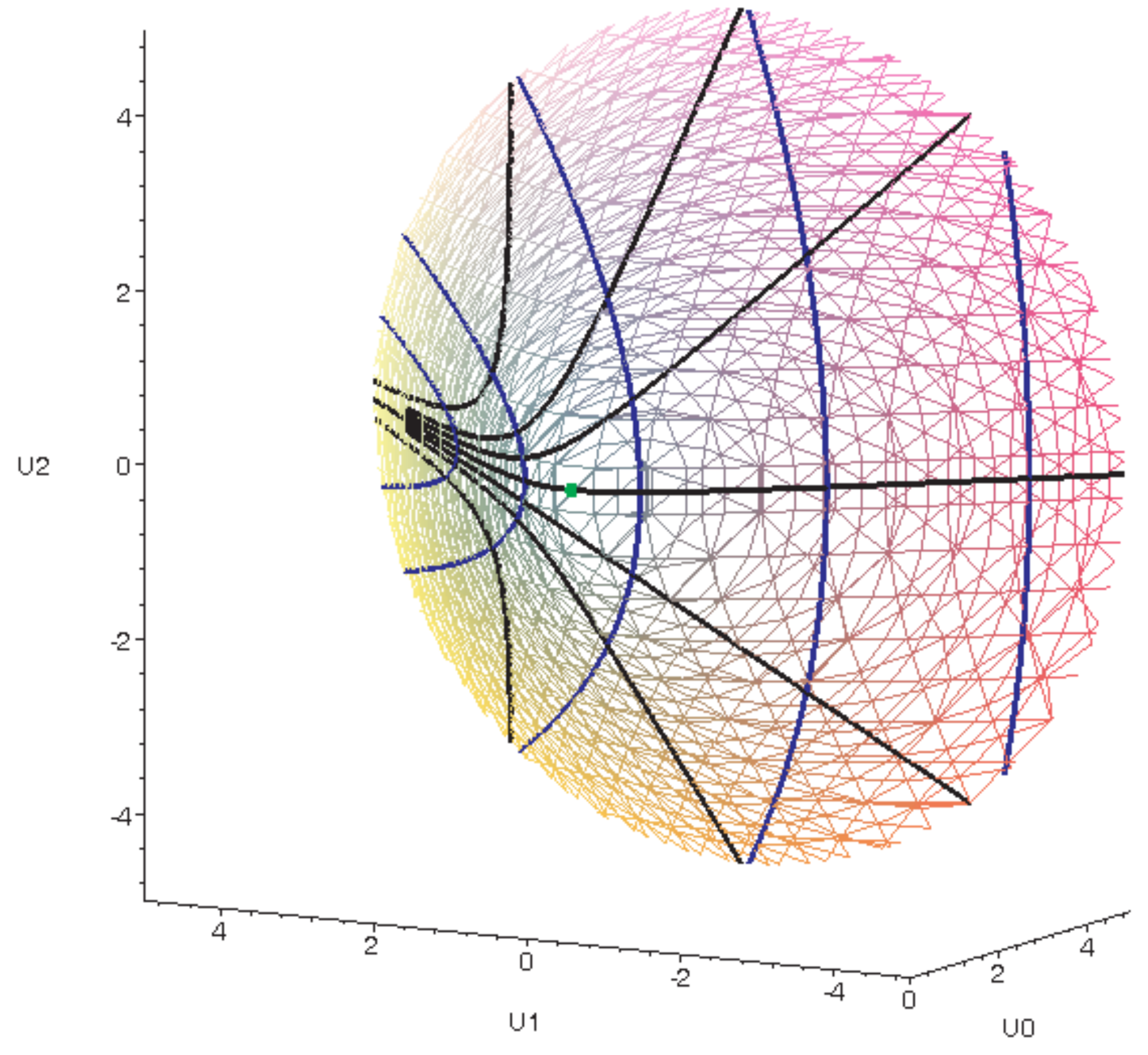}

\begin{center}
\caption{Horicyclic system}
\label{fig:6}
\end{center}
    \end{minipage}
    \hfill
    \begin{minipage}[t]{0.45\linewidth}
 \includegraphics[scale=0.3]{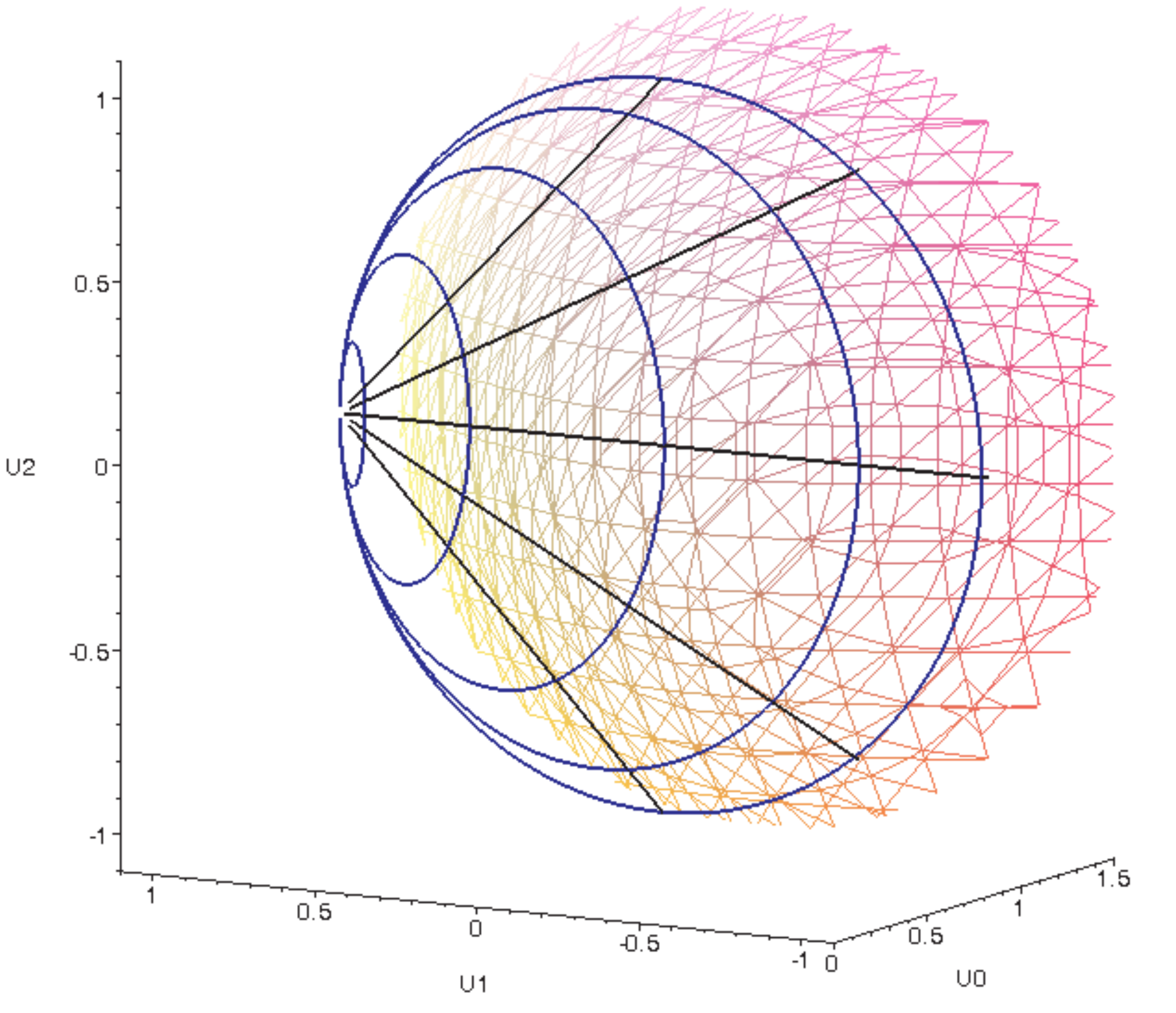}
\caption{Projective plane for horicyclic system is formed by straight lines passing through the boundary point and horicycles}
\label{fig:7}
    \end{minipage}
  \end{center}
 \end{figure}

{\bf 3a.} 
For the case of two-sheeted hyperboloid (for the upper sheet) we have for orthogonal separable coordinates ($\xi, \eta)$ (we take $g=0$)
\begin{eqnarray}
\xi = u_0 - u_1 > 0, 
\qquad  
\eta =  \frac{u_2}{u_1 - u_0}.
\end{eqnarray}
Taking into account (\ref{HYPER-01}) with $\epsilon = 1$ and putting $\eta = - \tilde{x}$, $\xi = R/\tilde{y}$ we obtain the orthogonal horicyclic system of coordinates (see Fig. \ref{fig:6}):
\be
\label{sys_hor}
u_0 = R \frac{\tilde{x}^2 + \tilde{y}^2 + 1} {2 \tilde{y} },
\quad
u_1 = R \frac{\tilde{x}^2 + \tilde{y}^2 - 1} {2 \tilde{y} },
\quad
u_2 = R \frac{\tilde{x}} {\tilde{y} },
\ee
where $\tilde{x}\in \mathbb{R}$, $\tilde{y} >0$.

For nonorthogonal $HO$ system we take $f(\xi,R)=R/\xi$, $g(\xi,R) = R(1-1/\xi)$. Denoting $\xi = \tilde{y}$, $\eta = - \tilde{x}$ we have:
\be
\label{sys_hor_no}
u_0 = R \frac{\left(\tilde{x} + \tilde{y} -1  \right)^2 + \tilde{y}^2 + 1} {2 \tilde{y} },
\quad
u_1 = R \frac{\left(\tilde{x} + \tilde{y} -1  \right)^2 + \tilde{y}^2 - 1} {2 \tilde{y} },
\quad
u_2 = R \frac{\tilde{x} + \tilde{y} -1} {\tilde{y} }.
\ee

\begin{figure}[htbp]
\begin{center}
    \begin{minipage}[t]{0.45\linewidth}
      \includegraphics[scale=0.35]{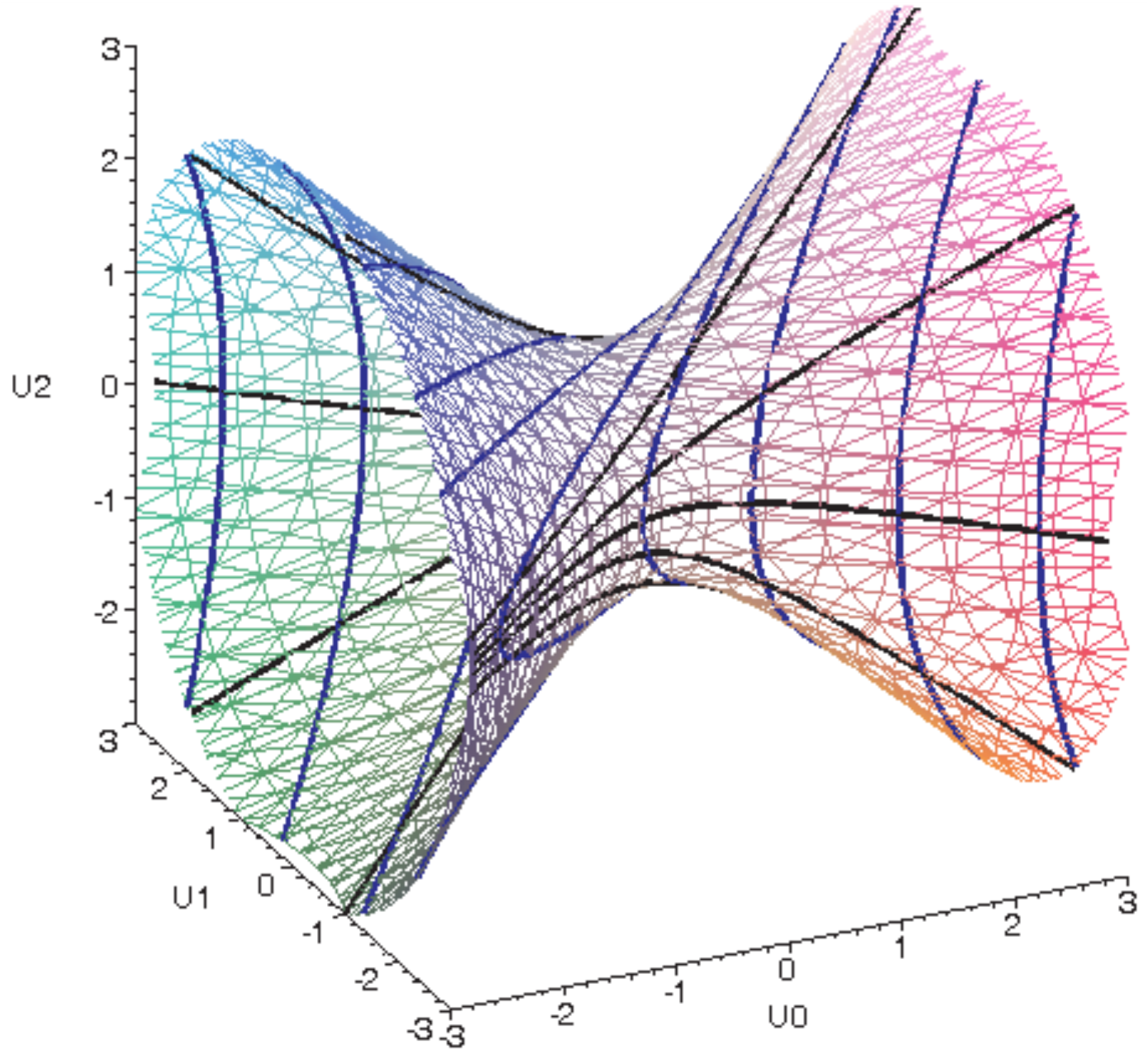}

\begin{center}
\caption{Horicyclic system}
\label{fig:23}
\end{center}
    \end{minipage}
    \hfill
    \begin{minipage}[t]{0.45\linewidth}
 \includegraphics[scale=0.35]{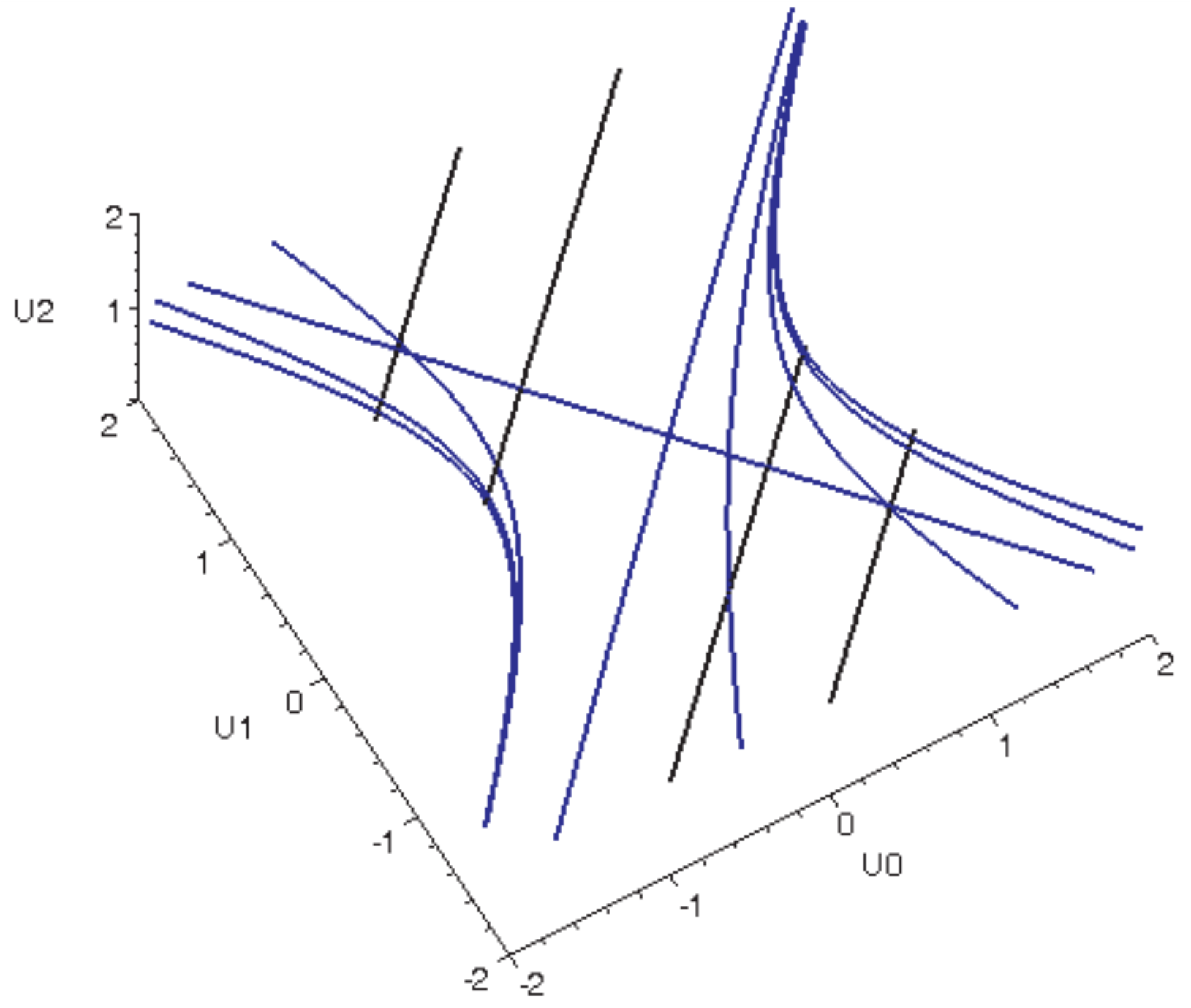}
\caption{Projective plane for horicyclic system}
\label{fig:24a}
    \end{minipage}
  \end{center}
\end{figure}

{\bf 3b.} 
In the case of one-sheeted hyperboloid ($\epsilon = -1$) we get the following form of 
horicyclic system of coordinates (see Fig. \ref{fig:23}):
\be
\label{sys_hor_1}
u_0 = R \frac{\tilde{x}^2 - \tilde{y}^2 + 1} {2 \tilde{y} },
\qquad
u_1 = R \frac{\tilde{x}^2 - \tilde{y}^2 - 1} {2 \tilde{y} },
\qquad
u_2 = R \frac{\tilde{x}} {\tilde{y} }.
\ee
where now  $\tilde{x}\in \mathbb{R}$, $\tilde{y} \in \mathbb{R}\setminus\{0\}$.

Also we consider nonorthogonal $HO$ system of the form ($f(\xi,R) = R\xi/2$, $g(\xi,R) = R$):
\be
\label{sys_hor_1_NO}
u_0 = \frac{R}{4}\left( \xi\eta^2 - 4\eta  + \xi\right),
\qquad
u_1 = \frac{R}{4}\left( \xi\eta^2 - 4\eta  - \xi\right),
\qquad
u_2 = R (1-\xi\eta/2).
\ee

It is clear that one can introduce the equivalent coordinate systems for (\ref{sys_hor}), (\ref{sys_hor_1}), (\ref{sys_hor_1_NO}) which diagonalize operator 
${S}^{(1)} = K_2 - L$ and are obtained by the permutation of coordinates $u_1$ and $u_2$.


\subsection{Classification of second-order symmetries} 
\label{subsection:2.2}

Now let us consider the second-order operator $S^{(2)}$ from  (\ref{FIRST-OPER-01})  of  the  enveloping  algebra  of  $so(2,1)$:
\begin{equation}
    S^{(2)} =  aK_1^2 + b\{K_1,K_2\} + cK_2^2 + d\{K_1,L\} + e\{K_2,L\} + fL^2,\label{S_2}
\end{equation}
where $\{X,Y\} = XY + YX$.

The quadratic form
\begin{equation}
M = \left( {{\begin{array}{*{20}c}
 a & b & d \\
 b & c & e \\
 d & e & f
\end{array} }} \right)  \label{F}
\end{equation}
corresponds to operator  (\ref{S_2}).

Under the automorphisms of  group $A_3$ the elements of form $M$ are transformed as follows:
  \begin{eqnarray}
&&  {M}_{K_1}^\prime = A_{K_1}^T M A_{K_1} = \label{AK1}\\
&&=
\left( {{\begin{array}{*{20}c}
 a & b \cosh a_1 + d \sinh a_1 & b \sinh a_1 + d \cosh a_1\\
 b \cosh a_1 + d \sinh a_1 & c \cosh^2 a_1 + e \sinh 2 a_1 + f \sinh^2 a_1&
 (c+f)/2 \sinh 2 a_1 + e\cosh 2 a_1 \\
 b \sinh a_1 + d \cosh a_1& (c+f)/2 \sinh 2 a_1 + e\cosh 2 a_1
 & c \sinh^2 a_1 + e \sinh 2 a_1 + f \cosh^2 a_1 \end {array}}}\right ), \nonumber \\
&& {M}_{K_2}^\prime = A_{K_2}^T M A_{K_2}, \label{AK2}\\
&&{M}_{L}^\prime = A_{L}^T M A_{L}, \label{ALF}
  \end{eqnarray}
where $A^T$ means transposed matrix. So, we have two hyperbolic rotations  (\ref{AK1}),  (\ref{AK2}) and  ordinary rotation  (\ref{ALF}).

We have to obtain the classification of matrices $M$ with respect to actions (\ref{AK1}) -- (\ref{ALF}) and the linear combination with the Casimir operator ${\cal C} = K_1^2 + K_2^2 - L^2$:
\begin{equation}
{M}_{\cal C}^\prime = \alpha_1 M  + \alpha_2 \left( {{\begin{array}{*{20}c}
 1 & 0 & 0 \\
 0 & 1 & 0 \\
 0 & 0 & -1
\end{array} }} \right)
= \alpha_1 M  + \alpha_2 A_{\cal C}, \label{Casimir}
\end{equation}
where $\alpha_1$ is non-zero constant.

The invariants of transformations (\ref{AK1}) -- (\ref{ALF}) are the following ones:
\[ I_1 = a + c - f, \quad I_2 = A + C - F, \quad I_3 = \det M,
\]
where $A$, $B$, $C$, ..., $F$ are the minors of the elements $a$, $b$, $c$, ..., $f$ of $M$ respectively.

In order to classify all $M$ forms let us consider the case when $I_3 = 0$. If it is not so, we can do it through transformation (\ref{Casimir}), namely ${M}_{\cal C}^\prime = M - \mu A_{\cal C}$, where $\mu$ is the real root of equation $\det(M - \mu A_{\cal C}) = 0$. If $\det M = 0$, then there are the following relations for minors of $M$:
\begin{equation}
    CF = E^2,\qquad AF = D^2,\qquad AC = B^2, \label{minors}
\end{equation}
and
\begin{eqnarray}
    aA - bB + dD = 0, \label{det1}\\
    -bB + cC - eE = 0, \label{det2}\\
    dD - eE + fF = 0. \label{det3}
\end{eqnarray}

It is easy to see that transformations (\ref{AK1}) -- (\ref{ALF}) act on the minors like on the elements of $M$, i.e. if we substitute $a$, $b$, $c$,... in (\ref{AK1}) -- (\ref{ALF}) by the corresponding minors $A$, $B$, $C$,... we obtain the same transformations for minors.

Acting by rotation $A_L$ on minor $B$ it can be reduced to zero. Since $I_3$ is the invariant of group $A_3$, then relations (\ref{minors}) -- (\ref{det3}) remain valid for the transformed minors. By virtue of (\ref{minors}) we have $AC=0$.

\textit{Case 1}. If $A=0$, then $D=0$ due to (\ref{minors}). There are two cases to consider here.
\begin{enumerate}
    \item $I_2 \neq 0$, i.e. $C-F \neq 0$, then by suitable choice of parameter $a_1 = \frac{1}{4}\ln\left(\frac{F+C-2E}{F+C+2E}\right)$, $E$ can be reduced to zero and minors $A$, $B$, $D$ stay equal to zero. Such value of the parameter exists because $(F+C-2E)(F+C+2E) = (C-F)^2>0$. Then from (\ref{minors}) we have $CF=0$. There are  two possible cases:
\begin{itemize}
    \item $C=0, F\neq 0$, then $f=0$ from (\ref{det3}), so $e=0$ from $A=0$ and $d=0$ from $C=0$. Then we can vanish element $b$ by rotation $A_L$ Finally, we have
    \[
M_1= \left.\left( {{\begin{array}{*{20}c}
 a & 0 & 0 \\
 0 & c & 0 \\
 0 & 0 & 0
\end{array} }} \right)\right|_{F = ac \neq 0}.
\]
    \item $C \neq 0, F = 0$, then $c=0$ from (\ref{det2}), so $e=0$ from $A=0$ and $b=0$ from $F=0$. Finally, we have
    \[
M_2= \left.\left( {{\begin{array}{*{20}c}
 a & 0 & d \\
 0 & 0 & 0 \\
 d & 0 & f
\end{array} }} \right)\right|_{C = af - d^2 \neq 0}.
\]
\end{itemize}
    \item $I_2 = 0$, i.e. $C = F$:
        \begin{itemize}
            \item $C = F = 0$, then $E = 0$ from (\ref{minors}) that is all the minors are equal to zero and the automorphisms can not change any minor, but they can transform the elements of $M$. By appropriate choice of rotation $A_L$ element $b$ can be reduced to zero. If $a\neq 0$, then $c=0$ from $F=0$, $e=0$ from $E=0$ and we have form $M_2$ under the condition $C=0$. If $a=0$, then $d=0$ from $C=0$ and through rotation $A_L$ with $a_3 = \pi/2$ we obtain the same form $M_2$.
            \item $C =F \neq 0$, then $f=c$ from (\ref{det2}), (\ref{det3}), so $e = \pm c$ from $A=0$ and $d = \pm b$ from $C=F$. Using reflections one can reduce the form with the higher sign elements to the form with the lower sign elements and finally we have form
    \[
M_3 = \left.\left( {{\begin{array}{*{20}c}
 a & b & -b \\
 b & c & -c \\
 -b & -c & c
\end{array} }} \right)\right|_{ ac \neq b^2}.
\]
        \end{itemize}
\end{enumerate}

\textit{Case 2}. If $A \neq 0$, then $C=0$ from (\ref{minors}), since $B=0$; and $E=0$ due to (\ref{minors}). There are two cases to consider here.
\begin{enumerate}
    \item $I_2 \neq 0$, i.e. $A-F \neq 0$, then by suitable choice of parameter $a_2 = \frac{1}{4}\ln\left(\frac{A+F-2D}{A+F+2D}\right)$ minor $D$ can be reduced to zero and minors $C$, $B$, $E$ stay equal to zero. Such parameter $a_2$ exists, because $(A+F-2D)(A+F+2D) = (A-F)^2>0$. Then from (\ref{minors}) we have $AF=0$, so $F=0$. All minors except $A$ are equal to zero, so $a=0$ from (\ref{det1}). Then $d=0$ from $C=0$; $b=0$ from $F=0$. The obtained form can be reduced to form $M_2$ through rotation $A_L$ with $a_3 = \pi/2$.
    \item $I_2 = A-F = 0$, so $A=F \neq 0$. Then $f = a$ from (\ref{det1}), (\ref{det2}); so $d = \pm a$ from $C=0$ and $b = \pm e$ from $A=F$. Using reflections one can reduce the form with the higher sign elements to the form with the lower sign elements and finally we have form
\begin{equation}
M_4 = \left.\left( {{\begin{array}{*{20}c}
 a & b & -a \\
 b & c & -b \\
 -a & -b & a
\end{array} }} \right)\right|_{ ac \neq b^2}.\label{ab}
\end{equation}
\end{enumerate}

Note, that

1. through rotation $A_L$ with parameter $a_3 = \pi/2$  form $M_3$ is reduced to form $M_4$;

2. form $M_1$ through (\ref{Casimir}) with $\alpha_1 = 1/c$, $\alpha_2 = -1$ is reduced to form $M_2$ with $a\neq -1$, $d=0$, $f=1$,

so there are two distinct inequivalent forms only: $M_2$ (without any condition) and $M_4$.

{\bf 1.} {$M_4.$}

If in (\ref{ab}) $I_1 = c\neq 0$, then by the composition $A_{L} \circ A_{K_1} \circ A_{L}$ with parameters $\sin a_3 = \sigma /\sqrt{\sigma^2 + 4}$, $\sinh a_1 = -\sigma\sqrt{\sigma^2 + 4}/2$, $\sigma = -b/c$ we reduce it to  form
    \[
    \left.\left( {{\begin{array}{*{20}c}
 a & 0 & -a \\
 0 & c & 0 \\
 -a & 0 & a
\end{array} }} \right)\right|_{ac \neq 0},
\]
then dividing it by $a$ and using reflections we obtain operators ($\gamma >0$):
\bea
\label{EP-SYSTEM}
S_{EP} = (K_1 + L)^2 + \gamma K_2^2, \qquad \text{if} \quad ac>0, 
\\[2mm]
\label{hP-SYSTEM}
S_{HP} = (K_1 + L)^2 - \gamma K_2^2, \qquad \text{if}  \quad ac<0.
\eea

If in (\ref{ab}) $I_1 = c = 0$, then  $b \neq 0$, by the same composition $A_{L} \circ A_{K_1} \circ A_{L}$ with parameter $\sigma = -a/(2b)$ we make $a=0$, and after applying reflections and the division by $b$ we have operator
\be
\label{SCP-SYSTEM}
S_{SCP} = \{K_1, K_2\} + \{K_2,L\}.
\ee

{\bf 2.} {$M_2.$}   \\
In this case the form looks like
\begin{equation}
M_2 = \left( {{\begin{array}{*{20}c}
 a & 0 & d \\
 0 & 0 & 0 \\
 d & 0 & f
\end{array} }} \right) \cong
\left( {{\begin{array}{*{20}c}
 0 & 0 & 0 \\
 0 & c & e \\
 0 & e & f
\end{array} }} \right) = \widetilde{M}_2
. \label{ae}
\end{equation}

Here we need to consider the value of the invariant of transformation (\ref{AK2}), namely $J = (a+f)^2 - 4d^2$ for $M_2$ (or the invariant for $A_{K_1}$: $\widetilde{J} = (c+f)^2 - 4e^2$ for $\widetilde{M}_2$).

1. $\widetilde{J} = (c+f)^2 - 4e^2 > 0$, then $c+f \neq 0$, and by the hyperbolic rotation (\ref{AK1}) with $\tanh 2a_1 = -2e/(c+f)$ we can vanish element $e$ in $\widetilde{M}_2$.

If $f = 0$, then we have operator
\be
\label{EQ-SYSTEM}
 S_{EQ}^{(2)} = K_2^2.
\ee

If $f \neq 0$, then dividing by $f$ we have form
\begin{equation}
    \left.\left( {{\begin{array}{*{20}c}
 0 & 0 & 0 \\
 0 & \overline{c} & 0 \\
 0 & 0 & 1
\end{array} }} \right)\right|_{\widetilde{J} = (\overline{c}+1)^2 >0,\: \overline{c}\neq -1}. \label{1c}
\end{equation}
    If $\overline{c}=0$ in (\ref{1c}), then we have operator
\be
\label{SPH-SYSTEM}
S_{SPH}^{(2)} = L^2.
\ee

If $\overline{c} \ne 0$, then from relation $\widetilde{J} = (\overline{c}+1)^2 >0$ we have $\overline{c}\in(-\infty,-1)\cup(-1,0)\cup(0,\infty)$.

a. If $\overline{c}>0$ in (\ref{1c}), then putting $\overline{c} = \sinh ^2 \beta$, $\beta \neq 0$ we have operator
\be
\label{E-SYSTEM}
S_E = L^2 + \sinh^2 \beta K_2^2, \qquad \beta \neq 0.
\ee

b. If $\overline{c} < -1$, then putting $\overline{c}^2 = -1/\sin^2\alpha$, $\sin^2\alpha \neq 0, \ne 1$ and multiplying the operator by $-\sin^2\alpha$ we have  operator
\be
\label{H-SYSTEM}
S_H = K_2^2 - \sin^2 \alpha \:L^2, \quad \sin^2\alpha \neq 0, \neq 1.
\ee

c. Finally, for values $-1 < \overline{c} < 0$, we can take $\overline{c} = -\tanh^2 \gamma$, $\gamma \neq 0$ and trough rotation (\ref{ALF}) with $a_3 = \pi/2$ we obtain
 \begin{equation}
    \left( {{\begin{array}{*{20}c}
 -\tanh^2 \gamma & 0 & 0 \\
 0 & 0 & 0 \\
 0 & 0 & 1
\end{array} }} \right). \label{c2}
\end{equation}
Then, acting by (\ref{Casimir}) over (\ref{c2}) with $\alpha_1 = \cosh^2\gamma$, $\alpha_2 = \sinh^2\gamma$ we reduce the above form to the case of $S_E$ and there is no new operator.

Let us note, that applying hyperbolic rotation (\ref{AK2}) with $a_2 = \beta$ to the operator $S_E$ and using (\ref{Casimir}) with $\alpha_1 = 1$, $\alpha_2 = \sinh^2 \beta$ we obtain the {\it rotated} operator
\be
\label{E1-SYSTEM}
S_{\tilde{E}} = \cosh 2\beta L^2 + 1/2 \sinh 2\beta \{K_1,L\}, \qquad \beta \neq 0.
\ee

2. $J = 0$, so $(a+f)^2 = 4d^2$, $d \neq 0$ (if $d=0$, then $a=-f$ and we have operator $S_{H}$ with $\sin^2 \alpha = 1$, but in this case $S_H - \Delta = K_2^2 - L^2 - (K_1^2 + K_2^2 - L^2) = - K_1^2 \sim K_2^2 = S_{EQ}$). So $| a+f | = 2|d|$, or dividing it by $d$: $|a+f|=2$.

a. If $I_1 = 0$, i.e. $a=f$ and $a=f=\pm 1$, using reflections we obtain

\be
\label{HO-SYSTEM}
S_{HO}^{(2)} = (K_1 + L)^2.
\ee

b. If $I_1 \neq 0$, i.e. $a\neq f$, and we can take $f=2-a$, if it's not so, then trough reflections we can do it, so we have
\begin{equation}
\left.\left( {{\begin{array}{*{20}c}
 a & 0 & 1 \\
 0 & 0 & 0 \\
 1 & 0 & 2-a
\end{array} }} \right)\right|_{a\neq 1}. \label{ae2}
\end{equation}

Then using (\ref{Casimir}) with $\alpha_2 = 1-a$, $\alpha_1 = 1$ we can reduce the form (\ref{ae2}) to
\begin{equation}
\left( {{\begin{array}{*{20}c}
 1 & 0 & 1 \\
 0 & \gamma & 0 \\
 1 & 0 & 1
\end{array} }} \right),
\end{equation}
where $\gamma= 1 - a \neq 0$ that corresponds to operator
$
\gamma K^2_2 + (K_1 + L)^2,
$ and it is equivalent to the form $M_4$ with $b=0$.

3. $J = (a+f)^2 - 4d^2 < 0$, $d \neq 0$, then by (\ref{AK2}) we can obtain $a = -f$,
so by (\ref{Casimir}) with $\alpha_1 = 1/d$, $\alpha_2 = - a/d$ we have
    \[
    \left( {{\begin{array}{*{20}c}
 0 & 0 & 1 \\
 0 & c & 0 \\
 1 & 0 & 0
\end{array} }} \right),
\]
that corresponds to operator $S_{SH}$, if we put $c = \sinh 2\beta$:
\be
\label{SH-SYSTEM}
S_{SH} = \sinh 2\beta K_2^2 + \{K_1, L\}.
\ee

Thus, we have obtained nine second order operators $S^{(2)}_\alpha$ of the enveloping algebra of $so(2,1)$. Each of them presents nonequivalent class with respect to the group of inner automorphisms $A_3$.


\subsection{Second-order symmetries and  non-subgroup coordinates on  two-dimensional hyperboloids}

Here we briefly describe the method of  simultaneous diagonalization  of   the  two second order  operators  $\Delta_{LB}$  and  $S^{(2)}_{\alpha}$ (see also \cite{PSW1}).   
We will use the analogy with classical mechanics  which  simplifies  our  discussion.  As it  is  known  on the space of constant curvature the  Schroedinger and Hamilton-Jacobi 
equation  separates in the same systems of coordinates. 

Let us take some local coordinate system $\xi=(\xi^1, \xi^2)$ that defines metric tensor (\ref{METRIC-01}) in space (\ref{HYPER-01}). For that system the classic Hamiltonian, 
describing a free motion, takes the form
\begin{equation}
\label{Hamiltonian-01}
{\cal H} (\xi, p) = g^{ik} p_i p_k ,  \qquad i,k = 1,2,
\end{equation}
where $p_i$ are the momenta classically conjugated to coordinates $\xi^i$. Let there exist the quadratic integral of motion ${\cal  S} (\xi, p)$ which is in involution with Hamiltonian (\ref{Hamiltonian-01}) 
\begin{equation}
\label{KILLING-01}
{\cal  S} (\xi, p) =  a^{ik} (\xi) p_i  p_k ,  \qquad a^{ik} = a^{ki},
\end{equation}
where tensor $a^{ik}$ is called Killing tensor of the second rank. Analogically, we can consider for the same space the motion of a free particle in quantum mechanics which is described simultaneously by two commuting  operators, that is, by Hamiltonian and by quadratic integral of motion:
\begin{equation}
\label{Hamiltonian-02}
H  = \Delta_{LB}  =  \frac{1}{\sqrt{g}}
\,  \frac{\partial
}{\partial \xi^i} \sqrt{g} \, g^{ik} \, \frac{\partial }{\partial
\xi^k},
\qquad\qquad
 S =   \frac{1}{\sqrt{g}}
 \,  \frac{\partial
}{\partial \xi^i} \sqrt{g} \, a^{ik} \, \frac{\partial }{\partial
\xi^k}, \qquad i,k = 1,2.
\end{equation}
The task is to find orthogonal coordinates $(\lambda_1, \lambda_2)$ which admit in classic mechanics the additive separation of variables related to equations  
${\cal H} (\xi, p) = \mathcal{E}$ and  ${\cal  S} (\xi, p)  = \mu$. In the case of quantum mechanics there is a multiplicative separation of variables related to the system of equations:  $H  \Psi  =  \mathcal{E} \Psi $ and  $S \Psi =  \mu \Psi$ where $\mu$ is a separation constant.

Such a coordinate system, generally speaking, can be determined from the condition of simultaneous reduction of quadratic forms ${\cal H} $ and  ${\cal  S} $ to canonic ones when matrices $g^{ik} $ and  $a^{ik} $ are of diagonal type. To make it, according to the general theory, one needs to solve equation 
\begin{equation}
\label{XARAKTER-01}
\det(a^{ik}  -  \lambda g^{ik})  =  0.
\end{equation}
Let us note that if one of the quadratic form, for example ${\cal H} (\xi, p)$ has the positive-definite associated matrix, then due to well known theorem from algebra, both roots $\lambda_1$ and $\lambda_2$ are real and distinct throughout space (\ref{HYPER-01}). Moreover, form ${\cal H} (\xi, p)$ is reduced to the normal form (all coefficients of quadratic form are equal to unit), but form ${\cal  S} (\xi, p) $ is diagonalized (is of canonic form, see \cite{KUROSH}, p. 219). In such a way it is possible to choose some new orthogonal coordinates as a functions of the roots  $\lambda_1$ and $\lambda_2$:  $u(\lambda_1)$  and  $v(\lambda_2)$ for which ${\cal H}$ and ${\cal  S}$ simultaneously take the following form:
\begin{equation}
\label{Hamiltonian-03}
{\cal H}  =  \frac{1}{\alpha(u) + \beta(v)}  \left ( p_{u}^2  +   p_{v}^2 \right) = \mathcal{E},
\qquad\qquad
{\cal  S}  =   \frac{1}{\alpha(u) + \beta(v)}  \left ( \alpha(u)  p_{u}^2  -  \beta(v) p_{v}^2 \right)  =  \mu.
\end{equation}
The form of Hamiltonian ${\cal H}$ in (\ref{Hamiltonian-03}) is called the Liouville form \cite{PERELOMOV}. Finally, the procedure of separation of variables for Hamilton-Jacobi equation leads to two equations:
\begin{equation}
\label{Hamiltonian-04}
 \beta  {\cal H}  +  {\cal  S} =  \beta  \mathcal{E}  + \mu ,
\qquad\qquad
 \alpha  {\cal H}  -  {\cal  S} =  \alpha  \mathcal{E}  - \mu.
\end{equation}
By the analogy, for Helmholtz equation we have
\begin{equation}
\label{Hamiltonian-05}
 (\beta  { H}  +  {S} )  \Psi  =  (\beta  \mathcal{E}  + \mu) \Psi ,
\qquad\qquad
 (\alpha  {H}  -  {S} ) \Psi =   (\alpha  \mathcal{E}  - \mu ) \Psi. 
\end{equation}
If none of quadratic forms is positive definite, then condition (\ref {XARAKTER-01}) does not guarantee the existence of the real or distinct eigenvalues. It may happen that $\lambda_1 = \lambda_2$ or $\lambda_1 $ and $\lambda_2$ are real ones only in some part of space and they do not provide the parametrization of whole hyperboloid (\ref{HYPER-01}). We can see this situation for the construction of coordinate system on one-sheeted hyperboloid.   

Now let us write down the direct algorithm to obtain the separable coordinate system corresponding to the one of symmetry forms determined in subsection \ref{subsection:2.2}. 

Let us take pseudo-spherical coordinates $(\tau, \varphi)$  (\ref{PSEUDO-SPHER-01}) as local ones $\xi=(\xi^1, \xi^2)$ on two-sheeted hyperboloid. Components $g_{ik}$ for this coordinate system are as follows: 
\begin{eqnarray}
\label{01-METRIX}
g_{ik} = R^2 \diag(1, \sinh^2\tau),
\qquad
g^{ik} = R^{-2} \diag(1, 1/\sinh^2\tau),
\qquad
\sinh^2\tau = R^{-2} (u_1^2 + u_2^2). 
\end{eqnarray}
The classical  free  Hamiltonian  (\ref{Hamiltonian-01})  is 
\begin{equation}
\label{Hamiltonian1}
{\cal H}  =  g^{ik} p_i p_k   =  \frac{1}{R^2}\left(p_{\tau}^2 + \frac{1}{\sinh^2\tau} p_{\varphi}^2\right) = 
\frac{1}{R^2}\left(p_{\tau}^2 + \frac{R^2}{u_1^2+u_2^2}  p_{\varphi}^2\right),
\end{equation}
and the symmetry  elements $K_1$, $K_2$, $L$ have the form  (in ambient space coordinates)
\begin{eqnarray}
\label{KLASSIK-01}
K_1 =  - \frac{u_2}{ \sqrt {u_1^2+u_2^2}  } p_\tau - \frac {u_0 u_1 }{u_1^2+u_2^2 } p_\varphi,
\qquad
K_2  =  -  \frac{u_1}{ \sqrt {u_1^2+u_2^2}  } p_\tau + \frac {u_0 u_2 }{u_1^2+u_2^2 } p_\varphi,
\qquad
L =  p_\varphi,
\end{eqnarray}
where $p_\tau$, $p_\varphi$ are classically conjugated to coordinates $(\tau, \varphi)$.  

In the case of one-sheeted hyperboloid $u_0^2 - u_1^2-u_2^2 = - R^2$ we use pseudo-spherical coordinates defined by (\ref{pspherical}). The components of metric tensor are as follows:
\begin{eqnarray}
\label{01-METRIX00}
g_{ik} = R^2 \diag(1,  - \cosh^2\tau),
\qquad
g^{ik} = R^{-2} \diag(1,  - 1/\cosh^2\tau).
\end{eqnarray}
Formulas (\ref{KLASSIK-01}) for operators $K_1$, $K_2$, $L$ are the same, but for classic Hamiltonian (\ref{Hamiltonian-01}) we obtain 
\begin{equation}
\label{Hamiltonian2}
{\cal H}  =  g^{ik} p_i p_k   =  \frac{1}{R^2}\left(p_{\tau}^2  - \frac{1}{\cosh^2 \tau} p_{\varphi}^2\right) = 
\frac{1}{R^2}\left(p_{\tau}^2  - \frac{R^2}{u_1^2+u_2^2}  p_{\varphi}^2\right).
\end{equation}
Using relations (\ref {KLASSIK-01}), we write down the polynomial $S^{(2)}_\alpha$ in the quadratic form
\begin{equation}
\label{01-CONSTANT}
S^{(2)}_\alpha = A  p_\tau^2 + 2B p_\tau p_\varphi + C p_\varphi^2,
\end{equation}
where in general the coefficients $A$, $B$, $C$ are the functions of the variables ($\tau, \varphi$) or ambient space variables ($u_0, u_1, u_2)$. Later we will work in the ambient coordinate system ($u_0, u_1, u_2$) as a  more universal one.

We need to diagonalize the matrix corresponding to form (\ref{01-CONSTANT}) by finding  the eigenvalues from equation (\ref{XARAKTER-01})
\be
\label{DETERMINANT-1}
\det(a^{ik} - \lambda g^{ik}) =     
\left| {{\begin{array}{*{20}c}
 A - \frac{\lambda}{g_{11}} &   B  \\
 B & C - \frac{\lambda}{g_{22}}
 \end{array} }} \right|=0,
\ee
which corresponds to the system of two algebraic equations   
\begin{equation}
\label{lambda}
\lambda_1 + \lambda_2 = A g_{11} + C g_{22}, \quad
\lambda_1\lambda_2 = (AC - B^2) g_{11} g_{22}
\end{equation}
and admits two different and real roots if the following condition is satisfied  
\begin{equation}
\label{01-lambda}
(A g_{11} - C g_{22})^2  +  4B^2 g_{11} g_{22} > 0.
\end{equation}
Resolving above system (\ref{lambda}) with the relation $u_0^2 - u_1^2-u_2^2= \epsilon R^2$,   $\epsilon = \pm 1$ we can determine separable orthogonal 
systems of coordinates on two- or one-sheeted hyperboloids. Let us note that in the case of two-sheeted hyperboloid we have $g_{11} g_{22} > 0$ and 
inequality (\ref{01-lambda}) is satisfied automatically. Thus, two different and real roots $\lambda_1$ and $\lambda_2$ exist. 
The corresponding problem on one-sheeted hyperboloid is more complicated. It may happen that  $\lambda_1$ and $\lambda_2$ are real only on 
a part of ${\tilde H}_2$ hyperboloid and, therefore, do not parametrize the whole space. 

We conserved the names given in article \cite{WINLUKSMOR} for every operator $S^{(2)}_\alpha$: Equidistant coordinates 
$S_{EQ} = K_2^2$, Spherical coordinates $S_{SPH} = L^2$, Horiciclic coordinates $S_{HO} = (K_1 + L)^2$, Semi-circular-parabolic coordinates $S_{SCP}$, 
Elliptic-parabolic coordinates $S_{EP}$, Hyperbolic-parabolic coordinates $S_{HP}$, Elliptic coordinates $S_{E}$ ($S_{\tilde{E}}$ for rotated elliptic coordinates), Hyperbolic coordinates $S_H$  and $S_{SH}$ for Semi-hyperbolic coordinates. Three first operators, namely: $S_{EQ}$, $S_{SPH}$ and $S_{HO}$ are the squares of the first order operators and provide the separation of variables in subgroup coordinate systems. Other six operators are non-subgroup ones. 

Five operators: $S_{EP}$, $S_{HP}$, $S_{E}$, ($S_{\tilde{E}}$),  $S_H$ and $S_{SH}$ depend on one dimensionless parameter, let us call it $\gamma$. The limit $\gamma \to 0$ (or $\gamma \to \infty$) simplifies significantly operators $S^{(2)}_\alpha$ and corresponding coordinates degenerate into a simpler subgroup systems (except the system for $S_{SCP}$).


\subsection{Semi-hyperbolic system of coordinates}
\label{SH-system}

{\bf 1.} \, Firstly let us consider two-sheeted hyperboloid. Writing down the operator $S_{SH} = \sinh 2\beta K_2^2 + \{ K_1, L\}$ as in equation 
(\ref{01-CONSTANT}) and using formula (\ref{01-METRIX}) we get that the algebraic system (\ref{lambda}) takes the form 
\begin{equation}
\label{lambda_SH}
\lambda_1 + \lambda_2 =2u_0 u_1 + c(u_1^2 - u_0^2), 
\qquad
\lambda_1\lambda_2 = -R^2(u_2^2 + 2c u_0 u_1)
\end{equation}
with $c=\sinh 2\beta$. Solution of (\ref{lambda_SH}) provides the algebraic form of semi-hyperbolic system of coordinates in terms  
of independent variables $\lambda_1$, $\lambda_2$: 
\begin{eqnarray}
\label{sys_sh_lambda}
u_0^2 &=& \frac{R^2}{2(c^2+1)}\left\{\sqrt{(c^2 + 1)\left(1 +
\frac{\lambda_1^2}{R^4}\right)\left(1 + \frac{\lambda_2^2}{R^4}\right)} +
1- \frac{\lambda_1}{R^2} \frac{\lambda_2}{R^2} -
c\left(\frac{\lambda_1}{R^2} + \frac{\lambda_2}{R^2}\right) \right\},
\nonumber\\[2mm]
u_1^2 &=& \frac{R^2}{2(c^2+1)}\left\{\sqrt{(c^2 + 1)
\left(1 + \frac{\lambda_1^2}{R^4}\right)\left(1 + \frac{\lambda_2^2}{R^4}\right)}
- 1 + \frac{\lambda_1}{R^2} \frac{\lambda_2}{R^2} +
c\left(\frac{\lambda_1}{R^2} + \frac{\lambda_2}{R^2}\right)\right\},
\\[2mm]
u_2^2 &=& - \frac{R^2}{c^2 + 1} \left(\frac{\lambda_1}{R^2} + c\right)\left(\frac{\lambda_2}{R^2} + c\right) .
\nonumber
\end{eqnarray}
where we choose that $\lambda_2/R^2 < - c  <  \lambda_1/R^2$. It is clear from (\ref{sys_sh_lambda}) for $SH$ coordinate system that the relation between ($\lambda_1, \lambda_2$) and Cartesian coordinates ($u_0, u_1, u_2$) is not ''one-to-one''. For every value of  $\lambda_1$, $\lambda_2$ there are eight corresponding points  ($\pm u_0, \pm u_1, \pm u_2$) in ambient space $E_{2,1}$. 

Geometrically the semi-hyperbolic system (\ref{sys_sh_lambda}) consists of two families of confocal semi-hyperbolas.  The distance between the semi-hyperbolas focus and the basis of its equidistances is equal to $2\beta R$. The dimensionless parameter $c = \sinh 2\beta$ defines the position of 
the semi-hyperbolas focus on the upper sheet of two-sheeted hyperboloid (Fig. \ref{fig:12}). One can obtain its coordinates making 
$\lambda_1/R^2 \to - c$ and $\lambda_2/R^2 \to - c$, then $F\left(u_0, u_1, u_2 \right) \equiv 
F\left( R\frac{\sqrt{\sqrt{c^2 + 1} + 1}}{\sqrt{2}}, \right.$ $\left. -R\frac{\sqrt{\sqrt{c^2 + 1} - 1}}{\sqrt{2}}, \, 0\right)$. 

There exist two simple particular cases of $SH$ system, namely: $c=0$ and $c=1$. For the first case the focus coordinates are $F\left(R, 0, 0\right)$ and symmetry operator takes the simple form $S_{SH} = \{ K_1, L\}$. Then system (\ref{sys_sh_lambda}) can be rewritten in new coordinates $\lambda_1/R^2 = \mu_1$ and $\lambda_2/R^2 = - \mu_2$:
\bea
\label{sys_sh_mu1}
u_0^2 &=& \frac{R^2}{2}\left\{ \sqrt{(1 +\mu_1^2)(1+\mu_2^2)}  +  \mu_1 \mu_2 + 1 \right\},
\nonumber\\[2mm]
u_1^2 &=& \frac{R^2}{2}\left\{ \sqrt{(1 +\mu_1^2)(1+\mu_2^2)}  -  \mu_1 \mu_2  - 1 \right\},
\\[2mm]
u_2^2 &=& R^2 \mu_1\mu_2,
\nonumber
\eea
where $\mu_1,  \mu_2 \ge 0$. For the second case  $c=1$ we have $S_{SH} = K_2^2 + \{ K_1, L\}$. Introducing dimensionless variables $\lambda_1/R^2 = \sinh\tau_1$ and  $\lambda_2/R^2 = \sinh\tau_2$ ($\sinh \tau_2  <  - 1  < \sinh\tau_1$) we get the trigonometric form of semi-hyperbolic coordinate system 
\begin{eqnarray}
\label{2SH_sys_sh_hyp}
u_0^2 &=& \frac{R^2}{4} \left[\sqrt{2}\cosh\tau_1\cosh\tau_2 - (\sinh\tau_1+1)(\sinh\tau_2+1) + 2\right],
\nonumber\\[2mm]
u_1^2 &=& \frac{R^2}{4}  \left[\sqrt{2}\cosh\tau_1\cosh\tau_2 + (\sinh\tau_1+1)(\sinh\tau_2+1) - 2\right],
\\[2mm]
u_2^2 &=& -\frac{R^2}{2}(\sinh\tau_1+1)(\sinh\tau_2+1).
\nonumber
\end{eqnarray}

For the large values of parameter $c \to \infty$ ($\beta \to \infty$) the focus goes to infinity and $SH$ system degenerates to the equidistant one. For the operator we have $S_{SH} /\sinh 2\beta  =   K_2^2  +   \{ K_1, L\} /\sinh 2\beta  \to  K_2^2$. Indeed, as $\beta \to \infty$ one can make that the variable $\lambda_2$ tends to $-\infty$ and at the same time the domain of $\lambda_1$  is expanded:  $\lambda_1 \in (-\infty, \infty)$. Putting $- \lambda_2/\sinh 2\beta = R^2 \cosh^2\tau_1$, $\lambda_1 = R^2 \sinh 2\tau_2$ in (\ref {sys_sh_lambda}) and taking $\beta \to \infty$ we come to equidistant system in form (\ref{sys_equi}).

Finally, note that putting $c = \frac{a-\gamma}{\delta}$, where $\alpha, \delta, \gamma$ are some constants and introducing new variables accordingly to the relations $\lambda_1 = R^2\left(\frac{a - \rho_1}{\delta} - c\right)$ and $\lambda_2 = R^2\left( \frac{a - \rho_2}{\delta} - c\right)$, semi-hyperbolic system of coordinates (\ref{sys_sh_lambda}) takes the well known form (see \cite{OLEV,WINLUKSMOR}). 

\noindent
{\bf 2.} \, In the case of one-sheeted hyperboloid instead of the algebraic equations (\ref{lambda_SH}) we have the following system      
\begin{equation}
\label{001-lambda}
\lambda_1 + \lambda_2 =2u_0 u_1 + c(u_1^2 - u_0^2), 
\qquad
\lambda_1\lambda_2 =  R^2(u_2^2 + 2c u_0 u_1).
\end{equation}
Indeed equations (\ref{lambda_SH}) and (\ref{001-lambda}) are connected by the transformation $R \to i R$. Therefore the roots $\lambda_1$, $\lambda_2$ are real and distinct when inequality (\ref{01-lambda}) is satisfied 
\begin{equation}
\label{01-cover}
|2u_0 u_1  +   c(u_1^2 - u_0^2)| > 2 R \sqrt{u_2^2 + 2c u_0 u_1}
\end{equation}
and are complex otherwise. Using now the substitutions $\lambda_1/R^2 = \sinh\tau_1$ and $ \lambda_2/R^2 = \sinh\tau_2$ 
we get the analog of trigonometric form of semi-hyperbolic coordinates (see Fig. \ref{fig:24}): 
\bea
\label{sys_sh_hyp}
u_0^2 &=& \frac{R^2}{2(c^2+1)} \left\{ \sqrt{c^2+1} \cosh\tau_1\cosh\tau_2 + (\sinh\tau_1-c)(\sinh\tau_2-c)-c^2-1\right\},
\nonumber\\[2mm]
u_1^2 &=& \frac{R^2}{2(c^2+1)} \left\{ \sqrt{c^2+1} \cosh\tau_1\cosh\tau_2 - (\sinh\tau_1-c)(\sinh\tau_2-c)+c^2+1\right\},
\\[2mm]
u_2^2 &=& \frac{R^2}{c^2+1}(\sinh\tau_1-c)(\sinh\tau_2-c),
\nonumber
\eea
where $\sinh\tau_1, \sinh\tau_2 \leq c$ ($SH$ of Type I) or $\sinh\tau_1, \sinh\tau_2 \geq c$ ($SH$ of Type II).
The above coordinate grid has two pairs of envelopes defined by equality  $|2u_0 u_1  +   c(u_1^2 - u_0^2)| = 2 R \sqrt{u_2^2 + 2c u_0 u_1}$. 
The points of its intersections have coordinates $\left(\pm R\frac{\sqrt{\sqrt{c^2 + 1} - 1}}{\sqrt{2}}, \right.$ $\left. \mp R\frac{\sqrt{\sqrt{c^2 + 1} + 1}}{\sqrt{2}}, \, 0\right)$. It is necessary to adjust the intervals for coordinate lines $\tau_i = \text{const}$. For example, one can take points of envelops as initial ones for family $\tau_1 = \text{const.}$ and as endpoints for the other family $\tau_2 = \text{const.}$

 Let us note that coordinate system (\ref{2SH_sys_sh_hyp}) 
can be constructed from (\ref{sys_sh_hyp}) by the transformations $R\to iR$, $\tau_1 \to \tau_1 - i\pi$ and $\tau_2 \to - \tau_2$.

In case of $c = 0$ the inequality (\ref{01-cover}) is equivalent to $|u_1|\geq  R$.  The change of variable 
$\mu_1 = \sinh\tau_1$, $\mu_2 = \sinh\tau_2$ leads to the following parametrization of coordinates:
\bea
\label{sys_sh_mu}
u_0^2 &=& \frac{R^2}{2}\left\{ \sqrt{(1 +\mu_1^2)(1+\mu_2^2)}  +  \mu_1 \mu_2 - 1 \right\},
\nonumber\\[2mm]
u_1^2 &=& \frac{R^2}{2}\left\{ \sqrt{(1 +\mu_1^2)(1+\mu_2^2)}  -  \mu_1 \mu_2  + 1 \right\},
\\[2mm]
u_2^2 &=& R^2 \mu_1\mu_2,
\nonumber
\eea
where $\mu_1, \mu_2  \geq 0$.  


\subsection{Semi-circular-parabolic system of coordinates}

\begin{figure}[htbp] 
\begin{center}
    \begin{minipage}[t]{0.45\linewidth}
      \includegraphics[scale=0.3]{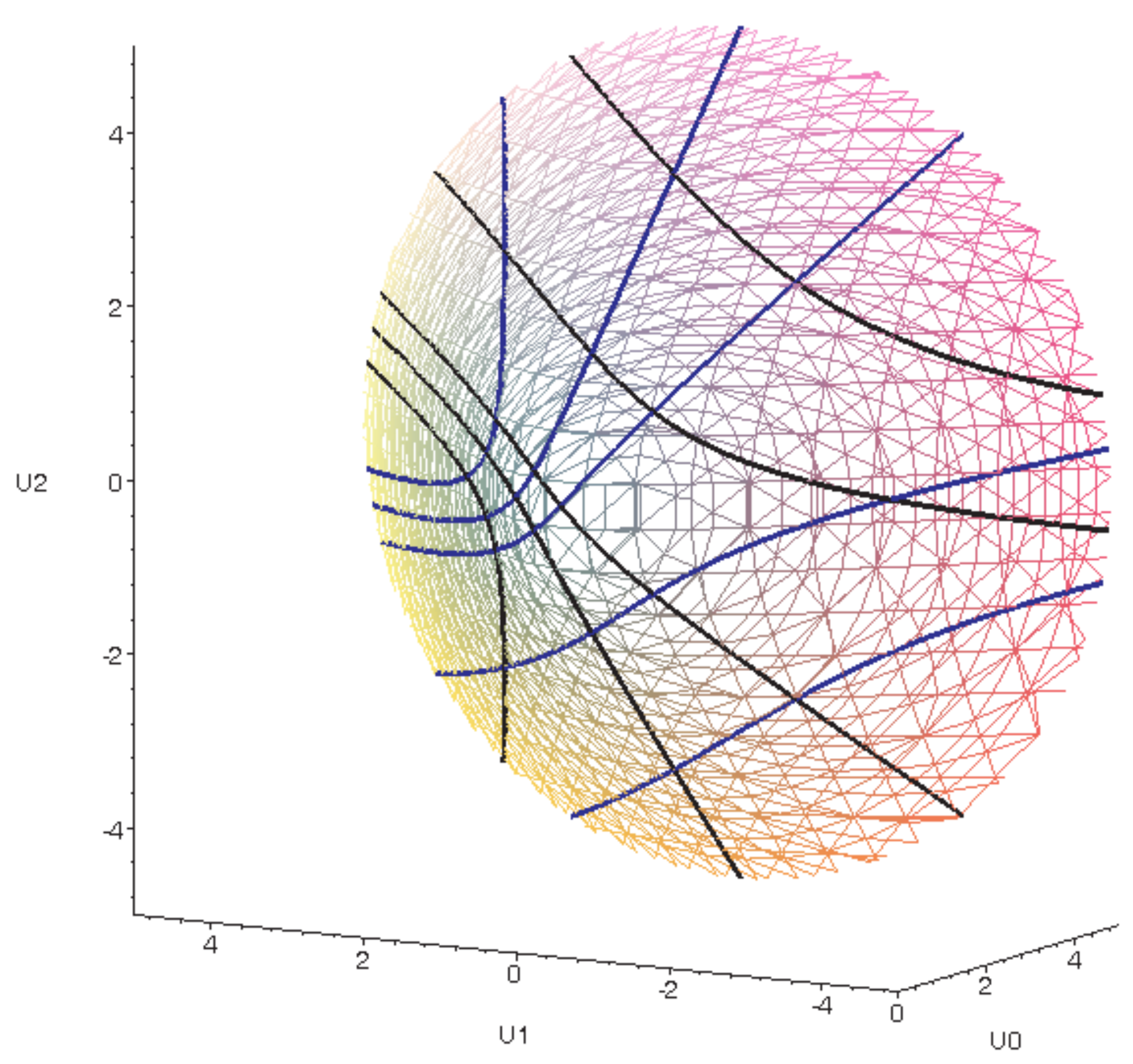}
\begin{center}
\caption{Semi-circular-parabolic system}
\label{fig:16}
\end{center}
    \end{minipage}
    \hfill
    \begin{minipage}[t]{0.45\linewidth}
 \includegraphics[scale=0.3]{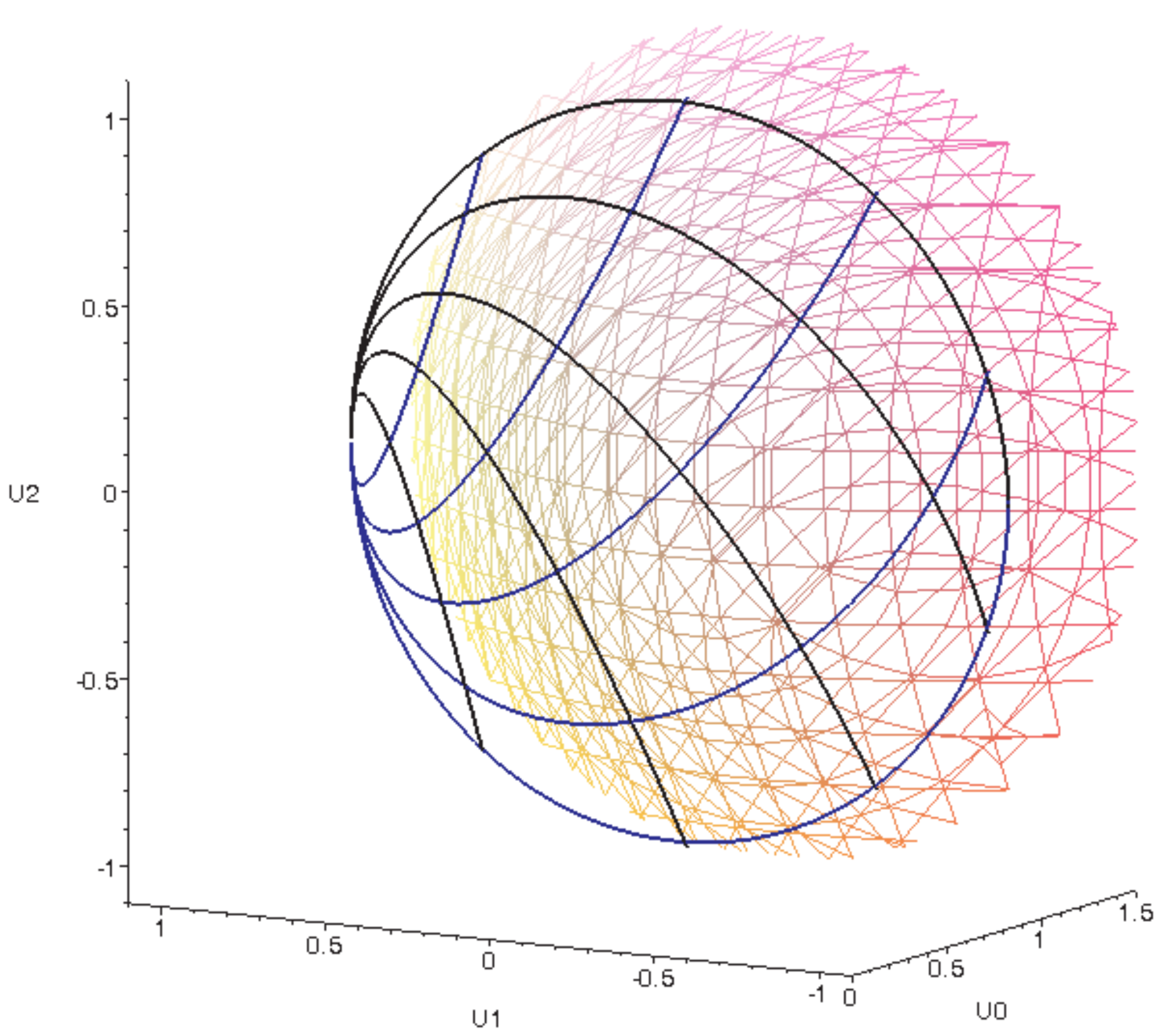}
\caption{Projective plane for semi-circular-parabolic system is formed by osculating parabolas}
\label{fig:17}
    \end{minipage}
  \end{center}
  \end{figure}

{\bf 1.} \, System (\ref{lambda}) for $S_{SCP} = \{K_1, K_2 \} + \{ K_2, L\}$ on two-sheeted hyperboloid
has the form
\begin{equation}
\lambda_1 + \lambda_2 =2 u_2 (u_1  - u_0), \qquad
\lambda_1\lambda_2 = -R^2 (u_1 - u_0)^2.
\label{lambdaSCP}
\end{equation}
Resolving system (\ref{lambdaSCP}), and making the change of variables $\lambda_1 = -R^2/\xi^2$,
$\lambda_2 = R^2/\eta^2$ we obtain the semi-circular-parabolic coordinate system (Fig. \ref{fig:16}):
\begin{equation}
\label{sys_scp_xi_eta}
u_0 = R\frac {\left(\eta^2 + \xi^2\right)^2 + 4}{8 \xi \eta}, \quad
u_1 = R\frac {\left(\eta^2 + \xi^2\right)^2 - 4}{8 \xi \eta},\quad
u_2 =  R{\frac {{\eta}^{2}-{\xi}^{2}}{2\xi\eta}},
\end{equation}
where $\xi$, $\eta >0$.

\begin{figure}[htbp]
\begin{center}
    \begin{minipage}[t]{0.45\linewidth}
             \includegraphics[scale=0.4]{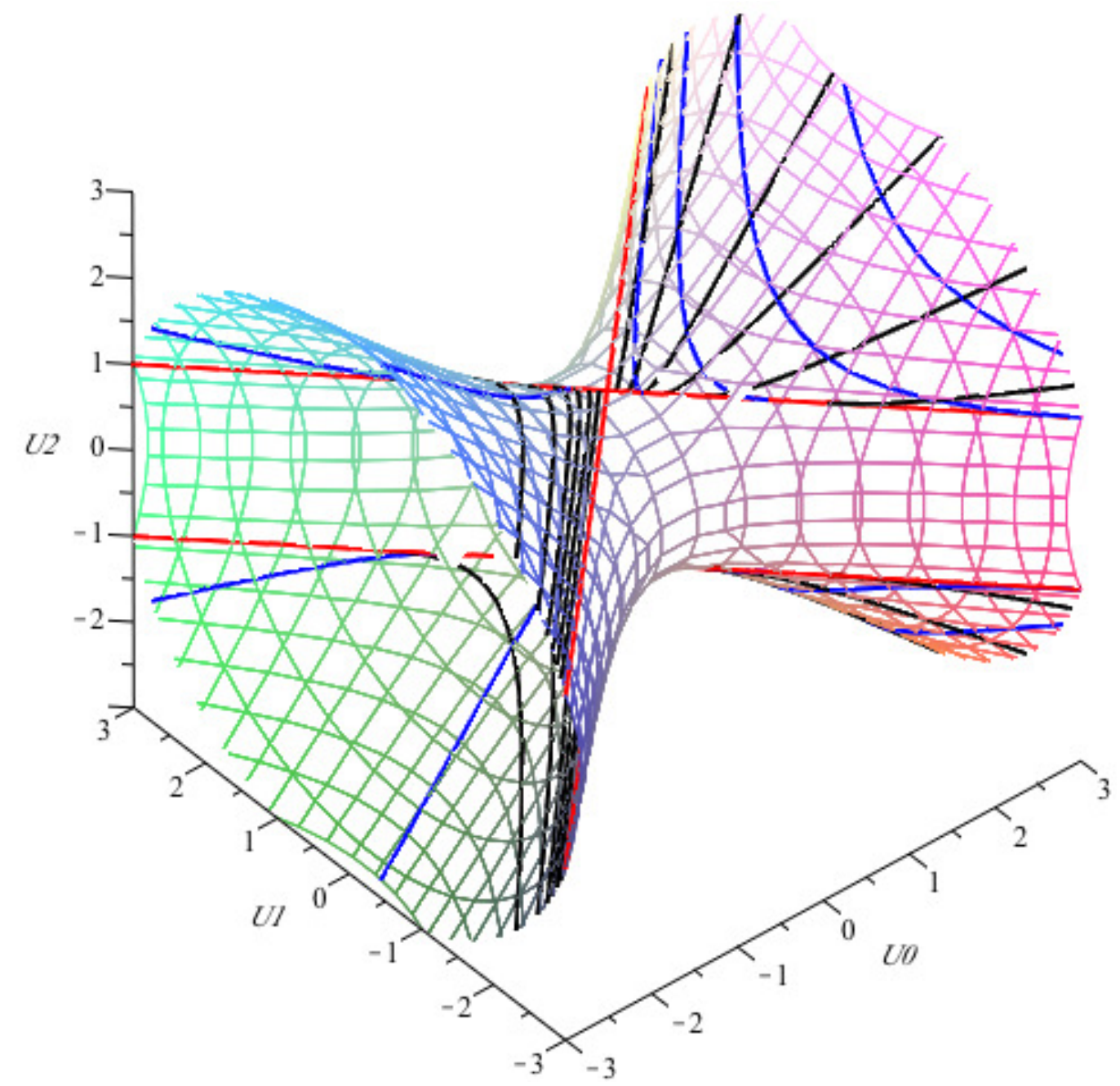}

\begin{center}
\caption{Semi-circular-parabolic system}
\label{fig:26}
\end{center}
    \end{minipage}
    \hfill
    \begin{minipage}[t]{0.45\linewidth}
 \includegraphics[scale=0.35]{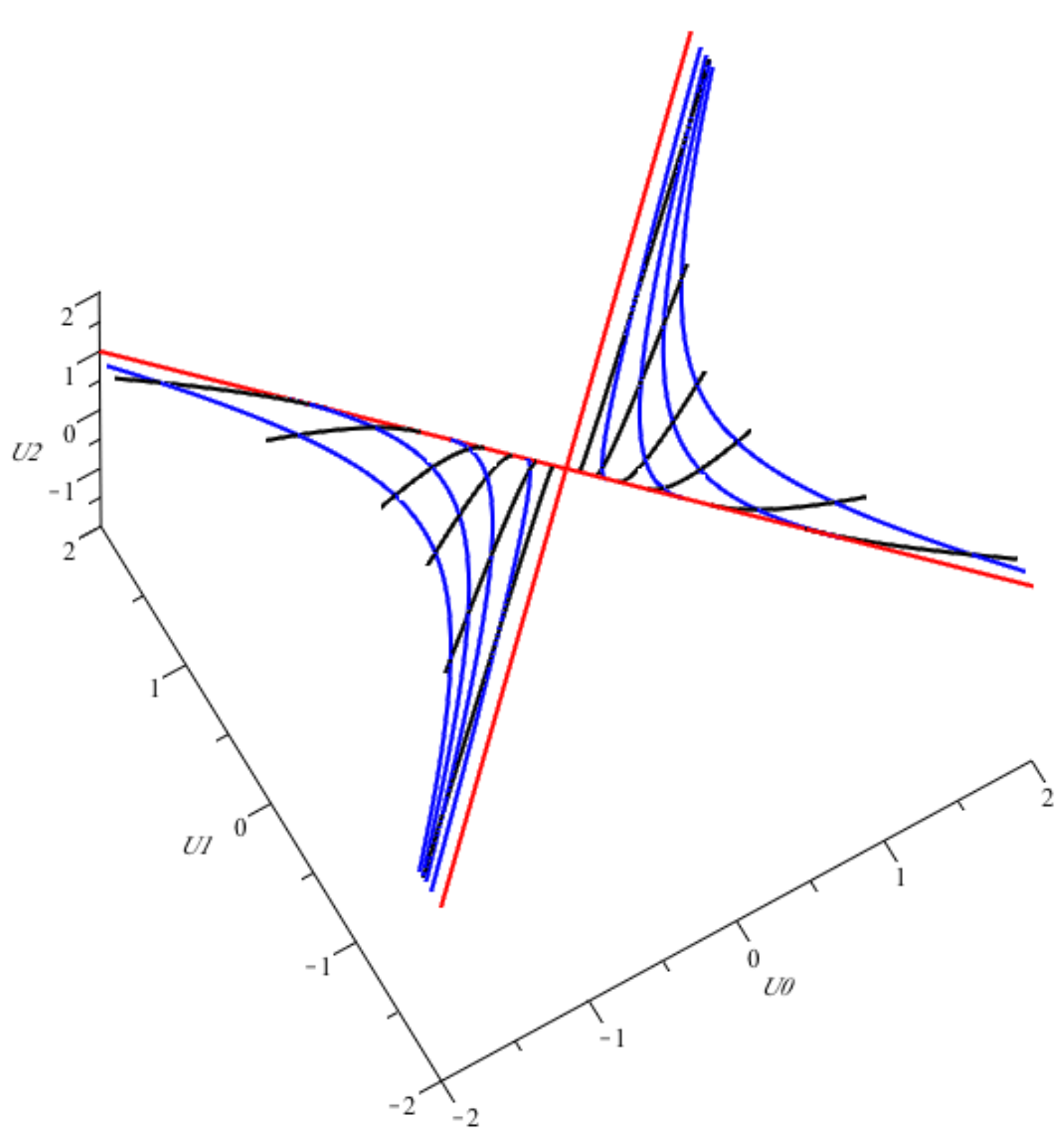}
\caption{Projective plane for semi-circular-parabolic system}
\label{fig:27}
    \end{minipage}
  \end{center}
\end{figure}

\noindent
{\bf 2.} \, In case of the one-sheeted hyperboloid we have 
\begin{equation}
\label{01-lambdaSCP}
\lambda_1 + \lambda_2 =2 u_2 (u_1  - u_0), 
\qquad
\lambda_1\lambda_2 = R^2 (u_1 - u_0)^2.
\end{equation}
In this case coordinate system covers  only  the  part of hyperboloid when 
$|u_2| > R$ (Fig. \ref{fig:26}) and has the following form
\begin{equation}
\label{sys_scp_xi_eta_1}
u_0 = R\frac {\left(\eta^2 - \xi^2\right)^2 + 4}{8 \xi \eta},
\quad
u_1 = R\frac {\left(\eta^2 - \xi^2\right)^2 - 4}{8 \xi \eta},
\quad
u_2 =  \pm R{\frac {{\eta}^{2}+{\xi}^{2}}{2\xi\eta}},
\end{equation}
where we introduce new variables $\lambda_1 = R^2/\xi^2$, $\lambda_2 = R^2/\eta^2$ and   $\xi > 0$ , $\eta \in \mathbb{R}\setminus\{0\}$. 
Sign $\pm$ for $u_2$ corresponds to the case $\lambda_1, \lambda_2<0$. Let us note that lines $u_0\pm u_1=0$, $|u_2|=R$ 
are envelopes for families of coordinate lines $\xi = \text{const.}$ (or $\eta = \text{const.}$).
To distinguish coordinate lines one needs to select the appropriate intervals. Thus, points of envelopes can be taken as initial ones. 
For example, for the fixed line $\xi=\xi_0$ the interval for $\eta$ is $(-\infty,-\xi_0) \cup (\xi_0,+\infty)$.  


\subsection{Elliptic-parabolic system of coordinates}
 
{\bf 1.}   Let us consider the operator $S_{EP} =  \gamma K_2^{2} + \left( K_1+L\right)^{2}$, $\gamma > 0$.
For such operator, system (\ref{lambda}) takes the form
\begin{equation}
\label{00-LAMBDA2}
\lambda_1 + \lambda_2 =  (u_0 - u_1)^2  + \gamma(u_0^2 - u_1^2), 
\qquad
\lambda_1\lambda_2 = \gamma R^2 (u_0 - u_1)^2,
\end{equation}
where we have preliminary changed the signs of $\lambda_i$:  $\lambda_i \to  - \lambda_i$.
Let us take the solution of system (\ref{00-LAMBDA2}) as follows 
\begin{eqnarray}
u_0&=& \frac{1}{2\sqrt{\gamma^3 \lambda_1\lambda_2}} \left(\frac{\gamma - 1}{R} \lambda_1\lambda_2  +  \gamma R (\lambda_1 + \lambda_2)\right),
\nonumber\\[2mm]
u_1 &=& \frac{1}{2\sqrt{\gamma^3 \lambda_1\lambda_2}} \left( - \frac{\gamma + 1}{R} \lambda_1\lambda_2 + \gamma R (\lambda_1 + \lambda_2)\right),
\\[2mm]
u_2^2 &=& \frac{R^2}{\gamma^2} \left(\frac{\lambda_1}{R^2} - \gamma\right) \left(\gamma - \frac{\lambda_2}{R^2}\right),
\nonumber
\end{eqnarray}
where  $0 < \lambda_2/R^2 \leq \gamma \leq \lambda_1/R^2$. If  we  choose  $\lambda_1 =  R^2 \gamma/ \cos^2\theta$ and   $\lambda_2 = R^2\gamma /\cosh^2 a$,  we obtain the elliptic-parabolic system of coordinates  in the trigonometric  form (Fig. \ref{fig:13})
\bea
\label{sys_ep_trig}
u_0 = \frac{R}{\sqrt{\gamma}} \frac{\cosh^2 a - \sin^2\theta + \gamma}{2\cos\theta\cosh a},  \quad
u_1 = \frac{R}{\sqrt{\gamma}} \frac{\cosh^2 a - \sin^2\theta - \gamma}{2\cos\theta\cosh a},   \quad
u_2 = R \tan\theta\tanh a,
\eea
where $a \ge 0$, $\theta\in\left(-\frac{\pi}{2}, \frac{\pi}{2}\right)$. Geometrically positive parameter $\gamma$ defines the position of the focus of elliptic parabolas lying on hyperboloid $H_2$. One can get its 
coordinates in limits $\cos\theta \to 1$, $\cosh a \to 1$. Thus we have 
\bea
\label{FOKUSES}
F\left(u_0, u_1, u_2 \right) \equiv F\left(R\frac{\gamma + 1}{2\sqrt{\gamma}}, \, R\frac{\gamma - 1}{2\sqrt{\gamma}}, \, 0\right). 
\eea

\noindent
{\bf 2.} For one-sheeted hyperboloid we obtain 
\begin{equation}
\label{01-LAMBDA}
\lambda_1 + \lambda_2 =   (u_0 - u_1)^2  +  \gamma(u_0^2 - u_1^2), 
\qquad
\lambda_1\lambda_2 =  -  \gamma R^2 (u_0 - u_1)^2
\end{equation}
and the corresponding system of coordinates takes the form  (see Fig. \ref{fig:28})
\begin{eqnarray}
u_0&=& \pm \frac{1}{2\sqrt{- \gamma^3 \lambda_1\lambda_2}} \left(\frac{1- \gamma}{R} \lambda_1\lambda_2  +  \gamma R (\lambda_1 + \lambda_2)\right),
\nonumber\\[2mm]
u_1 &=& \pm \frac{1}{2\sqrt{- \gamma^3 \lambda_1\lambda_2}} \left(\frac{1+ \gamma}{R} \lambda_1\lambda_2 + \gamma R (\lambda_1 + \lambda_2)\right),
\\[2mm]
u_2^2 &=& \frac{R^2}{\gamma^2} \left(\frac{\lambda_1}{R^2} + \gamma\right) \left(\frac{\lambda_2}{R^2} + \gamma \right),
\nonumber
\end{eqnarray}
where $-\gamma \leq \lambda_2/R^2 < 0 < \lambda_1/R^2$ (or $-\gamma \leq \lambda_1/R^2 < 0 < \lambda_2/R^2$). Putting 
$\lambda_2/(R^2 \gamma) = - 1/\cosh^2\tau_1$ and $\lambda_1/(R^2 \gamma) =  1/\sinh^2\tau_2$ we can rewrite the elliptic-parabolic system 
of coordinates on ${\tilde H}_2$ hyperboloid in form 
\bea
\label{sys_ep_trig1} 
u_0 = \frac{R}{\sqrt{\gamma}} \frac{\cosh^2 \tau_1 - \cosh^2\tau_2 + \gamma}{2\cosh \tau_1\sinh \tau_2},  \quad
u_1 = \frac{R}{\sqrt{\gamma}} \frac{\cosh^2 \tau_1 - \cosh^2\tau_2 - \gamma}{2\cosh \tau_1\sinh \tau_2},   \quad
u_2 = R \tanh \tau_1 \coth \tau_2,
\eea
where now $\tau_1 \in \mathbb{R}$, $\tau_2 \in  \mathbb{R}  \setminus \{0\}$. Let us note that on ${\tilde H}_2$  parameter $\gamma$ can be considered as a scale factor.

The elliptic-parabolic system as the one-parametric one includes two simple limiting cases, namely, the small and the large values of parameter 
$\gamma$, when the coordinates of focus (\ref{FOKUSES}) on $H_2$ move to the infinity. For $\gamma \sim 0$  we get that $S_{EP} \sim \left(K_1+L\right)^{2}$, 
whereas, for large $\gamma$: $S_{EP} /\gamma \sim  K_2^2$ and therefore $S_{EP}$ coordinates degenerate to the horicyclic  and equidistant system correspondingly. Thus, from equation (\ref{00-LAMBDA2}) we have 
\begin{equation}
\label{000-LAMBDA}
\lambda_{1,2} =  \frac{(u_0 - u_1)^2}{2}\left[1 +  \gamma \frac{u_0 + u_1}{u_0 - u_1} \pm 
\sqrt{1 + 2\gamma \frac{u_2^2 - R^2}{(u_0 - u_1)^2} + \gamma^2  \frac{(u_0 + u_1)^2}{(u_0 - u_1)^2}} \right],
\end{equation}
we get for $\gamma \sim 0$: $\cosh a \sim 1 + \gamma {\tilde x}^2/2$, $\cos\theta \sim \sqrt{\gamma} {\tilde y}$ and for large 
$\gamma \sim \infty$: $\cosh a \sim \sqrt{\gamma} e^{\tau_2}$, $\sin\theta \sim \tanh\tau_1$. Putting these results into equation (\ref{sys_ep_trig})
it is easy to obtain the limiting horicylic (\ref{sys_hor}) and equidistant (\ref{sys_equi}) systems of coordinates respectively.  By analogy one can prove that the elliptic-parabolic 
coordinates on one-sheeted hyperboloid degenerate to the same ones.  

It is also important to note that elliptic-parabolic coordinates introduced in (\ref{sys_ep_trig}) and (\ref{sys_ep_trig1}) cover the two- and one-sheeted hyperboloids completely.  


\subsection{Hyperbolic-parabolic system of coordinates}

{\bf 1.}                                                                                                                                                           
Let us consider operator $S_{HP} = - \gamma K_2^{2} + \left( K_1+L\right)^{2}$, $\gamma > 0$. In  this case we obtain (up to the 
change of sign  $\gamma  \to - \gamma$) the same algebraic equations as in (\ref{00-LAMBDA2}).
The solution looks as follows
\bea
u_0 &=& \frac{-1}{2\sqrt{-\gamma^3 \lambda_1\lambda_2}} \left(\frac{1+ \gamma}{R} \lambda_1\lambda_2
+ \gamma R(\lambda_1 + \lambda_2)\right),
\nonumber\\[2mm]
u_1 &=&  \frac{-1}{2\sqrt{-\gamma^3 \lambda_1\lambda_2}} \left( \frac{1 - \gamma}{R} \lambda_1\lambda_2
+ \gamma R(\lambda_1 + \lambda_2)\right),
\\[2mm]
u_2^2 &=& - \frac{R^2}{\gamma^2}\left(\frac {\lambda_1}{R^2} + \gamma\right)
\left(\frac{\lambda_2}{R^2} + \gamma \right),
\nonumber
\eea
where  $\lambda_2/R^2 \leq - \gamma < 0 < \lambda_1/R^2$. If  we  take $\lambda_1/R^2  =   \gamma /\sinh^2 b$,  
$\lambda_2/R^2  =  - \gamma /\sin^2\theta$, one can obtain the hyperbolic-parabolic coordinate system in the 
trigonometric form 
\bea
\label{sys_hp_trig}
u_0 = \frac{R}{\sqrt{\gamma}} \frac{\cosh^2 b - \sin^2\theta + \gamma}{2\sin\theta \sinh b},
\qquad
u_1 = \frac{R}{\sqrt{\gamma}} \frac{\cosh^2 b - \sin^2\theta - \gamma}{2\sin\theta \sinh b},
\qquad
u_2 = R \cot\theta \coth b,
\eea
where $b>0$, $\theta \in \left(0,\pi \right)$. 

\begin{figure}[htbp] 
\begin{center}
    \begin{minipage}[t]{0.45\linewidth}
      \includegraphics[scale=0.3]{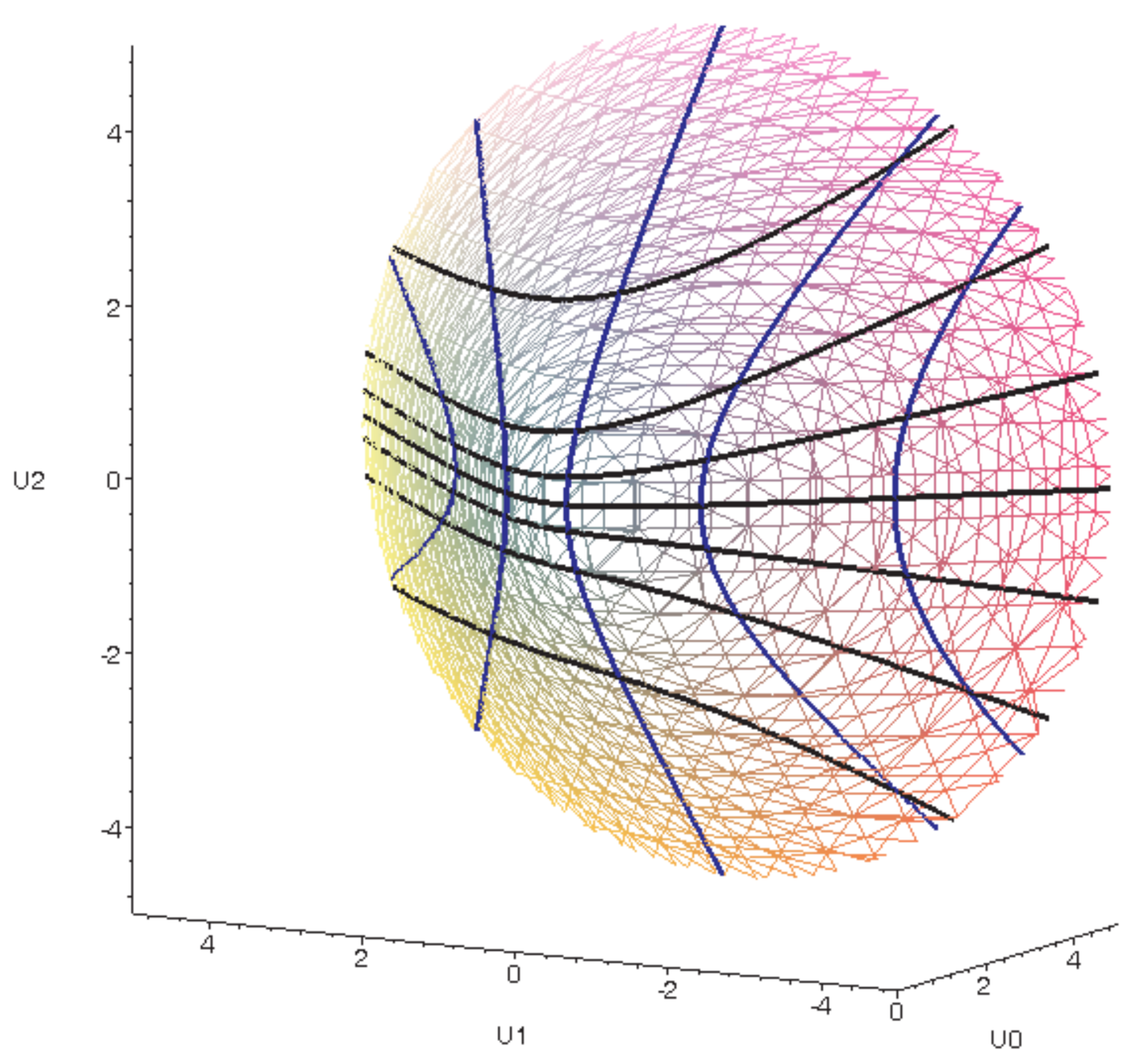}
\begin{center}
\caption{Hyperbolic-parabolic system}
\label{fig:15}
\end{center}
    \end{minipage}
    \hfill
    \begin{minipage}[t]{0.45\linewidth}
 \includegraphics[scale=0.3]{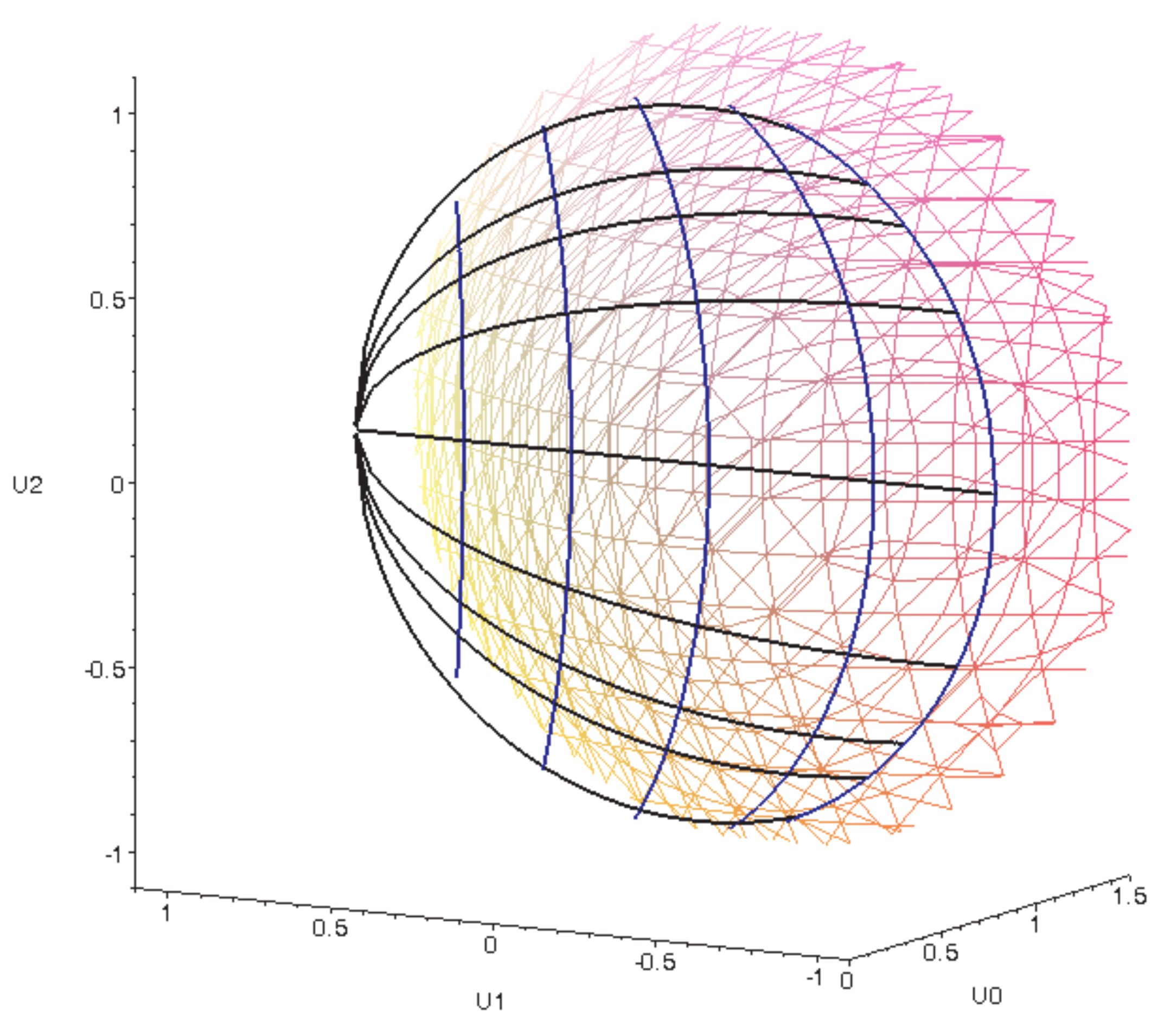}
\caption{Projective plane for hyperbolic-parabolic system is formed by concave hyperbolic
parabolas of two branches and concave hyperbolic parabolas of one branch}
\label{fig:16a}
    \end{minipage}
  \end{center}
\end{figure}

Geometrically, the $S_{HP}$ system is represented by the co-focal hyperbolic parabolas (see Fig. \ref{fig:15}). But in contrast to the case of 
elliptic-parabolic system the coordinates of the focus of hyperbolic parabolas are imaginary ones. Formally it can be seen if we make the change $\gamma \to - \gamma$ 
in formula (\ref{FOKUSES}). Thus we can consider parameter $\gamma$ as a scale factor. 

\noindent
{\bf 2.}
For  the one-sheeted hyperboloid  we obtain  that this coordinate system does not cover all  hyperboloid.
The covered part is defined by inequality (\ref{01-lambda})  
\be
\label{COVERED-01}
|u_0(1-\gamma) - u_1 (1+\gamma)| >  2R\sqrt{\gamma}.
\ee
The solution of equations 
\begin{equation}
\label{00-LAMBDA}
\lambda_1 + \lambda_2 =   (u_0 - u_1)^2   -  \gamma(u_0^2 - u_1^2), 
\qquad
\lambda_1\lambda_2 =  \gamma  R^2  (u_0 - u_1)^2
\end{equation}
gives us the hyperbolic-parabolic system of coordinates on $\tilde{H}_2$:  
\bea
\label{001-HYPER-PAR}
u_0 &=& 
\pm \frac{1}{2\sqrt{\gamma}} \left(\frac{ 1+\gamma}{\gamma R} \sqrt{\lambda_1 \lambda_2}
-   R  \frac{\lambda_1 + \lambda_2}{\sqrt{\lambda_1 \lambda_2}} \right),
\nonumber\\[2mm]
u_1 &=& 
\pm \frac{1}{2\sqrt{\gamma}} \left(\frac{ 1-\gamma}{\gamma R} \sqrt{\lambda_1 \lambda_2}
-   R  \frac{\lambda_1 + \lambda_2}{\sqrt{\lambda_1 \lambda_2}} \right),
\\[2mm]
u_2^2 &=&  \frac{R^2}{\gamma^2}\left(\frac {\lambda_1}{R^2} - \gamma\right)
\left(\frac{\lambda_2}{R^2} - \gamma \right).
\nonumber
\eea
System (\ref{001-HYPER-PAR}) splits the covered part of one-sheeted hyperboloid (\ref{COVERED-01}) 
into three different regions: (A) \, $0 < \gamma R^2 \leq \lambda_i$, (B) \, $0 <  \lambda_i \leq \gamma  R^2$ and 
(C) \, $\lambda_i < 0$ ($i=1,2$). Depending on the intervals of values for $\lambda_i$, the hyperbolic-parabolic 
system of coordinates can be parametrized in three different forms. 

(A). \, $0 < \gamma R^2 \leq \lambda_i$. Moving to new variables according to formulas $\lambda_1 =   \gamma R^2 / \sin^2\theta$,  
$\lambda_2 =  \gamma  R^2/ \sin^2\phi$ 
\bea
\label{sys_hp_trigTI}
u_0 =  \frac{R}{2 \sqrt{\gamma}} \frac{\cos^2 \theta - \sin^2\phi + \gamma}{\sin\theta  \sin\phi},
\quad
u_1 =  \frac{R}{2\sqrt{\gamma}} \frac{\cos^2 \theta - \sin^2\phi - \gamma}{\sin\theta  \sin\phi},
\quad
u_2 = R \cot\theta \cot \phi,
\eea
where  $\theta \in \left[- \frac{\pi}{2}, 0 \right) \cup \left(0, \frac{\pi}{2}\right]$, $\phi \in (0, \pi)$. We will call this system $S_{HP}$ system of Type I.

(B). \, $0 <  \lambda_i \leq \gamma  R^2$. If we take $\lambda_1  =  \gamma  R^2 \sin^2\theta$, $\lambda_2 = \gamma R^2 \sin^2\phi$, then we obtain $S_{HP}$ system of Type II: 
\bea
\label{sys_hp_trigTII}
u_0 &=&  \frac{R}{2\sqrt{\gamma}} \frac{\cos^2 \theta \cos^2\phi - 1 + \gamma \sin^2 \theta \sin^2\phi}{\sin\theta  \sin\phi},
\nonumber\\[3mm]
u_1 &=&  \frac{R}{2\sqrt{\gamma}} \frac{\cos^2 \theta \cos^2\phi - 1 - \gamma \sin^2 \theta \sin^2\phi}{\sin\theta  \sin\phi},
\\[3mm]
u_2 &=& R \cos\theta \cos \phi,
\nonumber
\eea
where $\theta \in \left[- \frac{\pi}{2}, 0 \right) \cup \left(0, \frac{\pi}{2}\right]$, $\phi \in (0, \pi)$. 

(C).  \, $\lambda_i < 0$.  The last  $S_{HP}$ system of Type III is 
\bea
\label{sys_hp_trigTIII}
u_0 &=&  \frac{R}{2\sqrt{\gamma}} \frac{\cosh^2 \theta \cosh^2\phi - 1 + \gamma \sinh^2 \theta \sinh^2\phi}{\sinh\theta \sinh\phi},
\nonumber\\[3mm]
u_1 &=&  \frac{R}{2\sqrt{\gamma}} \frac{\cosh^2 \theta \cosh^2\phi - 1 - \gamma \sinh^2 \theta \sinh^2\phi}{\sinh\theta \sinh\phi},
\\[3mm]
u_2 &=& \pm R \cosh \theta \cosh \phi,
\nonumber
\eea
where $\lambda_1 =  -  \gamma R^2 \sinh^2\theta$, $\lambda_2  =  - \gamma R^2 \sinh^2\phi$ and  $\theta  \in  \mathbb{R} \setminus \{0\}$, 
$\phi > 0$. 

Let us note that if $ (1+\gamma) u_1 < u_0 (1-\gamma) - 2R\sqrt{\gamma}$, then system $S_{HP}$ of Type I is defined in region $u_0 > -R(1-\gamma)/(2\sqrt{\gamma})$ and for  $S_{HP}$  of Type II, III we have $u_0 < -R(1-\gamma)/(2\sqrt{\gamma})$. 
If $(1+\gamma) u_1 > u_0 (1-\gamma) + 2R\sqrt{\gamma}$, then for  $S_{HP}$ Type I: $u_0 < R(1-\gamma)/(2\sqrt{\gamma})$ and for Type II, III 
we have $u_0 > R(1-\gamma)/(2\sqrt{\gamma})$. To avoid the intersection of coordinate lines of one family ($\theta = \text{const.}$ or $\phi = \text{const.}$) it is necessary to adjust intervals for the angles. For example, one can take points of the limiting lines
\be
\label{HP_evelopes}
|u_0(1-\gamma) - u_1 (1+\gamma)| =  2R\sqrt{\gamma}
\ee
as initial ones for $\phi$ and as endpoints for $\theta$ (see Fig. \ref{fig:30}).

As for the degeneration of hyperbolic-parabolic coordinates when $\gamma \sim 0$ or $\gamma \sim \infty$, the process is equivalent to elliptic-parabolic system. In limiting cases we obtain the horicyclic and equidistant systems.


\subsection{Elliptic system of coordinates}

\begin{figure}[htbp]
\begin{center}
    \begin{minipage}[t]{0.45\linewidth}
      \includegraphics[scale=0.3]{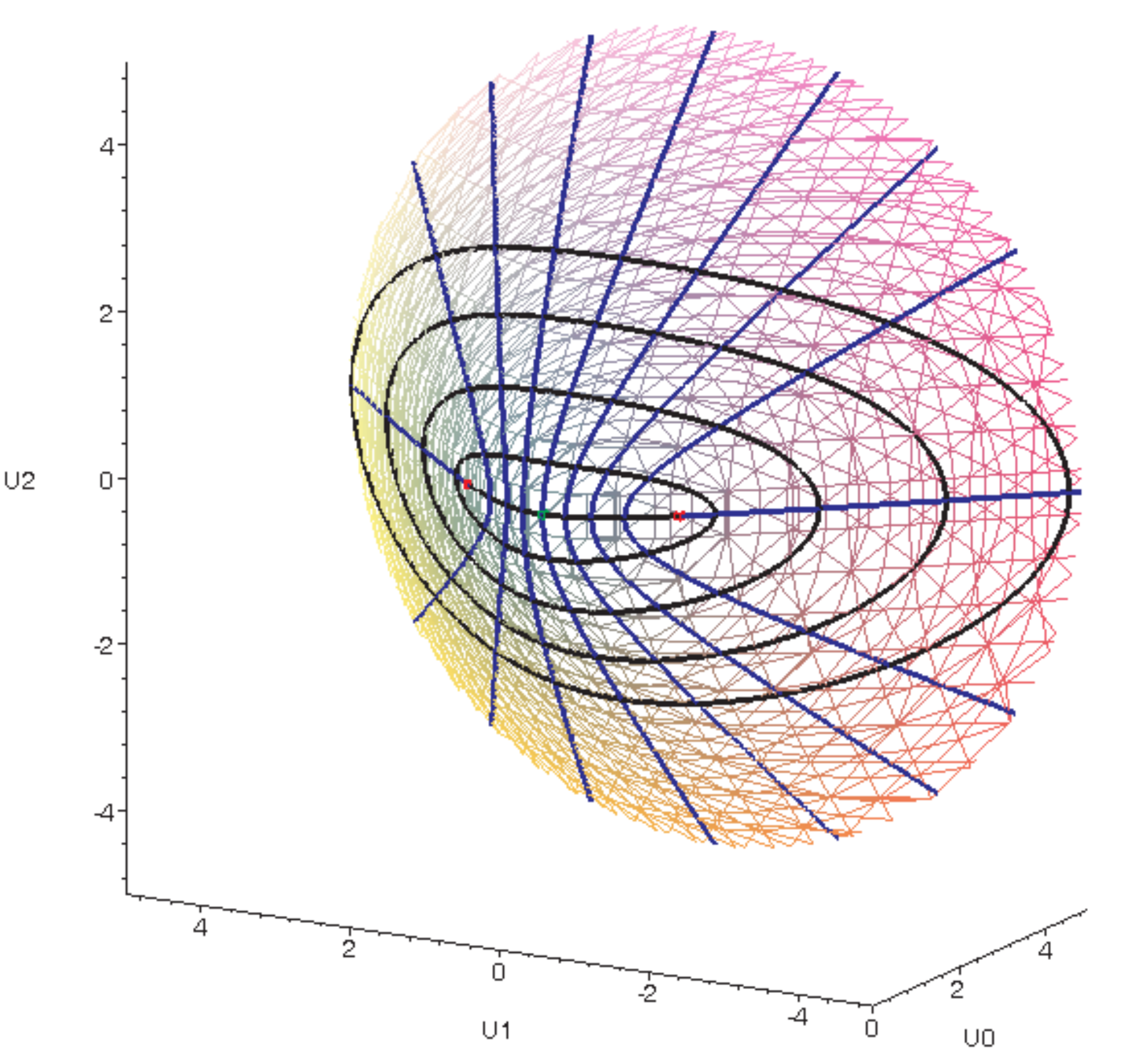}
\caption{Elliptic system}
\label{fig:8}
    \end{minipage}
    \hfill
    \begin{minipage}[t]{0.45\linewidth}
 \includegraphics[scale=0.3]{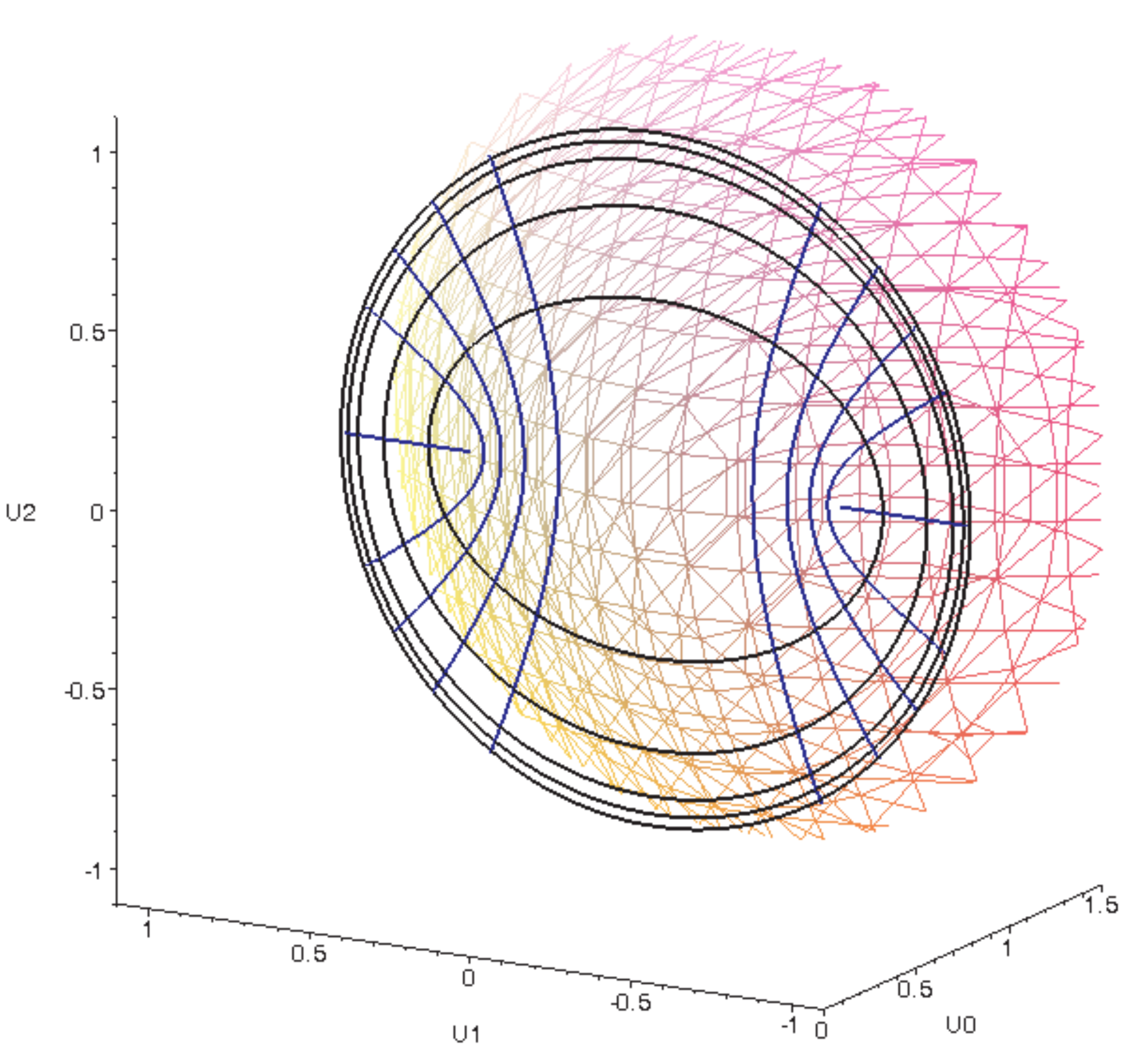}
\caption{Projective plane for elliptic system is formed by ellipses and convex hyperbolas}
\label{fig:9}
    \end{minipage}
  \end{center}
 \end{figure}

{\bf 1.} \, 
For operator  $S_{E} = L^2 + \sinh^2 \beta K_2^2$  algebraic  system  (\ref{lambda}) has the form 
\begin{equation}
\label{lambdaE}
\lambda_1 + \lambda_2 = - u_1^2 - u_2^2 \cosh^2 \beta - R^2 \sinh^2 \beta, 
\quad
\lambda_1\lambda_2 = R^2 u_1^2 \sinh^2 \beta.
\end{equation}
Its solution corresponds to the elliptic system of coordinates 
\bea
\label{01-sys_ell}
u_0^2 =  \frac{\left ({\lambda_1}-{R^2}\right )\left ({\lambda_2}-{R^2} \right )}{R^2 \cosh^2\beta}, 
\quad
u_1^2 =  \frac{\lambda_1 \lambda_2}{R^2  \sinh^2\beta},
\quad
u_2^2 =  -   \frac{ \left ({\lambda_1} + {R^2} \sinh^2\beta \right )\left ({\lambda_2}+ {R^2}\sinh^2\beta \right )}
{R^2 \sinh^2\beta \cosh^2\beta},
\eea
where  $\lambda_1 \leq - R^2 \sinh^2\beta \leq \lambda_2 \leq 0$.  To write  these coordinates  in  conventional form let us use  the  following  substitutions:  $\lambda_1/R^2  =  \frac{a_2-\rho_1}{a_2 - a_3}$, $\lambda_2 / R^2  =  \frac{a_2 - \rho_2}{a_2 - a_3}$
and $\sinh^2 \beta = (a_1 - a_2)/(a_2 - a_3)$, where  $a_i$, ($i=1,2,3$) are some constants so that $a_3 < a_2 < a_1$. 
In the new variables the elliptic system of coordinates is given by the following relations ($a_3 < a_2 \leq \rho_2 < a_1 \leq \rho_1$)  
\bea
\label{sys_ell}
u_0^2 = R^2{\frac {\left ({ \rho_1}-{ a_3}\right )\left ({ \rho_2}-{ a_3}
\right )}{\left ({ a_1}-{ a_3}\right )\left ({ a_2}-{ a_3}\right )}},
\quad
u_1^2 = R^2{\frac {\left ({ \rho_1}-{a_2}\right )\left ({ \rho_2}-{ a_2}
\right )}{\left ({ a_1}-{ a_2}\right )\left ({ a_2}-{ a_3}\right )}},
\quad
u_2^2 = R^2{\frac {\left ({ \rho_1}-{a_1}\right )\left (a_1 -  \rho_2 \right )}
{\left ({ a_1}-{ a_2}\right )\left ({ a_1}-{ a_3}\right )}},
\eea
and it coincides with the known from literature  definition of elliptic system of coordinates on hyperboloid $H_2$ \cite{GROP4,PSW1}. Parameters $a_i$ $(i=1,2,3)$ define the positions of  two  foci  for co-focal ellipses and hyperbolas on $H_2$ (see Fig. \ref{fig:8}). 
One can obtain its coordinates making $\rho_1 \to a_1$ and $\rho_2 \to a_1$, then 
$F_{1,2}\left(u_0,  \pm u_1, u_2 \right) \equiv F_{1,2}\left(R\frac{1}{k}, \pm R\frac{k^\prime}{k},0\right)$, 
where 
\begin{equation}
\label{MODULUS}
{k^\prime}^2 = \frac{a_1 - a_2}{a_1 - a_3}, 
\qquad
{k}^2 = \frac{a_2 - a_3}{a_1 - a_3},
\end{equation}
where $\sinh^2 \beta = \frac{{k^\prime}^2}{{k}^2}$ and $2\beta R$ is the distance between the foci.  Elliptic  coordinates (\ref{sys_ell})  are given in  algebraic form  and they actually
depend on the three parameters $a_i$.   Equivalently to this  form we can define the elliptic system in terms of  Jacobi  elliptic  functions 
which depend on just one  parameter.  It  frees  us  from  ambiguity in  determining  the position  of   a point  on hyperboloid $H_2$ in terms of elliptic coordinates.     
 
If we put  \cite{GROP4, PSW1}:
\begin{equation}
\label{Jef}
\rho_1 = a_1 -  (a_1 - a_3) \dn^2 (a, k),
\qquad
\rho_2 = a_1 - (a_1 - a_2) \sn^2(b, k^\prime),
\end{equation}
into expression  (\ref{sys_ell}),  where functions $\sn (a,k)$, $\cn (a,k)$ and $\dn (a,k)$ are Jacobi elliptic functions with modulus $k$ 
and $k^\prime$ \cite{BE3}  related by well known identities $\sn^2 (a,k) + \cn^2 (a,k) = 1$ and $k^2 \sn^2 (a,k) + \dn^2 (a,k) =1$, 
we obtain Jacobi form of the elliptic coordinates  
\bea
\label{sys_e_J}
u_0 = R \sn(a, k)\dn(b, k^\prime),\
u_1 = i R \cn(a, k) \cn(b, k^\prime), \
u_2 = i R \dn(a, k) \sn(b, k^\prime),
\eea
where $a \in \left( i K^\prime, i K^\prime + 2K\right)$, $b \in \left[0,4K^\prime \right)$, $k^2 + {k^\prime}^2 = 1$,
$K = K(k)$, $K^\prime = K(k^\prime)$  are complete  elliptic  integrals with $k$ and ${k^\prime}$, respectively. Jacobi elliptic functions degenerate into trigonometric or hyperbolic functions if one of the periods is infinite, 
i.e.  modulus $k^2$ is zero or one \cite{BE3}.  In the limiting case  $k^2 \to 1$, $k'^2 \to 0$ we have that $K \sim \infty$, 
$K' \sim \frac{\pi}{2}$ and  
\bea
\label{LIMIT-ELLIP-00A}
\sn(\mu, k) \to \tanh\mu, \qquad 
\dn(\mu, k) \to \frac{1}{\cosh\mu},
\eea
whereas for limit $k^2 \to 0$, $k'^2 \to 1$ we have $K \sim \frac{\pi}{2}$, $K' \sim \infty$ and  
\bea
\label{LIMIT-ELLIP-00B}
\sn(\mu, k) \to \sin\mu,  \qquad
\dn (\mu, k) \to  1.
\eea

\noindent
{\bf 2.} \, 
For the one-sheeted hyperboloid inequality (\ref{01-lambda}) holds over the whole space. 
Thus we can construct in the same way as in (\ref{sys_ell})  the elliptic system of coordinates (Fig. \ref{fig:32})
\bea
\label{sys_ell_1}
u_0^2 = R^2{\frac {\left ({ \rho_1}-{ a_3}\right )\left ( { a_3} - {\rho_2} \right )}{\left ({ a_1}-{ a_3}\right )\left ({ a_2}-{ a_3}\right )}}, 
\quad 
u_1^2 = R^2{\frac {\left ({ \rho_1}-{a_2}\right )\left ({ a_2} - { \rho_2} \right )}{\left ({ a_1}-{ a_2}\right )\left ({ a_2}-{ a_3}\right )}},
\quad
u_2^2 = R^2{\frac {\left ({a_1} - { \rho_1}\right )\left (a_1 -  \rho_2 \right )}{\left ({ a_1}-{ a_2}\right )\left ({ a_1}-{ a_3}\right )}},
\eea
where now  $\rho_2<a_3<a_2<\rho_1<a_1$.  Introducing Jacobi functions as in  (\ref{Jef}) we can rewrite coordinates (\ref{sys_ell_1}) in form 
\bea
\label{ONE-SHEETED-sys_e}
u_0 =  i R \sn(a, k) \dn(b, k^\prime),
\quad
u_1 =  -R \cn(a, k) \cn(b, k^\prime), 
\quad
u_2 =  - R \dn(a, k) \sn(b, k^\prime),
\eea
where now $a \in [K, K + i 4K^\prime)$ and $b \in (iK,  iK + 2K^\prime)$.

Let us finally note that the elliptic systems of coordinates are an one-parametric coordinate systems depending on $k$ and in the limit $k\to 0$ 
(respectively $k\to 1$) the subgroup coordinate systems arise. 
Indeed, in the case when $\beta \to 0$ (or $k \to 1$): $S_E \sim L^2$ and we get that the elliptic coordinates transform into the spherical ones, 
whereas for the large $\beta \sim \infty$ ($k \to 0$):  $S_E/\sinh^2\beta \sim K^2_2$ we obtain the equidistant coordinates. 
To track these limits directly on the level of coordinates, let us use some properties of Jacobi functions and complete elliptic integrals
in (\ref{LIMIT-ELLIP-00A}) and (\ref{LIMIT-ELLIP-00B}).  
Firstly let us consider the two-sheeted hyperboloid.   In the limiting case $k\to 1$, i.e., $k'\to 0$ we obtain
\bea
\begin{array}{ll}
\label{LIMIT-ELLIP-01}
\sn(a, k) \to \tanh(i\pi/2 + \mu) = \coth\mu \equiv \cosh\tau,
\quad
&\dn(a, k) \to \frac{1}{\cosh(i\pi/2 + \mu)} = - \frac{i}{\sinh\mu} \equiv \sinh\tau, 
\\[3mm]
\sn(b, k') \to \sin b \equiv \sin\varphi,  
\quad
&\dn (b, k') \to 1, 
\end{array}
\eea
where $\tau \in (0, \infty)$ and $\varphi \in [0, 2\pi)$. Therefore the elliptic coordinate system (\ref{sys_e_J}) on $H_2$ yields spherical coordinates (\ref{PSEUDO-SPHER-01}). 

To determine the second limit  $k^\prime \to 1$, $k\to 0$,  let us introduce new variables $(\nu, \mu)$ through the substitution 
$a = \nu + iK'$ and $b = \mu + K'$, where $\nu \in (0, 2K)$ and $\mu \in \left[-K^\prime, 3K^\prime\right)$. Using the following formulas \cite{BE3} 
\bea
\begin{array}{lll}
\label{LIMIT-ELLIP-00C}
\sn(\nu + iK', k) = \frac{1}{k \sn(\nu, k)}, 
\quad
&\cn(\nu + iK', k) = - \frac{i}{k} \frac{\dn(\nu, k)}{\sn(\nu, k)}, 
\quad
&\dn(\nu + iK', k) = - i  \frac{\cn(\nu, k)}{\sn(\nu, k)},
\\[3mm]
\sn(\mu + K', k') = \frac{\cn( \mu, k')}{\dn( \mu, k')},  
\quad
&\cn(\mu + K', k') =  - k \frac{\sn( \mu, k')}{\dn( \mu, k')},  
\quad
&\dn(\mu + K', k') = \frac{k}{\dn( \mu, k)},  
\end{array}
\eea
taking into account equations (\ref{LIMIT-ELLIP-00A}) and (\ref{LIMIT-ELLIP-00B}), and denoting 
$\tau_2=\mu$ and $\cosh\tau_1 = 1/\sin\nu$ it is easy to reproduce from the elliptic coordinates (\ref{sys_e_J}) 
the equidistant system of coordinates (\ref{sys_equi}).   

There is also an alternative way to find  the same results.  Let us consider equivalent  intervals for $a\in\left[K-2iK^\prime, K+2iK^\prime\right)$, $b\in\left(-iK, iK\right)$ 
(corresponding to order $a_3 < a_2 \leq \rho_1 < a_1 \leq \rho_2$) and making the change of variables $a = \pi/2 + i\tau_2$, $b = -i\arctan(\sinh\tau_1)$, from (\ref{sys_e_J}) ,
then  for  $k^\prime \to 1$, $k\to 0$ we obtain equidistant system (\ref{sys_equi}).

For one-sheeted hyperboloid, taking intervals $a\in \left(-iK^\prime, iK^\prime \right)$, $b\in\left[0,4K^\prime \right)$ (corresponding to $\rho_1<a_3<a_2<\rho_2<a_1$) and introducing new variables $a = \tanh^{-1} (-i\sinh \tau)$, $b = \varphi$ one can see that system (\ref{ONE-SHEETED-sys_e}) goes to pseudo  spherical one (\ref{pspherical}), when $k\to 1$, $k'\to 0$.

In the case when $k\to 0$, $k'\to 1$, introducing $a = \pi/2 + i\tau_2$, $b = \coth^{-1} (\cosh \tau_1) + i\pi/2$ and considering intervals $a \in [K, K + i 4K^\prime)$ and $b \in (iK,  iK + 2K^\prime)$ from (\ref{ONE-SHEETED-sys_e}) one can obtain equidistant system of Type I (\ref{sys_equi_1}), with $u_2 \leqslant -R$, $\tau_1\in \mathbb{R}$, $\tau_2\in \mathbb{R}^{+}$. To obtain the second part of equidistant system, when $u_2 \geqslant R$ one can consider the alternative interval $b\in (iK-2K^\prime, iK)$ taking $b = \coth^{-1}(- \cosh \tau_1) + i\pi/2$.

Finally, taking $a\in \left(-iK^\prime, iK^\prime \right)$, $b\in\left[0,4K^\prime \right)$ and $a = i\tau$, $b = \arccosh(-1/\sin\varphi)$ we obtain from (\ref{ONE-SHEETED-sys_e}) the equidistant coordinates of Type Ib (\ref{sys_equi_11}) with $\varphi \in [\pi, 2\pi)$ that corresponds to $u_1 \leqslant 0$.


\subsection{Rotated elliptic system of coordinates}

The  rotated  elliptic  system of coordinates  corresponding to operator  $S_{\tilde{E}} = \cosh 2\beta L^2 + 1/2 \sinh 2\beta \{K_1,L\}$  (see (\ref{E1-SYSTEM}) and (\ref{relation_coords_operators}))
can be obtained  from the elliptic system (\ref{sys_e_J})  through hyperbolic rotation with angle $\beta$  about axis $u_2$:
\begin{equation}
\label{sys_ell_r}
\left( {{\begin{array}{*{20}c}
 u^\prime_0 \\
 u^\prime_1 \\
 u^\prime_2
\end{array} }} \right) =
\left( {{\begin{array}{*{20}c}
 \cosh \beta & \sinh \beta & 0 \\
 \sinh \beta & \cosh \beta & 0 \\
 0 & 0 & 1
\end{array} }} \right)
\left( {{\begin{array}{*{20}c}
 u_0 \\
 u_1 \\
 u_2
\end{array} }} \right)
 =
\left( {{\begin{array}{*{20}c}
 u_0\cosh \beta + u_1\sinh \beta \\
 u_0\sinh \beta + u_1\cosh \beta \\
 u_2
\end{array} }} \right)
\end{equation}

Substituting now equation (\ref{sys_e_J}) in (\ref{sys_ell_r}) we obtain 
\bea
\label{sys_e_rot}
u^\prime_0 &=& \frac{R}{k} \left\{ \sn(a,k)\dn(b, k^\prime) + i k^\prime \cn(a,k)\cn(b, k^\prime)    \right\},
\nonumber\\[2mm]
u^\prime_1 &=& \frac{R}{k} \left\{ k^\prime \sn(a,k)\dn(b, k^\prime) +  i \cn(a,k)\cn(b, k^\prime)    \right\},
\\[2mm]
u^\prime_2 &=& i R \dn(a,k)\sn(b, k^\prime),
\nonumber
\eea
where intervals for $a$, $b$ are the same as in (\ref{sys_e_J}).  
This system of coordinates was introduced for the first time  in article \cite{GROP4} to provide the separation of variables 
in Schrodinger equation with Coulomb potential on two-dimensional two-sheeted hyperboloid.  

We do not introduce here the analog of rotated elliptic system of coordinates on the one-sheeted hyperboloid because 
this system does not lead to the new contraction on the pseudo-euclidean 
space $E_{1,1}$.


\subsection{Hyperbolic system of coordinates}

\begin{figure}[htbp] 
 \begin{center}
    \begin{minipage}[t]{0.45\linewidth}
      \includegraphics[scale=0.3]{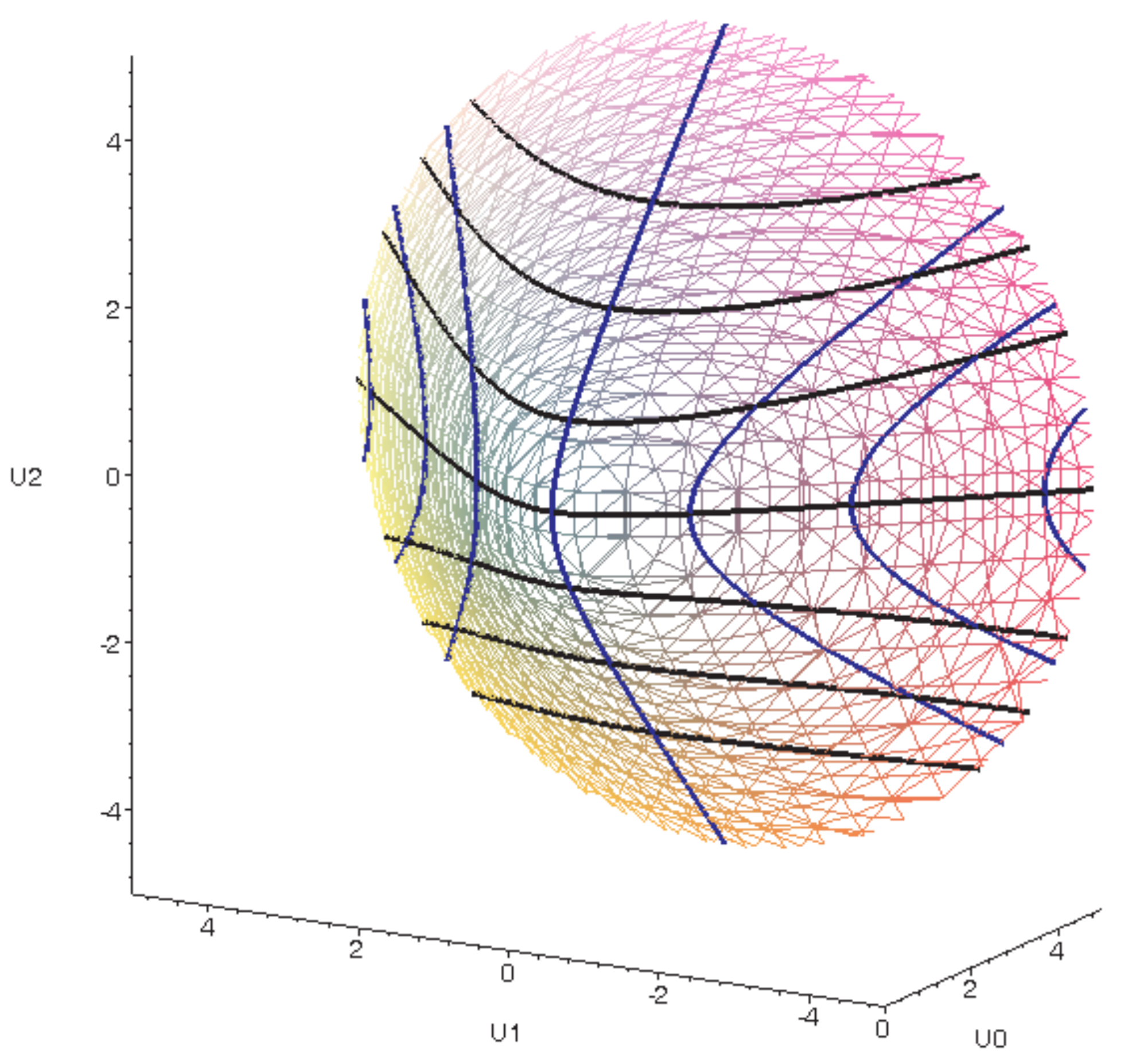}

\begin{center}
\caption{Hyperbolic system}
\label{fig:10}
\end{center}
    \end{minipage}
    \hfill
    \begin{minipage}[t]{0.45\linewidth}
 \includegraphics[scale=0.3]{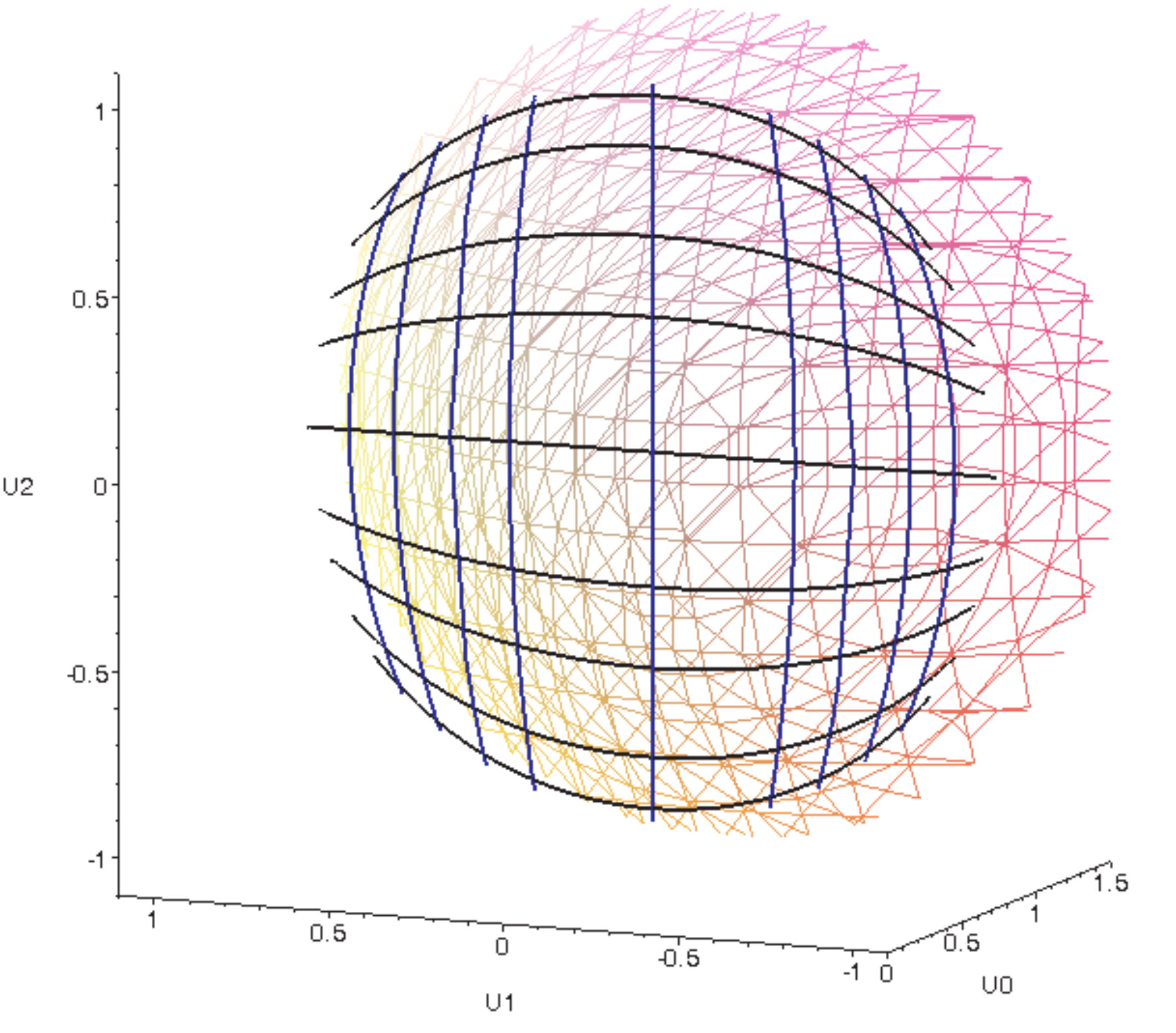}
\caption{Projective plane for hyperbolic system is formed by concave hyperbolas}
\label{fig:11}
    \end{minipage}
  \end{center}
\end{figure}

{\bf 1.}
For operator $S_{H} = K_2^2 - \sin^2 \alpha L^2$, $\sin^2\alpha \neq 0, \neq 1$ system (\ref{lambda}) looks as  follows
\bea
\label{lambdaH}
\lambda_1 + \lambda_2 =  u_1^2\sin^2\alpha - u_2^2 \cos^2 \alpha - R^2,
\quad
\lambda_1\lambda_2 = -R^2 u_1^2 \sin^2\alpha,
\eea
and it has the solution 
\bea
\label{01-sys_hyper}
u_0^2 = -  \frac{\left ({\lambda_1} + {R^2} \sin^2\alpha\right )\left ({\lambda_2}+ {R^2}\sin^2\alpha \right )}
{R^2 \sin^2\alpha \cos^2\alpha},
\quad
u_1^2 =  - \frac{\lambda_1 \lambda_2}{R^2  \sin^2\alpha},
\quad
u_2^2 =  - \frac{\left ({\lambda_1}+{R^2}\right )\left ({\lambda_2}+{R^2} \right )}{R^2 \cos^2\alpha}, 
\eea
where  $\lambda_1 \leq - R^2 < 0 \leq  \lambda_2$. Introducing  now new constants $a_3<a_2<a_1$ such that  
$\sin^2\alpha = (a_2-a_3)/(a_1 - a_3)$ and  making  the  change of variables: 
$\lambda_2/ R^2 = \frac{a_3-\rho_2}{a_1 - a_3}$, $\lambda_1/ R^2 = \frac{a_3 - \rho_1}{a_1 - a_3}$, we obtain the algebraic form  of  hyperbolic system  of  coordinates (Fig. \ref{fig:10}) 
\bea
\label{sys_hyp}
u_0^2 = R^2{\frac {\left ({\rho_1}-{a_2}\right )\left ({a_2 - \rho_2}\right )}
{\left ({a_1}-{a_2}\right )\left ({a_2}-{a_3}\right )}},
\quad
u_1^2 = R^2{\frac {\left ({\rho_1}-{a_3}\right )\left ({a_3 - \rho_2}\right )}
{\left ({a_1}-{a_3}\right )\left ({a_2}-{a_3}\right )}},
\quad
u_2^2 = R^2{\frac {\left ({\rho_1}-{a_1}\right )\left ({a_1 - \rho_2}\right )}
{\left ({a_1}-{a_2}\right )\left ({a_1}-{a_3}\right )}},
\eea
with $\rho_2<a_3<a_2<a_1<\rho_1$.

After  putting  formula  (\ref{Jef})  in  (\ref{sys_hyp}),  where moduli $k$ and $k'$ are the same as 
in formula (\ref{MODULUS}),   we obtain
\begin{equation}
\label{hyperbolic_Jacobi_2sh}
u_0 = - R \cn(a,k) \cn(b,k'), 
\qquad
u_1= i R\sn(a,k)\dn(b,k'),
\qquad
u_2= i R\dn(a,k)\sn(b,k').
\end{equation}
where  $a \in \left(iK, iK+ 2K \right)$, $b \in \left(iK, iK + 2K^\prime \right)$. 

\noindent
{\bf 2.}
The case of  one-sheeted hyperboloid is more complicated. From relation (\ref{01-lambda})  we obtain that  the covered parts of hyperboloid are defined by the following inequality
\begin{equation}
\label{surface-01}
|u_1^2 \sin^2\alpha - u_2^2 \cos^2\alpha  + R^2| > 2R |u_1\sin\alpha|.
\end{equation}
The same procedure as in the previous case  gives  us  the  hyperbolic  system of coordinates  on one-sheeted hyperboloid
in form (Fig. \ref{fig:34})
\bea
\label{sys_hyp_1}
u_0^2 = R^2{\frac {\left ({\rho_1}-{a_2}\right )\left ({\rho_2 - a_2}\right )}{\left ({a_1}-{a_2}\right )
\left ({a_2}-{a_3}\right )}},  
\quad
u_1^2 = R^2{\frac {\left ({\rho_1}-{a_3}\right )\left ({\rho_2 - a_3}\right )}{\left ({a_1}-{a_3}\right )
\left ({a_2}-{a_3}\right )}},
\quad
u_2^2 = R^2{\frac {\left ({\rho_1}-{a_1}\right )\left ({\rho_2 - a_1}\right )}{\left ({a_1}-{a_2}\right )
\left ({a_1}-{a_3}\right )}},
\eea
where it  is  necessary to select  two  cases:
\begin{itemize}
\item Hyperbolic Type I $H_I^A:$ $\rho_1$, $\rho_2<a_3<a_2<a_1$ ($H_I^B:$ $a_3<a_2<a_1<\rho_1$, $\rho_2$),
\item  Hyperbolic Type II $H_{II}^A:$ $a_3<\rho_1,\rho_2<a_2<a_1$ ($H_{II}^B:$ $a_3<a_2<\rho_1,\rho_2<a_1$).
\end{itemize}
It means that the covered part of one-sheeted hyperboloid (\ref{surface-01}) is splitting into several sub-parts corresponding to each type of the system as presented in Fig. \ref{fig:34}.  

It is more convenient for us to rewrite coordinates (\ref{sys_hyp_1}) in the form of Jacobi elliptic functions.  Introducing Jacobi functions as in (\ref{Jef}) with the same moduli $k$ and $k^\prime$, given by formula (\ref{MODULUS}),  we come to  
\begin{equation}
\label{TYPEII-02}
u_0 = - i R \cn(a,k)\cn(b, k^\prime), 
\qquad
u_1=  - R\sn(a,k) \dn(b, k^\prime),
\qquad
u_2=  - R\dn(a,k) \sn(b, k^\prime).
\end{equation}
It is easy to verify that angles $a$, $b$ run the intervals  
\begin{eqnarray}
\label{interval_a_b_1sh_hyp}
&&\text{for } H_I^A:\ a \in (-iK^\prime, iK^\prime), \quad b \in \left(iK, iK + 2K^\prime \right); \nonumber \\
&&\text{for } H_I^B:\ a \in (iK^\prime, iK^\prime + 2K), \quad b \in \left(-iK, iK \right); \\
&&\text{for } H_{II}^A:\ a \in \left[0, 4K\right), \quad b \in \left[K^\prime, K^\prime + i4K\right);\nonumber \\
&&\text{for } H_{II}^B:\ a \in [K, K + i4K^\prime), \quad b \in \left[0, 4K^\prime \right). \nonumber
\end{eqnarray}
Let us note that in general (except the particular case $k=k^\prime$) systems $H^B$ could not be obtained from $H^A$ through trigonometric rotation on $\pi/2$ over $u_0$.  One can observe that system $H_{I}^A$ is divided geometrically into four parts depending on the sign of $u_0$ and $u_2$, so it is necessary to choose the appropriate sign for these coordinates. The same is true for $H_I^B$. As for $H_{II}^A$ here we have two parts depending on the sign of $u_2$ and for $H_{II}^B$ one should select the sign of $u_1$.

Let us note, that families of coordinate curves $\rho_i = \text{const.}$ have straight lines as envelopes defined by the equalities 
\begin{equation}
\label{envelope_h}
|k u_1 \pm R|  = k^\prime |u_2|
\end{equation}
with intersections in the limit points $(0,\pm kR, \pm k^\prime R)$, $(\pm R k/ k^\prime, 0, \pm R/ k^\prime)$, $(\pm R k^\prime / k\pm R/k, 0)$. To avoid the intersection of coordinate curves on one family, it is necessary to take the points of envelops as initial ones for one family and as endpoints for the other family. For example, for system $H_{I}^A$ if we consider fixed $\rho_1 = \rho_{10} = \text{const.}$, then $\rho_2$ should be in the interval $\rho_2 \in (-\infty,\rho_{10})$ and for fixed $\rho_2 = \rho_{20}=\text{const.}$ the corresponding interval is $\rho_1 \in (\rho_{20},a_3)$.

In the limit case, when $k=\sin\alpha \to 0$ ($k^\prime \to 1$) operator $S_{H} = K_2^2 - \sin^2 \alpha L^2$ goes to equidistant one $K_2^2$. When $k^\prime \to 0$, considering $S_{SH} - {\cal C} = -K_1^2 + L^2 {k^\prime}^2$ we obtain the rotated equidistant operator $-K_1^2$. 

For system (\ref{hyperbolic_Jacobi_2sh}) on two-sheeted hyperboloid, taking $a \in \left(2K-iK^\prime, 2K+iK^\prime\right)$, $b\in (-iK, iK)$ and introducing variables $\alpha = \pi + i\tau_2$, $\beta =-i \arctan(\sinh\tau_1)$ in limit $k\to 0$ one can obtain equidistant system (\ref{sys_equi}).

Let us consider system $H_I^A$ (\ref{TYPEII-02}) with $a \in \left(2K-iK^\prime, 2K+iK^\prime\right)$, $b \in \left(iK, iK + 2K^\prime \right)$. Introducing $\alpha = \pi + i\tau_2$, $\beta = \tanh^{-1}(\cosh\tau_1) + i\pi/2$ we obtain equidistant system of  Type Ia (\ref{sys_equi_1}) when $k \to 0$. From $H_{II}^B$ with $\alpha = \pi/2 - i\tau$, $\beta = \tanh^{-1}(\cos\varphi)$ we come to equidistant system (\ref{sys_equi_11}) for $a \in \left[ K - 2iK^\prime, K+ 2iK^\prime\right)$, $b\in \left[-2K^\prime, 2K^\prime\right)$ with the same limit $k\to 0$. By analogy one can consider the contractions of the rest of the systems from (\ref{interval_a_b_1sh_hyp}) to corresponding equidistant systems on one-sheeted hyperboloid.

As in the case of rotated elliptic system on the two-sheeted hyperboloid, here we introduce the rotated hyperbolic system. Let us consider hyperbolic rotation (\ref{AK1}) through angle $a_1 = \arccosh\left(1/\cos\alpha\right)$ about axis $u_1$ (see (\ref{relation_coords_operators})):
\begin{equation}
\label{TYPEII-03}
\left( {{\begin{array}{*{20}c}
 u^\prime_0 \\
 u^\prime_1 \\
 u^\prime_2
\end{array} }} \right) =
\left( {{\begin{array}{*{20}c}
\frac{1}{\cos\alpha} & 0 & -\frac{\sin\alpha}{\cos\alpha} \\
0 & 1 & 0 \\
-\frac{\sin\alpha}{\cos\alpha} & 0 & \frac{1}{\cos\alpha}
\end{array} }} \right)
\left( {{\begin{array}{*{20}c}
 u_0 \\
 u_1 \\
 u_2
\end{array} }} \right)
 =
\left( {{\begin{array}{*{20}c}
 \frac{u_0 - u_2 \sin \alpha}{\cos\alpha} \\
 u_1 \\
 \frac{- u_0 \sin \alpha + u_2}{\cos\alpha}
\end{array} }} \right),
\end{equation}
then  
\begin{equation}
K_2 = \left(K_2^\prime + \sin\alpha L^\prime\right)/\cos\alpha,\quad L = \left(\sin\alpha K_2^\prime + L^\prime\right)/\cos\alpha
\end{equation}
and correspondingly operator $S_H $  transforms into rotated hyperbolic operator    
\bea
\label{rot_hyp-01}
S_{\tilde{H}} = c K_2^2 + \{K_2,L\}, \ c = \sin\alpha + 1/\sin\alpha,\ \sin^2\alpha \ne 0, 1.
\eea
For this system of coordinates we get from (\ref{TYPEII-02}) and (\ref{TYPEII-03}) (see Fig. \ref{fig:34a}):
\bea
\label{TYPEII-04}
u_0^\prime &=& \frac{R}{k^\prime}\left[ k \dn(a,k)\sn(b, k^\prime) - i \cn(a,k) \cn(b, k^\prime)\right], 
\nonumber\\[2mm]
u_1^\prime &=&  -R\sn(a,k) \dn(b, k^\prime),
\\[2mm]
u_2^\prime &=&  \frac{R}{k^\prime}\left[i k \cn(a,k)\cn(b, k^\prime) - \dn(a,k) \sn(b, k^\prime)\right],
\nonumber
\eea
with intervals for $a$ and $b$ as in (\ref{interval_a_b_1sh_hyp}). The reason for  introducing rotated hyperbolic system of coordinates is  the construction  of  another contraction limits showed later.


\section{Contractions on Two-Sheeted Hyperboloid}
\label{sec:3}

\subsection{Contraction of Lie algebra \texorpdfstring{$so(2,1)$}{so(2,1)} to \texorpdfstring{$e(2)$}{e(2)}}

To realize the contractions of Lie algebra $so(2,1)$ to $e(2)$ let us introduce Beltrami
coordinates on the hyperboloid $H_2$ in such a way
\begin{equation}
\label{Beltrami_H2}
x_{\mu} = R\frac{u_{\mu}}{u_0} = R\frac{u_{\mu}}{\sqrt{R^2+ u_1^2 + u_2^2}},
\qquad\qquad \mu = 1,2. 
\end{equation}
In terms of variables (\ref{Beltrami_H2}) generators (\ref{algebra_basis}) look like this
\bea
\label{generators_H2}
-\frac{K_1}{R} \equiv \pi_2 &=& \partial_{x_2} - \frac{x_2}{R^2}(x_1\partial_{x_1} + x_2 \partial_{x_2}),
\nonumber\\[2mm]
-\frac{K_2}{R} \equiv \pi_1 &=& \partial_{x_1} - \frac{x_1}{R^2}(x_1\partial_{x_1} + x_2 \partial_{x_2}),
\\[2mm]
L &=&  x_1\partial_{x_2} - x_2\partial_{x_1} = x_1\pi_2 - x_2\pi_1
\nonumber
\eea
and commutator relations of $so(2,1)$ take the form
\begin{equation}
    [\pi_1,\pi_2] = \frac{L}{R^2}, \ [\pi_1, L] = \pi_2, \ [L,\pi_2] = \pi_1. \label{comm_o21}
\end{equation}
Let us take the basis of $e(2)$ in the form
  \begin{equation}
    p_1 = \partial_{x_1}, \qquad p_2 = \partial_{x_2}, \qquad  M = x_2\partial_{x_1} - x_1\partial_{x_2},
     \label{basis_e2}
\end{equation}
with commutators
\begin{equation}
[p_1,p_2] = 0, 
\qquad [p_1, M] = -p_2 , 
\qquad
[M,p_2] = -p_1. \label{comm_e2}
\end{equation}
Then in limit $R^{-1} \to 0$ we have
\begin{equation}
\pi_1 \to p_1,
\qquad 
\pi_2 \to p_2,
\qquad 
L \to -M,
\end{equation}
relations (\ref{comm_o21}) contract to (\ref{comm_e2}), so algebra $so(2,1)$ contracts to $e(2)$.
Moreover, Laplace-Beltrami operator of $so(2,1)$ contracts to operator of $e(2)$:
\begin{equation}
\Delta_{LB} = \frac{1}{R^2}(K_1^2 + K_2^2 - L^2) = \pi_1^2 + \pi_2^2 - \frac{M^2}{R^2}
\to \Delta = p_1^2 + p_2^2.
\end{equation}


\begin{figure}[htbp]
\begin{center}
\begin{minipage}[c]{0.45\linewidth}
\includegraphics[scale=0.35]{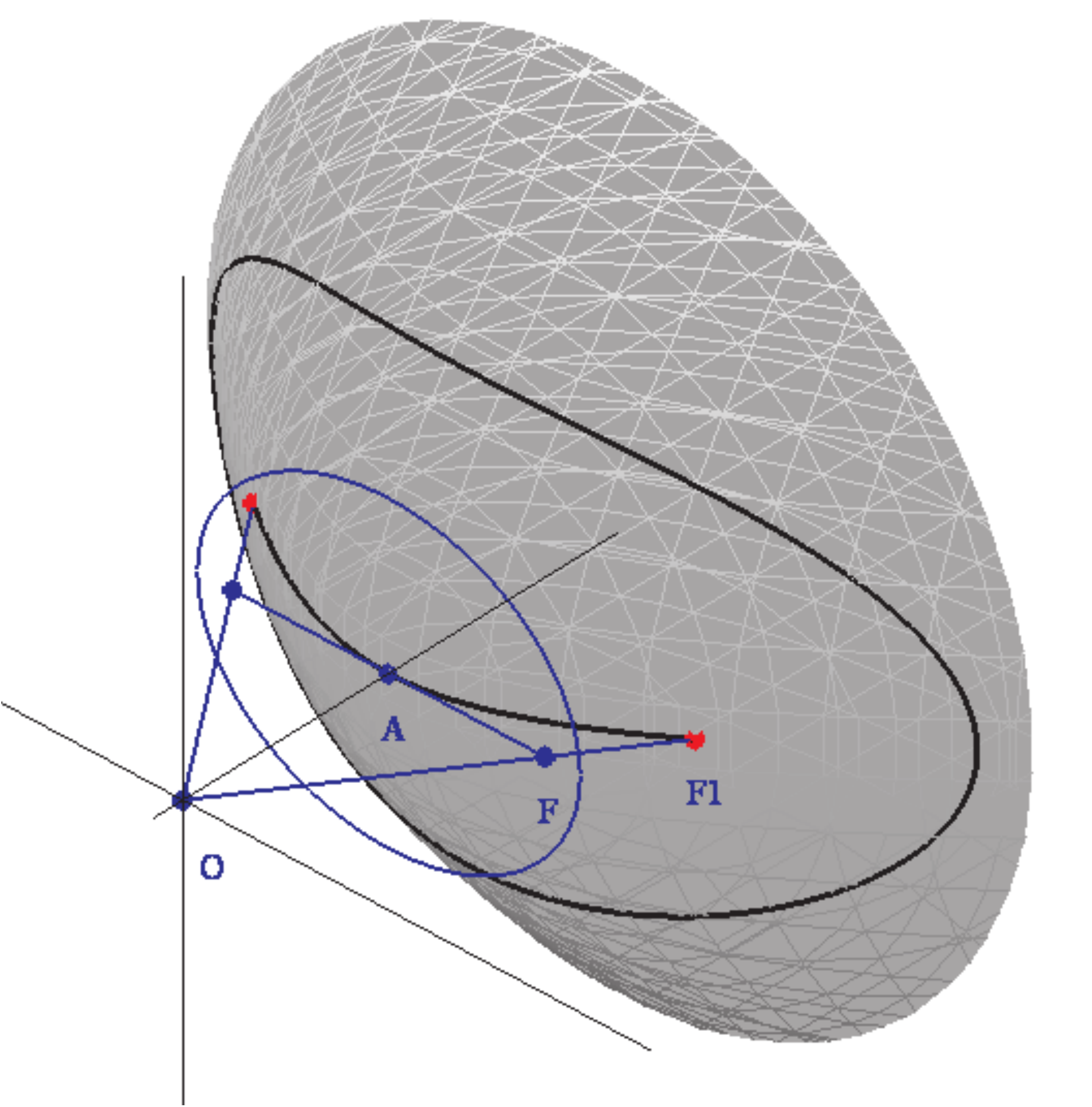}
\begin{center}
\caption{ Projective plane for two-sheeted hyperboloid}
\label{fig:1}
\end{center}
\end{minipage}
\end{center}
\end{figure}

\subsection{Contractions of the systems of coordinates}

In the case of hyperboloid $H_2$  Beltrami coordinates (\ref{Beltrami_H2}) are coordinates $(x_1, x_2)$  on   the  projective  plane  $u_0 = R$  in  the  interior of  circle
$x_1^2 + x_2^2 = R^2$ (inhomogeneous or projective coordinates, see Fig. \ref{fig:1}).
In contraction limit $R\to \infty$ this circle is transformed to euclidean 
plane $E_2$. It allows to relate the coordinate systems on $H_2$ to the corresponding ones on $E_2$.

The metric for projective plane $(x_1, x_2)$ induced by the metric on two-sheeted hyperboloid, has the form 
\begin{equation}
ds^2 = \left(1-\frac{x_1^2 + x_2^2}{R^2}\right)^{-2} \left[ \left( 1 - \frac{x_2^2}{R^2} \right) dx_1^2 + 2\frac{x_1 x_2}{R^2} dx_1 dx_2 + \left(1-\frac{x_1^2}{R^2}\right) dx_2^2 \right]
\end{equation}
and contracts to the metric on euclidean plane $ds^2 = dx_1^2 + dx_2^2 \sim dx^2 + dy^2$.

\subsection{Contractions of nonorthogonal systems}

{\bf 1.} Considering nonorthogonal pseudo-spherical system (\ref{PSEUDO-SPHER-01_NO}) for the fixed geodesic parameter $r=\tau R$, we obtain that $\tau \sim r/R$ when $R\to \infty$. Beltarmi coordinates contract as follows:
\begin{eqnarray*}
x_1 = R \frac{u_1}{u_0} = R \tanh \tau \cos\left(\varphi + \frac{R \tau}{\alpha}\right) \rightarrow  r \cos\left(\varphi + \frac{r}{\alpha}\right), \\
x_2 = R \frac{u_2}{u_0} = R \tanh \tau \sin\left(\varphi + \frac{R \tau}{\alpha}\right) \rightarrow  r \sin\left(\varphi + \frac{r}{\alpha}\right),
\end{eqnarray*}
so we get nonorthogonal polar coordinates on euclidean plane $E_2$ (see Table \ref{tab:7}). For corresponding operator we obtain $-S_{SPH} = -L \to M = y\partial_x -x\partial_y = X_S$.

\noindent {\bf 2.} For nonorthogonal equidistant system (\ref{sys_equi_NO}), taking  $\tau_1 \sim y'/R$, $\tau_2 \sim x'/R$, $\alpha \sim R$ we obtain:
\begin{eqnarray*}
x_1 = R \frac{u_1}{u_0} = R\tanh(R\tau_1/\alpha + \tau_2)  \rightarrow x' + y', \qquad
x_2 = R \frac{u_2}{u_0} = R\frac{ \tanh (R\tau_1/\alpha)}{\cosh\tau_2} \rightarrow  y',
\end{eqnarray*}
where $(x', y')$ are nonorthogonal Cartesian coordinates on $E_2$ (Table \ref{tab:7}). For symmetry operator we have $- S_{EQ}/R = -K_2/R \to p_1 = X_C$.

\noindent {\bf 3.} Let us consider nonorthogonal horicyclic system (\ref{sys_hor_no}) in permuted form $u_1 \leftrightarrow u_2$. Then in contraction limit one can obtain nonorthogonal Cartesian coordinates $(x', y')$, if $\tilde{x} \sim y'/R$, $\tilde{y} \sim 1 + x'/R$. For corresponding Beltrami coordinates we have:
\[
x_1 \to x' + y', \qquad x_2 \to x'
\]
and $-S_{\bar{HO}}/R = (-K_2 + L)/R =  \pi_1 + L/R \to p_1 = X_C$. 

\subsection{Pseudo-spherical to polar}

For pseudo-spherical coordinate system (\ref{PSEUDO-SPHER-01}) we fix the geodesic parameter
$r=\tau R$. Then, if $R^{-1}$ tends to zero, then $\tau \to 0$ $\tanh \tau \simeq \tau  \simeq \frac{r}{R}$.
In the limit for Beltrami coordinates we have
\begin{eqnarray*}
x_1 = R \frac{u_1}{u_0} = R \tanh \tau \cos\varphi \rightarrow  r \cos\varphi, \qquad
x_2 = R \frac{u_2}{u_0} R \tanh \tau \sin\varphi  \rightarrow  r \sin\varphi.
\end{eqnarray*}
Thus  the pseudo-spherical coordinates on $H_2$ contract into polar coordinates $(r, \varphi)$ on the euclidean plane  $E_2$ 
(see Table \ref{tab:7}).   For the  corresponding  pseudo-spherical operator we obtain  $S_{SPH}^{(2)} = L^2  \to M^2 = X^2_S$.  

\subsection{Equidistant to Cartesian}

For equidistant system of coordinates (\ref{sys_equi}) Beltrami coordinates (\ref{Beltrami_H2})
look like
\begin{eqnarray*}
 x_1 = R\tanh\tau_2,\qquad  x_2 = R\tanh\tau_1/\cosh\tau_2.
\end{eqnarray*}
Taking limit $R^{-1} \to 0$, $\tau_1 \to 0$, $\tau_2 \to 0$ and  putting  $\sinh \tau_1 \simeq y/R$,
$\sinh \tau_2 \simeq x/R$ we get  
\begin{eqnarray*}
x_1\rightarrow x, \qquad  x_2\rightarrow y
\end{eqnarray*}
and  
\begin{eqnarray*}
\frac{S_{EQ}^{(2)} }{R^2} =  \frac{K_2^2}{R^2}\to p_1^2  = X_C^2,
\end{eqnarray*}
where $(x,y)$ are the orthogonal Cartesian coordinates on euclidean plane $E_2$ (see Table \ref{tab:7}). 


\subsection{Horicyclic to Cartesian}

For horicyclic  variables $\tilde{x}$, $\tilde{y}$ (\ref{sys_hor}) we have
\begin{eqnarray*}
\tilde x = \frac{u_2}{u_0 - u_1} , 
\qquad\qquad 
\tilde y = \frac{R}{u_0 - u_1}. 
\end{eqnarray*}
In contraction limit $R \to \infty$  we obtain 
\begin{eqnarray*}
\tilde x \to \frac{y}{R},  \qquad\qquad  
\tilde y  \to 1 + \frac{x}{R},
\end{eqnarray*}
and Beltrami coordinates are transforming to Cartesian ones:
\[
x_1 = R \frac{\tilde{x}^2 + \tilde{y}^2 - 1}{\tilde{x}^2 + \tilde{y}^2 + 1} \to x,
\qquad
x_2 = R \frac{2 \tilde{x}}{\tilde{x}^2 + \tilde{y}^2 + 1} \to y.
\]

For the corresponding  operator $S_{HO}^{(2)}$  we  get
\[
\frac{S_{HO}^{(2)}}{R^2} = \frac{(K_1+L)^2}{R^2} = \pi_2^2 + \frac{L^2}{R^2} - \frac{1}{R}\{\pi_2,L\}\to p_2^2 \simeq X_C^2.
\]


\subsection{Elliptic coordinates to elliptic, Cartesian, polar and parabolic ones}

On the projective plane the ellipse foci of (\ref{sys_ell}) have coordinates $F\left(\pm R \tan \alpha, 0\right)$
where $\tan^2 \alpha  = (a_1 - a_2)/(a_1 - a_3) $ and $\alpha$ is angle $AOF$ (see Fig. \ref{fig:1}).
The same points are the foci for hyperbolas. We must distinguish three cases for limit $R \to \infty$, 
namely:  when the length of  $AF$  is  fixed  and  $\alpha \to 0$,  when  $AF$ and  $\alpha$ go  to  zero,
and when  $\alpha$ is  fixed  and  $AF \to \infty$.  The latter limiting procedure involves two additional 
cases when one or two foci go to infinity with pseudo-radius $R$.  

\subsubsection{Elliptic to elliptic}

\label{sec:EllepticToElliptic}
Let us introduce parameter $D = \sqrt{a_1 - a_2}$ that defines the focal distance for elliptic coordinate system on  $E_2$ plane. 
The value of $D$ is the limit one for the length of arc $AF_1$, as well for the length of $AF$, when $R\to\infty$  (see Fig. \ref{fig:1}).
Therefore,  for  the fixed  values of  $D$  in  contraction  limit  $-a_3 \simeq R^2 \to \infty$  we get  
\begin{eqnarray}
\label{rad_ell1}
\sinh^2\beta  =  \frac{a_1 - a_2}{a_2 - a_3}  \simeq  \frac{D^2}{R^2} \to 0, 
\end {eqnarray}
and  
\begin{eqnarray*}
S_E = L^2 + \sinh^2\beta K_2^2 = L^2 + R^2 \sinh^2\beta \pi_1^2  \to M^2 + D^2 p_1^2 = X_E.
\end {eqnarray*}
Introducing new variables $\xi$ and $\eta$  as
\begin{eqnarray*}
\sinh^2 \xi = \frac{\rho_1 - a_1}{a_1 - a_2}, 
\qquad 
\cos^2 \eta = \frac{\rho_2 - a_2}{a_1 - a_2},
\qquad
\xi \in [0, \infty), 
\quad
\eta \in [0, 2\pi),
\end{eqnarray*}
we can rewrite the elliptic system of coordinates (\ref{sys_ell}) in the form   
\begin{eqnarray*}
u_0^2 &=& R^2 \left[1+ \frac{a_1-a_2}{a_1-a_3}\sinh^2\xi\right] 
\left[1+\frac{a_1-a_2}{a_2-a_3}\cos^2\eta \right],
\\
u_1^2 &=& \frac{R^2 D^2}{a_2-a_3} \cos^2\eta \cosh^2\xi,
\\
u_2^2 &=& \frac{R^2 D^2}{a_1-a_3} \sin^2\eta\sinh^2\xi.
\end{eqnarray*}
Taking  now limit  $-a_3 \simeq R^2 \to \infty$  and  using  relation   (\ref{rad_ell1})  it is easy to see  
that Beltrami coordinates  
\begin{eqnarray*}
x_1 =  R  \frac{u_1}{u_0} \to  D \cos\eta \cosh\xi,
\quad
x_2 = R \frac{u_2}{u_0}  \to  D \sin\eta\sinh\xi
\end{eqnarray*}
contract  to the ordinary elliptic coordinates on euclidean plane $E_2$ (see Table \ref{tab:7}).

\subsubsection{Elliptic to polar}

Let us consider the case when $a_1 - a_2  \simeq 1/R^2$, $a_2 - a_3 \simeq R^2$, then
\[
\sinh^2 \beta  = \frac{a_1-a_2}{a_2-a_3} \simeq \frac{1}{R^4}.
\]
Geometrically it means that foci coincide as $R\to \infty$.  Thus for  operator $S_E$ we have 
\[
S_E \simeq L^2 +  \frac{1}{R^2}  \pi_1^2 \to M^2 = X_S^2.
\]
Let us  introduce new variables
\[
r^2 = \rho_1 - a_2,\qquad \cos^2 \varphi = \frac{\rho_2 - a_2}{a_1 - a_2}. 
\]
Then Beltrami coordinates take the form
\[
x_1^2 = \frac{(a_1 - a_3)}{R^2}    \frac{r^2 \cos^2\varphi}{(1+r^2/R^2)(1+\cos^2\varphi/R^4)}, 
\qquad
x_2^2 = \frac{(a_2 - a_3)}{R^2}   \frac{\sin^2\varphi (r^2 -1/R^2)}{(1+r^2/R^2)(1+\cos^2\varphi/R^4)}
\]
and  in limit  $(a_2 - a_3)  \simeq   (a_1 - a_3)  \simeq  R^2 \to \infty$  we  obtain 
\[
x_1^2 \to r^2 \cos^2\varphi, \qquad x_2^2 \to r^2 \sin^2\varphi
\]
that coincide  with the orthogonal polar coordinates on euclidean plane $E_2$ (Table \ref{tab:7}).

\subsubsection{Elliptic to Cartesian}

Let us  fix  angle $\alpha$ (i.e.  the length  of  $AF$ tends to infinity  when $R\to\infty$) and $a_1 - a_2 = a_2 - a_3$ (or $k^2 = k^{{\prime}2} = 1/2$). Then we get 
\[
\frac{S_E}{R^2} = \frac{1}{R^2} (L^2 + K_2^2) =   \frac{L^2}{R^2}  +  \pi_1^2  \to  p_1^2  
=   X_C^2.
\]
Taking into account that from (\ref{sys_e_J})
\begin{eqnarray*}
- \cn^2 a = \frac{1}{2}\left\{ \frac{u_0^2+u_2^2}{R^2}
+ \sqrt{\left(\frac{u_0^2+u_2^2}{R^2}\right)^2 - \frac{4u_1^2}{R^2}}\right\},
\qquad
\cn^2 b = \frac{1}{2}  \left\{ \frac{u_0^2+u_2^2}{R^2} 
-  \sqrt{\left(\frac{u_0^2+u_2^2}{R^2}\right)^2 - \frac{4u_1^2}{R^2}}\right\},
\end{eqnarray*}
we obtain for the large $R$:  
\begin{eqnarray*}
\dn\, a \to -i\frac{y}{R},\ \cn\, a \to -i\left(1+\frac{y^2}{R^2}\right), \ \sn \, a \to \sqrt{2}\left(1+\frac{y^2}{4R^2}\right);\\
\dn\, b \to \frac{1}{\sqrt{2}}\left(1+ \frac{x^2}{2R^2}\right),\ \cn\, b \to \frac{x}{R}, \ \sn \, b \to 1-\frac{x^2}{2R^2}.
\end{eqnarray*}
Thus, in contraction limit  $R \to \infty$ we get that  Beltrami  coordinates 
(\ref{Beltrami_H2}) take Cartesian form: 
\begin{eqnarray*}
x_1 = R\frac{i \cn\, a\, \cn\, b}{ \sn\, a \, \dn\, b} \to x,
\qquad 
x_2 = R \frac{i \dn\, a\, \sn\, b}{\sn\, a\, \dn\, b}  \to y.
\end{eqnarray*}

\subsubsection{Rotated elliptic to parabolic}

\begin{figure}[htbp]
 \begin{center}
    \begin{minipage}[t]{0.45\linewidth}
      \includegraphics[scale=0.33]{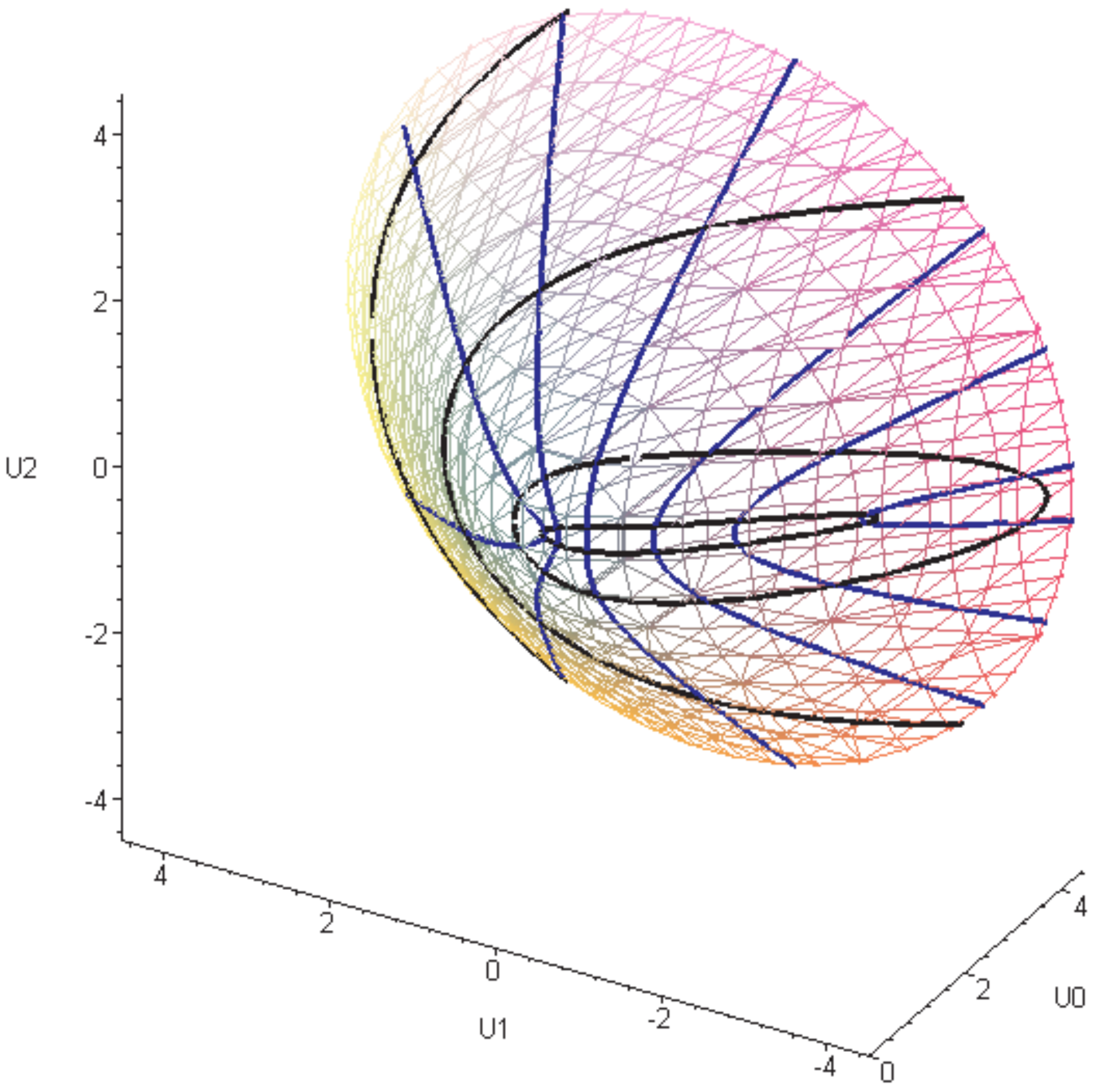}

\begin{center}
\caption{Rotated elliptic system}
\label{fig:8a}
\end{center}
    \end{minipage}
    \hfill
    \begin{minipage}[t]{0.45\linewidth}
 \includegraphics[scale=0.37]{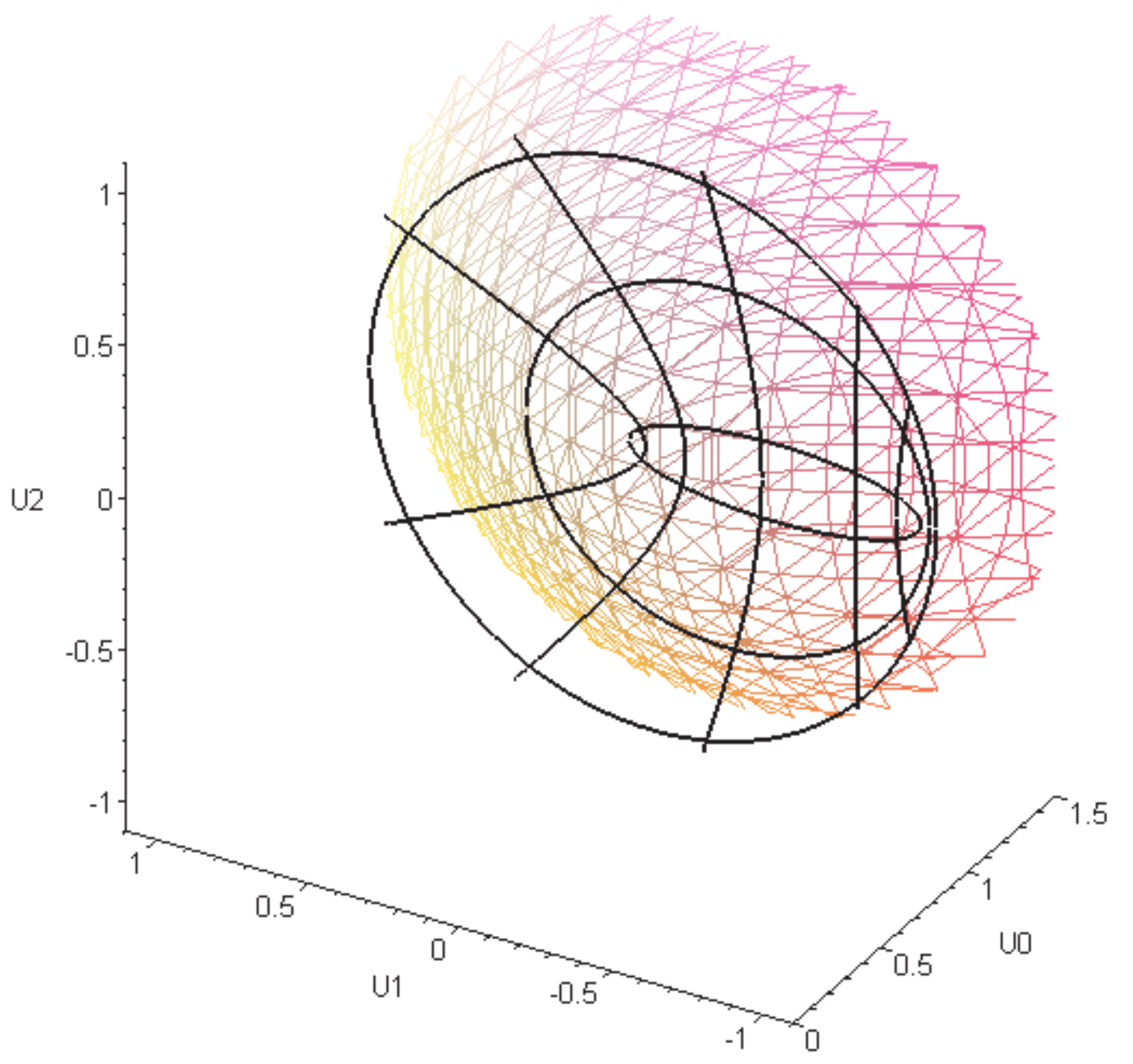}
\caption{Projective plane for rotated elliptic system}
\label{fig:9a}
    \end{minipage}
  \end{center}
\end{figure}

The geometrical difference between elliptic system and the rotated elliptic one is the positions of the foci of ellipses. 
For the rotated elliptic system (see Fig. \ref{fig:8a}) one of the foci is in $(R^2,0,0)$. Let us fix constants $a_i$ in such a way: $a_1 - a_2 = a_2 - a_3$. Then rotating elliptic coordinates (\ref{sys_e_rot}) 
take the form 
\bea
\label{new-sys_e_rot}
u^\prime_0 = R \left\{\sqrt{2}\, \sn\, a\, \dn\, b + i\, \cn\, a\,  \cn\, b    \right\},
\quad
u^\prime_1 = R \left\{ \sn\, a\, \dn\, b +  i \sqrt{2}\, \cn\, a\, \cn\, b    \right\},
\quad
u^\prime_2 =  i R\, \dn\, a\, \sn\, b,
\nonumber
\eea
with moduli $k = k^\prime=1/\sqrt{2}$ for all Jacobi functions. For the large $R$ we obtain
\begin{equation}
\begin{array}{ll}
\i \cn a = \frac{1}{2}\sqrt{\left( 1 + \sqrt{2} \frac{u^\prime_1}{R} - \frac{u^\prime_0}{R}\right)^2 + 2\frac{{u^\prime_2}^2}{R^2}}
- \frac{1}{2} \sqrt{\left( 1 - \sqrt{2} \frac{u^\prime_1}{R} + \frac{u^\prime_0}{R}\right)^2 + 2 \frac{{u^\prime_2}^2}{R^2}  } 
\simeq -1 + \frac{\sqrt{2}}{2}\frac{u^2}{R},
\\[3mm]
\cn b = \frac{1}{2}\sqrt{\left( 1 + \sqrt{2} \frac{u^\prime_1}{R} - \frac{u^\prime_0}{R}\right)^2 + 2\frac{{u^\prime_2}^2}{R^2}}
+ \frac{1}{2} \sqrt{\left( 1 - \sqrt{2} \frac{u^\prime_1}{R} + \frac{u^\prime_0}{R}\right)^2 + 2\frac{{u^\prime_2}^2}{R^2}  } 
\simeq 1 + \frac{\sqrt{2}}{2}\frac{v^2}{R}.
\end{array}
\end{equation}
For limit $R\to\infty$ we have that Beltrami coordinates go into the parabolic ones: 
\[
x_1\to \frac{u^2 - v^2}{2}, 
\qquad 
x_2 \to uv. 
\]

For the symmetry operator we find
\[
 \frac{S_{\tilde{E}}} {\sqrt{2} R} = 
  \frac{1} {\sqrt{2} R} \left[ \cosh 2\beta L^2 + 1/2 \sinh 2\beta \{K_1,L\}\right] 
=    \frac{3}{\sqrt{2} R} L^2  -  \{\pi_2, L\}  \to  \{p_2, M\}  = X_P 
\]
which corresponds to the parabolic coordinates on $E_2$ plane (see Table \ref{tab:7}).

\subsection{Hyperbolic to Cartesian}

The hyperbolic system of coordinates (\ref{sys_hyp}) is defined by three parameters
$a_1$, $a_2$, $a_3$, which fixe the position of  the hyperbola foci on hyperboloid. Considering orthogonal
projections of coordinates to plane $u_0 = R$, we obtain the families of hyperbolas.
Fixing $\rho_2$ for the first family of hyperbolas its minimal focal distance is equal to
$2d_1$, where $d_1 = R\sqrt{\frac{a_1 - a_3}{a_2 - a_3}} = \frac{R}{k}$.
The minimal focal distance for the second family of hyperbolas (when $\rho_1$ is a constant)
is equal to $2d_2$, where $d_2 = R\sqrt{\frac{a_1 - a_3}{a_1 - a_2}} = \frac{R}{k^\prime}$.

If we denote $F_1(R, d_1, 0)$,  $F_{21}(R, 0, d_2)$ and $F_{22}(R, 0, -d_2)$ then parameter
$\sin^2 \alpha = \frac{a_2 - a_3}{a_1 - a_3} = k^2$ where $2\alpha$ is angle $F_{21} F_1 F_{22}$.
Note, that $\sin^2\alpha \ne 0, \ne 1$ and one can say that this parameter plays the role of a scale factor for
the projective coordinates.

For simplicity let us take $a_1 - a_2 = a_2 - a_3$. Then $k^2 = k^{{\prime}2}= \sin^2\alpha = 1/2$ and 
\[
\frac{S_H}{R^2} = \frac{1}{R^2} \left (K_2^2 - \frac{1}{2} L^2\right) = \pi_1^2 - \frac{L^2}{2R^2}
\to p_1^2 = X_C^2.
\]
From equation (\ref{hyperbolic_Jacobi_2sh}) we have 
\begin{eqnarray*}
 \cn^2 a = \frac{1}{2}\left\{ \frac{u_1^2-u_2^2}{R^2}
- \sqrt{\left(\frac{u_1^2-u_2^2}{R^2}\right)^2 + \frac{4u_0^2}{R^2}}\right\},
\qquad
- \cn^2 b = \frac{1}{2}  \left\{ \frac{u_1^2-u_2^2}{R^2} 
+  \sqrt{\left(\frac{u_1^2-u_2^2}{R^2}\right)^2 + \frac{4u_0^2}{R^2}}\right\}.
\end{eqnarray*}
For the large $R$ we get that 
\begin{eqnarray*}
\dn\, a \to -\frac{i y}{R\sqrt{2}}, \ \cn\, a \to -i\left(1 + \frac{y^2}{2R^2}\right), \ \sn\, a \to \sqrt{2} \left(1+\frac{y^2}{4R^2}\right);\\
\dn\, b \to -\frac{i x}{R\sqrt{2}}, \ \cn\, b \to -i\left(1 + \frac{x^2}{2R^2}\right), \ \sn\, b \to \sqrt{2} \left(1+\frac{x^2}{4R^2}\right),
\end{eqnarray*}
and Beltrami coordinates in contraction limit $R \to \infty$ go into Cartesian ones:
\begin{equation}
x_1 = -i R \frac{\sn\, a\ \dn\, b}{\cn\, a\ \cn\, b} \to x,
\qquad
x_2 = -i R \frac{\dn\, a\ \sn\, b}{\cn\, a\ \cn\, b} \to y.
\end{equation}


\subsection{Semi-hyperbolic coordinates to Cartesian and parabolic ones}

\subsubsection{Semi-hyperbolic to Cartesian}

\begin{figure}[htbp] 
\begin{center}
    \begin{minipage}[t]{0.45\linewidth}
      \includegraphics[scale=0.3]{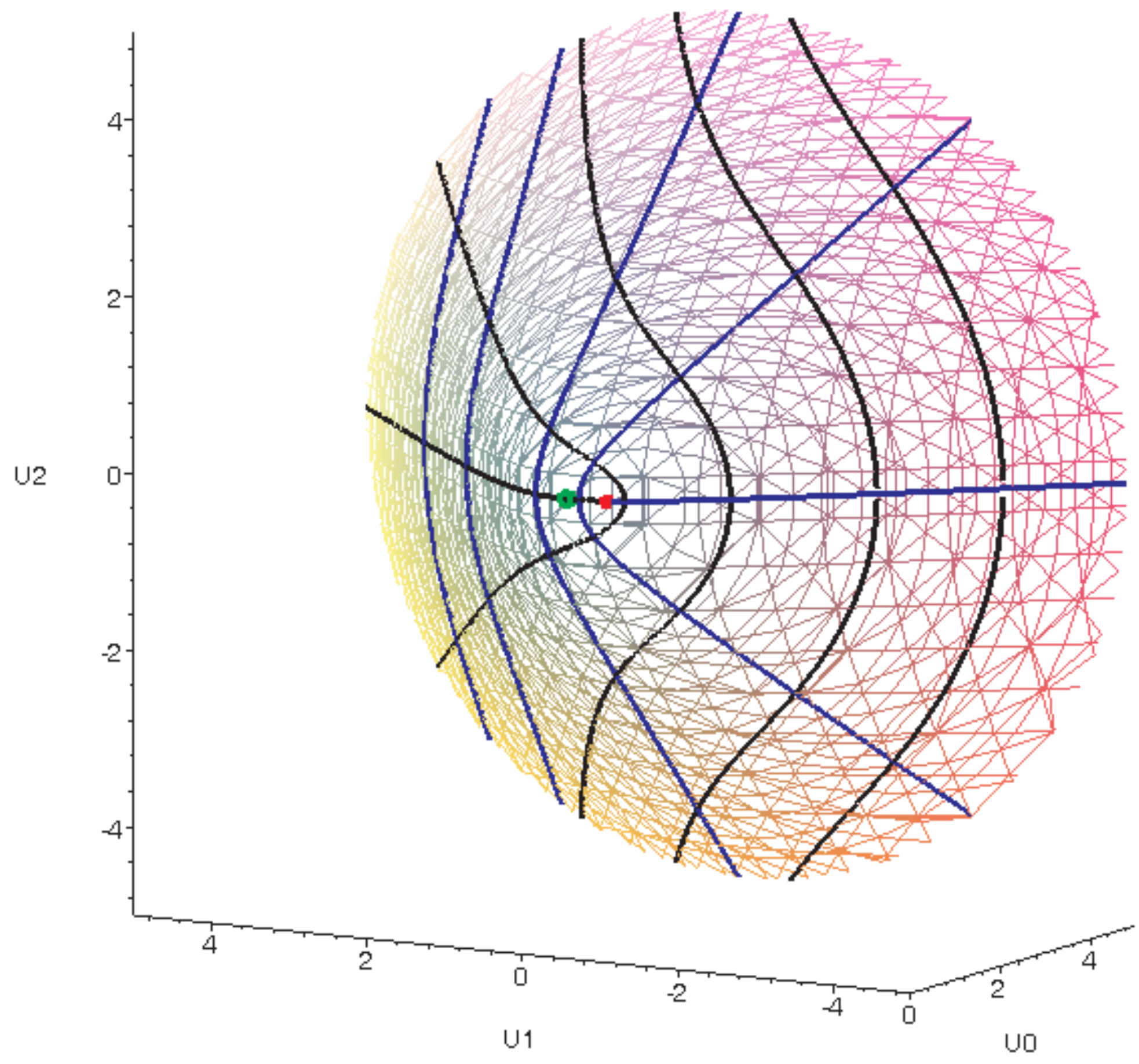}

\begin{center}
\caption{Semi-hyperbolic system}
\label{fig:12}
\end{center}
    \end{minipage}
    \hfill
    \begin{minipage}[t]{0.45\linewidth}
 \includegraphics[scale=0.3]{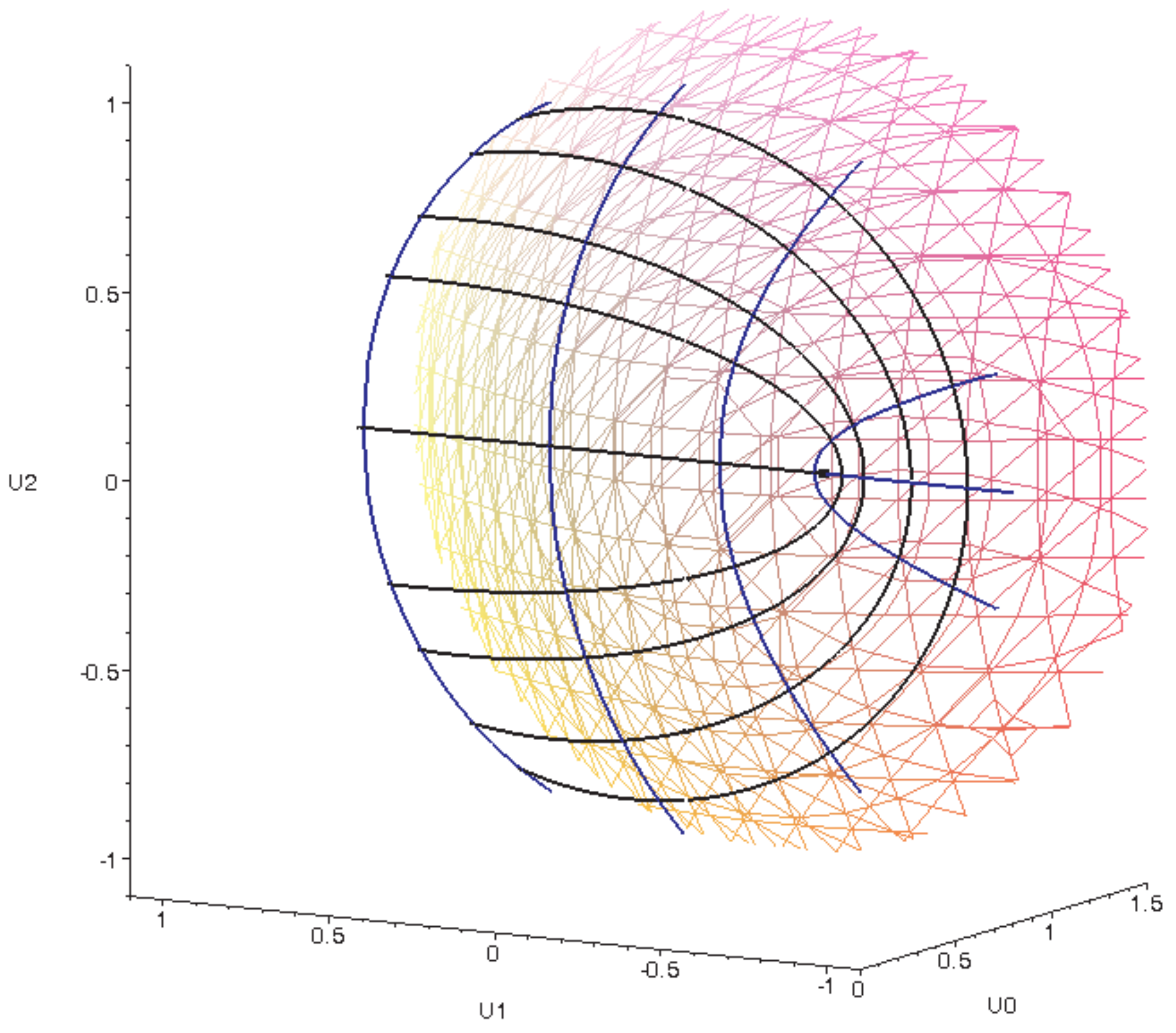}
\caption{Projective plane for semi-hyperbolic system is formed by semihyperbolas where the focus has coordinates
$\left( - R \sqrt{\frac{\sqrt{c^2 + 1}-1}{\sqrt{c^2 + 1}+1}}, \, 0\right)$.}
\label{fig:13a}
    \end{minipage}
  \end{center}
\end{figure}

Let us fix parameter $c = 1$ in the semi-hyperbolic system of coordinates (\ref{2SH_sys_sh_hyp}).  
Then, the coordinates of the focus (see Fig.\ref{fig:13a}) are moved to infinity as $R\to \infty$ and the semi-hyperbolic system of coordinates contracts into Cartesian one.  
Indeed, if we write out the coordinates $\tau_1$, $\tau_2$ from (\ref{2SH_sys_sh_hyp}) as 
\begin{equation}
\sinh \tau_{1,2} = -\frac{u_0 u_1}{R^2} - \frac{1}{2}\left(\frac{u_2^2}{R^2} + 1\right) \pm
\sqrt{ \left( \frac{u_0 u_1}{R^2} + \frac{1}{2}\left(\frac{u_2^2}{R^2} + 1\right)\right)^2
+ \frac{u_2^2}{R^2} - 2\frac{u_0u_1}{R^2}}.
\end{equation}
we get for the large $R$ 
\begin{equation}
\sinh\tau_1 \to - 2\frac{x}{R}, \qquad
\sinh\tau_2 \to - 1 - 2\frac{y^2}{R^2}.
\end{equation}
Hence, it is easy to see that at contraction limit $R \to \infty$  Beltrami coordinates go into 
Cartesian ones: $x_1 \to x$, $x_2 \to y$.  For the corresponding symmetry  operator   
we obtain
\[
\frac{S_{SH}}{R^2} = \frac{1}{R^2} \left ( K_2^2 + \{K_1,L\}\right) \simeq \pi_1^2 - \frac{1}{R}\{\pi_2,L \} 
\to p_1^2 = X_C^2.
\]

\subsubsection{Semi-hyperbolic to parabolic}

In case of $c = 0$, the coordinates of focus $F$ are fixed on the projective plane (see Fig.\ref{fig:13a}) 
and in the contraction limit $R \to \infty$ the semi-hyperbolic coordinates tend to parabolic ones. Indeed, 
for variables $\mu_{1,2}$ in (\ref{sys_sh_mu1}) we have at the large $R$ 
\begin{equation}
\mu_1 = \sqrt{ \frac{u_0^2 u_1^2}{R^4} + \frac{u_2^2}{R^2}} + \frac{u_0 u_1}{R^2} \to \frac{u^2}{R},
\qquad
\mu_2 = \sqrt{ \frac{u_0^2 u_1^2}{R^4} + \frac{u_2^2}{R^2}} - \frac{u_0 u_1}{R^2} \to \frac{v^2}{R}.
\end{equation}
Therefore Beltrami coordinates in limit $R \to \infty$ contract to parabolic ones:
\begin{equation}
x_1 \to \frac{u^2 - v^2}{2},
\qquad
x_2 \to uv.
\end{equation}
For symmetry operator $S_{SH}$ we obtain   
\[
 \frac{S_{SH}}{R} =  \frac{1}{R} \{K_1,L\} \to \{p_2,M\} = X_P.
\]


\subsection{Elliptic-parabolic to Cartesian and parabolic}

\begin{figure}[htbp] 
\begin{center}
    \begin{minipage}[t]{0.45\linewidth}
      \includegraphics[scale=0.3]{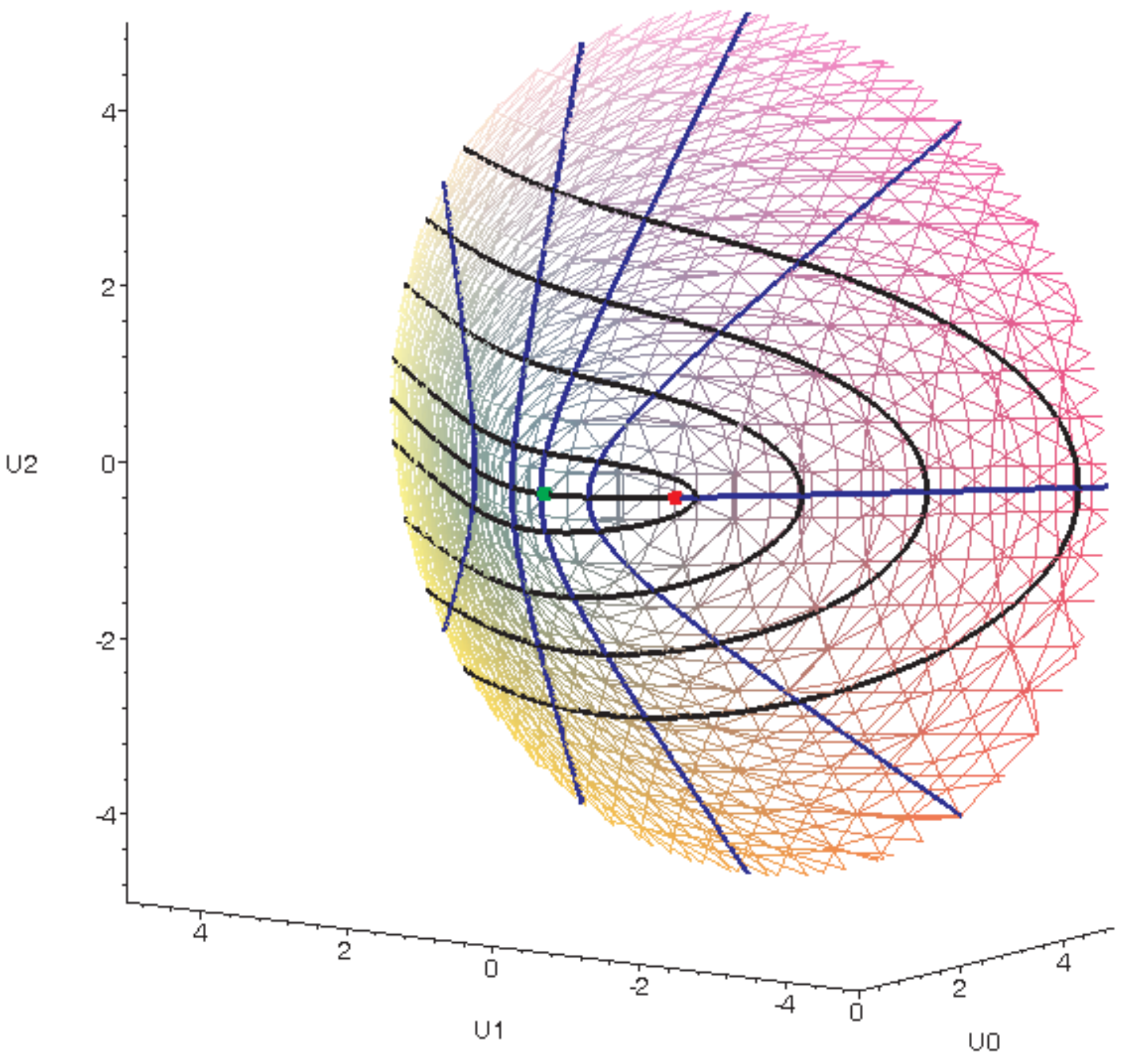}
\begin{center}
\caption{Elliptic-parabolic system}
\label{fig:13}
\end{center}
    \end{minipage}
    \hfill
    \begin{minipage}[t]{0.45\linewidth}
 \includegraphics[scale=0.3]{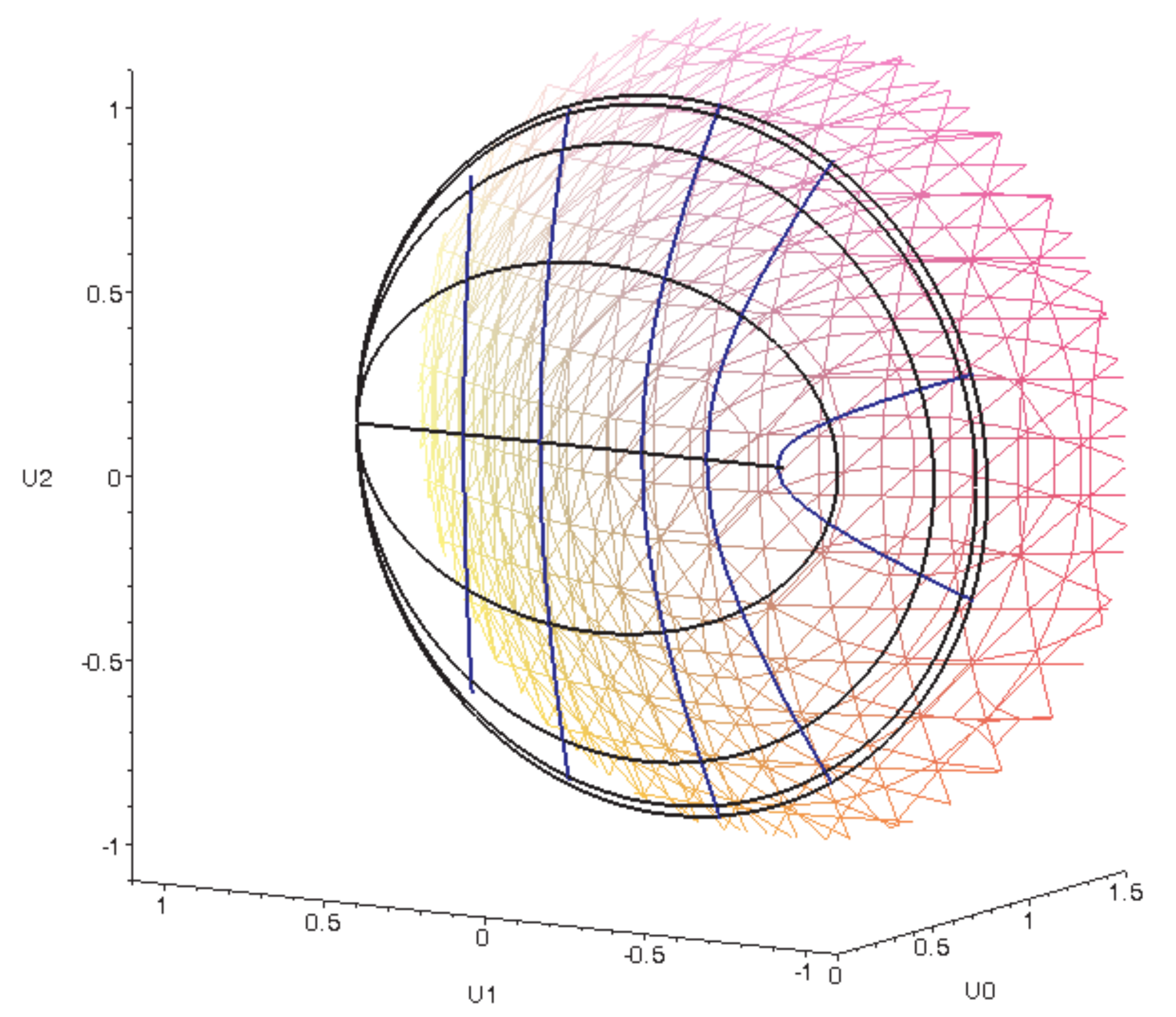}
\caption{Projective plane for elliptic-parabolic system is formed by elliptic parabolas and convex hyperbolic parabolas. The coordinates of the focus 
on the projective plane is $\left(R\frac{\gamma - 1}{\gamma +1},0\right)$.}
\label{fig:14}
    \end{minipage}
  \end{center}
\end{figure} 

\subsubsection{Elliptic-parabolic to Cartesian}

We start with the case when parameter $\gamma \ne 1$. Then the focus on projective plane 
(see Fig. \ref{fig:14}) moves to infinity as pseudo-radius $R\to \infty$. 
For variables $\xi_{1} =  \cos^2\theta$ and  $\xi_{2} =  \cosh^2 a$ from (\ref{sys_ep_trig}) we have 
\begin{equation}
\xi^2_{1,2} = \frac{u_0(\gamma + 1) + u_1(\gamma -1) \mp
\sqrt{ [u_0(\gamma + 1) + u_1(\gamma - 1)]^2 - 4R^2\gamma  } }{2(u_0-u_1)}.
\end{equation}
For the calculation of the behavior of variables $\xi_{1,2}$ at the large $R$, we must distinguish two cases of parameter $\gamma$.   
For the case of $\gamma\in(0,1)$ we obtain:
\begin{equation}
\cos\theta \to \sqrt{\gamma} \left(1 + \frac{x}{R}\right), \qquad
\sinh a \to \sqrt{ \frac{\gamma}{1-\gamma}} \frac{y}{R}, 
\label{sys_ep_xi}
\end{equation}
whereas for $\gamma >1$, we get 
\begin{equation}
\sin\theta \to \sqrt{\frac{\gamma}{\gamma - 1}} \frac{y}{R},
\qquad
\cosh a \to \sqrt{\gamma} \left(1 + \frac{x}{R}\right).
\label{sys_ep_xi_2}
\end{equation}
It is easy to see that in both cases Beltrami coordinates in contraction limit $R\to \infty$ 
go into Cartesian ones: $x_1 \to x$, $x_2 \to y$. For symmetry operator $S_{EP}$ we obtain
\[
\frac{1}{(\gamma-1) R^2}\left[S_{EP} - R^2\Delta_{LB}\right]  = \pi_1^2  - \frac{1}{(\gamma-1) R^2}
\left[R \{\pi_2,L\} - 2 L^2 \right] \to p_1^2 = X_C^2.
\]

\subsubsection{Elliptic-parabolic to parabolic}

Let us take now $\gamma = 1$, then the coordinates of the focus on projective plane are fixed at point $(0,0)$. 
For the large $R$ we have 
\begin{equation}
\sin^2\theta = \frac{- u_1 + \sqrt{u_0^2 - R^2}}{u_0 - u_1}\to \frac{v^2}{R},
\qquad
\sinh^2 a = \frac{u_1 + \sqrt{u_0^2 - R^2}}{u_0 - u_1}\to \frac{u^2}{R},
\end{equation}
and hence  Beltrami coordinates contract into the parabolic ones:
\begin{equation}
x_1 \to \frac{u^2 - v^2}{2},
\qquad
x_2 \to uv.
\end{equation}
Symmetry operator $S_{EP}$ transforms as follows
\[
 \frac{S_{EP}}{R} - R \Delta_{LB}  =   -\{\pi_2,L\}  + 2\frac{L^2}{R} \to \{p_2,M\} = X_P.
\]


\subsection{Hyperbolic-parabolic to Cartesian}

Variables $\xi_1 = \sin \theta$, $\xi_2 = \sinh b$ from (\ref{sys_hp_trig}) are defined by the following formulas:
\begin{equation}
\xi_{1,2}^2 = \frac{ \pm u_0(1-\gamma) \mp u_1 (\gamma +1) +
\sqrt{ [u_0(\gamma - 1) + u_1(\gamma + 1)]^2 + 4R^2 \gamma } }
{ 2(u_0 - u_1) }.
\end{equation}
For the large $R$ we have
\begin{equation}
\cos\theta \to  \sqrt{\frac{\gamma}{\gamma + 1}} \frac{y}{R},
\qquad
\sinh b \to \sqrt{\gamma} \left(1 + \frac{x}{R}\right).
\end{equation}
There is no any special value of $\gamma > 0$ (in contrast to the case of elliptic-parabolic system, 
where contracted variables (\ref{sys_ep_xi}) or (\ref{sys_ep_xi_2}) have some singularity).  Here $\gamma$ is just a scale factor, 
and Beltrami coordinates in the limit $R\to\infty$  go into Cartesian ones:
$x_1 \to x$, $x_2 \to y$.

In this case we have for the symmetry operator: 
\[
- \frac{1}{(\gamma + 1) R^2}\left[S_{HP} - R^2\Delta_{LB}\right]  = \pi_1^2 + \frac{1}{\gamma + 1}
\left[ \{\pi_2, L/R\} - 2L^2\right] \to  p_1^2  =  X_C^2.
\]

\subsection{Semi-circular parabolic to Cartesian}

Write out the semi-circular parabolic coordinates (\ref{sys_scp_xi_eta}) $\eta$ and $\xi$ in form 
\begin{equation}
\eta^2 = \frac{\sqrt{R^2 + u_2^2} + u_2}{u_0 - u_1},
\qquad
\xi^2 = \frac{\sqrt{R^2 + u_2^2} - u_2}{u_0 - u_1}.
\end{equation}
Then in the limit $R \to \infty$ we get
\begin{equation}
\label{PREDEL-SCP-01}
\eta^2 \to 1 + \frac{x+y}{R},
\qquad
\xi^2 \to 1 + \frac{x-y}{R}.
\end{equation}
Correspondingly for symmetry operator 
\be
\label{PREDEL-SCP-02}
\frac{S_{SCP}}{R^2} = \frac{1}{R^2}\left(\{K_1, K_2\} + \{K_2, L\}\right) = 
\{\pi_2,\pi_1\} - \frac{1}{R} \{\pi_1, L\} \to  2 p_2p_1 \sim X_C^2.
\ee
Thus, it can be seen from equation (\ref{PREDEL-SCP-01}) that in limit $R\to \infty$, semi-circular parabolic coordinates $\eta$ and $\xi$ are "entangled" with Cartesian coordinates
$x$, $y$ on euclidean plane $E_{2}$, namely each of the variables $\eta$ and $\xi$ depend directly 
on two coordinates $x$, $y$. To unravel the existing relationship between the contracting semi-circular 
parabolic and Cartesian coordinates we can rotate system (\ref{sys_scp_xi_eta}) about axis $u_0$ through angle $a_3=-\pi/4$ (see (\ref{relation_coords_operators})). It takes the form 
\bea
\label{sys_scp_xi_eta_rot}
u_0^\prime &=& u_0 = R\frac {\left(\eta^2 + \xi^2\right)^2 + 4}{8 \xi \eta},
\nonumber
\\[2mm]
u_1^\prime &=& \frac{u_1 + u_2}{\sqrt{2}} = \frac{R}{\sqrt{2}}
\left(\frac {\left(\eta^2 + \xi^2\right)^2 - 4}{8 \xi \eta} +  \frac {{\eta}^{2}-{\xi}^{2}}{2\xi\eta} \right),
\\[2mm]
u_2^\prime &=& \frac{-u_1 + u_2}{\sqrt{2}} = \frac{R}{\sqrt{2}}
\left( - \frac {\left(\eta^2 + \xi^2\right)^2 - 4}{8 \xi \eta} + \frac {{\eta}^{2}-{\xi}^{2}}{2\xi\eta} \right).
\nonumber
\eea
The corresponding generators of  $SO(2,1)$ group have transformed as
\be
K_1 = \frac{K_2^\prime + K_1^\prime}{\sqrt{2}}, 
\qquad
K_2 = \frac{K_2^\prime - K_1^\prime}{\sqrt{2}}, 
\qquad 
L = L^\prime,
\ee
so that the symmetry operator is given by 
\be
S^\prime_{SCP} = {K_2^\prime}^2 - {K_1^\prime}^2 - \frac{1}{\sqrt{2}}\{K_1^\prime, L^\prime \} 
+  \frac{1}{\sqrt{2}}\{K_2^\prime, L^\prime \}.
\ee
Now from equation (\ref{sys_scp_xi_eta_rot}) we obtain  
\begin{equation}
\eta^2 = \frac{\sqrt{2R^2 + (u^\prime_1 + u^\prime_2)^2} + u^\prime_1  + u^\prime_2}{\sqrt{2} u^\prime_0 - u^\prime_1 +  u^\prime_2}, 
\qquad
\xi^2 =  \frac{\sqrt{2R^2 + (u^\prime_1 + u^\prime_2)^2} - u^\prime_1  - u^\prime_2}{\sqrt{2} u^\prime_0 - u^\prime_1 +  u^\prime_2}.
\end{equation}
Taking the limit $R \to \infty$ we have 
\begin{equation}
\eta^2 \to 1 + \sqrt{2}\frac{x}{R}, 
\qquad
\xi^2 \to 1 - \sqrt{2}\frac{y}{R},
\end{equation}
and hence Beltrami coordinates in limit  $R\to\infty$  go into Cartesian ones: $x_1 \to x$, $x_2 \to  y$.
For the symmetry operator we have 
\be
\frac{S^\prime_{SCP}}{R^2} + \Delta_{LB} = \frac{1}{R^2} \left[ 
{K_2^\prime}^2 - {K_1^\prime}^2 - \frac{1}{\sqrt{2}}\{K_1^\prime, L^\prime \} +  \frac{1}{\sqrt{2}}\{K_2^\prime, L^\prime \}  + \Delta_{LB}\right]
\to  2 p_1^2 \sim X_C^2.
\ee

\section{Contractions on One-Sheeted Hyperboloid}
\label{sec:4}

\subsection{Contraction of Lie algebra \texorpdfstring{$so(2,1)$}{so(2,1)} to \texorpdfstring{$e(1,1)$}{e(1,1)}}

To realize contraction of Lie algebra $so(2,1)$ to $e(1,1)$ let us introduce Beltrami
coordinates on hyperboloid ${\tilde H_2}$ in such a way
\begin{equation}
y_{\mu} = R\frac{u_{\mu}}{u_2} = R\frac{u_{\mu}}{\sqrt{R^2 + u_0^2 - u_1^2}},
\qquad
\mu = 0,1.
\label{Beltrami_1}
\end{equation}
In terms of variables (\ref{Beltrami_1}) generators (\ref{algebra_basis}) look like
\bea
\label{generators_H1}
-\frac{K_1}{R} &\equiv& \pi_0 = \partial_{y_0} - \frac{y_0}{R^2}(y_0\partial_{y_0} + y_1 \partial_{y_1}),
\nonumber\\[3mm]
-K_2 & = & y_1\pi_0 + y_0\pi_1 = y_0\partial_{y_1} + y_1\partial_{y_0},
\\[3mm]
-\frac{L}{R} &\equiv& \pi_1 = \partial_{y_1} + \frac{y_1}{R^2}(y_0\partial_{y_0} + y_1 \partial_{y_1}),
\nonumber
\eea
and commutator relations of $so(2,1)$ take the form
\begin{equation}
    [\pi_0,\pi_1] = -\frac{K_2}{R^2}, 
    \qquad 
    [\pi_0, K_2] = - \pi_1 , 
    \qquad
    [K_2,\pi_1] = \pi_0. \label{comm_o12}
\end{equation}
Let us take the basis of $e(1,1)$ in the form
\begin{equation}
    p_0 = \partial_{y_0}, \qquad p_1 = \partial_{y_1}, \qquad  
    N = y_0\partial_{y_1} + y_1\partial_{y_0},
     \label{basis_e11}
\end{equation}
with commutators
\begin{equation}
    [p_0, p_1] = 0, \qquad [p_0, N] = p_1 , \qquad [N, p_1] = - p_0. 
    \label{comm_e11}
\end{equation}
Then in limit $R^{-1} \to 0$ we have
\begin{equation}
\label{H1_to}
    \pi_0 \to p_0, \qquad \pi_1 \to p_1, \qquad  K_2 \to - N,
\end{equation}
relations (\ref{comm_o12}) contract to (\ref{comm_e11}), so algebra $so(2,1)$ contracts to $e(1,1)$.
Moreover, $so(2,1)$ Laplace-Beltrami operator contracts to the $e(1,1)$ one:
\begin{equation}
    \Delta_{LB} = \frac{1}{R^2}(K_1^2 + K_2^2 - L^2) = \pi_0^2 + \frac{N^2}{R^2} - \pi_1^2
    \to \Delta = p_0^2 - p_1^2.
\end{equation}



\subsection{Contractions for systems of coordinates}
\label{sec:9}

\begin{figure}[htbp] 
\begin{center}
\begin{minipage}[c]{0.5\linewidth}
      \includegraphics[scale=0.3]{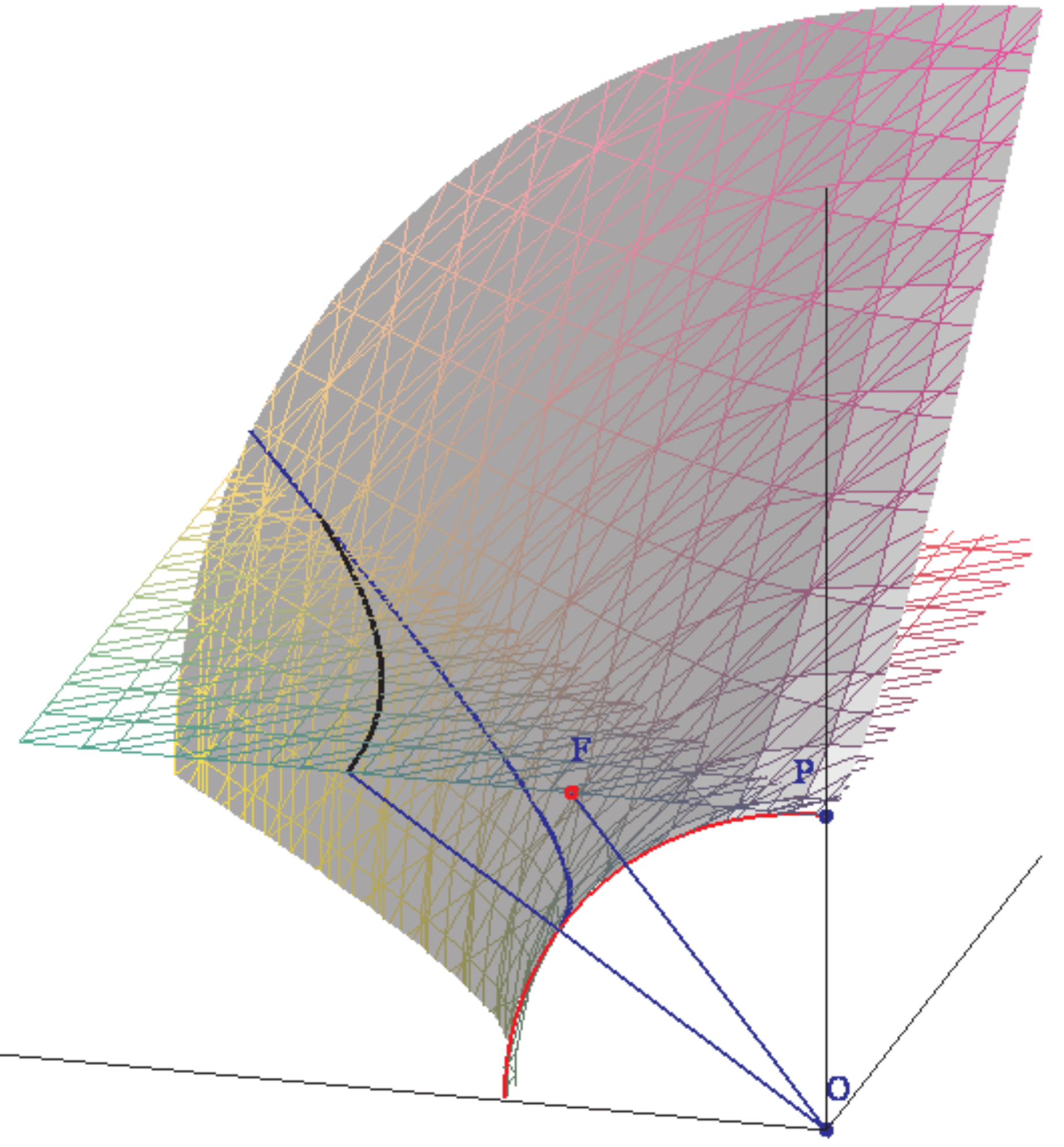}

\begin{center}
\caption{Projective plane for one-sheeted hyperboloid}
\label{fig:36}
\end{center}
    \end{minipage}
\end{center}
\end{figure}

In the geometric sense Beltrami coordinates $(y_0, y_1)$ in equation (\ref{Beltrami_1}) mean a mapping of a point of one-sheeted hyperboloid ${\tilde H_2}$ to the tangent or projection plane $u_2 = R$ by the straight line passing trough the origin of coordinates (see Fig. \ref{fig:36}). Let us note that the diametrically opposite points of ${\tilde H_2}$ are moved to the same point on projective plane.
 
\begin{figure}[htbp] 
\begin{center}
    \begin{minipage}[t]{0.45\linewidth}
      \includegraphics[scale=0.3]{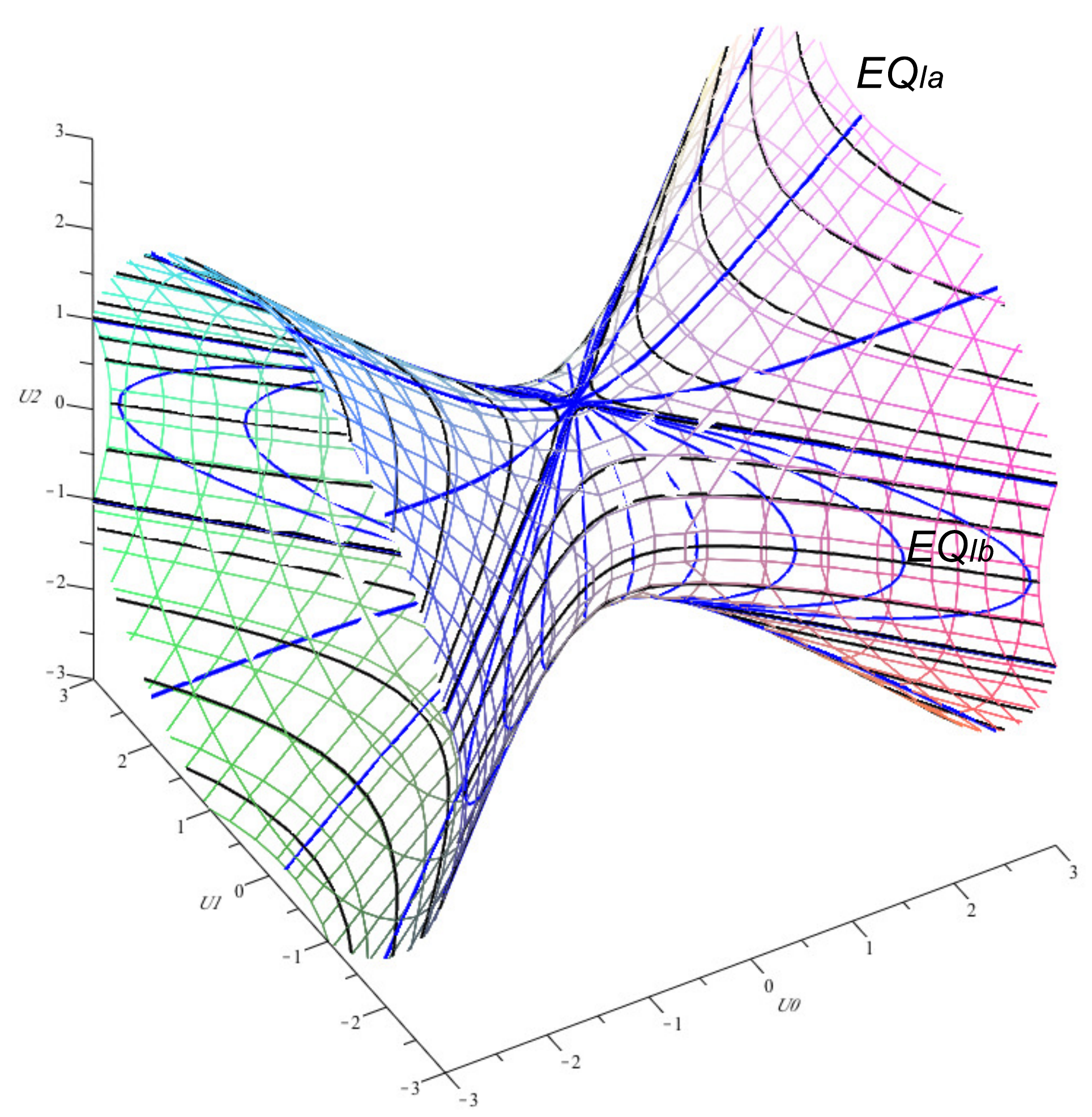}

\begin{center}
\caption{Equidistant systems of Type Ia ($|u_2|\geq R$) and Type Ib ($|u_2|\leq R$)}
\label{fig:21}
\end{center}
    \end{minipage}
    \hfill
    \begin{minipage}[t]{0.45\linewidth}
 \includegraphics[scale=0.35]{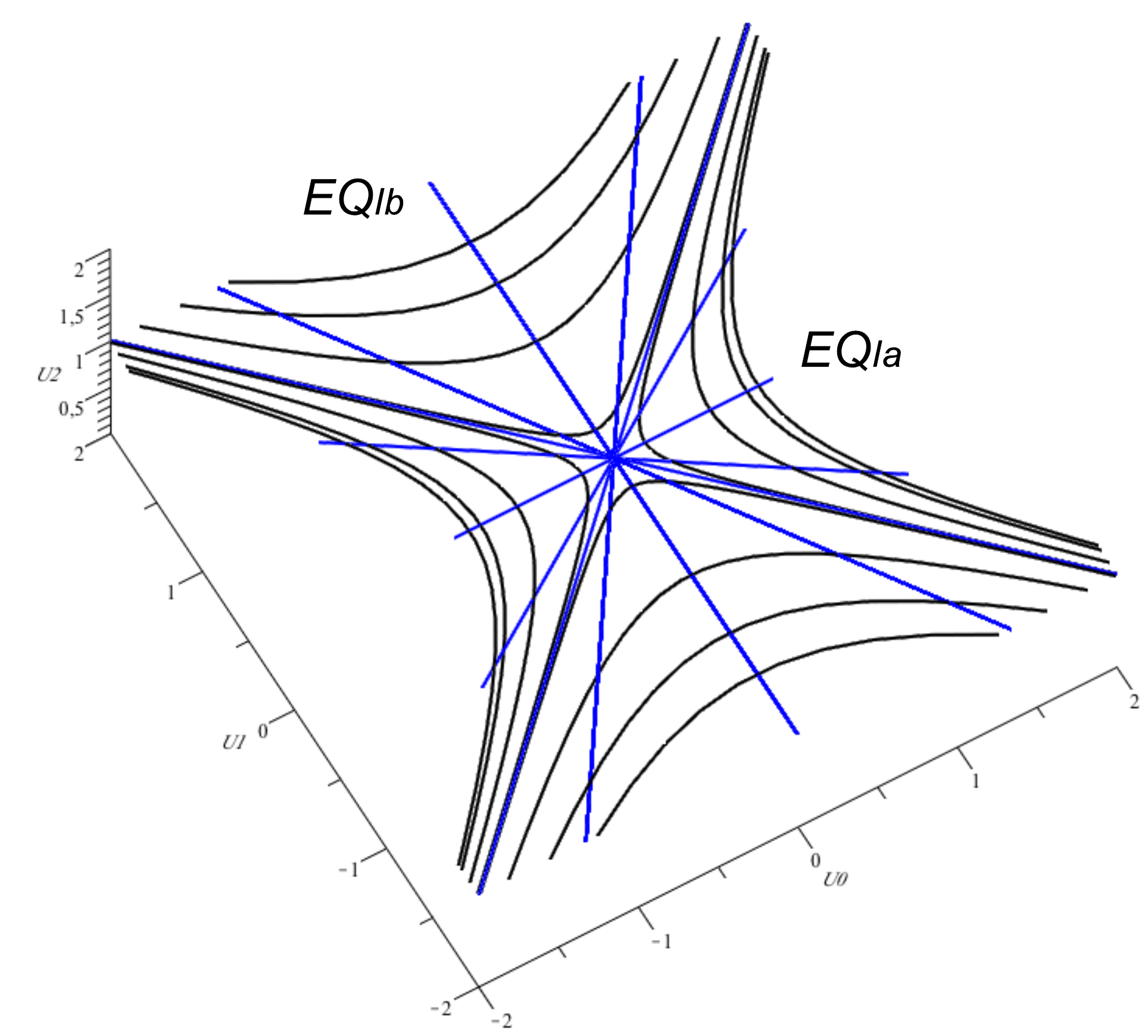}
\caption{Projective plane for equidistant systems of Type Ia and Type Ib}
\label{fig:22}
    \end{minipage}
  \end{center}
\end{figure}

The projective plane, unlike the case of the two-sheeted hyperboloid, is bounded by two hyperbolas  $y_0 = \pm \sqrt{y_1^2 + R^2}$ as
shown in Figure \ref{fig:22}. The asymptotes $|y_1| = |y_0|$ of hyperbola $y_0^2 - y_1^2 = R^2$  divide the projective plane into 
four segment $|y_1| > |y_0|$ and $|y_0| > |y_1|$.  Herewith the parts of hyperboloid with $0 < u_2 \leq R$ and $u_2 \geq R$ are displayed to the 
regions $|y_1| \geq |y_0|$ and $|y_0| \geq |y_1|$ respectively. In contraction limit $R\to\infty$ projective plane $(y_0, y_1)$
transforms into Cartesian coordinates $t$, $x$ ($|t|>|x|$) on pseudo-euclidean space $E_{1,1}$ (or into Cartesian coordinates $\tilde{t}$, $\tilde{x}$, $|\tilde{x}|>|\tilde{t}|$). Simultaneously the image of coordinate system 
on the one-sheeted hyperboloid contracts to the one of the systems on $E_{1,1}$. 
    
To determine the contraction of coordinate system on ${\tilde H_2}$ in explicit form, one needs to define the "geodesic" system. It is easy to see that the equidistant system can be taken as the geodesic one. Indeed, the grid of this system on the projective plane is formed by hyperbolas and straight-lines passing trough the origin (Fig. \ref{fig:22}). Obviously these straight-lines are the geodesic ones and are transformed to geodesic lines on  plane $E_{1,1}$ in contraction limit. We use this fact to determine the asymptotic behavior of independent variables for systems on ${\tilde H_2}$ as $R \to \infty$.
 
The one-sheeted hyperboloid has an interesting property of violence of "symmetry" in contraction limit for equivalent coordinates: different but corresponding to the same operator systems have different contraction limits. It is easy to see for the contractions of $K_1$ and $K_2$ in (\ref{generators_H1}). The reason is that  projective plane $u_2=R$ has no preference with respect to plane $u_1=R$ (in contrast to the case of two-sheeted hyperboloid with projective plane $u_0=R$). Taking plane $u_1=R$ as projective one one can obtain a  different contraction result. This procedure can be considered as a projection on $u_2=R$ of permuted system (see the note after (\ref{relation_coords_operators})). Such a system is obtained by permutation of coordinates $u_1 \leftrightarrow u_2$ and for operators: $K_1 \leftrightarrow K_2$, $L\to -L$.

The metric for coordinates (\ref{Beltrami_1}) of projective plane $(y_0, y_1)$ has the form
\begin{equation}
ds^2 = \left(1+ \frac{y_1^2 - y_0^2}{R^2}\right)^{-2} \left[\left(1+\frac{y_1^2}{R^2}\right) dy_0^2 - 2\frac{y_0 y_1}{R^2} dy_0 dy_1 + \left(\frac{y_0^2}{R^2}-1\right) dy_1^2 \right]
\end{equation}
and in contraction limit it goes to pseudo-euclidean metric $ds^2 = dy_0^2 - dy_1^2 \sim dt^2 - dx^2$.

\subsection{Contractions of nonorthogonal systems}

{\bf 1.} For nonorthogonal pseudo-spherical system (\ref{pspherical_NO}) in contraction limit $R\to \infty$, $\tau \sim -t'/(2R)$, $\varphi \sim - x'/R$, $\alpha\sim R$ we have for Beltrami coordinates:
\be
y_0 = R\frac{\tanh \tau}{\cos(\varphi + R\tau/\alpha)} \sim - \frac{ t' }{2},\qquad  y_1 = -R \tan(\varphi + R\tau/\alpha) \sim \frac{t'}{2} + x',
\ee
where $(t',x')$ are nonorthogonal Cartesian coordinates of Type III (see Tab. \ref{tab:6}). Symmetry operator $S_{SPH} = L $ contracts as follows
\be
- \frac{L}{R} = \pi_1 \to p_1 = X_C^{III}.
\ee

\noindent {\bf 2.} {\bf a.} Considering nonorthogonal equidistant coordinates Ia (\ref{sys_equi_1_NO}) in contraction limit  $\tau_1 \sim r/R$, $\tau_2 \sim \tau$, we have for Beltrami coordinates:
\be
y_0\sim \frac{1}{2} \left(\alpha e^\tau + r^2 e^{-\tau}/\alpha\right), \qquad y_1\sim \frac{1}{2} \left(\alpha e^\tau - r^2 e^{-\tau}/\alpha \right),
\ee
where $(r,\tau)$ are semi-hyperbolic nonorthogonal coordinates ($|t|>|x|$) on $E_{1,1}$ plane (see Table \ref{tab:6}).

{\bf b.} For nonorthogonal equidistant coordinates Ib (\ref{sys_equi_11_NO}) we have $\varphi \sim r/R$ and Beltrami coordinates contract:
\be
y_0\sim \frac{1}{2} \left(\alpha e^\tau - r^2 e^{-\tau}/\alpha\right), \qquad y_1\sim \frac{1}{2} \left(\alpha e^\tau + r^2 e^{-\tau}/\alpha \right),
\ee
where $(r,\tau)$ are semi-hyperbolic nonorthogonal coordinates ($|t|<|x|$).

Symmetry operator contracts as follows
\be
- K_2 \to  N = X_S.
\ee

{\bf c.} For nonorthogonal $EQ$ system of type IIb (\ref{sys_equi_11b_NO}) in contraction limit we get $\varphi \sim t'/(2R)$, $\tau \sim x'/R$, $\alpha \sim R$ and corresponding Beltrami coordinates go like this:
\be
y_0 = R \tanh(\tau + R \varphi/\alpha) \sim x' + \frac{t'}{2},\qquad y_1 = -R \frac{\tan \varphi }{\cosh (\tau + R \varphi/\alpha)} \sim -\frac{t'}{2},
\ee
where $(t', x')$ are ''permuted'' nonorthogonal Cartesian coordinates of Type III. Symmetry operator contracts as follows: $-K_1/R \to p_0 \sim X_C^{III}$. 

Let us note, that for nonorthogonal $EQ$ system of type IIa ($|u_1|>R$) we obtain $|y_0| = R |u_0/u_2| = R |\coth(\tau_2 - g(\tau_1,R))| > R$ for any form of $g$. It means that such a system does not contract to Cartesian III system on $E_{1,1}$ as $R\to \infty$.

\noindent {\bf 3.} Let us consider nonorthogonal horicyclic system (\ref{sys_hor_1_NO}). Taking $\xi \sim t'/R$, $\eta \sim -x'/R$ for corresponding Beltrami coordinates we have:
\begin{equation*}
y_0  \rightarrow x' + t'/4,\qquad
y_1  \rightarrow x'-t'/4,
\end{equation*}
where $(x', t')$ are nonorthogonal Cartesian coordinates of the Type II (see Table \ref{tab:6}). For symmetry operator we obtain: $-S_{EQ}/R = -(K_1 + L)/R = \pi_0 + \pi_1 \to p_0 + p_1 = X_C^{II}$.

\begin{figure}[htbp] 
\begin{center}
\begin{minipage}[c]{0.4\linewidth}
      \includegraphics[scale=0.35]{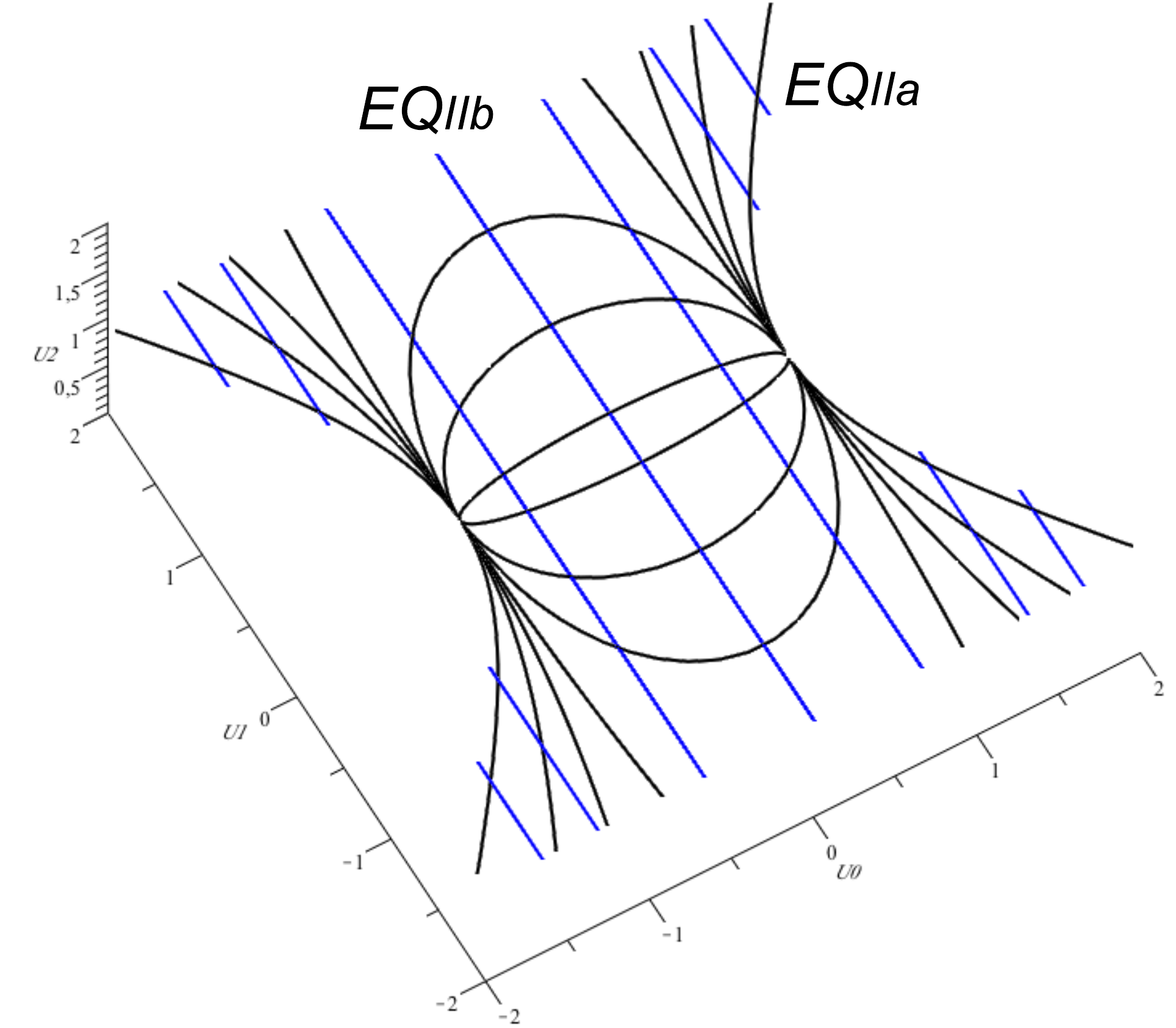}

\begin{center}
\caption{Projective plane for permuted equidistant systems $EQ^*$ of Type IIa and Type IIb}
\label{fig:37}
\end{center}
    \end{minipage}
\end{center}
\end{figure}

\subsection{Equidistant coordinates of Type Ia and Type Ib to pseudo-polar ones}

Let us fix in equidistant coordinate system of Type Ia  (\ref{sys_equi_1}) the geodesic parameter $r= \tau_1 R$  (see Fig. \ref{fig:22}).
Then for the large $R$ angle $\tau_1$ tends to zero and $\tanh\tau_1 \simeq \tau_1 \simeq \frac{r}{R}$. In  contraction limit 
$R \to \infty$ the Beltrami coordinates (\ref{Beltrami_1}) transform accordingly to 
\begin{eqnarray*}
y_0 = R \frac{u_0}{u_2} = R\tanh\tau_1\cosh\tau_2 \rightarrow t = r\cosh\tau_2,
\\[1mm]
y_1 = R \frac{u_1}{u_2} = R\tanh\tau_1\sinh\tau_2 \rightarrow x = r\sinh\tau_2,
\end{eqnarray*}
where variables $(r, \tau_2)$   present the pseudo-polar coordinate system that covers only part 
$|t| > |x|$  of pseudo-euclidean plane $E_{1,1}$ (see Fig. \ref{fig:22}). 

In case of equidistant coordinate system of Type Ib [see equation (\ref{sys_equi_11})] for the fixed geodesic parameter $r= \varphi R$ 
we get for $R^{-1}\to 0$ that $\tan\varphi \simeq \frac{r}{R}$. In contraction limit $R \to \infty$ Beltrami coordinates  
go into pseudo-polar ones: 
\begin{eqnarray*}
y_0 = R \frac{u_0}{u_2} = R\tan\varphi \sinh\tau \rightarrow \tilde{t} = r\sinh\tau,
\\[1mm]
y_1 = R \frac{u_1}{u_2} = R\tan\varphi \cosh\tau \rightarrow \tilde{x} = r\cosh\tau.
\end{eqnarray*}
and cover the remainder part $|\tilde{x}| > |\tilde{t}|$ of plane $E_{1,1}$ (see Fig. \ref{fig:22}). 
For the symmetry operator  we obtain $S_{EQ}^{(2)}= K_2^2 \to N^2 = X_{S}^2 $ (see Table \ref{tab:6}).

\subsection{Equidistant coordinates of Type IIb to Cartesian ones}

Let us use the equivalent  form  of symmetry operator $\bar{S}_{EQ}^{(2)} = K_1^2$.   Two equidistant systems correspond to this operator, namely for  $|u_1| \geq R$  and for $|u_1| \leq R$,
which we call of  Type IIa, IIb  respectively (see Fig. \ref{fig:37}).  The  equidistant  system  of  Type IIb can be obtained from (\ref{sys_equi_11}) by permutation $u_1 \leftrightarrow u_2$ and it looks as follows  
\be
\label{sys_equi_11_rot}
u_0 = R \sin \varphi \sinh \tau,
\quad
u_1 = R \cos \varphi,
\quad
u_2 = R \sin \varphi \cosh \tau.
\ee
Then  for  the  large $R$  we obtain
\be
\cot \varphi = \frac{u_1}{\sqrt{u_2^2 - u_0^2}} \simeq  \frac{x}{R}, 
\qquad 
\tanh\tau = \frac{u_0}{u_2}\simeq  \frac{t}{R}.
\ee
Therefore in contraction limit  $R \to \infty$  Beltrami  coordinates  go  into Cartesian ones:
\be
y_0 =  R\tanh\tau\to t, 
\qquad 
y_1 =  R  \frac{\cot \varphi}{\cosh\tau} \to x.
\ee
For symmetry operator  in the contraction limit we  obtain 
\[
\frac{\bar{S}_{EQ}^{(2)}}{R^2} = \pi_0^2  \to p_0^2 =  X_C^{I}.
\]
As for the equidistant system of Type IIa: 
\be
\label{sys_equi_1_rot}
u_0 = R \sinh \tau_1 \cosh\tau_2,
\quad
u_1 = R \cosh \tau_1,
\quad
u_2 = R \sinh \tau_1 \sinh \tau_2,
\ee
here  we  get  $|y_0|= R |u_0/u_2| = R|\coth\tau_2| > R$,  it means  that  the contraction limit  
$R\to\infty$  does  not  exist  for  this  system of coordinates.

\begin{figure}[htbp]
\begin{center}
    \begin{minipage}[t]{0.45\linewidth}
      \includegraphics[scale=0.35]{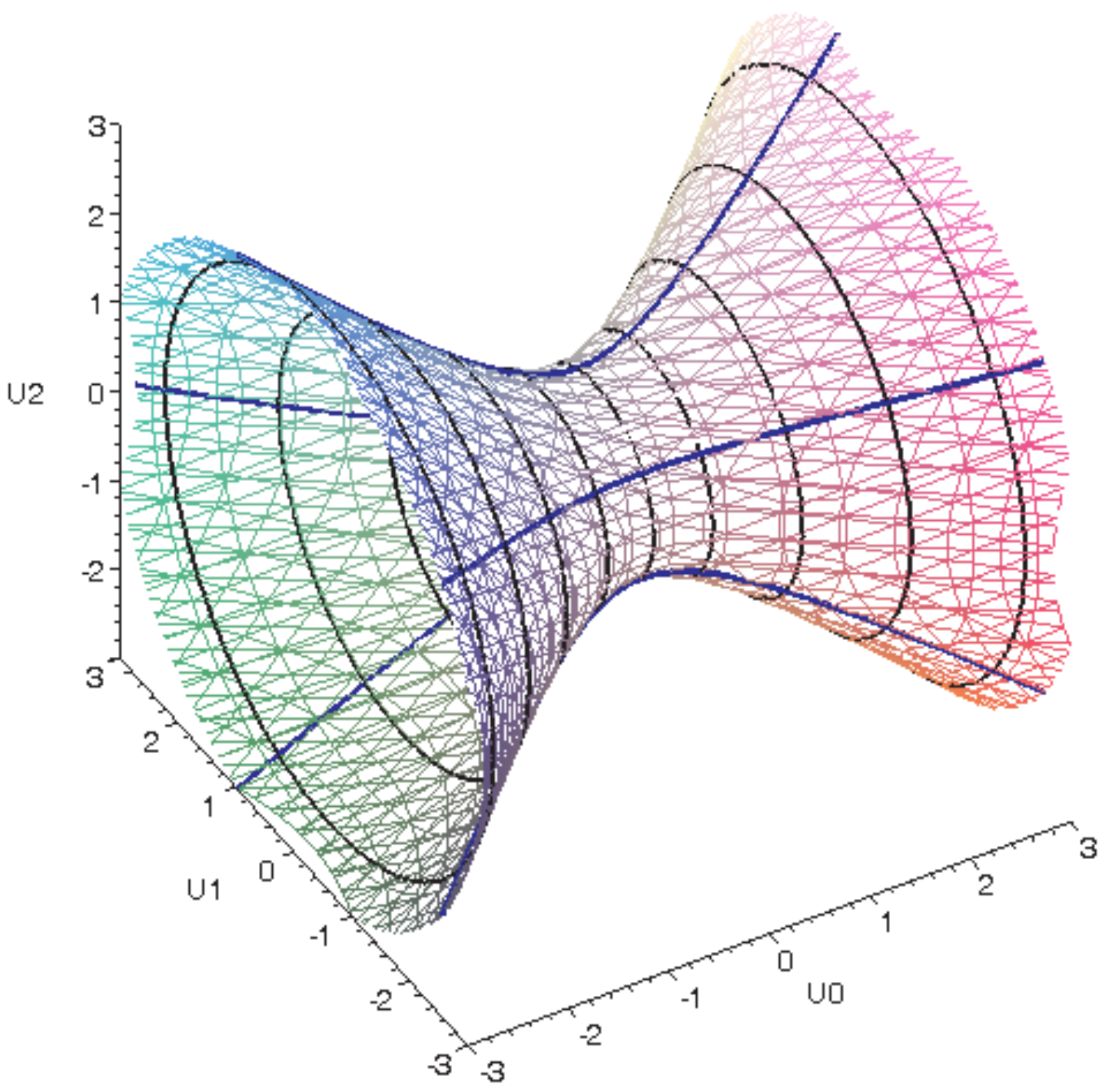}

\begin{center}
\caption{Pseudo-spherical system}
\label{fig:19}
\end{center}
    \end{minipage}
    \hfill
    \begin{minipage}[t]{0.45\linewidth}
 \includegraphics[scale=0.35]{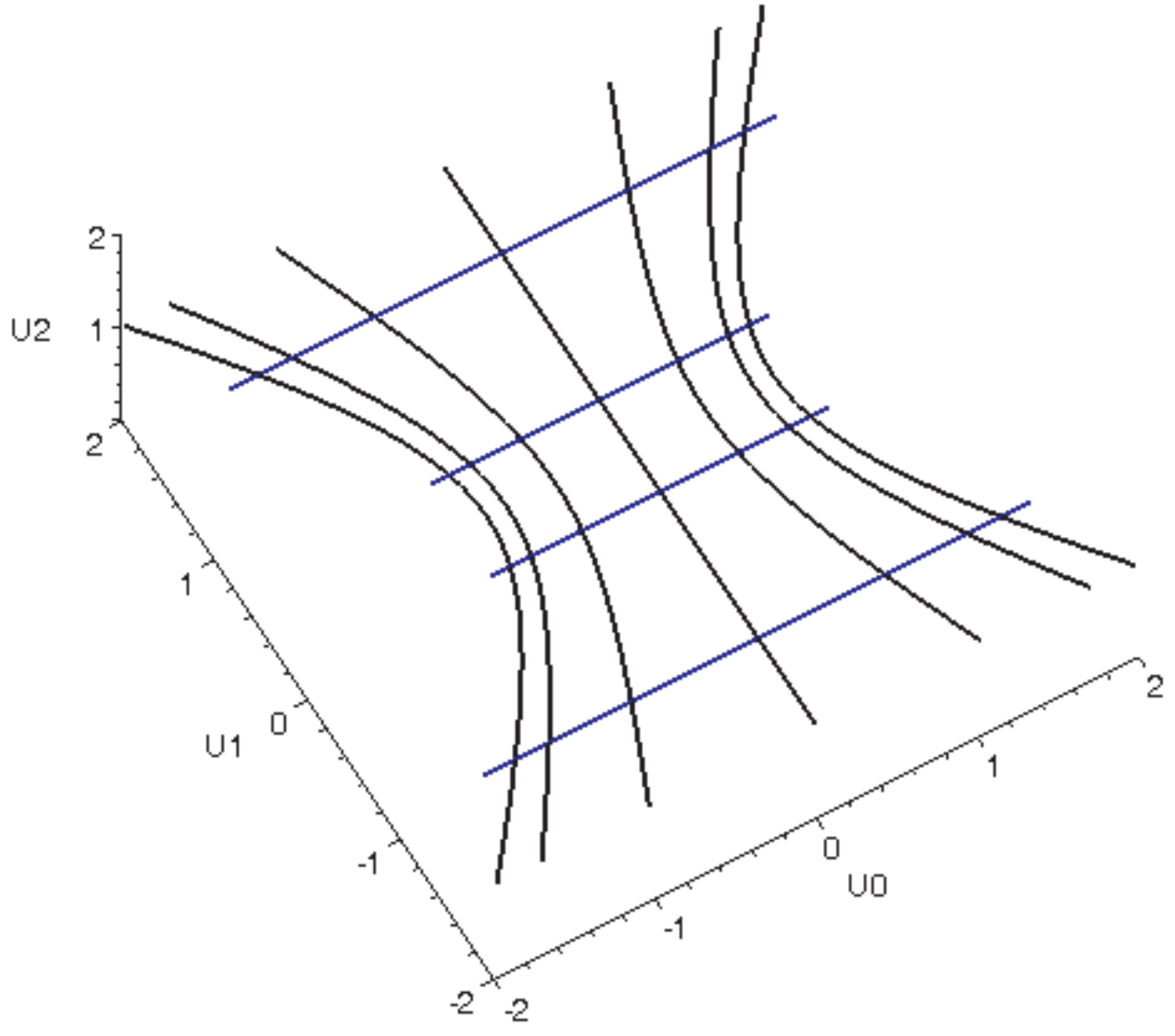}
\caption{Projective plane for pseudo-spherical system}
\label{fig:20}
    \end{minipage}
  \end{center}
\end{figure}

\subsection{Pseudo-spherical coordinates to Cartesian ones}

For pseudo-spherical system of  coordinates (\ref{pspherical})  we have (see Fig. \ref{fig:20})
\begin{eqnarray*}
y_0 = \frac{u_0}{u_2} =  R \frac{\tanh\tau}{\sin\varphi},
\qquad
y_1 =  \frac{u_1}{u_2}  = R\cot\varphi.
\end{eqnarray*}
In limit $R \to \infty$,  we  obtain   $\tanh\tau \simeq t/R$ and  $\cot\varphi \simeq x/R$ 
hence the Beltrami coordinates $(y_0, y_1)$  go into Cartesian ones   $y_0 \to t$,  \, $y_1 \to x$.
In this  case  the symmetry operator takes the form: 
\[
\frac{S_{SPH}^{(2)}}{R^2} = \frac{L^2}{R^2} = \pi_1^2 \to p_1^2  =  X_C^{I}.
\]

\subsection{Horicyclic coordinates to Cartesian ones}
\label{sec:HO_to_C}

\begin{figure}[htbp] 
\begin{center}
\begin{minipage}[c]{0.4\linewidth}
      \includegraphics[scale=0.35]{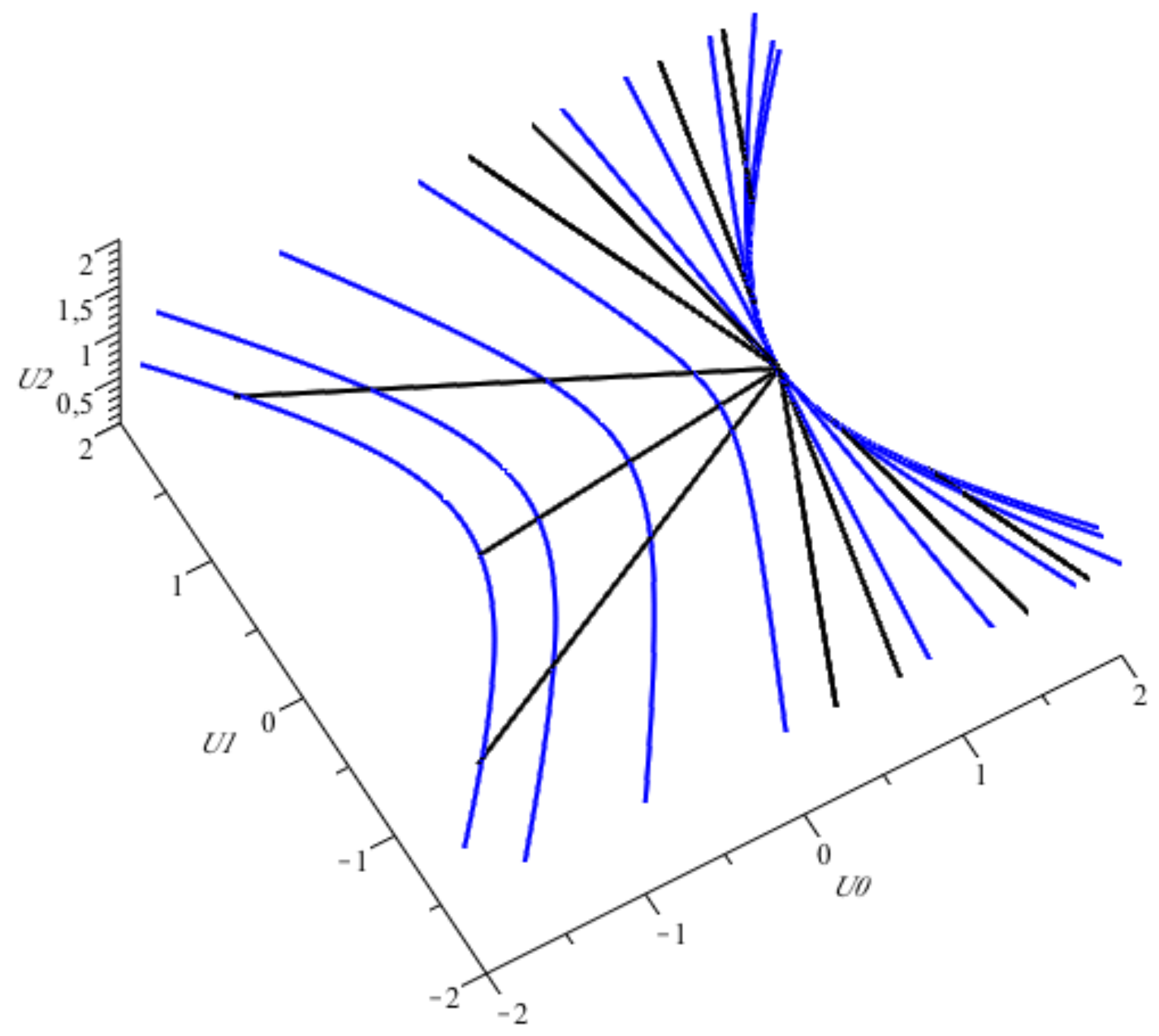}

\begin{center}
\caption{Projective plane for permuted horicyclic system $\bar{S}_{HO}$}
\label{fig:24b}
\end{center}
    \end{minipage}
\end{center}
\end{figure}

We start with horiciclyc coordinates  (\ref{sys_hor_1}) but  interchange  the coordinates 
$u_1 \leftrightarrow u_2$, i.e. put (see Fig. \ref{fig:24b})
\be
\label{HOtoC}
u_0 = R \frac{\tilde{x}^2 - \tilde{y}^2 + 1} {2 \tilde{y} },
\qquad
u_1 = R \frac{\tilde{x}} {\tilde{y} },
\qquad
u_2 = R \frac{\tilde{x}^2 - \tilde{y}^2 - 1} {2 \tilde{y} }.
\ee
Then the corresponding symmetry operator has the form   
\be
\bar{S}_{HO}^{(2)} = K_2^2 - \{K_2, L\} + L^2.
\ee
For  variables $\tilde{x}$, $\tilde{y}$ we  obtain
\begin{equation}
\tilde{x} = \frac{u_1}{u_0 - u_2}, \qquad \tilde{y} = \frac{R}{u_0-u_2}.
\end{equation}
In the limit  $R \to \infty$ we get  $\tilde{x} \to -  \frac{x}{R}$ and  $\tilde{y}  \to  -(1 + \frac{t}{R})$
and Beltrami coordinates go into Cartesian ones 
\bea
y_0 =  R\frac{ \tilde{x}^2 - \tilde{y}^2 + 1 }{ \tilde{x}^2 - \tilde{y}^2 - 1} \to t,
\qquad
y_1 =  2R\frac{\tilde{x}}{\tilde{x}^2 - \tilde{y}^2 - 1} \to x.
\eea
For symmetry operator we have 
\[
\frac{\bar{S}_{HO}^{(2)}}{R^2} = \frac{K_2^2}{R^2} + \frac{1}{R} \{K_2, \pi_1 \} + \pi_1^2  
\to  p_1^2 = X_C^{I}.
\]
Note that contraction for coordinates  (\ref{HOtoC}) is much more simple then for (\ref{sys_hor_1}),  
since in this case there is no entanglement of horiciclic  and Cartesian coordinates for the large $R$.

\subsection{Elliptic coordinates to elliptic I and Cartesian ones}

\begin{figure}[htbp]
\begin{center}
    \begin{minipage}[t]{0.45\linewidth}
      \includegraphics[scale=0.35]{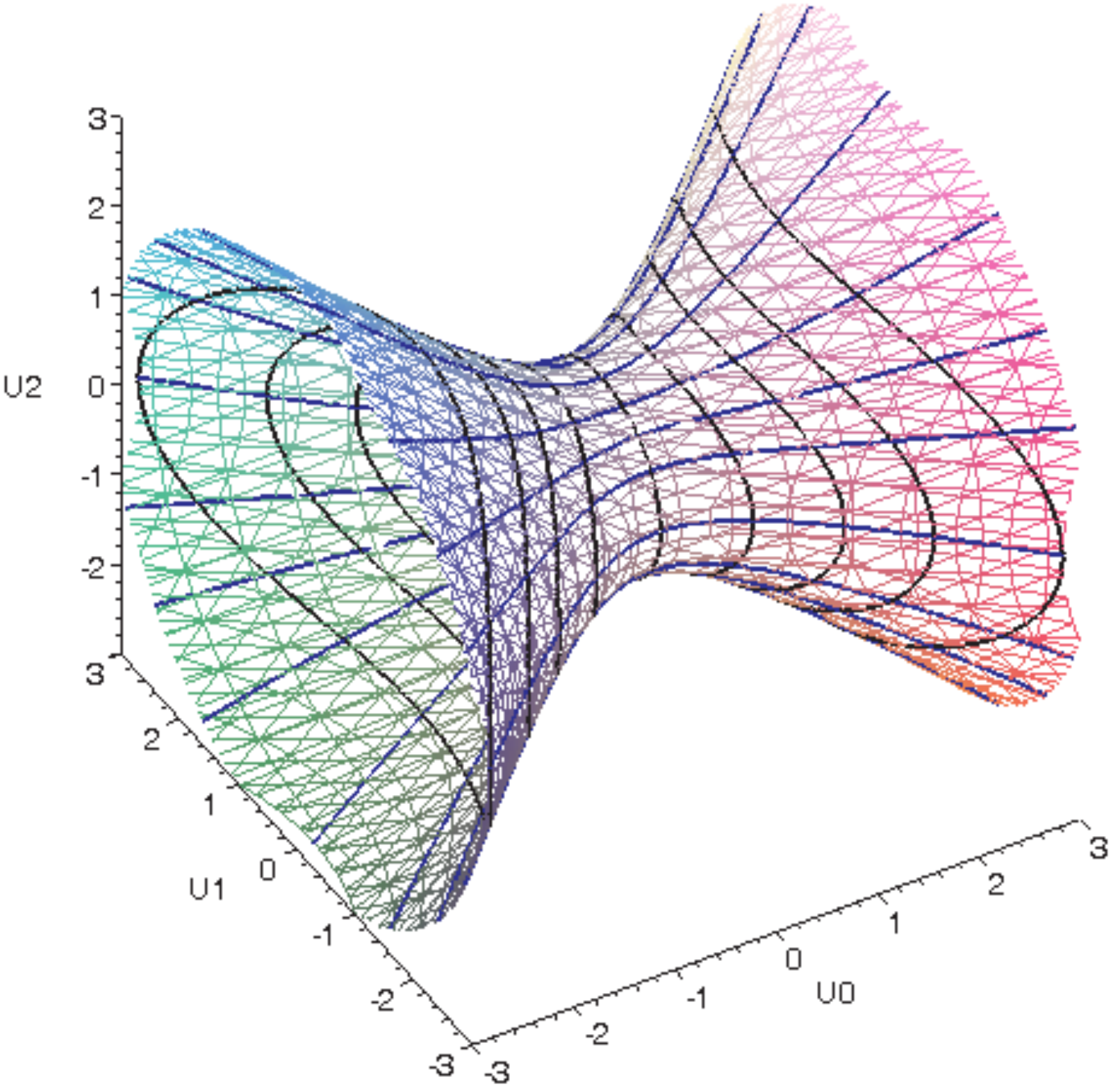}

\begin{center}
\caption{Elliptic system}
\label{fig:32}
\end{center}
    \end{minipage}
    \hfill
    \begin{minipage}[t]{0.45\linewidth}
 \includegraphics[scale=0.35]{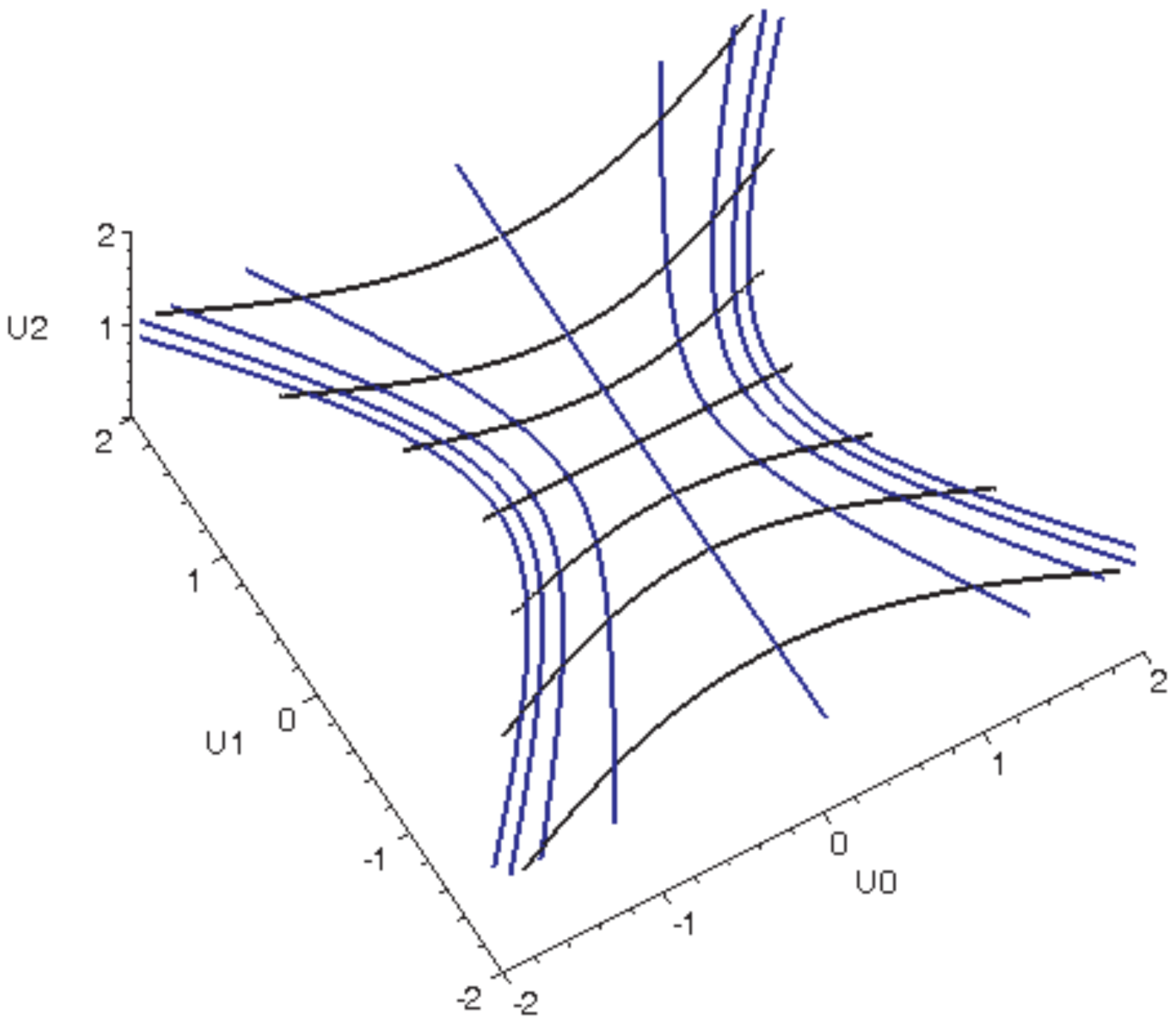}
\caption{Projective plane for elliptic system}
\label{fig:33}
    \end{minipage}
  \end{center}
\end{figure}

Elliptic system (\ref{sys_ell_1}) has three parameters $a_1$, $a_2$ and $a_3$, that define the
position of foci on hyperboloid. On the projective plane the hyperbola foci have coordinates
$F\left(0,R\sqrt{\frac{2\rho_1 - a_2 - a_3}{a_1-\rho_1}}\right)$. Then the minimal focus distance for
corresponding projective hyperbolas is $|FP| = R\sqrt{\frac{a_2-a_3}{a_1-a_2}} = \frac{R}{\sinh\beta}$,
or $\sinh\beta = R/|FP| = \cot \alpha$, where $\alpha$ is angle $FOP$, $|OP| = R$ (see Figs. \ref{fig:32} 
and \ref{fig:33}).
There are two interesting limiting cases in the sense of contractions. First, when the minimal focus distance 
$|FP|$ is fixed, so $\sinh\beta \sim R$ and second, when $\sinh\beta$ is fixed and therefore 
$|FP| \sim R$.

\subsubsection{Elliptic to elliptic}

Let us introduce parameter $D = \sqrt{a_2 - a_3}$, which is the limit for $|FP|$ as $R\to\infty$, 
$a_1 \simeq R^2$ and it represents the minimal focal distance to the focus point for both families of hyperbolas on $E_{1,1}$ plane.

 In new variables $\xi$ and $\eta$:
\begin{eqnarray*}
\sinh^2 \eta = \frac{\rho_1 - a_2}{a_2 - a_3}, \qquad \cosh^2 \xi = \frac{a_2 - \rho_2}{a_2 - a_3},
\end{eqnarray*}
we can rewrite the elliptic system  (\ref{sys_ell_1})  in the  form
\bea
\label{sys_ell1_1}
u_0 &=&  \frac { R D}{\sqrt{a_1 - a_3}} \cosh\eta \sinh\xi,
\nonumber\\[1mm]
u_1 &=& \frac { R D}{\sqrt{a_1 - a_2}} \sinh \eta \cosh\xi ,
\\[1mm]
u_2^2 &=& R^2 \left[1 -  \frac { a_2 -  a_3}{a_1 - a_2} \sinh^2\eta\right]  
\left[1 +  \frac { a_2 -  a_3}{a_1 - a_3} \sinh^2\xi\right] . 
\nonumber
\eea
In contraction limit  $a_1 \simeq R^2  \to  \infty$ and  Beltrami  coordinates  
tend into the elliptic ones of Type I (see Table \ref{tab:6})  on $E_{1,1}$ plane:
\begin{eqnarray*}
y_0 \to t = D \cosh\eta \sinh \xi,
\qquad
y_1 \to x = D \sinh\eta \cosh \xi.
\end{eqnarray*}
For the symmetry operator  in limit  $a_1 \simeq R^2  \to  \infty$  we have
\[
\frac{D^2}{R^2} \, S_E =  \frac{D^2}{R^2} \left[L^2  +   \sinh^2\beta  K_2^2 \right] 
= D^2 \pi_1^2 +  \frac{a_1 - a_2}{R^2}  K_2^2  \to  N^2 +  D^2 p_1^2 = X_{E}^I .
\]

\subsubsection{Elliptic to Cartesian}

Let us  fix  angle $\alpha$ (or the same $\sinh\beta$) and choose  for simplicity  $a_1 - a_2 = a_2 - a_3$.  
Then  $k^2 = k^{\prime{2}} = 1/2$,  $\sinh\beta = \cot\alpha = 1$ and 
we have  $S_E = L^2 + K_2^2$.   From  equation  (\ref{ONE-SHEETED-sys_e})  we get   
\begin{equation}
    - \cn^2 a  = \sqrt{\left(\frac{u_0^2 + u_2^2}{2R^2}\right)^2
    + \frac{u_1^2}{R^2}} -  \frac{u_0^2 + u_2^2}{2R^2},  
		\quad 
		- \cn^2 b   =  \sqrt{\left(\frac{u_0^2 + u_2^2}{2R^2}\right)^2 + \frac{u_1^2}{R^2}} + \frac{u_0^2 + u_2^2}{2R^2} .
\end{equation}
Taking  now  limit  $R\to \infty $ we have  
\begin{equation}
\cn\, a \to \frac{ix}{R}, \qquad \dn\, b \to  -\frac{it}{R},  
\end{equation}
and hence Beltrami  coordinates (\ref{Beltrami_1})  go  into  Cartesian ones :  $y_0 \to t$  and  $y_1 \to x$.
For the  symmetry  operator we have
\[
\frac{S_E}{R^2} = \pi_1^2 + \frac{K_2^2}{R^2} \to p_1^2  = X_C^{I}.
\]

\subsection{Hyperbolic coordinates to elliptic II, III, Cartesian, parabolic I and pseudo-polar ones}

\begin{figure}[htbp]
\begin{center}
    \begin{minipage}[t]{0.45\linewidth}
      \includegraphics[scale=0.23]{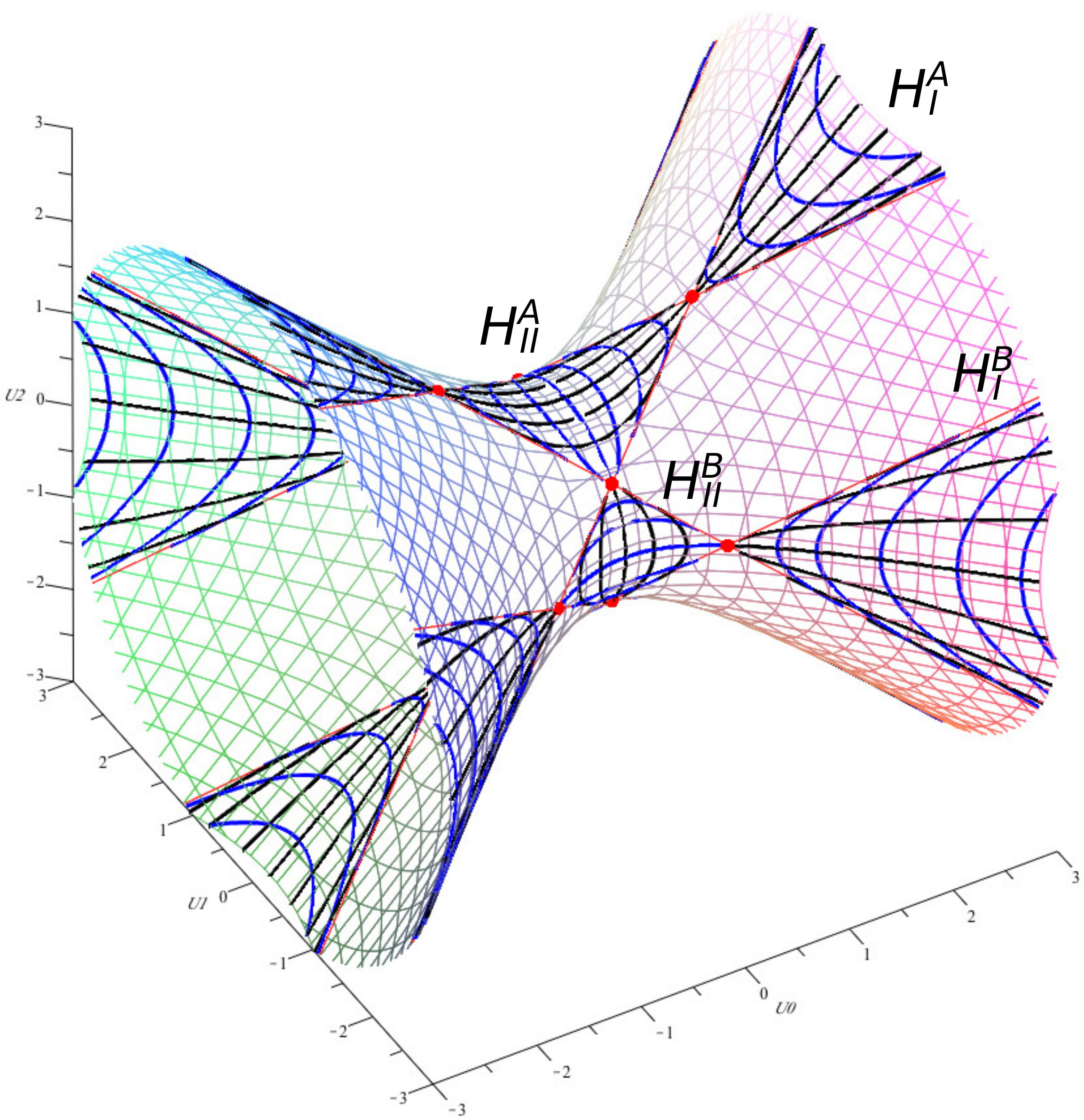}

\begin{center}
\caption{Hyperbolic system}
\label{fig:34}
\end{center}
    \end{minipage}
    \hfill
    \begin{minipage}[t]{0.45\linewidth}
 \includegraphics[scale=0.28]{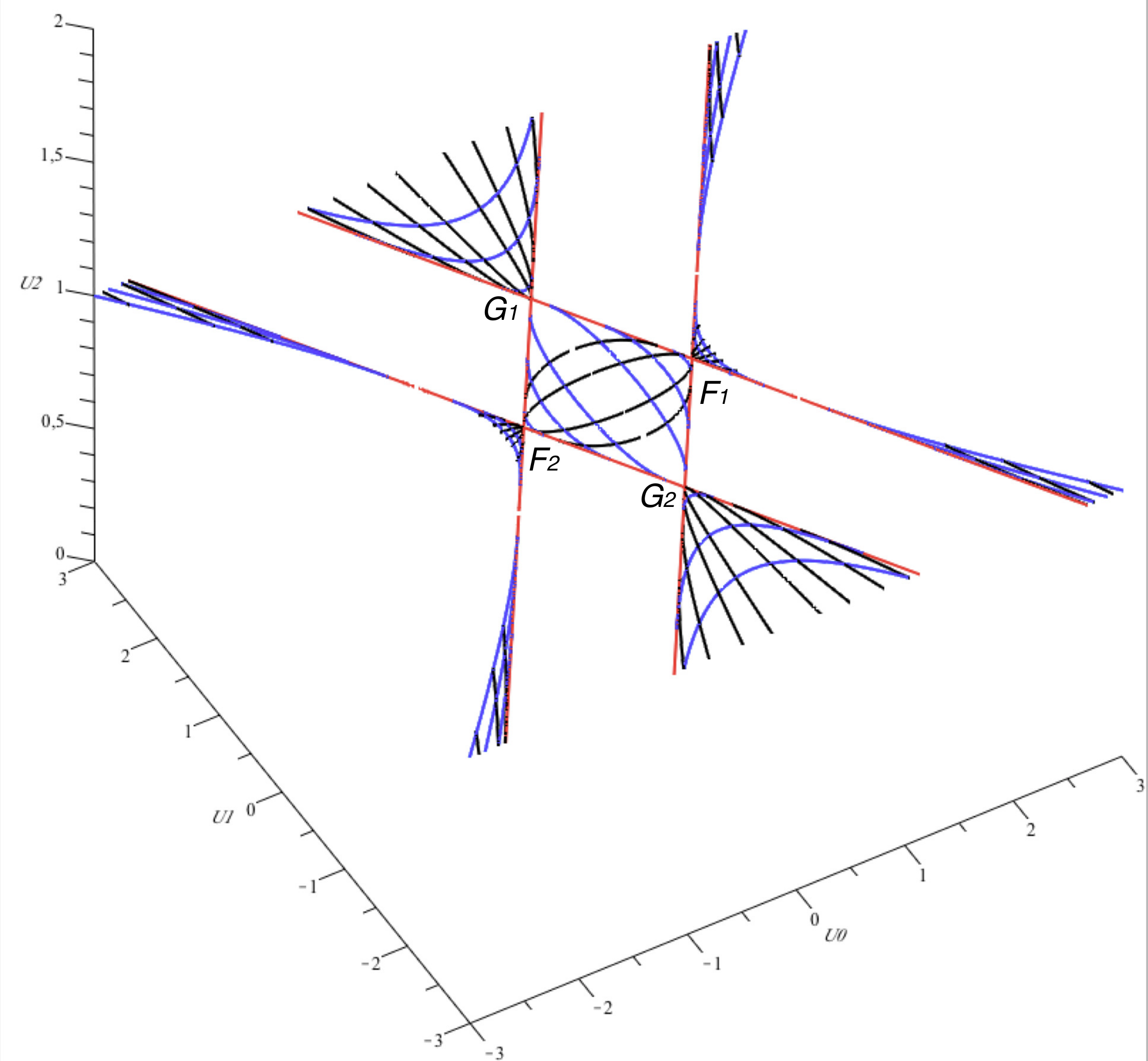}
\caption{Projective plane for hyperbolic system}
\label{fig:35}
    \end{minipage}
  \end{center}
  \end{figure}

The hyperbolic system of  coordinates (\ref{sys_hyp_1})  in algebraic form  is  determined  by three  parameters 
$a_1$, $a_2$ and $a_3$, which define the points of  intersection of  envelopes.  On the projective plane  the  points  have  
coordinates  $F_{1,2}\left(\pm R \sin\alpha, 0\right)$ and  $G_{1,2}\left(0, \pm R \tan \alpha\right)$. 
Distance $|F_1 F_2|$ is equal to $2f = 2 R \sqrt{\frac{a_2-a_3}{a_1-a_3}} = 2 R \sin \alpha$ (see Figs. \ref{fig:34} and \ref{fig:35}).
Therefore  at the contraction limit  $R  \to \infty$ we  must  distinguish  two cases,  namely   when parameter  
$f$  or angle  $\alpha$ are fixed.  

\subsubsection{Hyperbolic to elliptic}

We start with the case  when  $f \sim  R/\sqrt{a_1}$  is  fixed. Consider firstly the  hyperbolic  system of  coordinates  $H_I^A$  
(when $\rho_1$, $\rho_2<a_3<a_2<a_1$).   Introducing  the  new  coordinates $\xi$ and $\eta$ by
\begin{eqnarray*}
    \cosh^2 \eta = \frac{a_2 - \rho_1}{a_2 - a_3},   
		\qquad 
		\cosh^2 \xi = \frac{a_2 - \rho_2}{a_2 - a_3},
\end{eqnarray*}
and taking limit $a_1 \simeq R^2 \to \infty$,  we obtain  for  Beltrami coordinates  (\ref{Beltrami_1})    
\begin{eqnarray*}
    y_0   \to  t = d \cosh\eta \cosh \xi,
    \qquad
    y_1  \to  x = d \sinh\eta \sinh \xi.
\end{eqnarray*}
Here  $(\xi, \eta)$ are the elliptic coordinates  of  Type II(\textit{i}) (see Table \ref{tab:6}) and  $d = \sqrt{a_2 - a_3}$  
is  the  minimal  focus  distance for hyperbolas on $E_{1,1}$ plane. 

In the case of  hyperbolic system $H_{II}^A$  (when $a_3<\rho_1,\rho_2<a_2<a_1$)  using  the  new   coordinates: 
\begin{eqnarray*}
\cos^2 \eta = \frac{a_2 - \rho_1}{a_2 - a_3}, 
\qquad 
\cos^2 \xi = \frac{a_2 - \rho_2}{a_2 - a_3}, 
\end{eqnarray*}
it  is  easy  to  see   that  Beltrami  coordinates  (\ref{Beltrami_1})   
take  the  form  of  the  elliptic  coordinates of  the Type II(\textit{ii})   (see Table \ref{tab:6}) on $E_{1,1}$ plane:
\begin{eqnarray*}
    y_0  \to t = d \cos \eta  \cos \xi,
    \qquad
    y_1  \to x  = d \sin\eta \sin \xi,
\end{eqnarray*}
where $d$ means now  the  maximal  focus  distance  for  ellipses.   For the symmetry  operator  we have
\be
S_H = K_2^2 - \sin^2\alpha L^2 = K_2^2 - {R^2}\sin^2{\alpha}\, \pi_1^2  \to N^2 - d^2 p_1^2 = X_E^{II}.
\ee
Let us note that  elliptic coordinates of Types II(\textit{i}) and II(\textit{ii}) do not cover completely pseudo-euclidean plane 
$E_{1,1}$ (see Table \ref{tab:6} and \cite{Kalnins1}).

\subsubsection{Hyperbolic to Cartesian}

Let us fix angle $\alpha$ and  choose  $a_1 - a_2 = a_2 - a_3$ for the simplicity, then  $\sin \alpha = 1/\sqrt{2}$  and  $f = R/\sqrt{2} \to \infty$, when $R\to\infty$. As one can see from Figs. \ref{fig:34}, \ref{fig:35} only system $H^A_{II}$ on projective plane covers the origin of coordinates, therefore we proceed to contract this system with $u_2 = R\, \dn\,a \,\sn\,b \geq R$. From equation (\ref{TYPEII-02})  we obtain (considering $ - \cn^2 b \geq \cn^2 a$):
\begin{equation}
\cn^2 a  =  \frac{u_2^2 - u_1^2}{2R^2} - \sqrt{\left(\frac{u_2^2 - u_1^2}{2R^2}\right)^2
    - \frac{u_0^2}{R^2}},
\qquad
-  \cn^2 b  =  \frac{u_2^2 - u_1^2}{2R^2} + \sqrt{\left(\frac{u_2^2 - u_1^2}{2R^2}\right)^2
    - \frac{u_0^2}{R^2}}. 
\end{equation}
Therefore  for  the large  $R$  we have
\begin{equation}
    \cn\, a \to \frac{t}{R}, 
		\qquad 
		\dn\, b \to  \frac{x}{R},
\end{equation}
and Beltrami coordinates contract to Cartesian ones $y_0 \to t$ and  $y_1 \to x$.
In  this  case  the  symmetry  operator  takes  the form 
\[
-  \frac{2}{R^2}  S_H  =  -  \frac{2}{R^2}  \left(K_2^2 - \sin^2\alpha L^2\right) =  
\pi_1^2  - \frac{2K_2^2}{R^2} \to  p_1^2  =  X_C^{I}.
\]

\subsubsection{Rotated hyperbolic to parabolic I}
\begin{figure}[htbp]
\begin{center}
    \begin{minipage}[t]{0.45\linewidth}
      \includegraphics[scale=0.25]{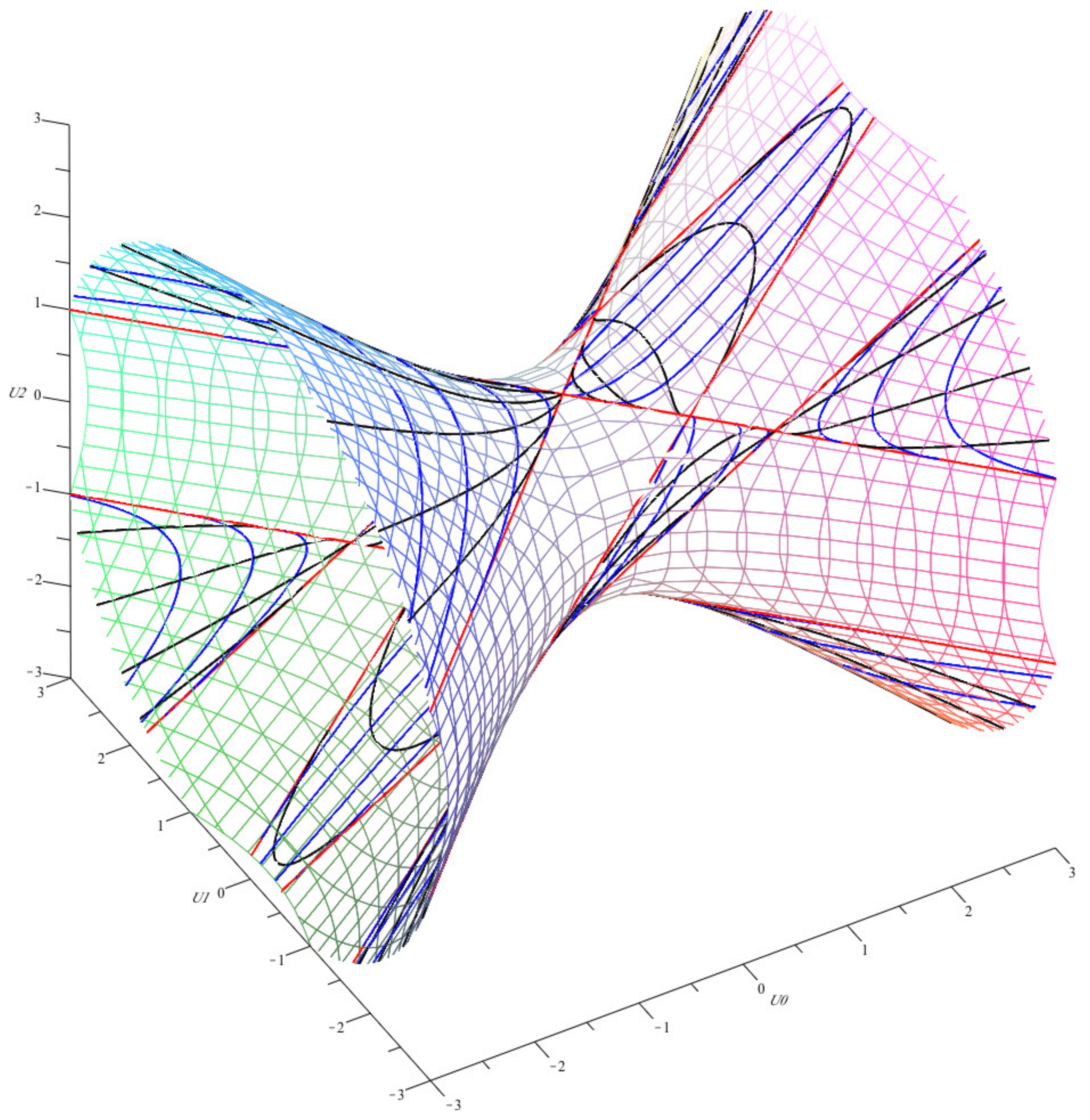}

\begin{center}
\caption{Rotated hyperbolic system}
\label{fig:34a}
\end{center}
    \end{minipage}
    \hfill
    \begin{minipage}[t]{0.45\linewidth}
 \includegraphics[scale=0.32]{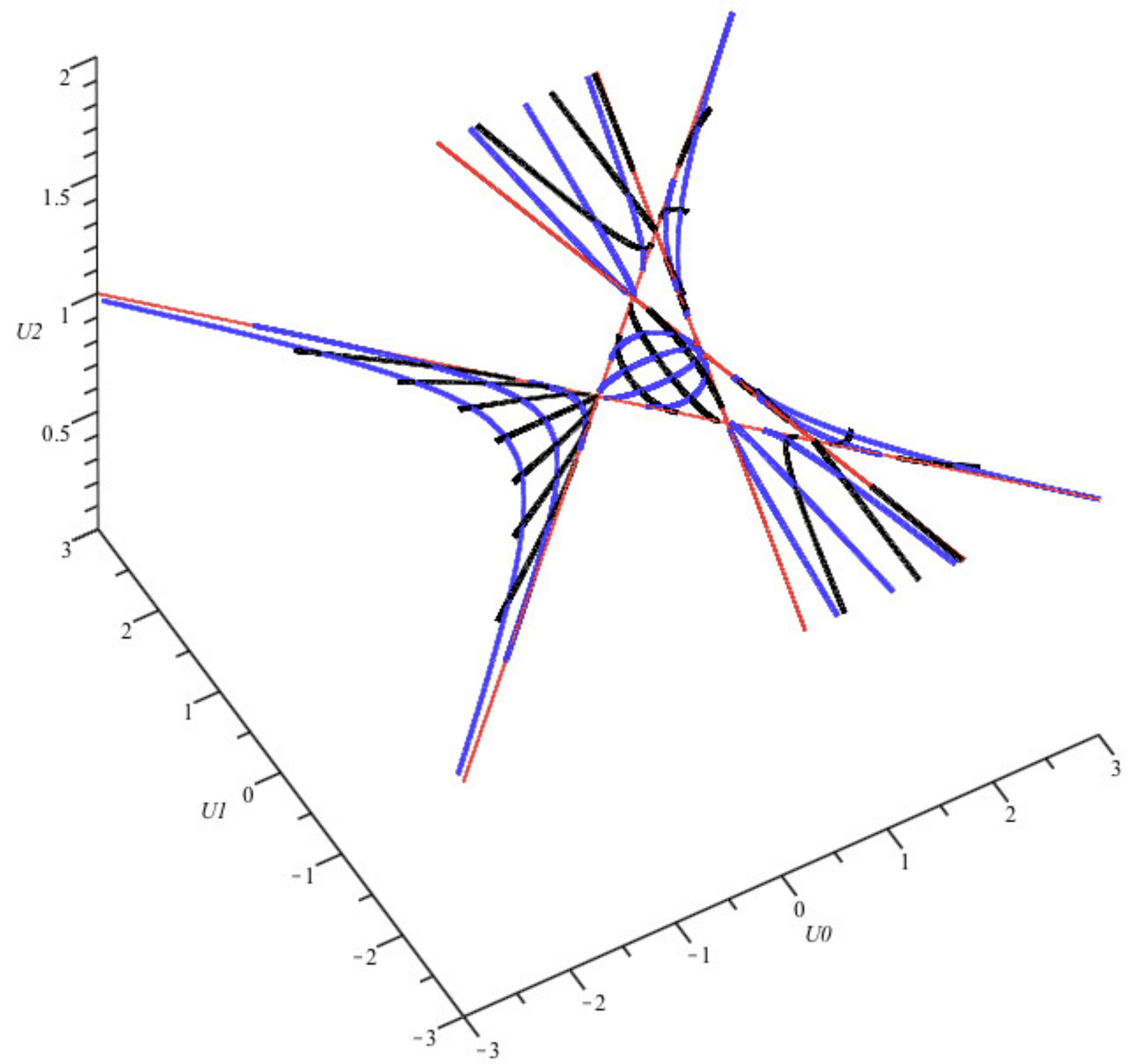}
\caption{Projective plane for rotated hyperbolic system}
\label{fig:35a}
    \end{minipage}
  \end{center}
\end{figure}

The principal difference between hyperbolic system and the rotated one is a position of the limit points 
(points of intersections of envelopes). For the rotated hyperbolic system one of the limit points is fixed in $(0,0,R)$ 
(see Fig. \ref{fig:34a} and Fig. \ref{fig:35a}) and it does not depend on parameter $\alpha$ that plays the critical part for the contraction.

Let us  fix parameter $k$ (or  parameter $k'$).  Then  for the large  $R$  we obtain 
\begin{equation}
\label{LIMIT-PAR-01}
\sn(a,k) \to  \frac{i v}{\sqrt{kR}},  
\qquad 
\dn(b,k^\prime) \to i u \sqrt{\frac{k}{R}},
\end{equation}
where we take into account the following expressions 
\begin{eqnarray*}
\label{rho_1rho_2_h_rotated}
  k^2  \sn^2(a,k) = \frac{1}{2R^2}\left\{(u_1^\prime)^2 k^2 - (ku_0^\prime+u_2^\prime)^2 + R^2
	\sqrt{\left[(u_1^\prime)^2 k^2 - (ku_0^\prime+u_2^\prime)^2 + R^2\right]^2 - 4R^2 k^2(u_1^\prime)^2}\right\},
	\\
  \dn^2(b,k^\prime) = \frac{1}{2R^2}\left\{(u_1^\prime)^2 k^2 - (ku_0^\prime+u_2^\prime)^2 + R^2
	- \sqrt{\left[(u_1^\prime)^2 k^2 - (ku_0^\prime+u_2^\prime)^2 + R^2\right]^2 - 4R^2 k^2(u_1^\prime)^2}\right\},
\end{eqnarray*}
and $v, u$  are  the  parabolic  coordinates  (see Table \ref{tab:6}).   Here  we  use  the  coordinate  system   for  region  $H_I^A$  
(with  $u_2 \geq R$, $u_0 > 0$):
\begin{equation}
\label{TYPEII-002}
u_0 =  R \cn(a,k) \, i \cn(b, k^\prime), 
\qquad
u_1=  R\, i \sn(a,k)\, i \dn(b, k^\prime),
\qquad
u_2=   R\dn(a,k) \sn(b, k^\prime),
\end{equation}
which  after  hyperbolic rotation   (\ref{TYPEII-03}) has  the  form:   
\bea
\label{H_IA_rotated}
u_0^\prime &=& \frac{R}{k^\prime}\left[ - k \dn(a,k)\sn(b, k^\prime)  +  i \cn(a,k) \cn(b, k^\prime)\right], 
\nonumber\\[2mm]
u_1^\prime &=&  R\, i \sn(a,k) \, i \dn(b, k^\prime),
\\[2mm]
u_2^\prime &=&  \frac{R}{k^\prime}\left[ - k \cn(a,k)\, i\cn(b, k^\prime) + \dn(a,k) \sn(b, k^\prime)\right].
\nonumber
\eea
Finally,  in contraction limit  $R\to \infty$  the   rotated  hyperbolic  system  (\ref{H_IA_rotated})   goes into the following one
\begin{equation}
\label{PARAB-01}
    y_0  \to t = \frac{u^2+v^2}{2},  
    \quad
    y_1 \to  x = uv,
\end{equation}
which  coincides  with  Parabolic I  system of coordinates (see Table \ref{tab:6}). For the symmetry operator we obtain  ($c = k + 1/k$):
\[
\frac{S_{\tilde{H}}}{R} = \frac{1}{R} \left( c K_2^2 +  \{K_2, L\} \right) \to  \{N, p_1\} = X_P^I.
\]
Let us note that parabolic coordinates  (\ref{PARAB-01})  cover only part  of  $t > 0$ and $|t| > |x|$.  To describe the  
missing part  of  the  pseudo-euclidean  plane  $t < 0$  it is enough  to  change  $u_0 \to - u_0$  in  rotated  system  (\ref{H_IA_rotated}). 

\subsubsection{Rotated hyperbolic to pseudo-polar}

Let us take $ k \sim R^{-2}$,  then $k^\prime \sim 1$ and $c\sim R^2$. Using  the system  of  coordinates (\ref{H_IA_rotated}),  for the  large  $R$, 
instead of  (\ref{LIMIT-PAR-01})  we get 
\begin{equation}
\label{LIMIT-PAR-02}
    \sn(a,k) \to i \sinh\tau_2, \quad \cn(b,k^\prime) \to  -  i \frac{r}{R},
\end{equation}
and  the  rotated  hyperbolic system goes  into  pseudo-polar one
\begin{equation}
    y_0 \to t  =  r \cosh \tau_2,
    \quad
    y_1 \to x  =  r \sinh \tau_2.
\end{equation}
For the  symmetry  operator  we  have
\[
\frac{S_{\tilde{H}}}{R^2} = \frac{c}{R^2} K_2^2 +   \frac{1}{R^2} \{ K_2,  L \}   \to   N^2  =  X_S^2.
\]
Finally let us note that the interchange of coordinates $u_1 \leftrightarrow u_2$ does not lead to new contractions 
of hyperbolic system of coordinates.

\begin{figure}[htbp]
\begin{center}
    \begin{minipage}[t]{0.45\linewidth}
      \includegraphics[scale=0.3]{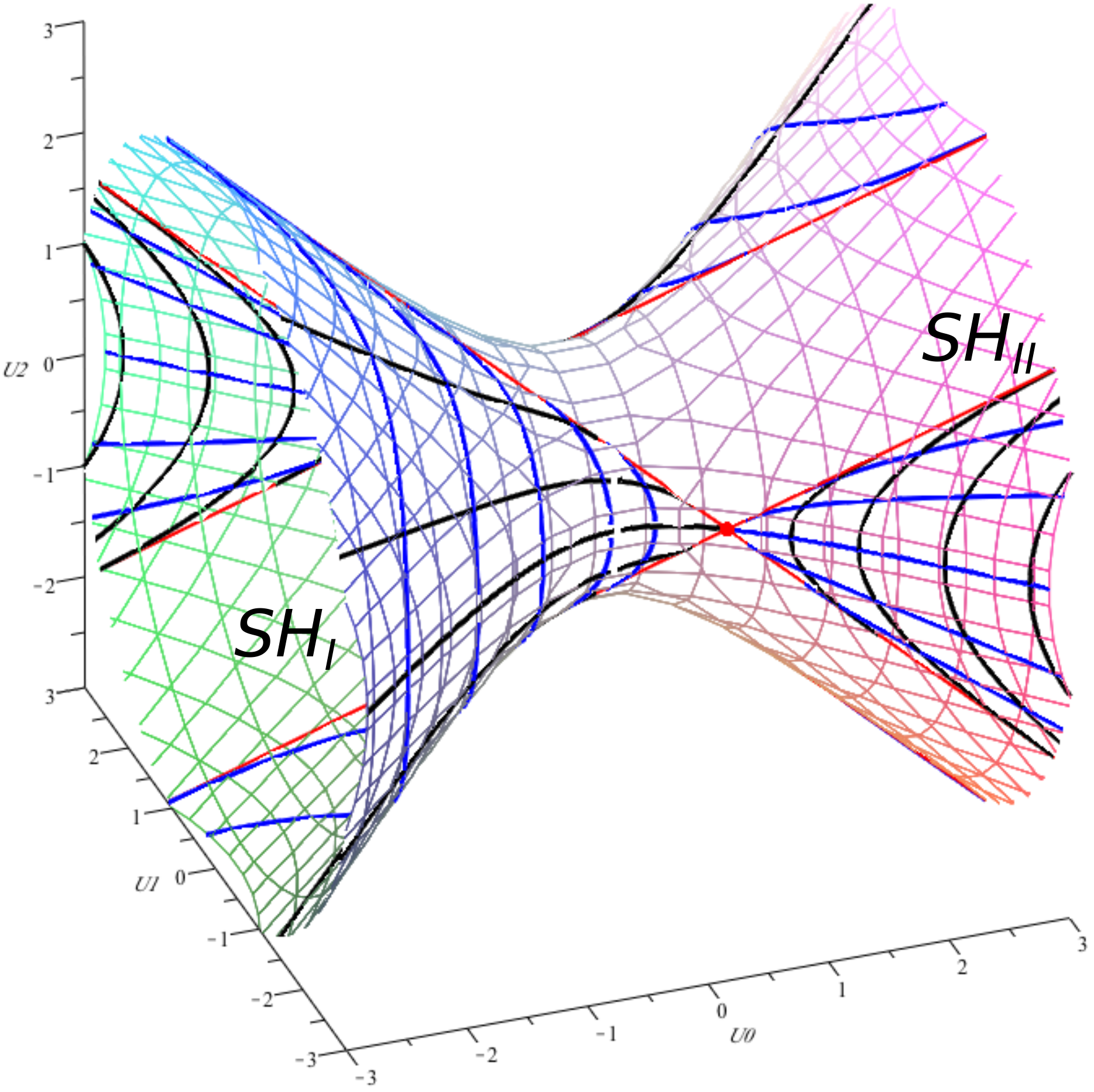}

\begin{center}
\caption{Semi-hyperbolic system 
}
\label{fig:24}
\end{center}
    \end{minipage}
    \hfill
    \begin{minipage}[t]{0.45\linewidth}
 \includegraphics[scale=0.35]{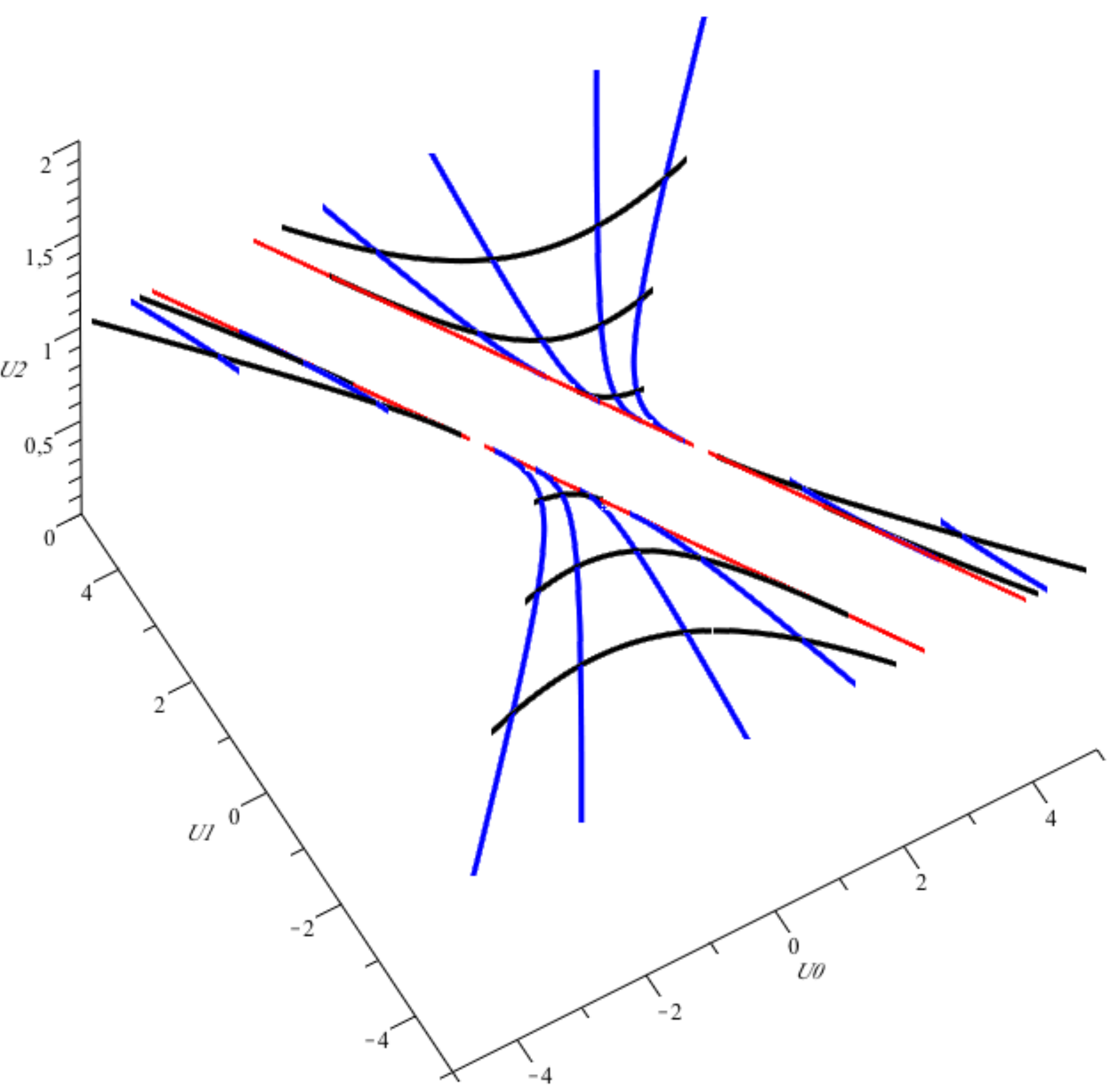}
\caption{Projective plane for semi-hyperbolic system}
\label{fig:25}
    \end{minipage}
  \end{center}
\end{figure}

\subsection{Semi-hyperbolic coordinates to Cartesian, Hyperbolic I and Parabolic I ones}

As we have shown  in paragraph  \ref{SH-system} semi-hyperbolic  system  of  coordinates  (\ref{sys_sh_hyp})  depends   
on the  dimensionless  parameter $c = \sinh 2\beta$.  It  does not cover the whole surface of  one-sheeted hyperboloid  
(see (\ref{01-cover})).  The parameter  $c$, as  shown  in Fig. \ref{fig:24} and Fig. \ref{fig:25}, 
splits semi-hyperbolic  system  of coordinates  into  two  parts  with  $\sinh\tau_1, \sinh\tau_2 \leq c$ ($SH_{I}$)  
and  $\sinh\tau_1, \sinh\tau_2 \geq c$ ($SH_{II}$).

\subsubsection{Semi-hyperbolic to hyperbolic}
\label{subsec:sh_to_C}

For coordinates  $\xi_{1,2} = \sinh\tau_{1,2}$ we have (assuming $\sinh \tau_1 > \sinh \tau_2$)
\begin{equation}
\label{xi12_SH}
\xi_{1,2} = \frac{u_0u_1}{R^2} + \frac{c}{2}\frac{u_1^2 - u_0^2}{R^2} \pm
\sqrt{\left\{\frac{u_0u_1}{R^2} + \frac{c}{2}\frac{u_1^2 - u_0^2}{R^2}\right\}^2 -
\frac{u_2^2 + 2c u_0 u_1}{R^2}}.
\end{equation}
Let  us  take now  $c = 2 R^2/l^2$  ($l$ is some constant). We  will  only  consider here system  $SH_{I}$ 
(for  $SH_{II}$ angles $\tau_{1,2} \to \infty$). Equation (\ref{xi12_SH}) denotes the limiting procedure.  
Indeed we obtain  $ \xi_{1,2} \to \sinh \zeta_{1,2}$ as $R \to \infty$ and for Beltrmi coordinates we get  
\begin{equation}
y_0 \to t = \frac{l}{2}\left[\cosh\frac{\zeta_1 - \zeta_2}{2} - \sinh\frac{\zeta_1 + \zeta_2}{2}\right],
\qquad
y_1 \to  x = \frac{l}{2}\left[\cosh\frac{\zeta_1 - \zeta_2}{2} + \sinh\frac{\zeta_1 + \zeta_2}{2}\right],
\end{equation}
i.e the  semi-hyperbolic coordinate  contract  into  hyperbolic coordinates of Type I  (see Table \ref{tab:6}). 
For the corresponding symmetry operator we get 
\[
\frac{l^2}{2 R^2} S_{SH} =  K_2^2  + \frac{l^2}{2} \{\pi_0, \pi_1\} \to N^2 +  l^2 p_0 p_1 = X_H^I.
\]

\subsubsection{Semi-hyperbolic coordinates to Cartesian and Parabolic I ones}

Let us  fix  the value of $\sinh2\beta$. Then symmetry operator $S_{SH}$ contracts into Cartesian one
\[
\frac{S_{SH}}{2R^2} =  \frac{\sinh2\beta}{2R^2} K_2^2  +  \frac{1}{2}\{\pi_0, \pi_1\} \to  p_0 p_1 = X_C^{I}.
\]
In spite of this, it is easy to see that the origin of coordinates on projective plane is not covered by $SH$ system and for large $R$ the radical expression in formula (\ref{xi12_SH}) 
becomes negative (because of $u_2 \sim R$). Hence, the semi-hyperbolic coordinates in form (\ref{sys_sh_hyp}) do  
not contract to Cartesian ones on $E_{1,1}$ plane. 

{\bf 1.} Let us consider  system (\ref{sys_sh_hyp})  but now we interchange coordinates   $u_1 \leftrightarrow u_2$. 
Then new symmetry operator  is  $\bar{S}_{SH} = \sinh2\beta K_1^2  -  \{K_2,L\}$.
We assume that parameter $\sinh 2\beta > 0$.  In  contraction limit  $R\to\infty$ we have 
\[
\frac{\bar{S}_{SH}}{ \sinh2\beta R^2} \to  \ p_0^2  \simeq X_C^I.
\]
Further,  from  equation  (\ref{xi12_SH}), taking $u_1 \leftrightarrow u_2$,  we obtain  for  variables  $\xi_{1,2} = \sinh\tau_{1,2}$: 
\begin{equation}
\xi_{1,2} = \frac{u_0u_2}{R^2} + \frac{c}{2}\frac{u_2^2 - u_0^2}{R^2} \pm \sqrt{\left\{\frac{u_0u_2}{R^2}
+ \frac{c}{2}\frac{u_2^2 - u_0^2}{R^2}\right\}^2 - \frac{u_1^2 + 2c u_0 u_2}{R^2}}.
\label{SH_coordinates-01}
\end{equation}
Taking limit  $R\to \infty$  in  (\ref{SH_coordinates-01})  we get  that   
\begin{equation}
\xi_1 \to c - \frac{c^2+1}{c}\frac{x^2}{R^2},
\qquad
\xi_2 \to  2\frac{t}{R}, 
\label{SH_contraction_limit}
\end{equation}
and  hence $y_0 \to t$ and $y_1 \to x$, i.e.  the  permuted semi-hyperbolic coordinates  $SH$ contract  into 
Cartesian ones on pseudo-euclidean plane $E_{1,1}$. Note that one needs to take $SH_I$ or $SH_{II}$ depending on the sign of $c$ to cover the origin of coordinates on projective plane.

\vspace{0.3cm}

{\bf 2.} Consider now the case  when  $\sinh 2\beta = 0$.   Then  in  contraction  limit  $R\to\infty$  we have
\[
- \frac{\bar{S}_{SH}}{R} = - \{K_2,\pi_1\}  \to \{N, p_1\} = X_P^I,
\]
which  corresponds to  the  parabolic  coordinates  of  Type I   in  $E_{1,1}$  plane.
From  equation  (\ref{SH_coordinates-01})  for  the  large  $R$  we  get   
\begin{equation}
\xi_1 \to \frac{u^2}{R}, 
\qquad 
\xi_2 \to \frac{v^2}{R},
\end{equation}
and Beltrami coordinates   go  into  parabolic  ones  of  Type I  (see Table \ref{tab:6}):
\[
y_0 \to t = \frac{u^2 + v^2}{2},
\qquad
y_1 \to x = uv.
\]

\begin{figure}[htbp]
\begin{center}
    \begin{minipage}[t]{0.45\linewidth}
      \includegraphics[scale=0.35]{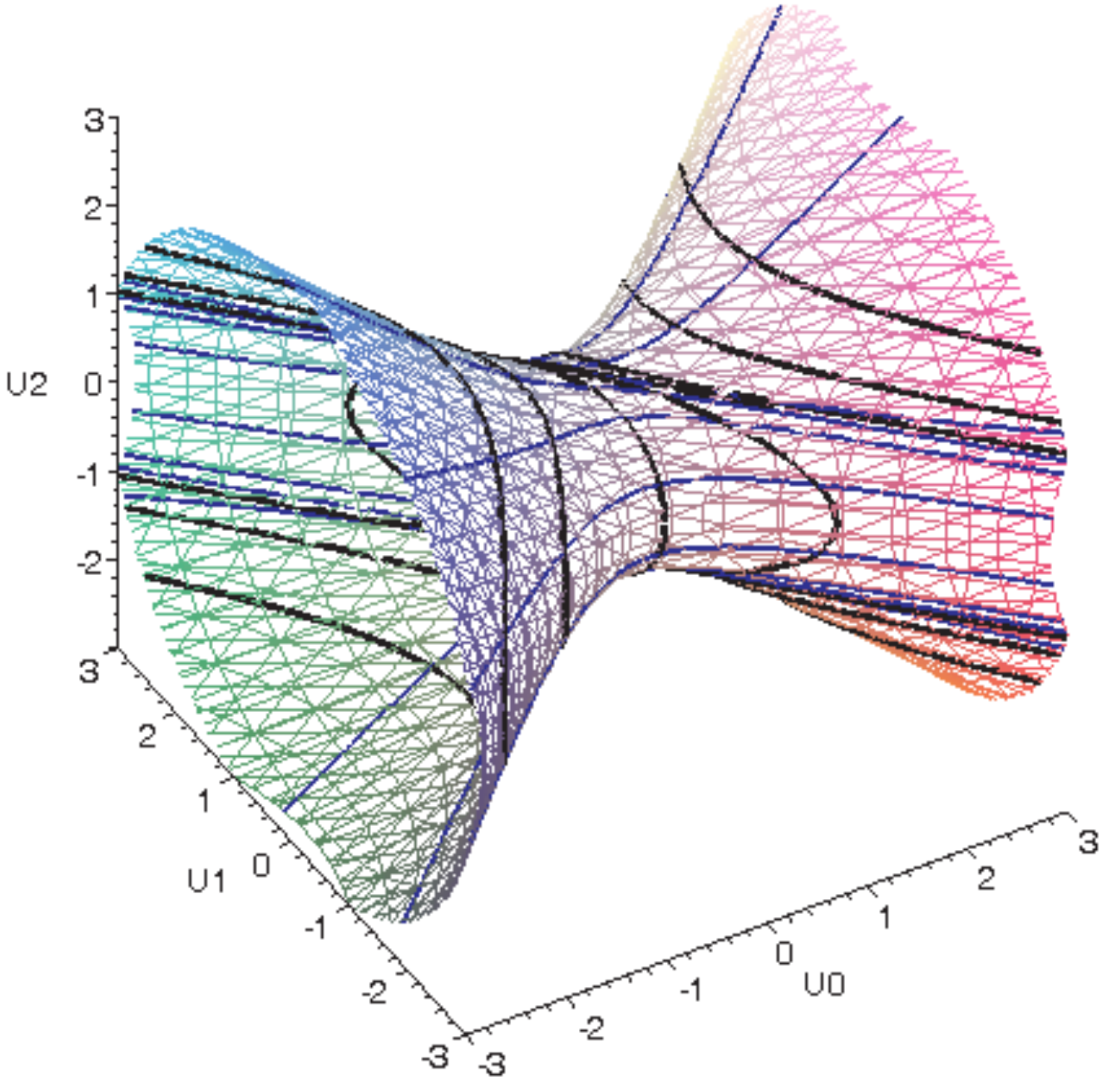}

\begin{center}
\caption{Elliptic-parabolic system}
\label{fig:28}
\end{center}
    \end{minipage}
    \hfill
    \begin{minipage}[t]{0.45\linewidth}
 \includegraphics[scale=0.35]{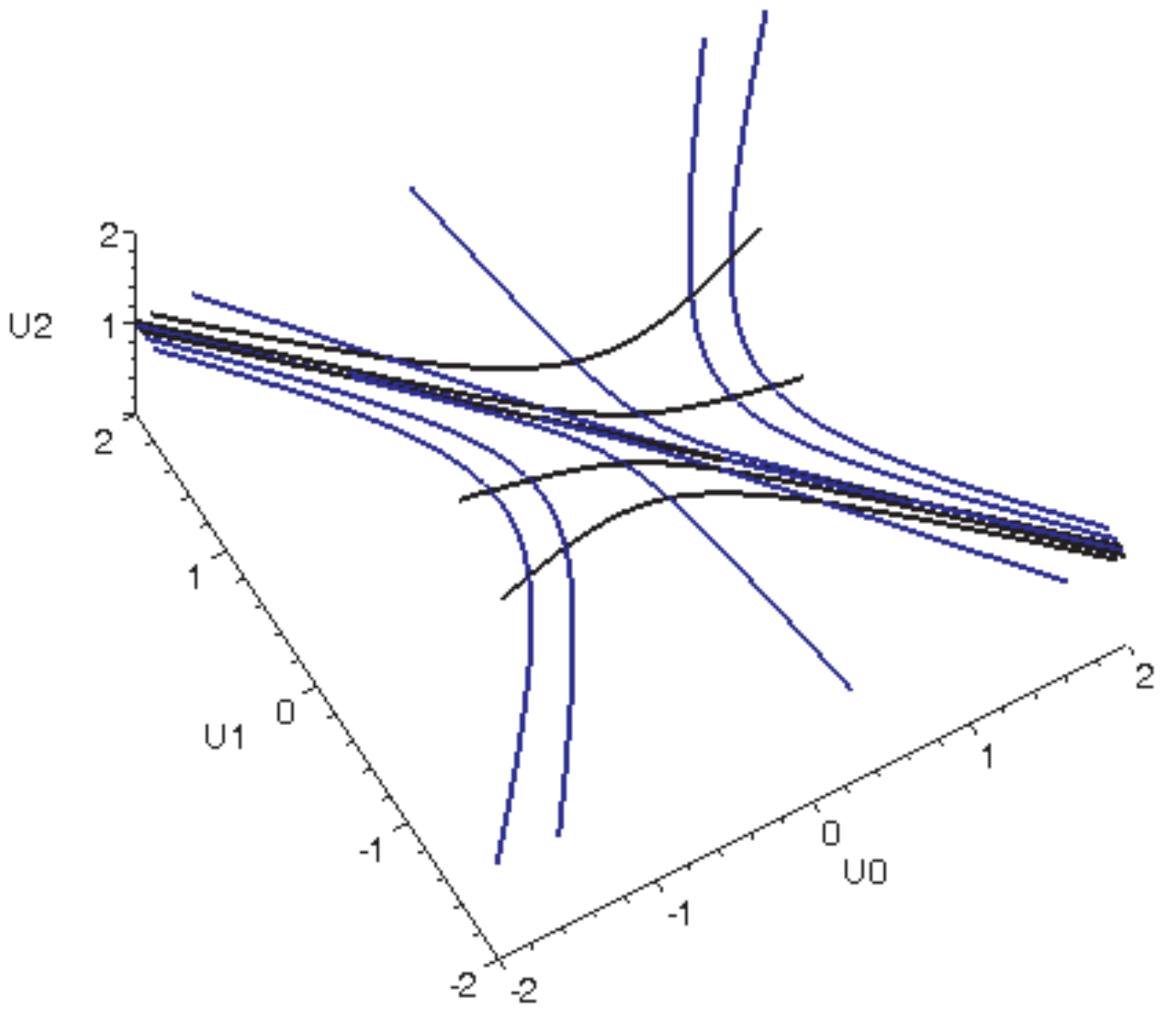}
\caption{Projective plane for elliptic-parabolic system}
\label{fig:29}
    \end{minipage}
  \end{center}
\end{figure}

\subsection{Elliptic-parabolic coordinates to hyperbolic II and Cartesian ones}

The  elliptic-parabolic system  of  coordinates is determined  by parameter  $\gamma > 0$, which is included in the definition 
of this system (see Fig. \ref{fig:28}). The projective plane for this system is presented in Fig. \ref{fig:29}.   

\subsubsection{Elliptic-parabolic to hyperbolic II}

We consider elliptic-parabolic  coordinates (\ref{sys_ep_trig1})  with parameter  $\gamma \sim R^2/l^2$ ($l>0$ is some constant)
and put  $\xi_1 = -\gamma/\sinh^2\tau_2$ and  $\xi_2 = \gamma/ \cosh^2\tau_1$.  From  equation  (\ref{sys_ep_trig1}) we have  ($\xi_1<0<\xi_2\leq\gamma$): 
\begin{equation}
\label{EP-001}
\xi_{1,2} = -\frac{1}{2R^2} \left\{(u_0 - u_1)^2 + \gamma(u_0^2 - u_1^2) \pm 
\sqrt{\left[ (u_0-u_1)^2 +\gamma(u_0^2-u_1^2)  \right]^2 +  4R^2\gamma(u_0 - u_1)^2}\right\}.
\end{equation}
Then in contraction limit we obtain 
\begin{equation}
\xi_{1,2} \to
 \mp 2 e^{2\zeta_{1,2}},
\end{equation}
and for Beltrami coordinates  we have  ($t>x$)   
\begin{equation}
  y_0  \to t=  l\left[\sinh(\zeta_2 - \zeta_1) + e^{\zeta_1 + \zeta_2}\right],  
	\qquad
  y_1  \to x = l\left[\sinh(\zeta_2 - \zeta_1) - e^{\zeta_1 + \zeta_2}\right].
\end{equation}
where $\zeta_1$ and  $\zeta_2$  are  the  hyperbolic coordinates of Type II (see Table \ref{tab:6}). 
For symmetry operator  in  the  same limit  we have 
\[
\frac{l^2}{R^2}  S_{EP}  =  \frac{l^2}{R^2} \left[ \gamma K_2^2 + (K_1 + L)^2 \right]  =   K_2^2 +  l^2 (\pi_0 + \pi_1)^2   
\to  N^2 +  l^2 (p_0 + p_1)^2  =  X_H^{II}.
\]

\subsubsection{Elliptic-parabolic to Cartesian}
\label{971}

We will start from coordinates   (\ref{sys_ep_trig1})  but  we will interchange coordinates   $u_1 \leftrightarrow u_2$.   
Then we  have 
\begin{equation}
\label{EP-002}
\bar{S}_{EP} = \gamma K_1^2 + (K_2-L)^2.
\end{equation}
Let us fix parameter $\gamma > 0$.   Introducing the new variables  by $\xi_1 = \cosh \tau_1$, $\xi_2 = \sinh \tau_2$
and  using equation   (\ref{sys_ep_trig1})  it is easy to find  that (we consider $u_2 > u_0$, $\tau_2 < 0$ to cover the origin of coordinates):
\begin{equation}
\label{EP-003}
\xi^2_{1,2} = \frac{ \pm u_0(\gamma + 1) \pm u_2(\gamma -1) - \sqrt{ \left[u_0(\gamma + 1) + u_2(\gamma -1) \right]^2 + 4R^2\gamma  } }{2(u_0 - u_2)}.
\end{equation}

From formula  (\ref{EP-003})  we have for the large $R$ 
\begin{equation}
\sinh\tau_{1}  \to   \sqrt{\frac{\gamma}{\gamma+1} } \frac{x}{R},
\qquad
\sinh\tau_{2}  \to  -\sqrt{\gamma} \left(1 + \frac{t}{R}\right),
\label{EP_to_C}
\end{equation}
and Beltrami  coordinates  go  to  Cartesian ones: $y_0 \to t$ and $y_1 \to -x$.   
For symmetry operator  $\bar{S}_{EP}$ we  get 
\[
\frac{\bar{S}_{EP}}{R^2} = \gamma \pi_0^2 + \left(\frac{K_2}{R} + \pi_1 \right)^2
\to \gamma p_0^2 + p_1^2 \simeq X_C^{I}.
\]
Finally  let us  note  that  we  do not  use the elliptic-parabolic system of coordinates in  form  (\ref{sys_ep_trig1})  because  
for the large $R$ the  variables  $\xi_1 = \cosh \tau_1$ and $\xi_2 = \sinh \tau_2$ are expressed as the combination of Cartesian coordinates $t$ and $x$.

\begin{figure}[htbp]
\begin{center}
    \begin{minipage}[t]{0.45\linewidth}
      \includegraphics[scale=0.4]{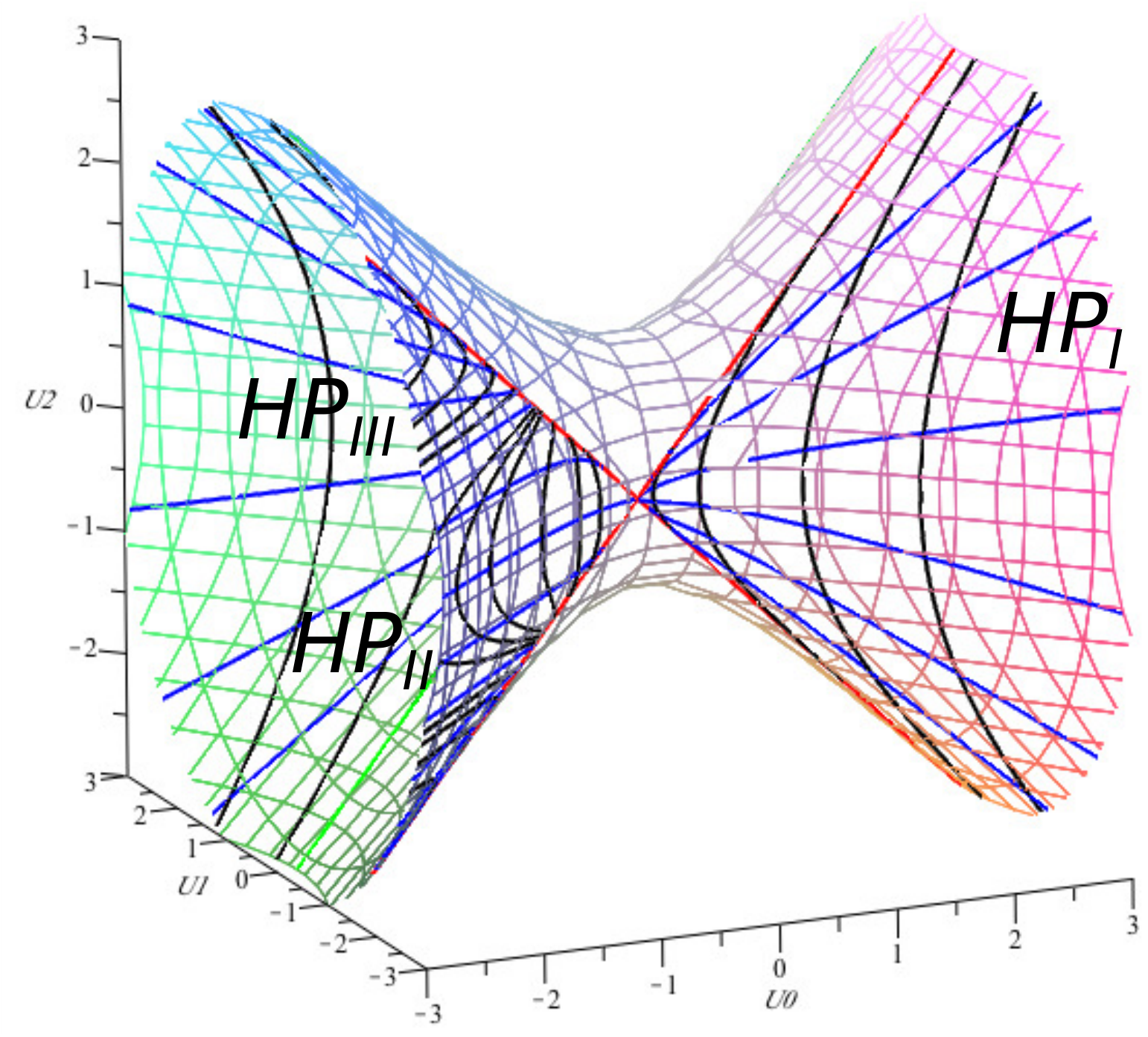}

\begin{center}
\caption{Hyperbolic-parabolic system}
\label{fig:30}
\end{center}
    \end{minipage}
   \hfill
    \begin{minipage}[t]{0.45\linewidth}
 \includegraphics[scale=0.4]{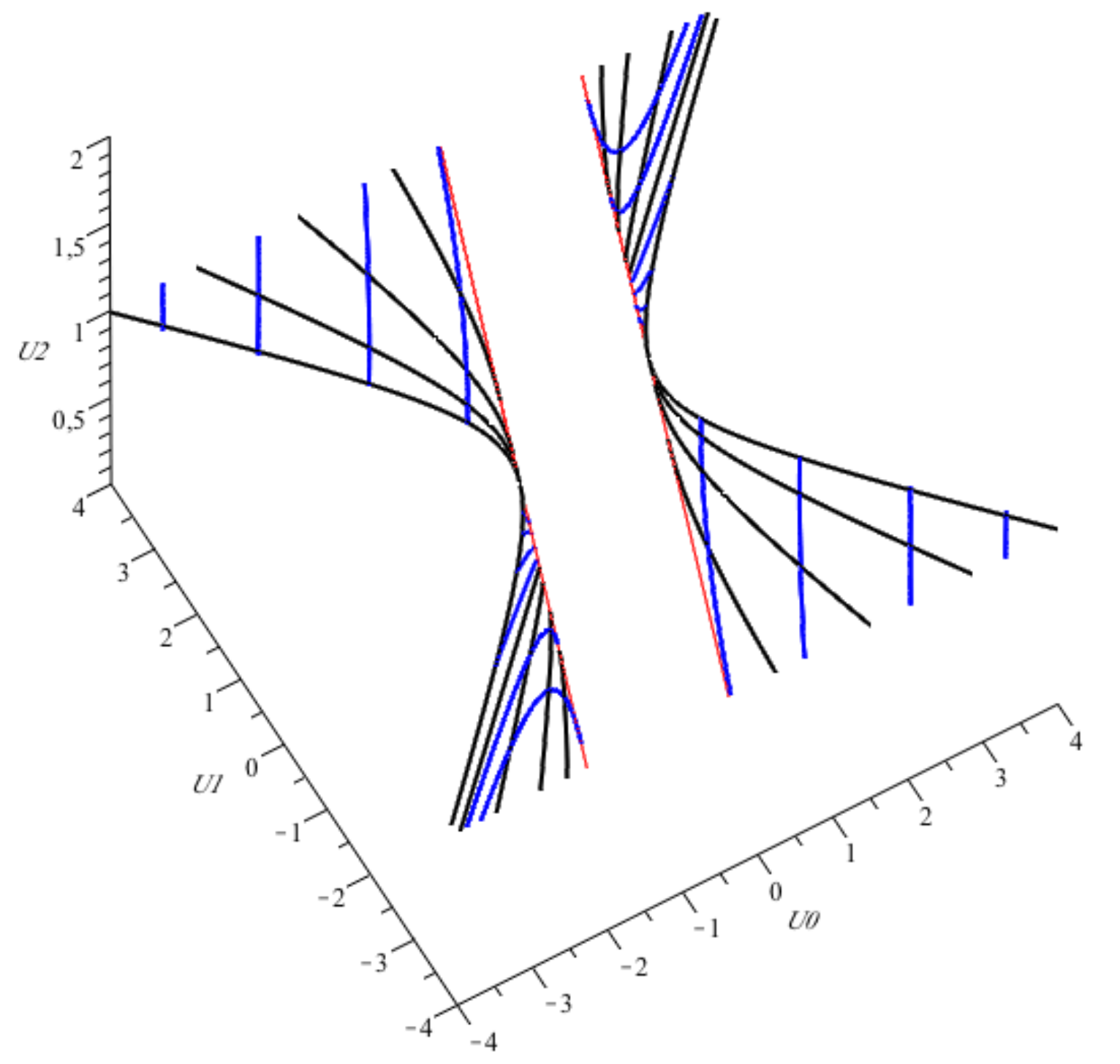}
\caption{Projective plane for hyperbolic-parabolic system}
\label{fig:31}
    \end{minipage}
  \end{center}
\end{figure}

\subsection{Hyperbolic-parabolic to hyperbolic III, Cartesian and Parabolic I}

The  hyperbolic-parabolic system of coordinates depends on parameter $\gamma > 0$. This parameter defines the points of intersection of envelopes (\ref{HP_evelopes}) on hyperboloid with coordinates $\left(\pm R\frac{1-\gamma}{2\sqrt{\gamma}}, \pm R\frac{1+\gamma}{2\sqrt{\gamma}}, 0\right)$ (see Fig. \ref{fig:30}).

\subsubsection{Hyperbolic-parabolic to Hyperbolic III}

In  this section we shall always assume that  $\gamma = R^2/l^2$ ($l$ is some constant).   
We introduce notation  $\xi_1= \sin\theta$ and  $\xi_2 = \sin\phi$.  Then,  for  the  coordinates  of   
Type I (\ref{sys_hp_trigTI})  we obtain 
\begin{equation}
\xi^2_{1,2} = \frac{  u_0(1 - \gamma) - u_1(\gamma + 1) \mp 
\sqrt{ (u_0(1-\gamma) - u_1(1+\gamma))^2 - 4 l^2 } }{2(u_0-u_1)}
\label{hpI}
\end{equation}
whereas  for coordinates  of Type II (\ref{sys_hp_trigTII}) 
\begin{equation}
\xi^2_{1,2} = \frac{u_0-u_1}{2 R^2 \gamma} \left\{  u_0(1 - \gamma) - u_1(\gamma + 1) \pm
\sqrt{ \left[u_0(1-\gamma) - u_1(1+\gamma)\right]^2 - 4 l^2}  \right\}.
\label{hpII}
\end{equation}
Let us put now   $\xi^{-1}_{1,2} =  \frac{\sqrt{2} l}{R}e^{\zeta_{2,1}}$ for coordinates  of Type I and    
${\xi_{1,2}}  =  \frac{\sqrt{2} l}{R}e^{\zeta_{2,1}}$ for Type II respectively.  It  is  easy  to see  that  
in  contraction  limit  $R\to \infty$   Beltrami  coordinates  $y_0 $ and  $y_1 $ of the system   
of  Type I, II   contract  into  hyperbolic coordinates  of Type III, that cover part  $x>t$ on 
pseudo-euclidean plane $E_{1,1}$  (see Table \ref{tab:6}):
\begin{equation}
y_0 \to  t = l\left[\cosh(\zeta_1 - \zeta_2) - e^{\zeta_1 + \zeta_2}\right],  
\qquad
y_1 \to  x = l\left[\cosh(\zeta_1 - \zeta_2) + e^{\zeta_1 + \zeta_2}\right].
\end{equation}
For hyperbolic-parabolic coordinates (\ref{sys_hp_trigTIII})  of  Type III  we  obtain 
\begin{equation}
\xi^2_{1,2} = \frac{u_0-u_1}{2 R^2 \gamma} \left\{  u_0(\gamma - 1) + u_1(\gamma + 1) \pm
\sqrt{ \left[u_0(\gamma - 1) + u_1(1+\gamma)\right]^2 - 4R^2\gamma  }  \right\}
\label{hpIII}
\end{equation}
here we use notation $\xi_1= \sinh\theta$ and  $\xi_2 = \sinh\phi$.  Choosing  now  ${\xi_{1,2}}  =  \frac{\sqrt{2}l}{R}e^{\zeta_{1,2}}$  
we  get  that in contraction  limit   $R\to \infty$   the system of  Type III  contracts   into  hyperbolic coordinates  of  Type III  that cover part  $t>x$ on  pseudo-euclidean plane $E_{1,1}$  (see Table \ref{tab:6}):      
\begin{equation}
y_0 \to  t = l\left[\cosh(\zeta_1 - \zeta_2) + e^{\zeta_1 + \zeta_2}\right],  
\qquad
y_1 \to  x = l\left[\cosh(\zeta_1 - \zeta_2) - e^{\zeta_1 + \zeta_2}\right].
\end{equation}
For  the  corresponding  symmetry  operator  in  the  same  limit  we get    
\[
-  \frac{l^2}{R^2} S_{HP} =   K_2^2   -    \frac{l^2}{R^2} (K_1 + L)^2   \to  N^2 -  l^2 (p_0 + p_1)^2  = X_H^{III}.
\]
Let us note that there is  uncovered part  of  pseudo-Euclidean plane,  namely $|t+x|<2| l |$.   This result  is  a  consequence  of  splitting 
the  hyperbolic-parabolic coordinates  as shown in Fig. \ref{fig:31}.

\subsubsection{Hyperbolic-parabolic to  Cartesian}

As in the previous section,  we look at all three types of hyperbolic-parabolic coordinates  but  interchange coordinates $u_1 \leftrightarrow u_2$.  Then  we arrive to  the  equivalent symmetry operator    
$\bar{S}_{HP} = - \gamma K_1^2 + (K_2-L)^2$,  where $\gamma>0$ and  $\gamma \neq 1$ (this case is considered later).   
In contraction limit  $R \to \infty$   we  have
\[
\frac{\bar{S}_{HP}}{R^2} = -\gamma \pi_0^2 + \left(\frac{K_2}{R} + \pi_1\right)^2
\to -\gamma p_0^2 + p_1^2 \simeq X_C^{I}.
\]

\begin{figure}[htbp]
\begin{center}
    \begin{minipage}[t]{0.45\linewidth}
      \includegraphics[scale=0.4]{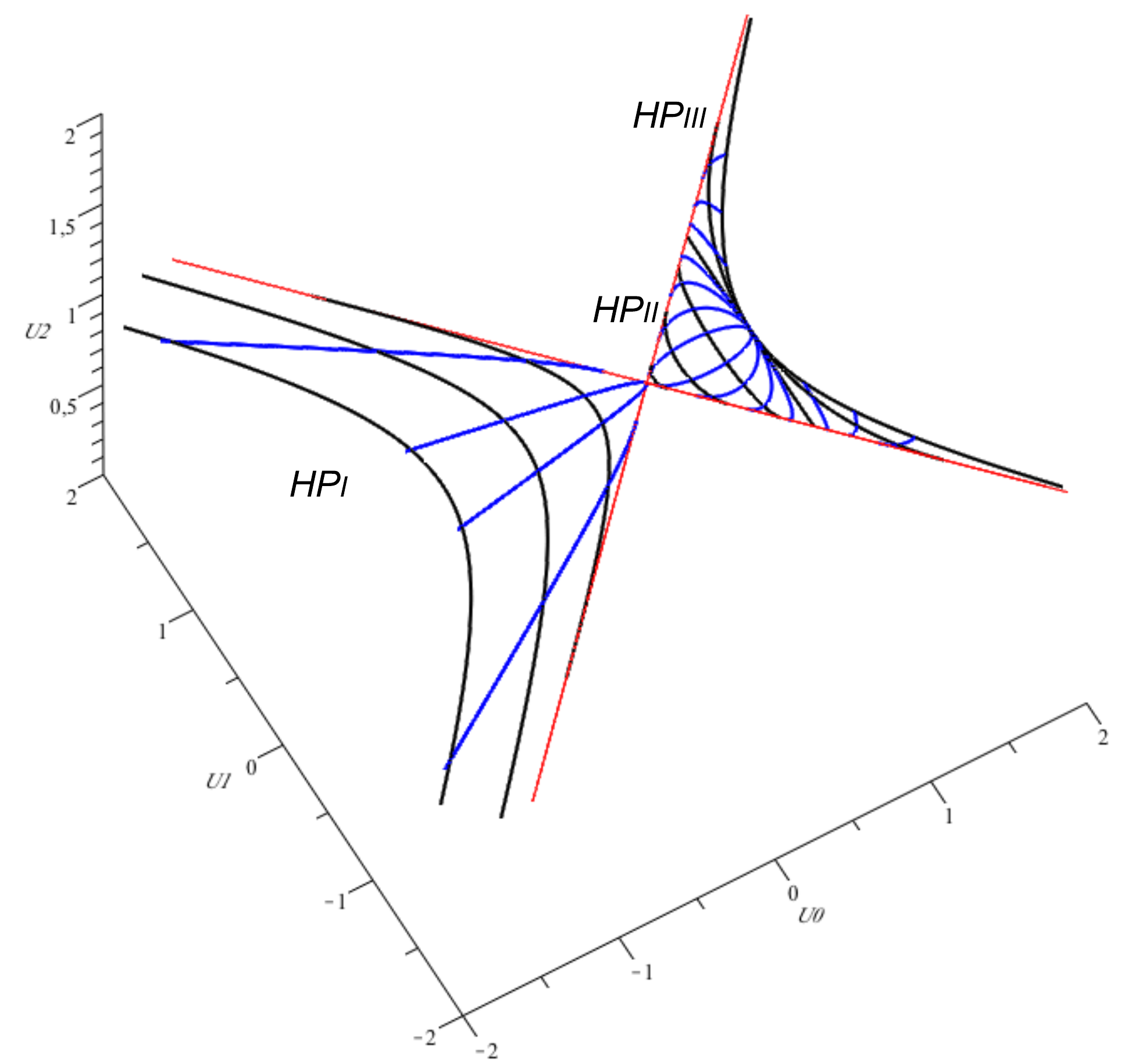}

\begin{center}
\caption{Projective plane for permuted system $\bar{S}_{HP}$, $0< \gamma < 1$ }
\label{fig:42b}
\end{center}
    \end{minipage}
   \hfill
    \begin{minipage}[t]{0.45\linewidth}
 \includegraphics[scale=0.4]{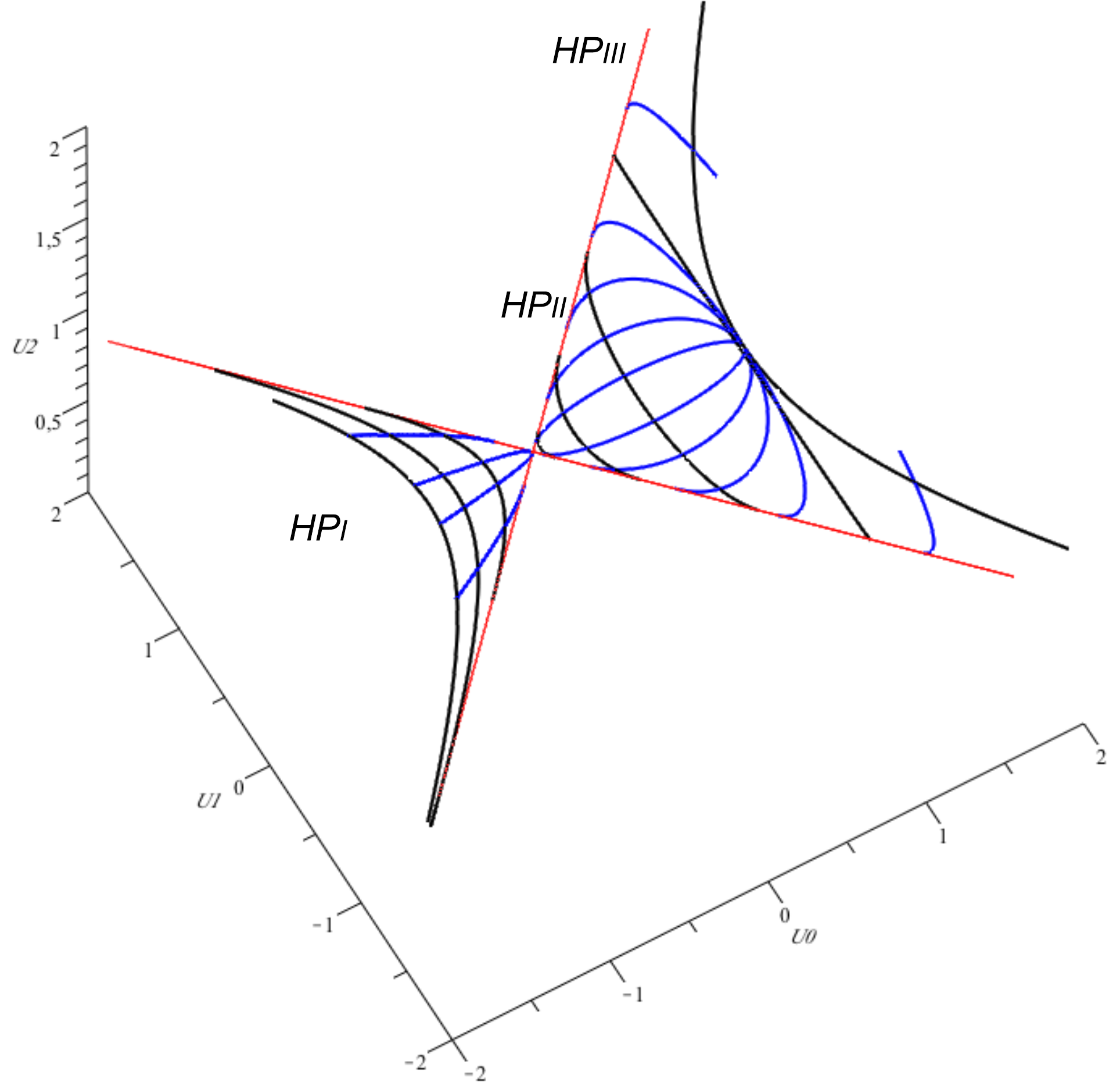}
\caption{Projective plane for permuted system $\bar{S}_{HP}$, $\gamma > 1$}
\label{fig:42c}
    \end{minipage}
  \end{center}
\end{figure}

The point of intersection of envelopes on projective plane is in $\left(\frac{1-\gamma}{1+\gamma}, 0 \right)$. Therefore the origin of coordinates is covered by different types of $HP$ system, depending on the value of $\gamma$.

From relations (\ref{hpI}) for $HP_I$, after interchanging $u_1 \leftrightarrow u_2$ we obtain  
\begin{equation}
\label{HPI_GAMMA}
\xi^2_{1,2} = \frac{  u_0(1 - \gamma)  -  u_2(\gamma + 1) \mp \sqrt{ \left[u_0(1 - \gamma) - u_2(\gamma + 1)\right]^2 - 4R^2\gamma  } }{2(u_0 - u_2)}
\end{equation}
and hence, for the  large  $R$  
\begin{equation}
\cos \theta  \to \sqrt{\frac{\gamma}{1 - \gamma}} \frac{x}{R},
\qquad
\sin \phi \to \sqrt{\gamma}\left(1 + \frac{t}{R}\right).
\label{HPI_to_C}
\end{equation}
Here we take $0< \gamma <1$, because only in this case system $HP_I$ on the projective plane covers the origin $(0,0)$ and goes to Cartesian coordinates (see Fig. \ref{fig:42b} with $\gamma=1/2$).

In the same way for permuted system $HP_{II}$ from (\ref{hpII}) we find 
\begin{equation}
\label{HPII_GAMMA}
\xi^2_{1,2} = \frac{u_0 - u_2}{2\gamma R^2}\left\{u_0(1-\gamma) - u_2(\gamma +1)  
\pm 
\sqrt{\left[ u_0(1-\gamma)  - u_2(\gamma +1) \right]^2 - 4\gamma R^2} \right\}
\end{equation}
and for the large $R$ 
\begin{equation}
\sin\theta \to \frac{1}{\sqrt{\gamma}}\left(1 - \frac{t}{R}\right) ,
\qquad
\cos\phi \to \sqrt{\frac{\gamma}{\gamma-1}} \frac{x}{R},
\label{HPII_to_C}
\end{equation}
where now $\gamma>1$, because only in this case permuted system $HP_{II}$ on the projective plane covers the origin $(0,0)$ and contracts to Cartesian system (see Fig. \ref{fig:42c} with $\gamma =2$).

The permuted system obtained from $HP_{III}$ does not cover $(0,0)$ for any value of $\gamma$ and in contraction limit does not contract to Cartesian coordinates.

\subsubsection{Hyperbolic-parabolic to Parabolic I}
We will consider here the case  when parameter $\gamma = 1$.  In this case the point of intersection of envelopes is in $(0,0)$ of projective plane and $HP$ systems do not cover this point.

For  symmetry operator  we  obtain  
\[
\bar{S}_{HP} =  - K_1^2 + (K_2 - L)^2 = 2 K_2^2 - \{K_2,L\}  - R^2 \Delta_{LB}
\]
and in contraction  limit  $R\to \infty$  we get 
\[
-\frac{\bar{S}_{HP}}{R}-R\Delta_{LB} = -2 \frac{K_2^2}{R} - \{K_2,\pi_1\} \to \{N,p_1\} = X_P^I.
\]

For coordinates  $\xi_{1,2}^2 =  \left(- u_2 \mp \sqrt{u_2^2 - R^2}\right)/{(u_0 - u_2)}$  of  Type I  (\ref{HPI_GAMMA})  we have:
\begin{equation}
\label{HPI_GAMMA1}
\xi_1^2 \to 1 - \frac{u^2}{R},  
\qquad
\xi_2^2 \to 1 - \frac{v^2}{R};
\end{equation}
for coordinates $\xi_{1,2}^2 = (u_0  - u_2)\left(-u_2\pm\sqrt{u_2^2-R^2}\right)/R^2$  of  Type II  (\ref{HPII_GAMMA})  we obtain
\begin{equation}
\label{HPII_GAMMA2}
\xi_1^2 \to 1 - \frac{v^2}{R}, 
\qquad 
\xi_2^2 \to 1 - \frac{u^2}{R};
\end{equation}
and  for  coordinates $\xi_{1,2}^2 = (u_0  - u_2)\left(u_2\pm\sqrt{u_2^2-R^2}\right)/R^2$ of Type III (here $\xi_1 =\sinh\theta$, $\xi_2 = \sinh\phi$) we get
\begin{equation}
\label{HPIII_GAMMA3}
\xi_1^2 \to 1 + \frac{v^2}{R}, 
\qquad 
\xi_2^2 \to 1 + \frac{u^2}{R}.
\end{equation}
In limit $R\to \infty$ we obtain
\[
y_0 \to t = \frac{1}{2} \left(u^2 + v^2\right),
\qquad
y_1 \to x = u v,
\]
which coincides with Parabolic I coordinates in pseudo-euclidean plane $E_{1,1}$  (see Table \ref{tab:6}).

\subsection{Semi-circular-parabolic  to Cartesian}

In contraction limit operator $S_{SCP}$ gives
\be
\label{NONSEP-OPERATOR}
\frac{S_{SCP}}{R} = \frac{1}{R}\{K_1, K_2\} + \frac{1}{R} \{K_2,L\} \to \left\{p_0, N \right\} + \left\{ p_1, N \right\},
\ee
which was already well known as the operator that does not generate a coordinate system with separation of variables (see \cite{Kalnins1}). The contraction limit of SCP system also does not correspond to any separable system on $E_{1,1}$. 

Nevertheless,   it  is  possible  to  construct   the  equivalent  form of  semi-circular-parabolic  system which  contracts 
to Cartesian  coordinates.   Let us  make  two  consecutive  hyperbolic  rotations. The first  one  through   the angle $a_1 \not= 0$  with  
respect  to axis  $u_1$ and the second  one   through angle $a_2 \not= 0$  with  respect  to axis  $u_2$.  
The  resulting  rotation is given by matrix  
\begin{equation}
\label{00-TYPEII-03}
\left( {{\begin{array}{*{20}c}
 u^\prime_0 \\
 u^\prime_1 \\
 u^\prime_2
\end{array} }} \right) =
\left( {{\begin{array}{*{20}c}
\cosh a_2 \cosh a_1 & \sinh a_2 &  \cosh a_2 \sinh a_1 \\
\sinh a_2 \cosh a_1 & \cosh a_2 &  \sinh a_2  \sinh a_1 \\
\sinh a_1 & 0 & \cosh a_1
\end{array} }} \right)
\left( {{\begin{array}{*{20}c}
 u_0 \\
 u_1 \\
 u_2
\end{array} }} \right).
\end{equation}
To simplify  all  subsequent  formulas  we choose  $a_1 = \ln \frac{1}{\sqrt{2}}$ and  
$a_2 = - \tanh^{-1} (\cosh a_1 + \sinh a_1) =  -  \tanh^{-1} \frac{1}{\sqrt{2}}$.  Then  the new symmetry 
operator takes the form  
\be
\tilde{S}_{SCP} = \frac{1}{4} K_1^2 + \frac{1}{4}  \left\{K_1, K_2\right\}  - \frac{3}{4} K_2^2  +
\frac{1}{\sqrt{2}}  \left\{K_2, L\right\}   -  \frac{1}{2} L^2.
\label{SCP_rot_hyp}
\ee
In contraction limit $R\to \infty$ we obtain 
\be 
\frac{4}{R^2}  \left(\frac{R^2}{2} \Delta_{LB}  -  \tilde{S} _{SCP} \right)   \to  p_0^2  =  X_C^{I}.
\ee
For coordinates $\xi$ and $\eta$,  after some long  and tedious  calculations we  get 
\bea
\label{SCP_rot3}
\xi \to 2\left(1 + \frac{x}{R}\right), 
\qquad
\eta \to \sqrt{2} -  \frac{2t}{R},
\eea
and  Beltrami coordinates  (\ref{Beltrami_1})  contract  into the Cartesian  ones:  
$y_0 \to t$, $y_1 \to x$.

\section{Conclusion and Discussion}
\label{sec:5}
 From the algebraic point of view on the problem of variables
separation for the two-dimensional Helmholtz equation the set of
the first and second order operators have been constructed. These
operators do not commute between themselves, but they commute with
Laplace-Beltrami operator and correspond to separation of variables
in Helmholtz equation on two-dimensional hyperboloid embedded
in three-dimensional Minkovski space $E_{2,1}$.
The results of the classification of second order operators are essentially
in agreement with those of \cite{WINLUKSMOR}.
We have shown that the procedure of the construction of separable systems
of coordinates, based on the first order
operator is not unique and leads not only to the orthogonal subgroup type
coordinates but describes also nonorthogonal ones. We have found
three such type systems of coordinates. The detailed description of all
possible orthogonal systems of coordinates, corresponding to each
operator sets $\left\{\Delta_{LB}, S^{(2)}_{\alpha}\right\}$ is given as
for two-sheeted (upper sheet) $H_2$, either for one-sheeted
${\tilde H}_2$ hyperboloids.
       
All coordinate systems on $H_2$ parametrize completely the surface of hyperboloid. As a consequence there exists an intimate 
connection  between  symmetry operators and systems of coordinates at the contraction limit from two-sheeted hyperboloid
into euclidean plane $E_{2}$. All coordinate systems in Euclidean plane $E_{2}$ can be obtained as the limit of space $H_2$ under
the contraction procedure $R\to \infty$. We have presented here some new contractions that were unnoticed in the previous papers \cite{IZM3,PSW1} in addition to already known ones. The contraction of the nonorthogonal systems of coordinates on $H_2$ into $E_{2}$ ones also are given.

We have shown that differently from $H_2$ hyperboloid, only five, namely: pseudo-spherical, equidistant (splitting into two parts), 
horiciclic,  elliptic-parabolic  and elliptic systems of coordinates completely cover one-sheeted hyperboloid. As a reflection of 
this fact, the contraction limit of symmetry operator does not guarantee the existence of the contractions of the corresponding system of 
coordinates on ${\tilde H}_2$ hyperboloid into the pseudo-euclidean plane $E_{1,1}$. It occurs with the semi-circular-parabolic, 
semi-hyperbolic and hyperbolic coordinate systems. Nevertheless, one can consider the contraction limit $R\to \infty$ in these 
coordinates using the complex ones. We can explain this situation by example of semi-circular-parabolic system of coordinates. 
Let us consider operator $\bar{S}_{SCP}$ 
\[
\bar{S}_{SCP} = \{K_1, K_2\} - \{K_1, L\},
\]
which corresponds to semi-circular-parabolic system (\ref{sys_scp_xi_eta_1}) with permutation $u_1 \leftrightarrow u_2$.  For limit $R\to\infty$ it is easy to see that it contracts into Cartesian operator
on $E_{1,1}$:  
\[
-\frac{\bar{S}_{SCP}}{2R^2} = -\frac{1}{2R}\{\pi_0, N \} +  \frac{1}{2} \{\pi_0,\pi_1\}  \to  p_0 p_1  =  X_C^I.
\]
Now we are going to check what happens to the semi-circular-parabolic system at the contraction limit $R\to\infty$. Let us consider system (\ref{01-lambdaSCP}) wherein previously we make the change $u_1 \leftrightarrow u_2$. The result is  
\be
\lambda_1 + \lambda_2 =  2 u_1 (u_2  - u_0),  
\qquad 
\lambda_1  \lambda_2  =  R^2 (u_2  - u_0)^2,
\ee
that gives three possibilities for the values of $\lambda_i$: 1. $\lambda_i \geqslant 0$; 2. $\lambda_i \leqslant 0$; 
3. $\lambda_1 = a + ib$, $\lambda_2 = a - ib$, $a,b\in \mathbb{R}$. For the first and the second ones the coordinate 
system covers only the part of hyperboloid when $|u_1| \geqslant R$, because in these cases
\be
u_1^2 = R^2 \frac{\left(\lambda_1 + \lambda_2\right)^2}{4\lambda_1 \lambda_2} \geqslant R^2.
\ee
Geometrically in contraction limit $R\to \infty$ the coordinate grid for permuted SCP system on the projective plane results to be empty. 
To avoid it let us formally consider the third case with $a = A R^2$, $b = B R^2$, $A\in\mathbb{R}$, $B>0$. 

Then coordinate system (nonorthogonal and non separable) looks as follows ($|u_1| < R$ and we consider $u_2 > 0$):
\bea
u_0 &=& \frac{R}{2}\left( \sqrt{A^2 + B^2} - \frac{B^2}{\left(A^2 + B^2\right)^{3/2}}\right),
\nonumber \\
u_1 &=& R \frac{A}{\sqrt{A^2 + B^2}}, \\
u_2 &=& \frac{R}{2}\left( \sqrt{A^2 + B^2} + \frac{B^2}{\left(A^2 + B^2\right)^{3/2}}\right).
\nonumber
\eea
For the above system variables $A$ and $B$ contract in such a way
\be
 A = \frac{u_1(u_0 + u_2)}{R^2} \to -\frac{x}{R}, 
 \qquad 
 B = \frac{u_0 + u_2}{R} \sqrt{1- \frac{u_1^2}{R^2}} \to 1 + \frac{t}{R}
\ee
and finally one obtains Cartesian coordinates $y_0 \to t$, $y_1 \to x$.

Thus we have investigated all forms of {\bf real} analytic contractions from one-sheeted hyperboloid ${\tilde H}_2$ including the equivalent systems of coordinates. Some of them (equivalent systems of coordinates), as equidistant ones, give us different contraction limits on $E_{1,1}$. We have also considered the contraction of the nonorthogonal systems of coordinates on ${\tilde H}_2$ and have shown that they transform 
into nonorthogonal ones on $E_{1,1}$ classified earlier by Kalnins in \cite{Kalnins1}. 

We  shed  the  light  on  the  mysterious  existence of an invariant 
operator (\ref{NONSEP-OPERATOR}) which does not correspond to any separation  system of coordinates for the Helmholtz equation 
on pseudo-euclidean space $E_{1,1}$. We have observed that the relation between symmetry operators and separation systems 
of coordinates on the euclidean or pseudo-euclidean spaces can  sometimes be better understood in the framework of 
contractions of the spaces of constant curvature as spheres or hyperboloids into these "flat" spaces. 

We have shown that except one, namely parabolic coordinates of type II, all orthogonal systems of coordinates on  
pseudo-euclidean plane $E_{1,1}$ could be obtained by contraction limit $R\to\infty$ from a system on the one-sheeted hyperboloid 
${\tilde H}_2$. 

To find the counterpart of parabolic coordinates of type II on ${\tilde H}_2$ hyperboloid, which we have not found among 
the nine orthogonal systems of coordinates, let us make the following steps.   
Let us consider formula (\ref{S_2}) describing the general form of operator $S^{(2)}$, choosing in  (\ref{S_2}) $b=e=1$, $c=0$ 
and $a = f = - d = \frac{\alpha}{R}$, where $\alpha$ is a constant. Then we obtain the "minimal"  operator
\be
\label{S2_to_X_P_II}
S^{(2)} = \{ K_1, K_2 \} + \{ L, K_2 \} + \frac{\alpha}{R} \left(L - K_1 \right)^2
\ee
which contracts into $X_P^{II}$:
\be
\frac{S^{(2)}}{R}  \to  \{p_0, N\} + \{p_1, N\} + \alpha (p_0 - p_1)^2  = X_P^{II}.
\ee
The characteristic equation (\ref{lambda}) corresponding to operator (\ref{S2_to_X_P_II}) gives the 
solution
\bea
\lambda_1 + \lambda_2 = \frac{\alpha}{R} \left(u_0 + u_1\right)^2 + 2 u_2 \left(u_0 - u_1\right), 
\quad
\frac{\lambda_1 \lambda_2}{R^2} =  (u_0 - u_1)^2 - \frac{4 \alpha}{R} u_2 (u_0 + u_1).
\label{lambda_discuss}
\eea
Since we are interested in contractions only, we consider the transition for large $R$ directly for system of equations (\ref{lambda_discuss}). From (\ref{lambda_discuss}) it is easy to see that 
\bea
\frac{\lambda_1 + \lambda_2}{R}  \simeq 2 (t-x), 
\qquad
\frac{\lambda_1 \lambda_2}{R^2}  \simeq (t-x)^2 - 4\alpha (t+x),
\eea
where $(t,x)$ are Cartesian coordinates on $E_{1,1}$.  Introducing new variables $\xi = - \lambda_1/(4R)$ and 
$\eta = - \lambda_2/(4R)$ and expressing $t$, $x$, from the above system we obtain the parabolic coordinates 
of type II 
\be
\label{Parabolic_II_coordinates}
t = \frac{1}{2\alpha} \left(\eta - \xi \right)^2 - \left(\eta + \xi \right),\qquad
x = \frac{1}{2\alpha} \left(\eta - \xi \right)^2 + \left(\eta + \xi \right).
\ee
Thus, the contractions to the parabolic coordinates of type II can also be found within a compound 
system of coordinates on one-sheeted hyperboloid corresponding to symmetry operator (\ref{S2_to_X_P_II}).     

In the nearest future we are planning to continue our investigation on the contractions of basis functions
and interbases expansions associated with orthogonal separation systems of coordinates for Helmholtz equation
on one- and two-sheeted hyperboloids.

\section{Acknowledgments}

Authors thank E. Kalnis for the useful discussion on the subject of this article. 
The research by G.S. Pogosyan was partially supported by the Armenian national grant 13-1C288 and the Armenian-Belarus grant Nr. 13RB-035. We appreciate the program PRO-SNI 2015 (UdeG).



\begin{landscape}

\begin{table}[htbp]

\renewcommand{\arraystretch}{1.1}
\renewcommand\tabcolsep{2pt}

\caption{Coordinate Systems on the Two-sheeted Hyperboloid  ($^*$ means permutation $u_1 \leftrightarrow u_2$) }
\label{tab:1}
\hfuzz=32pt
\begin{center}
\begin{tabular}{| l | l | l |} 
\hline
Coordinate System, Operator&
Coordinates & Contracted System on $E_2$\\
\hline
$   \begin{array}{ll}
    \text{Ia.~(pseudo)-Spherical (SPH)} \\  L^2 \\ \tau > 0, 0 \leqslant \varphi < 2\pi
    \end{array}
$
&
$   \begin{array}{ll}
    u_0 = R \cosh \tau \\ u_1 = R \sinh \tau \cos \varphi \\ u_2 = R \sinh \tau \sin \varphi
    \end{array}
$
&
Polar
\\
\hline
$   \begin{array}{ll}
    \text{Ib.~Spherical (} \mathit{nonorthogonal} \text{)}  \\  L \\ \tau > 0, 0 \leqslant \varphi < 2\pi
    \end{array}
$
&
$   \begin{array}{ll}
    u_0 = R \cosh \tau \\ u_1 = R \sinh \tau \cos (\varphi + R\tau/\alpha) \\ u_2 =R \sinh \tau \sin (\varphi + R\tau/\alpha)
    \end{array}
$
&
Polar nonorthogonal
\\
\hline
$   \begin{array}{ll}
    \text{IIa.~Equidistant (EQ)} \\ K_2^2 \\ \tau_1, \tau_2 \in \mathbb{R}
    \end{array}
$
&
$   \begin{array}{ll}
   u_0 = R \cosh \tau_1 \cosh \tau_2 \\ u_1 = R \cosh \tau_1 \sinh \tau_2 \\ u_2 = R \sinh \tau_1
    \end{array}
$
&
Cartesian
\\
\hline
$   \begin{array}{ll}
    \text{IIb.~Equidistant (} \mathit{nonorthogonal} \text{)} \\ K_2 \\ \tau_1, \tau_2 \in \mathbb{R}
    \end{array}
$
&
$   \begin{array}{ll}
   u_0 = R \cosh \tau_1 \cosh (\tau_2 + R \tau_1/\alpha) \\ u_1 = R \cosh \tau_1 \sinh (\tau_2 + R\tau_1/\alpha) \\ u_2 = R \sinh \tau_1
    \end{array}
$
&
Cartesian nonorthogonal
\\
\hline
$   \begin{array}{ll}
    \text{IIIa.~Horicyclic (HO)} \\ (K_1 + L)^2 \\ \tilde{x} \in \mathbb{R}, \tilde{y} >0 
    \end{array}
$
&
$   \begin{array}{ll}
  u_0 = R \frac{\tilde{x}^2 + \tilde{y}^2 + 1} {2 \tilde{y} } \vspace{1mm} \\ u_1 = R \frac{\tilde{x}^2 + \tilde{y}^2 - 1} {2 \tilde{y} } \vspace{1mm} \\ u_2 = R \frac{\tilde{x}} {\tilde{y} }
    \end{array}
$
&
Cartesian
\\
\hline
$   \begin{array}{ll}
    \text{IIIb.~Horicyclic}^{*} \text{(} \mathit{nonorthogonal} \text{)} \\ K_2 - L \\ \tilde{x} \in \mathbb{R}, \tilde{y} >0 
    \end{array}
$
&
$   \begin{array}{ll}
u_0 = R \frac{\left(\tilde{x} + \tilde{y} -1  \right)^2 + \tilde{y}^2 + 1} {2 \tilde{y} }
\vspace{1mm} \\
u_1 = R \frac{\tilde{x} + \tilde{y} -1} {\tilde{y} }
\vspace{1mm} \\
u_2 = R \frac{\left(\tilde{x} + \tilde{y} -1  \right)^2 + \tilde{y}^2 - 1} {2 \tilde{y} } 
    \end{array}
$
&
Cartesian nonorthogonal
\\
\hline
$   \begin{array}{ll}
    \text{IV.~Elliptic-Parabolic (EP)} \\ (K_1 + L)^2 + \gamma K_2^2,  \gamma>0 \\ a \geq 0, \theta\in\left(-\frac{\pi}{2}, \frac{\pi}{2}\right)
    \end{array}
$
&
$   \begin{array}{ll}
  u_0 = \frac{R}{\sqrt{\gamma}} \frac{\cosh^2 a - \sin^2\theta + \gamma}{2\cos\theta\cosh a} \vspace{1mm}\\ 
  u_1 = \frac{R}{\sqrt{\gamma}} \frac{\cosh^2 a - \sin^2\theta - \gamma}{2\cos\theta\cosh a} \vspace{1mm}\\
  u_2 = R \tan\theta\tanh a 
    \end{array}
$
&
$  \begin{array}{ll}
    \text{Cartesian } (\gamma \ne 1)\\
    \hdashline\\
    \text{Parabolic } ( \gamma = 1)
    \end{array}
$
\\
\hline
$   \begin{array}{ll}
    \text{V.~Hyperbolic-Parabolic (HP)} \\ (K_1 + L)^2 - \gamma K_2^2,  \gamma>0 \\ b>0, \theta \in \left(0,\pi \right)
    \end{array}
$
&
$   \begin{array}{ll}
  u_0 =\frac{R}{\sqrt{\gamma}}\frac{\cosh^2 b - \sin^2\theta + \gamma}{2\sin\theta \sinh b} \vspace{1mm} \\
  u_1 =\frac{R}{\sqrt{\gamma}}\frac{\cosh^2 b - \sin^2\theta - \gamma}{2\sin\theta \sinh b} \vspace{1mm} \\
  u_2 = R \cot\theta \coth b
    \end{array}
$
&
Cartesian
\\
\hline
$   \begin{array}{ll}
    \text{VIa.~Semi-Circular-Parabolic (SCP)} \\ \{K_1, K_2\} + \{K_2,L\} \\ \xi, \eta >0
    \end{array}
$
&
$   \begin{array}{ll}
 u_0 = R\frac {\left(\eta^2 + \xi^2\right)^2 + 4}{8 \xi \eta} \vspace{1mm}\\
 u_1 = R\frac {\left(\eta^2 + \xi^2\right)^2 - 4}{8 \xi \eta} \vspace{1mm} \\
 u_2 =  R{\frac {{\eta}^{2}-{\xi}^{2}}{2\xi\eta}}
    \end{array}
$
&
Cartesian
\\
\hline

\end{tabular}

\end{center}
\end{table}
\end{landscape}


\begin{landscape}

\begin{table}[htbp]

\renewcommand{\arraystretch}{1.1}
\renewcommand\tabcolsep{2pt}

\label{tab:2}
\hfuzz=32pt
\begin{center}
\begin{tabular}{| l | l | l |} 
\hline
Coordinate System, Operator&
Coordinates & Contracted System on $E_2$ \\
\hline
$   \begin{array}{ll}
    \text{VIb.~Semi-Circular-Parabolic (SCP')} \\ K_2^2 - K_1^2 - \{K_1, L\}/\sqrt{2} + \{K_2,L\}/\sqrt{2} \\ \xi, \eta >0
    \end{array}
$
&
$   \begin{array}{ll}
 u_0 = R\frac {\left(\eta^2 + \xi^2\right)^2 + 4}{8 \xi \eta} \vspace{1mm}\\
 u_1 + u_2 = R \frac {{\eta}^{2}-{\xi}^{2}}{\sqrt{2}\xi\eta} \vspace{1mm}\\
 u_1 - u_2 = R\frac {\left(\eta^2 + \xi^2\right)^2 - 4}{4\sqrt{2} \xi \eta} 
    \end{array}
$
&
Cartesian
\\
\hline
$   \begin{array}{ll}
    \text{VIIa.~Elliptic (E)} \\ L^2 + \sinh^2\beta K_2^2,\  \sinh^2 \beta = \frac{a_1 - a_2}{a_2 - a_3} \\ a_3<a_2\leq \rho_2<a_1\leq \rho_1,\ k^2 = \frac{a_2-a_3}{a_1-a_3},\\
    a \in \left( i K^\prime, i K^\prime + 2K\right), b \in \left[0,4K^\prime \right)
    \end{array}
$
&
$   \begin{array}{ll}
u_0^2 = R^2{\frac {\left ({ \rho_1}-{ a_3}\right)\left ({ \rho_2}-{ a_3} \right )}{\left ({ a_1}-{ a_3}\right)\left ({ a_2}-{ a_3}\right )}} \vspace{2mm}\\
u_1^2 = R^2{\frac {\left ({ \rho_1}-{a_2}\right )\left ({ \rho_2}-{ a_2}\right )}{\left ({ a_1}-{ a_2}\right )\left ({ a_2}-{ a_3}\right)}} \vspace{2mm} \\
u_2^2 = R^2{\frac {\left ({\rho_1}-{a_1}\right )\left (a_1 -  \rho_2 \right )}{\left ({a_1}-{ a_2}\right )\left ({ a_1}-{ a_3}\right )}}
    \end{array}
$
$   \begin{array}{ll}
u_0 = R \sn(a, k)\dn(b, k^\prime) \vspace{2mm}\\
u_1 = i R \cn(a, k) \cn(b, k^\prime) \vspace{2mm} \\
u_2 = i R \dn(a, k) \sn(b, k^\prime)
   \end{array}
$
&
$   \begin{array}{ll}
  \text{Elliptic} \\ (a_1-a_2 = D^2, -a_3 \simeq R^2)\\
 \hdashline
\text{Polar} \\(a_1-a_2\simeq R^{-2}, a_2-a_3\simeq R^2) \\
 \hdashline
 \text{Cartesian} \\ (a_1 - a_2 = a_2 - a_3)
    \end{array}
$
\\
\hdashline
$   \begin{array}{ll}
    \text{VIIb.~Elliptic (\~E) (h. rotation of E)} \\ \cosh 2\beta L^2 + 1/2 \sinh 2\beta \{K_1,L\} \\ 
    \end{array}
$
&
$   
\begin{array}{ll}
u_0 = \frac{R}{k} \left\{ \sn(a,k)\dn(b, k^\prime) + i k^\prime \cn(a,k)\cn(b, k^\prime)    \right\}
\vspace{2mm} \\
u_1 = \frac{R}{k} \left\{ k^\prime \sn(a,k)\dn(b, k^\prime) +  i \cn(a,k)\cn(b, k^\prime)    \right\}
\vspace{2mm} \\
u_2 = i R \dn(a,k)\sn(b, k^\prime)
\end{array}
$
&
Parabolic
\\
\hline
$   \begin{array}{ll}
    \text{VIII.~Hyperbolic (H)} \\ K_2^2 - \sin^2\alpha \:L^2 \\ \sin^2\alpha = \frac{a_2 - a_3}{a_1 - a_3} = k^2 \\ \rho_2<a_3<a_2<a_1<\rho_1; k^2 + {k^\prime}^2 = 1\\
     a \in \left(iK, iK+ 2K \right),\  b \in \left(iK, iK + 2K^\prime \right)
    \end{array}
$
&
$   \begin{array}{ll}
  u_0^2 = R^2{\frac {\left ({\rho_1}-{a_2}\right )\left ({a_2 - \rho_2}\right )}{\left ({a_1}-{a_2}\right )\left ({a_2}-{a_3}\right )}} \vspace{2mm}\\
  u_1^2 = R^2{\frac {\left ({\rho_1}-{a_3}\right )\left ({a_3 - \rho_2}\right )}{\left ({a_1}-{a_3}\right )\left ({a_2}-{a_3}\right )}} \vspace{2mm}\\
  u_2^2 = R^2{\frac {\left ({\rho_1}-{a_1}\right )\left ({a_1 - \rho_2}\right )}{\left ({a_1}-{a_2}\right )\left ({a_1}-{a_3}\right )}}
       \end{array}
$
$   \begin{array}{ll}
u_0 = - R \cn(a,k) \cn(b,k') 
\vspace{2mm} \\
u_1= i R\sn(a,k)\dn(b,k')
\vspace{2mm} \\
u_2= i R\dn(a,k)\sn(b,k')
       \end{array}
$

&
Cartesian 
\\
\hline
$   \begin{array}{ll}
    \text{IXa.~Semi-Hyperbolic (SH-I)} \\ c K_2^2+ \{K_1, L\}, c =\sinh 2\beta\ne 0 \\ \sinh \tau_2  <  - c  < \sinh\tau_1
    \end{array}
$
&
$   \begin{array}{ll}
 u_0^2 + u_1^2 = \frac{R^2}{\sqrt{c^2+1}} \cosh\tau_1\cosh\tau_2 \vspace{2mm} \\
  u_0^2 - u_1^2 = \frac{R^2}{{c^2+1}}\left[c^2+1 - (\sinh\tau_1+c)(\sinh\tau_2+c) \right] \vspace{2mm} \\
 u_2^2 = -\frac{R^2}{c^2+1}(\sinh\tau_1+c)(\sinh\tau_2+c)
       \end{array}
$
&
 Cartesian ($c \ne 0$)
 \\
\hdashline
$   \begin{array}{ll}
    \text{IXb.~Semi-Hyperbolic (SH-II)} \\ \{K_1, L\} \\ \mu_1, \mu_2 \geqslant 0
    \end{array}
$
&
$   \begin{array}{ll}
u_0^2 = \frac{R^2}{2}\left\{ \sqrt{(1 +\mu_1^2)(1+\mu_2^2)}  +  \mu_1 \mu_2 + 1 \right\}
\vspace{2mm} \\
u_1^2 = \frac{R^2}{2}\left\{ \sqrt{(1 +\mu_1^2)(1+\mu_2^2)}  -  \mu_1 \mu_2  - 1 \right\}
 \vspace{2mm}\\
u_2^2 = R^2 \mu_1\mu_2
    \end{array}
$
&
Parabolic
\\
\hline
\end{tabular}

\end{center}
\end{table}
\end{landscape}




\begin{landscape}

\begin{table}[htbp]

\renewcommand{\arraystretch}{1.1}
\renewcommand\tabcolsep{2pt}

\caption{Coordinate Systems on the One-sheeted Hyperboloid  ($^*$ means permutation $u_1 \leftrightarrow u_2$)}
\label{tab:3}
\hfuzz=32pt
\begin{center}
\begin{tabular}{| l | l | l | l |}
\hline
Coordinate System, Operator&
Coordinates & Contracted System on $E_{1,1}$, Operator & Comments\\
\hline
$   \begin{array}{ll}
    \text{Ia.~Equidistant (EQ-Ia)}, K_2^2\\ \tau_1, \tau_2 \in \mathbb{R}
    \end{array}
$&
$   \begin{array}{ll}
u_0 = R \sinh \tau_1 \cosh \tau_2\\
u_1 = R \sinh \tau_1 \sinh \tau_2\\
u_2 = \pm  R \cosh \tau_1
    \end{array}
$
&
$   \begin{array}{ll}
\text{Pseudo Polar}, N^2 \\
  |t| > |x|
 \end{array}
$
&
$|u_2| > R$
\\
\hdashline
$   \begin{array}{ll}
    \text{Ib.~Equidistant (EQ-Ia)}, K_2\\
    \tau_1, \tau_2 \in \mathbb{R}
    \end{array}
$&
$   \begin{array}{ll}
u_0 = R \sinh \tau_1 \cosh (\tau_2 - \ln [R\tau_1/\alpha])\\
u_1 = R \sinh \tau_1 \sinh (\tau_2 - \ln [R\tau_1/\alpha])\\
u_2 = \pm  R \cosh \tau_1
    \end{array}
$
&
$   \begin{array}{ll}
\text{Semihyperpolic}, N \\
  |t| > |x|
 \end{array}
$
&
$   \begin{array}{ll}
|u_2| > R
\\
  \text{nonorthogonal}
\end{array}
$
\\
 \hdashline
$   \begin{array}{ll}
    \text{Ic.~Equidistant (EQ-Ib)}, K_2^2 \\ \tau \in \mathbb{R}, \, \varphi \in [0,2\pi)
    \end{array}
$&
$ \begin{array}{ll}
u_0 = R \sin \varphi \sinh \tau \\
u_1 = R \sin \varphi \cosh \tau \\
u_2 =  R \cos \varphi
   \end{array}
$
&
$   \begin{array}{ll}
\text{Pseudo Polar}, N^2 \\
|t| < |x|
\end{array}
$
&
$|u_2| < R$
\\
\hdashline
$   \begin{array}{ll}
    \text{Id.~Equidistant (EQ-Ib)}, K_2 \\ \tau \in \mathbb{R}, \, \varphi \in [0,2\pi)
    \end{array}
$&
$ \begin{array}{ll}
u_0 = R \sin \varphi \sinh (\tau - \ln [R\varphi/\alpha]) \\
u_1 = R \sin \varphi \cosh (\tau - \ln [R\varphi/\alpha]) \\
u_2 =  R \cos \varphi
   \end{array}
$
&
$   \begin{array}{ll}
\text{Semihyperbolic}, N \\
|t| < |x|
\end{array}
$
&
$   \begin{array}{ll}
|u_2| < R
\\
  \text{nonorthogonal}
\end{array}
$
\\
\hdashline
$   \begin{array}{ll}
    \text{Ie.~Equidistant (EQ-IIb)}^*, K_1^2 \\ \tau \in \mathbb{R}, \, \varphi \in [0,2\pi)
    \end{array}
$&
$ \begin{array}{ll}
u_0 = R \sin \varphi \sinh \tau \\
u_1 =  R \cos \varphi\\
u_2 =  R \sin \varphi \cosh \tau
   \end{array}
$
&
$   \begin{array}{ll}
\text{Cartesian}^{I}, p_0^2 \\
\end{array}
$
&
$|u_1| < R$
\\
\hdashline
$   \begin{array}{ll}
    \text{If.~Equidistant (EQ-IIb)}^*,  K_1 \\  \tau \in \mathbb{R}, \, \varphi \in [0,2\pi)
    \end{array}
$&
$ \begin{array}{ll}
u_0 = -R \cos \varphi \sinh (\tau+ R\varphi/\alpha) \\
u_1 =  R \sin \varphi\\
u_2 =  - R \cos \varphi \cosh (\tau+R \varphi\alpha)
   \end{array}
$
&
$   \begin{array}{ll}
\text{Cartesian}^{III}, p_0
\end{array}
$
&
$   \begin{array}{ll}
|u_1| < R
\\
  \text{nonorthogonal}
\end{array}
$
\\
\hline
$   \begin{array}{ll}
     \text{IIa.~Pseudo-Spherical (SPH)}, L^2 \\
      \tau \in \mathbb{R}, \, \varphi \in [0,2\pi)
    \end{array}
$&
$   \begin{array}{ll}
u_0 = R \sinh \tau\\
u_1 = R \cosh \tau \cos \varphi\\
u_2 = R \cosh \tau \sin \varphi
     \end{array}
$
&
$   \begin{array}{ll}
\text{Cartesian}^{I}, p_1^2
\end{array}
$
&
\\
\hdashline
$   \begin{array}{ll}
     \text{IIb.~Pseudo-Spherical (SPH)}, L \\
      \tau \in \mathbb{R}, \, \varphi \in [0,2\pi)
    \end{array}
$&
$   \begin{array}{ll}
u_0 = R \sinh \tau\\
u_1 = - R \cosh \tau \sin(\varphi + R \tau/\alpha)\\
u_2 = R \cosh \tau \cos (\varphi + R\tau/\alpha)
     \end{array}
$
& 
$ \begin{array}{ll}
\text{Cartesian}^{III}, p_1
    \end{array}
$
&
nonorthogonal
\\
\hline
$   \begin{array}{ll}
    \text{IIIa.~Horicyclic (HO)}^*, (K_2-L)^2 \\ \tilde{x} \in \mathbb{R}, \tilde{y} \in \mathbb{R}\setminus\{0\}
    \end{array}
$&
$   \begin{array}{ll}
u_0 = R \frac{\tilde{x}^2 - \tilde{y}^2 + 1} {2 \tilde{y} } \vspace{1mm} \\
u_1 =  R \frac{\tilde{x}} {\tilde{y} } \vspace{1mm} \\
u_2 = R \frac{\tilde{x}^2 - \tilde{y}^2 - 1} {2 \tilde{y} }
    \end{array}
$
& 
$ \begin{array}{ll}
\text{Cartesian}^{I}, p_1^2
    \end{array}
$
&
\\
\hdashline
$   \begin{array}{ll}
    \text{IIIb.~Horicyclic (HO)}, K_1  + L  \\  \xi,\eta \in \mathbb{R}
    \end{array}
$&
$   \begin{array}{ll}
u_0 = R (\xi\eta^2 - 4\eta + \xi)/4 \\
u_1 =  R (\xi\eta^2 - 4\eta - \xi)/4 \\
u_2 = R (1-\xi\eta/2)
    \end{array}
$
& 
$   \begin{array}{ll}
\text{Cartesian}^{II}, p_0 + p_1
    \end{array}
$
&
nonorthogonal
\\
\hline
\end{tabular}
\end{center}
\end{table}

\end{landscape}


\begin{landscape}

\begin{table}[htbp]

\renewcommand{\arraystretch}{1.1}

\renewcommand\tabcolsep{2pt}

\label{tab:4}
\hfuzz=32pt
\begin{center}
\begin{tabular}{| l | l | l | l |}
\hline
Coordinate System&
Coordinates & Contracted System, Operator&Comments\\
\hline
$   \begin{array}{ll}
    \text{IVa.~Elliptic-Parabolic (EP)}\\
    \gamma K_2^2+(K_1+L)
    \\
    \tau_1 \geq 0 , \tau_2 \in \mathbb{R}\setminus\{0\}, \gamma>0
    \end{array}
$&
$   \begin{array}{ll}
u_0 = \frac{R}{\sqrt{\gamma}} \frac{\cosh^2 \tau_1 - \cosh^2\tau_2 + \gamma}{2\cosh \tau_1\sinh \tau_2} \vspace{1mm}\\
u_1 = \frac{R}{\sqrt{\gamma}} \frac{\cosh^2 \tau_1 - \cosh^2\tau_2 - \gamma}{2\cosh \tau_1\sinh \tau_2} \vspace{1mm}\\
u_2 = R^2 \tanh \tau_1 \coth\tau_2
    \end{array}
$
& $ \begin{array}{ll}
 \text{Hyperbolic}^{II},\\N^2 + l^2(p_0+p_1)^2,\
 \gamma \text{ is fixed}
    \end{array}
    $
&
 \\
 \hdashline
$   \begin{array}{ll}
    \text{IVb.~Elliptic-Parabolic (EP)}^*\\  \gamma K_1^2+(K_2-L) \\
    \tau_1 \geq 0 , \tau_2 \in \mathbb{R}\setminus\{0\}, \gamma>0
    \end{array}
$&
$   \begin{array}{ll}
u_0 = \frac{R}{\sqrt{\gamma}} \frac{\cosh^2 \tau_1 - \cosh^2\tau_2 + \gamma}{2\cosh \tau_1\sinh \tau_2} \vspace{1mm}\\
u_1 = R^2 \tanh \tau_1 \coth\tau_2 \vspace{1mm}\\
u_2 = \frac{R}{\sqrt{\gamma}} \frac{\cosh^2 \tau_1 - \cosh^2\tau_2 - \gamma}{2\cosh \tau_1\sinh \tau_2}
    \end{array}
$
& $ \begin{array}{ll}
 \text{Cartesian}^{I},\\\gamma p_0^2+p_1^2, \qquad
 \gamma \simeq \frac{R^2}{l^2}
    \end{array}
    $
&
 \\
\hline
$   \begin{array}{ll}
  \text{V.~Elliptic (E)}\\ L^2 + \sinh^2 \beta K_2^2 \\
  \rho_2<a_3<a_2<\rho_1<a_1
  \\
  a \in [K, K + i 4K^\prime), \ b \in (iK,  iK + 2K^\prime)
    \end{array}
$
&
$   \begin{array}{ll}
u_0^2 = R^2{\frac {\left ({ \rho_1}-{ a_3}\right )\left ( { a_3} - {\rho_2} \right )}{\left ({ a_1}-{ a_3}\right )\left ({ a_2}-{ a_3}\right )}} \vspace{1mm}\\
u_1^2 = R^2{\frac {\left ({ \rho_1}-{a_2}\right )\left ({ a_2} - { \rho_2} \right )}{\left ({ a_1}-{ a_2}\right )\left ({ a_2}-{ a_3}\right )}} \vspace{1mm}\\
u_2^2 = R^2{\frac {\left ({a_1} - { \rho_1}\right )\left (a_1 -  \rho_2 \right )}{\left ({ a_1}-{ a_2}\right )\left ({ a_1}-{ a_3}\right )}}
    \end{array}
$
$
\begin{array}{ll}
u_0 =  i R \sn(a, k) \dn(b, k^\prime) \vspace{1mm} \\
u_1 =  -R \cn(a, k) \cn(b, k^\prime) \vspace{1mm} \\
u_2 =  - R \dn(a, k) \sn(b, k^\prime)
   \end{array}
$
&
$ \begin{array}{ll}
  \text{Elliptic}^{I},\qquad N^2 + D^2 p_1^2, \\
  D = \sqrt{a_1 - a_2},\qquad a_1 \simeq R^2
  \\
  \hdashline
  \\
 \text{Cartesian}^{I},\qquad  p_1^2,\\
  \beta \text{ is fixed: }a_1 - a_2 = a_2 - a_3
    \end{array}
    $
& $\sinh^2\beta = \frac{a_1-a_2}{a_2-a_3}$
 \\
\hline
$   \begin{array}{ll}
    \text{VIa.~Semi-Circular-Parabolic (SCP)}\\ 
    \{K_1,K_2 \} + \{K_2,L\}\\
    \xi>0,\ \eta \in \mathbb{R}\setminus\{0\}
    \end{array}
$&
$   \begin{array}{ll}
    u_0 = R\frac {\left(\eta^2 - \xi^2\right)^2 + 4}{8 \xi \eta} \vspace{1mm} \\
    u_1 = R\frac {\left(\eta^2 - \xi^2\right)^2 - 4}{8 \xi \eta} \vspace{1mm} \\
    u_2 = \pm R{\frac {{\eta}^{2}+{\xi}^{2}}{2\xi\eta}}
    \end{array}
$
&
$   \begin{array}{ll}
\text{there are no} \\
\text{separated coordinates}\\
\{p_0, N\} + \{p_1,N\}
\end{array}
$
&
$|u_2|>R$ \\
\hdashline
$   \begin{array}{ll}
    \text{VIb.~SCP (h. rotation )}\\
     (K_1^2 + \{K_1,K_2\} - 3 K_2^2)/4 +\\
      \{K_2,L\}/\sqrt{2} - L^2/2 \\
    \xi\eta \ne 0
    \end{array}
$&
$   \begin{array}{ll}
  u_0 = R \frac{(\eta^2 - \xi^2)^2 - 4\xi^2 - 4\eta^2 + 20}{16\xi\eta} \vspace{1mm} \\  
u_1 = R \frac{(\eta^2 - \xi^2)^2 + 4\xi^2 + 4\eta^2 - 28}{16\sqrt{2} \xi\eta}  \vspace{1mm}\\
u_2 = - R \frac{(\eta^2 - \xi^2)^2 - 12\xi^2 - 12\eta^2 + 4}{16\sqrt{2} \xi\eta}
    \end{array}
$
&
$   \begin{array}{ll}
\text{Cartesian}^I, p_0^2
\end{array}
$
&
 \\
\hline
$   \begin{array}{ll}
  \text{VIIa.~Hyperbolic-Parabolic (HP-I)}\\
  -\gamma K_2^2 + (K_1+L)^2\\
  \theta\in[-\pi/2,\pi/2]\setminus\{0\}, \phi \in (0, \pi)
  \\
  \hdashline
  \\
    \text{VIIb.~Hyperbolic-Parabolic (HP-II)}\\
  \theta\in[-\pi/2,\pi/2]\setminus\{0\}, \phi \in (0, \pi)
   \\
   \\
  \hdashline
  \\
    \text{VIIc.~Hyperbolic-Parabolic (HP-III)}\\
  \theta \in \mathbb{R}\setminus\{0\}, \phi>0 
    \end{array}
$
&
$   \begin{array}{ll}
u_0 =  \frac{R}{2\sqrt{ \gamma}} \frac{\cos^2 \theta - \sin^2\phi + \gamma}{\sin\theta  \sin\phi}\vspace{1mm}\\
u_1 =  \frac{R}{2\sqrt{ \gamma}} \frac{\cos^2 \theta - \sin^2\phi - \gamma}{\sin\theta  \sin\phi}\vspace{1mm}\\
u_2 = R \cot\theta \cot \phi
\\
\hdashline
u_0 =  \frac{R}{2\sqrt{\gamma}} \frac{\cos^2 \theta \cos^2\phi - 1 + \gamma \sin^2 \theta \sin^2\phi}{\sin\theta  \sin\phi} \vspace{1mm}\\
u_1 =  \frac{R}{2\sqrt{\gamma}} \frac{\cos^2 \theta \cos^2\phi - 1 - \gamma \sin^2 \theta \sin^2\phi}{\sin\theta  \sin\phi}\vspace{1mm}\\
u_2 = R \cos\theta \cos \phi
\\
\hdashline
u_0 =  \frac{R}{2\sqrt{\gamma}} \frac{\cosh^2 \theta \cosh^2\phi - 1 + \gamma \sinh^2 \theta \sinh^2\phi}{\sinh\theta \sinh\phi} \vspace{1mm}\\
u_1 =  \frac{R}{2\sqrt{\gamma}} \frac{\cosh^2 \theta \cosh^2\phi - 1 - \gamma \sinh^2 \theta \sinh^2\phi}{\sinh\theta \sinh\phi}\vspace{1mm} \\
u_2 = \pm R\cosh \theta \cosh \phi
    \end{array}
$
& 
$\begin{array}{ll}
 \text{Hyperbolic}^{III} \text{ for all HP systems}, \\ N^2 - l^2(p_0 + p_1)^2,
    \gamma \simeq R^2/l^2
    \\
  \hdashline
  \\
  \text{Cartesian}^{I}, -\gamma p_0^2 + p_1^2,\\
   \gamma \text{ is fixed }\\
   \text{for permuted HP}_I \text{ if } 0 < \gamma < 1\\
   \text{for permuted HP}_{II} \text{ if } \gamma > 1
    \end{array}
$
&
$\begin{array}{ll}
|u_0 (1-\gamma) + \\ u_1 (1+\gamma)| > 2R\sqrt{\gamma}
\end{array}
$
 \\
\hline
\end{tabular}
\end{center}
\end{table}
\end{landscape}



\begin{landscape}

\begin{table}[htbp]

\renewcommand{\arraystretch}{1.1}
\renewcommand\tabcolsep{3pt}

\hfuzz=32pt
\begin{center}
\begin{tabular}{| l | l | l | l |}
\hline
Coordinate System&
Coordinates & Contracted System, Operator&Comments\\
\hline
$   \begin{array}{ll}
    \text{VIId.~Hyperbolic Parabolic}\\ \text{(HP)}^*\\
    -K_1^2 + (K_2-L)^2 
    \end{array}
$&
\text{All permuted HP systems}
&
$ \begin{array}{ll}
\text{Parabolic}^I, \{N, p_1\}\end{array}
$
&
$   \gamma = 1$
\\
\hline
$   \begin{array}{ll}
  \text{VIIIa.~Hyperbolic I}\\
  K_2^2 -\sin^2 \alpha L^2 \\
  \sin^2\alpha = \frac{a_2-a_3}{a_1-a_3}=k^2 \\[1mm]
  H_I^A: \rho_i <a_3<a_2<a_1\\
   a \in (-iK^\prime, iK^\prime), \\ b \in \left(iK, iK + 2K^\prime \right)\\[2mm]
  H_I^B: a_3<a_2<a_1<\rho_i \\
   a \in (iK^\prime, iK^\prime + 2K), \\ b \in \left(-iK, iK \right)\\
  \hdashline
  \\
   \text{VIIIb.~Hyperbolic II}\\
  H_{II}^A: a_3< \rho_i <a_2<a_1\\
  a \in \left[0, 4K\right), \\ b \in \left[K^\prime, K^\prime + i4K\right)\\[2mm]
  H_{II}^B: a_3<a_2<\rho_i< a_1\\
  a \in [K, K + i4K^\prime), \\ b \in \left[0, 4K^\prime \right)
    \end{array}
$&
$   \begin{array}{ll}
u_0^2 = R^2{\frac {\left ({\rho_1}-{a_2}\right )\left ({\rho_2 - a_2}\right )}{\left ({a_1}-{a_2}\right )\left ({a_2}-{a_3}\right )}}\\[2mm]
u_1^2 = R^2{\frac {\left ({\rho_1}-{a_3}\right )\left ({\rho_2 - a_3}\right )}{\left ({a_1}-{a_3}\right )\left ({a_2}-{a_3}\right )}}\\[2mm]
u_2^2 = R^2{\frac {\left ({\rho_1}-{a_1}\right )\left ({\rho_2 - a_1}\right )}{\left ({a_1}-{a_2}\right )\left ({a_1}-{a_3}\right )}}
    \end{array}
$
$  \begin{array}{ll}
u_0 = - i R \cn(a,k)\cn(b, k^\prime)\\[2mm]
u_1=  - R\sn(a,k) \dn(b, k^\prime)\\[2mm]
u_2=  - R\dn(a,k) \sn(b, k^\prime)
 \end{array}
$
&
$\begin{array}{ll}
H^A_{II} \text{ to Cartesian}^{I},\\
p_1^2, \\ a_1-a_2 = \\
a_2 - a_3
\end{array}
$
\vline
$
  \begin{array}{ll}
   H^A_I \text{ to Elliptic}^{II}(i),\\ N^2 - d^2 p_1^2, \\ d = \sqrt{a_2 - a_3}
   \\[2mm]
    \hdashline
    \\
   H^A_{II} \text{ to Elliptic}^{II}(ii), \\N^2 - d^2 p_1^2, \\      a_1 \simeq R^2
    \end{array}
$
&
$
\begin{array}{ll}
|u_1^2 \sin^2 \alpha - u_2^2 \cos^2\alpha \\+ R^2|
> 2R | u_1 \sin\alpha |
\end{array}
$
 \\
\hdashline
$   \begin{array}{ll}
  \text{VIIIc.~Hyperbolic (H-III) } \\
  (\text{h. rotation of } H_I^A)\\
  c K_2^2 + \{K_2,L\}\\
    c = k+1/k
     \end{array}
$&
$   \begin{array}{ll}
u_0 = \frac{R}{k^\prime}\left[ - k \dn(a,k)\cn(b, k^\prime)  +  i \cn(a,k) \sn(b, k^\prime)\right], 
\\[2mm]
u_1 =  R\, i \sn(a,k) \, i \dn(b, k^\prime),
\\[2mm]
u_2 =  \frac{R}{k^\prime}\left[ - k \cn(a,k)\, i\cn(b, k^\prime) + \dn(a,k) \sn(b, k^\prime)\right]
    \end{array}
$
&
$
  \begin{array}{ll}
   \text{Pseudo Polar},\ N^2,\ k \simeq R^{-2} \\
   \\
    \hdashline
   \\
   \text{Parabolic}^{I}, \ \{N,p_1\},  k \text { is fixed}
    \end{array}
$
&
 \\
\hline
$   \begin{array}{ll}
    \text{IX.~Semi-Hyperbolic}\\
    \sinh 2\beta K_2^2 + \{K_1,L\}\\
    \text{SH}_I: \tau_i \leq \arcsinh\, c,\\
    \text{SH}_{II}: \tau_i \geq \arcsinh\, c\\
    c = \frac{a-\gamma}{\delta}=\sinh 2\beta
    \end{array}
$&
$   \begin{array}{ll}
u_0^2 + u_1^2 = \frac{R^2}{\sqrt{c^2+1}}  \cosh\tau_1\cosh\tau_2 
\nonumber\\[2mm]
u_0^2 - u_1^2 = \frac{R^2}{2(c^2+1)} \left\{ (\sinh\tau_1-c)(\sinh\tau_2-c)-c^2-1\right\}
\\[2mm]
u_2^2 = \frac{R^2}{c^2+1}(\sinh\tau_1-c)(\sinh\tau_2-c)
\nonumber
    \end{array}
$
&
$ \begin{array}{ll}
SH_I \text{ to Hyperbolic}^I, N^2+l^2p_0 p_1, \\ c\simeq 2 R^2/l^2, l=\text{const.} \\
\hdashline
\\
 SH^* \text{ to Cartesian}^I, p_0^2,\\ c \ne 0 \text{ is fixed}
\\
\hdashline
\\
 SH^* \text{ to Parabolic}^I, \{N,p_1\},\\ c = 0
\end{array}
$
&
\\
\hline
\end{tabular}
\label{systemH1r}
\end{center}
\end{table}

\end{landscape}


\begin{landscape}

\begin{table}[htbp]

\renewcommand{\arraystretch}{1.1}

\renewcommand\tabcolsep{2pt}

\caption{Coordinate systems
 on pseudo-Euclidean plane $E_{1,1}$}
\label{tab:6}
\hfuzz=32pt
\begin{center}
\begin{tabular}{|l|l|l|c|l|}\hline
Coordinate System&Integral of Motion $X$&Coordinates&Comments\\
\hline
I.~a. Cartesian, type I    &$X^I_C = p_0 p_1$ \qquad \ \ or &$t,$  & \\
    &$X^{I}_C = (p_0 + p_1)^2$ \ or & $x$ & orthogonal \\
     &$X^{I}_C = p_1^2$ or $X^{I}_C = p_0^2$&$ $  $ $ & \\
\hdashline
b. Cartesian, type II     &$X^{II}_C = p_0 + p_1 $&$t = x' + t'/4$,\qquad $x = x' - t'/4$ & nonorthogonal\\
\hdashline
c. Cartesian, type III     &$X^{III}_C = p_1 $&$t = -t'/2 $,\qquad $x = t'/2 + x'$ & nonorthogonal\\
\hline
IIa.~Pseudo-polar &$X_S^2=N^2$                        &$t=r \cosh \tau_2$ &\\
$r \geq 0$, $-\infty<\tau_2<\infty$ &     &$x=r \sinh \tau_2$& $t > |x|$\\
\hdashline
IIb.~Pseudo-polar &$X_S^2=N^2$                        &$\tilde{t}=r \sinh \tau$ &\\
$r \geq 0$, $-\infty<\tau<\infty$ &     &$\tilde{x}=r \cosh \tau$& $|\tilde{t}| < \tilde{x}$\\
\hline
III.~Parabolic type I. &$X_P^{I}= \{p_1, N\}$&$t=\frac{1}{2}(u^2+v^2)$&\\
$v \geq 0$, $-\infty<u<\infty$          &&$x= uv$  &\\
\hline
IV.~Parabolic type II. &$X_P^{II}=\{p_0, N\}+\{p_1, N\}$
&$t=\frac{1}{2\alpha}(\eta-\xi)^2-(\eta+\xi)$& \\
$-\infty < \eta,\xi < \infty$&
$ + \alpha(p_0-p_1)^2$&$x= \frac{1}{2\alpha} (\eta-\xi)^2 + (\eta+\xi)$&\\
\hline
V.~Hyperbolic type I.  &$X_H^I=N^2+l^2 p_0p_1$
&$t=\frac{l}{2}\left(\cosh\frac{\zeta_1+\zeta_2}{2}
- \sinh\frac{\zeta_1-\zeta_2}{2}\right)$&\\
$-\infty<\zeta_1,\zeta_2 <\infty$&
&$x=\frac{l}{2}\left(\cosh\frac{\zeta_1+\zeta_2}{2}
+\sinh\frac{\zeta_1-\zeta_2}{2}\right)$&\\
\hline
VI.~Hyperbolic type II.  &$X_H^{II}=N^2+l^2 (p_0+p_1)^2$
&$t= l\left(\sinh(\zeta_1-\zeta_2) + e^{\zeta_1+\zeta_2}\right)$&\\
$-\infty<\zeta_1,\zeta_2 <\infty$
&&$x= l\left(\sinh(\zeta_1-\zeta_2) - e^{\zeta_1+\zeta_2}\right)$&\\
\hline
VII.~Hyperbolic type III.  &$X_H^{III}=N^2-l^2 (p_0+p_1)^2$
&$t= l\left(\cosh(\zeta_1-\zeta_2) \pm e^{\zeta_1+\zeta_2}\right)$& 
uncovered part
\\
$-\infty<\zeta_1,\zeta_2 <\infty$   &
&$x= l\left(\cosh(\zeta_1-\zeta_2) \mp e^{\zeta_1+\zeta_2}\right)$&  $|t+x|<2|l|$\\
\hline
VIII.~Elliptic, type I.  &$X_E^{I}=N^2+D^2 p_1^2$
&$t= D\sinh \xi \cosh \eta$&\\
$-\infty<\eta,\xi <\infty$ &
&$x= D\cosh \xi \sinh \eta$&\\
\hline
IX.~Elliptic, type II.&$X_E^{II}=N^2-d^2 p_1^2$
&$(i)$ \,$t= d\cosh \eta \cosh \xi$& 
\\
$(i)\, -\infty<\eta <\infty$, $\xi \geq 0$&
&\,\,\,$x= d\sinh \eta \sinh \xi;$&\\
$(ii)\,  0<\eta<2\pi, 0\leq \xi < \pi$& &$(ii)$ \,
$t= d\cos \eta \cos \xi$&\\
&&\,\,\,$x= d\sin \eta \sin \xi$&\\
\hline
X.~Semi-Hyperbolic  &$X_S=N$
&$(i)\, 2t = e^\tau + r^2 e^{-\tau}  $&nonorthogonal\\
$r>0, -\infty < \tau < +\infty $ &
&$2x= e^\tau - r^2 e^{-\tau};  $& $(i)\, |t| > |x|$\\
&&$(ii)\, 2t = e^\tau - r^2 e^{-\tau}  $&\\
 &
&$2x= e^\tau + r^2 e^{-\tau}  $& $(ii)\, |t| < |x|$\\
\hline
\end{tabular}
\end{center}
\end{table}  



\begin{table}[htbp]

\renewcommand{\arraystretch}{1.1}

\renewcommand\tabcolsep{2pt}

\caption{Systems of coordinates
 on Euclidean plane $E_{2}$}
\label{tab:7}
\hfuzz=32pt
\begin{center}
\begin{tabular}{|l|l|l|c|l|}\hline
Coordinate System&Integral of Motion $X$&Coordinates&Comments\\
\hline
I.~Cartesian  &$X^2_C = p_1^2$&$x$ & orthogonal\\
$-\infty < x, y < \infty$ & &$y$  &\\
\hline
Ia.~Cartesian  &$X_C = p_1$&$x=x' + y'$  & nonorthogonal\\
$-\infty < x', y' < \infty$ & &$y=y'$ &\\
\hline
II.~Polar &$X^2_S=M^2$                        &$x=r \cos \varphi$ & orthogonal\\
$r \geq 0$, $0 \leqslant \varphi  < 2\pi$ &     &$y=r \sin \varphi$&  \\
\hline
IIa.~Polar &$X_S=M$                       &$x=r \cos(\varphi + r/\alpha)$ & nonorthogonal\\
$r \geq 0$, $0 \leqslant \varphi  < 2\pi$ &     &$y=r \sin(\varphi + r/\alpha)$& $\alpha \ne 0$  \\
\hline
III.~Parabolic &$X_P= \{p_2, M\}$&$x=\frac{1}{2}(u^2 - v^2)$&orthogonal\\
$u  \geq 0$, $-\infty< v <\infty$          &&$y= uv$  &\\
\hline
IV.~Elliptic  &$X_E =M^2 + D^2 p_1^2$
&$x= D\cosh \xi \cos \eta$&orthogonal\\
$ 0 \leqslant \xi < \infty$, $ 0 \leqslant \eta < 2 \pi $ &
&$y= D\sinh \xi \sin \eta$&\\
\hline
\end{tabular}
\end{center}
\end{table}

\end{landscape}

\end{document}